\documentclass[aps,prx,reprint,showpacs,floatfix,superscriptaddress]{revtex4-2}

\usepackage[T1]{fontenc}
\usepackage{times}

\usepackage{amsmath,amssymb}
\usepackage{bm}
\usepackage{dsfont}
\usepackage{braket}

\usepackage{graphicx}
\usepackage[table]{xcolor}
\usepackage{booktabs}
\usepackage{multirow}
\usepackage{tabularray}

\usepackage[title]{appendix}
\usepackage[shortlabels]{enumitem}
\usepackage{verbatim}
\usepackage[normalem]{ulem} 

\definecolor{lblue}{RGB}{51,71,158}
\usepackage[colorlinks=true,citecolor=blue,linkcolor=blue,urlcolor=lblue]{hyperref}
\usepackage{orcidlink}

\makeatletter
\AtBeginDocument{\let\@elt\relax}
\makeatother

\newcommand{\op}[2]{\ket{#1}\!\bra{#1}}
\newcommand{\norm}[1]{\left\lVert #1 \right\rVert}
\newcommand{\Tr}{\mathrm{Tr}}

\begin{document}

\title{Restricted typicality in non-equilibrium quantum many-body systems}

\author{Konrad Pawlik\orcidlink{0009-0009-6304-2578}}
\email{konrad.pawlik@doctoral.uj.edu.pl}
\affiliation{Szkoła Doktorska Nauk Ścisłych i Przyrodniczych, Uniwersytet Jagielloński, ulica Stanisława \L{}ojasiewicza 11, PL-30-348 Krak\'ow, Poland}
\affiliation{Instytut Fizyki Teoretycznej,  Wydział Fizyki, Astronomii i Informatyki Stosowanej, Uniwersytet Jagielloński, ulica Stanisława \L{}ojasiewicza 11, PL-30-348 Krak\'ow, Poland }

\author{Piotr Sierant\orcidlink{0000-0001-9219-7274}}
\email{piotr.sierant@bsc.es}
\affiliation{   Barcelona Supercomputing Center, Barcelona 08034, Spain}

\author{Jakub Zakrzewski\orcidlink{0000-0003-0998-9460}}
\email{jakub.zakrzewski@uj.edu.pl}
\affiliation{Instytut Fizyki Teoretycznej,  Wydział Fizyki, Astronomii i Informatyki Stosowanej, Uniwersytet Jagielloński, ulica Stanisława \L{}ojasiewicza 11, PL-30-348 Krak\'ow, Poland }
\affiliation{Mark Kac Complex
    Systems Research Center, Jagiellonian University in Krakow, PL-30-348 Krak\'ow,
    Poland. }

\date{\today}

\begin{abstract}
    Characterizing  the time evolution of generic quantum many-body systems is a fundamental challenge, as representing the exact state requires exponentially scaling computational resources. While hydrodynamics and statistical mechanics successfully simplify this task by predicting the expectation values of local observables, these macroscopic frameworks provide no information about nonlinear characteristics of the quantum state.
    In this work, we demonstrate that for systems exhibiting a timescale separation, with dynamics governed by the transport of conserved charges, this lost information can be systematically recovered.
    By constraining the maximum-entropy Scrooge ensemble solely by the system's slow modes, we accurately reconstruct complex, nonlinear quantum properties of the global time-evolved state, including the half-chain entanglement entropy and the participation entropy in the computational basis. Our results generalize the paradigm of canonical typicality, revealing that a hydrodynamically bottlenecked system behaves as a typical pure state within a dynamically restricted submanifold of the Hilbert space—a phenomenon we term global restricted typicality.
\end{abstract}

\maketitle

\section{Introduction}

Typicality arguments have been deeply embedded in the study of quantum mechanics since its inception. Early works by Schr\"odinger~\cite{Schrodinger27} and von Neumann~\cite{vonNeumann29,vonNeumann10} posited that the local properties of individual pure states, as well as the expectation values of macroscopic observables, can be remarkably well described by statistical ensemble averages. Modern extensions of these foundational ideas have uncovered fundamental structural properties of high-dimensional Hilbert spaces. Building upon Schr\"odinger's insights, the discovery of canonical typicality~\cite{Popescu06, Goldstein06} demonstrated that for almost all pure states residing in a large microcanonical subspace, small local subsystems are indistinguishable from a thermal Gibbs density matrix. Concurrently, modern reevaluations of von Neumann's quantum ergodic theorem have established the concept of normal typicality~\cite{Goldstein10N}, extending this framework to show that the macroscopic observables of typical states within an energy shell similarly equilibrate to their ensemble averages. These typicality principles are underlined by the phenomenon of measure concentration~\cite{Popescu06, Gemmer09, Lloyd13}. As the system size increases, the statistical distribution of relevant observables sharply peaks around the ensemble mean, ensuring that the properties of a complex many-body quantum state can be faithfully captured by sampling only a small number of states from the underlying ensemble.

The utility of ensembles of quantum states in describing many-body systems extends far beyond static local subsystems. For instance, the mean energy ensemble~\cite{Brody98, Bender05, Brody07, Muller11, Ji11, Fresch13} is constructed from random vectors constrained to a fixed expectation value of a given Hamiltonian, exhibiting typicality under specific physical conditions~\cite{Muller11}. Another foundational concept is the random-phase ensemble~\cite{Reimann07}, which demonstrates typicality provided that a macroscopically large number of energy levels are populated. Recently, this ensemble has been utilized to formalize the phenomenon of Hilbert space ergodicity~\cite{Mark24}, revealing that the unitary evolution of pure states explores the dynamically available Hilbert space uniformly over time~\cite{Nakata12, Pilatowsky-Cameo23,Pilatowsky-Cameo24}. Furthermore, the paradigm of dynamical typicality~\cite{Bartsch09} establishes that pure initial states sharing similar macroscopic expectation values will typically yield indistinguishable time evolution for those macroscopic observables. Beyond their theoretical significance, these typicality principles offer powerful practical advantages for numerical simulations~\cite{White09me,Sugiura12,Wietek21}; by replacing intractable mixed density matrices with classical distributions of pure states, simulating just a few typical wavefunctions is sufficient to evaluate finite-temperature observables with high precision.

While those random ensembles provide a rigorous foundation for evaluating linear expectation values and local reduced density matrices, i.e., quantities determined strictly by the first statistical moment of the distribution, they remain agnostic to the non-linear quantum correlations that characterize complex many-body states. The physical necessity of investigating these higher statistical moments is deeply rooted in the replica formalism~\cite{Calabrese04, Daley12, Islam15, Brydges19, Bittel2026complete, Deneris26comparing, Sierant26comm, Tang26witness}.
Any quantity nonlinear in the state — such as a bipartite Rényi entanglement entropy or the moments of measurement-outcome probabilities — can be expressed as the expectation value of an operator acting on $q$ replicas of the system, $\mathcal{H}^{\otimes q}$. The ensemble average of such a quantity is therefore determined by the $q$-th statistical moment of the pure-state distribution, and, whenever the distribution concentrates, this average faithfully characterizes typical individual states.

Access to these higher-order moments has enabled the discovery of phenomena such as deep thermalization~\cite{Cotler23, Claeys22, Ippoliti22, Choi23, Ippoliti23, Lucas23, Wilming22, Bhore23, Chan24, Varikuti24, Vairogs25, Shrotriya25, Liu26}. By analyzing the projected ensemble, defined as the collection of pure states arising from projective measurements on an extensive subsystem, one finds that its higher moments form quantum state $k$-designs, demonstrating that thermalization extends far beyond the first statistical moment of the emergent ensemble. Furthermore, the dynamics of these ensembles can be rigorously formulated as a problem of probability transport~\cite{Anza26}. Similarly, for unitary evolution of isolated systems, evaluating higher statistical moments along a continuous Hilbert space trajectory $\ket{\psi(t)}$, which form the basis of the temporal ensemble, provides the required mathematical structure to capture information scrambling, out-of-time-ordered correlations, and the precise boundaries of Hilbert-space ergodicity~\cite{Nakata12, Pilatowsky-Cameo23, Pilatowsky-Cameo24, Mark24}.
Quantum resource theories~\cite{Chitambar19} provide resource measures that quantify nonclassical aspects of quantum states and are, typically, nonlinear functions of the state, studied recently in equilibrium and non-equilibrium quantum many-body systems~\cite{Bera25syk, Odavic24, Fux24, Jasser25, Odavic25, Lami26, Magni25anti, Falcao25u1, Tirrito2025universal, Lami25mps, Aditya25mpemba, Xiao26non, Falcao25mbl, Turkeshi25spreading, Sierant26faf, Aditya26a, Trigueros26unitary, Aditya26coherence, Sierant26magic, Falcao26faf, Nava26, Aditya26equivalence}.
The theoretical framework to predict and manipulate these non-linear physical properties has garnered intense recent interest~\cite{Calabrese04, Sekino08, Nakata12, Daley12, Islam15,Kaufman16, Nahum18,Elben18, Elben19,Brydges19, Huang20, Mi21, Claeys22, Ho22, Ippoliti22, Wilming22, Cotler23, Choi23, Ippoliti23, Lucas23, Bhore23, Elben23, Pilatowsky-Cameo23, Chan24, Mark24, Pilatowsky-Cameo24, Varikuti24, Vairogs25, Shrotriya25} driven by state-of-the-art quantum simulation experiments~\cite{Islam15, Brydges19, Mi21, Rosenberg24, Wienand24}  that are now capable of directly measuring these higher-order correlations in the laboratory.

Historically, the Haar ensemble~\cite{Haar33,Halmos50} has served as the foundational statistical baseline for predicting higher-order quantum properties, famously yielding the Page curve for bipartite entanglement~\cite{Page93}. However, its inherent lack of spatial and energetic structure systematically fails for physical many-body states, which are intrinsically restricted by macroscopic dynamical constraints~\cite{Vidmar17, Roberts17, Cotler17, Lu19, Bianchi22, Tirrito25}. Among statistical distributions capable of explicitly respecting these physical restrictions, the Scrooge ensemble~\cite{Jozsa94, Goldstein06S,Reimann08, Goldstein16,Mark24,Liu24,Teufel25, Manna25, Chang25,McGinley25,Mok26,Xu26,Liu26,Wu26,Igelspacher26,Liu26nonlocal,Sarma26} has recently emerged as a particularly powerful theoretical tool. Defined as the unique pure-state distribution that maximizes accessible information entropy subject to a fixed first statistical moment $\rho$, it provides a natural generalization of the Haar measure. It has been successfully utilized to describe nonlinear functions of quantum state $\rho$  in thermal equilibrium, in both local~\cite{Goldstein06S} and extensive~\cite{Xu26,Sarma26} subsystems. Recent insights reveal that pure-state ensembles subject to physical constraints naturally follow the maximum entropy principle~\cite{Mark24}; encoding those constraints into a density matrix $\rho$ uniquely yields the Scrooge measure. This fundamental connection establishes the Scrooge ensemble as the governing statistical distribution for non-linear phenomena such as finite-temperature deep thermalization and Hilbert-space ergodicity. Importantly, while previous studies have largely explored the Scrooge ensemble within abstract quantum information settings, including random unitary circuits or projected ensembles, its construction as a maximally agnostic measure constrained solely by $\rho$ makes it an ideal statistical baseline for characterizing the time evolution of realistic, interacting many-body systems.

In this work, we show that for quantum many-body systems featuring a set of macroscopic slow modes whose relaxation is fundamentally bottlenecked by the underlying Hamiltonian dynamics, the Scrooge ensemble provides an accurate description of the global properties of individual time-evolved states. We argue that once the fast degrees of freedom scramble, the expectation values of the slow modes act as effective dynamical constraints on the continuous time evolution, which, via the maximum entropy principle~\cite{Mark24}, uniquely fixes a Scrooge ensemble. While established measure concentration for the Scrooge ensemble~\cite{Teufel25,Igelspacher26} does not strictly guarantee the concentration of global, higher-order functions of quantum state, we demonstrate that the ensemble faithfully captures specific complex non-linear quantities of physical interest at sufficiently long evolution times. Specifically, we derive analytical formulas for the annealed averages of both the entanglement and participation entropies within the Scrooge ensemble, providing an exact closed-form characterization of these non-linear quantities. We then validate this physical correspondence within the paradigmatic setting of quantum hydrodynamics~\cite{Kadanoff63, Pines66, Spohn91}. While standard hydrodynamics is inherently limited to predicting the expectation values of local observables, the application of our framework reveals that the analytic ensemble predictions accurately reproduce the exact microscopic dynamics, including effective hydrodynamical scaling exponents of nonlinear quantities. These findings reveal that generic quantum states obtained by time evolution behave, even in their global features, like typical members of a dynamically restricted manifold of states, a phenomenon we term \textit{global restricted typicality}. Finally, we highlight the broad universality of this phenomenon, discussing its natural applicability to a wide array of physical systems, including the generalized hydrodynamics (GHD) of integrable models~\cite{Castro-Alvaredo16, Bertini16, Doyon17, Bulchandani18, DeNardis18, Doyon20, Essler23, Doyon25} and regimes of weak Hilbert space fragmentation~\cite{Sala20, Khemani20, Guardado20, Moudgalya22,Kohlert23, Wang25}.

This work is organized as follows. In Sec.~\ref{sec:scrooge}, we recall the theoretical framework of the Scrooge ensemble, and we show how knowledge of the system's slow modes enables the systematic retrieval of nonlinear functions of quantum state. In Sec.~\ref{sec:numerics}, we derive analytic baselines for the entanglement and participation entropies of the Scrooge ensemble. We then numerically validate our framework, showing that the ensemble accurately reconstructs the complex, nonlinear entropies of a time-evolved state, and we carefully discuss the physical limitations of this approach. In Sec.~\ref{sec:integrable_main}, we analytically apply our formalism to integrable systems via GHD, revealing that the Scrooge measure exactly recovers the formula for entanglement in inhomogeneous quenches when supplemented with the causal constraints of the semiclassical quasiparticle picture. Finally, in Sec.~\ref{sec:typical_manifold}, we formally establish the phenomenon of global restricted typicality. By introducing the canonical and microcanonical Scrooge ensembles, we prove that the underlying pure state undergoes exact geometric measure concentration onto a dynamically restricted hydrodynamic manifold.

\begin{figure*}[ht!]
    \centering
    \includegraphics[width=0.95\linewidth]{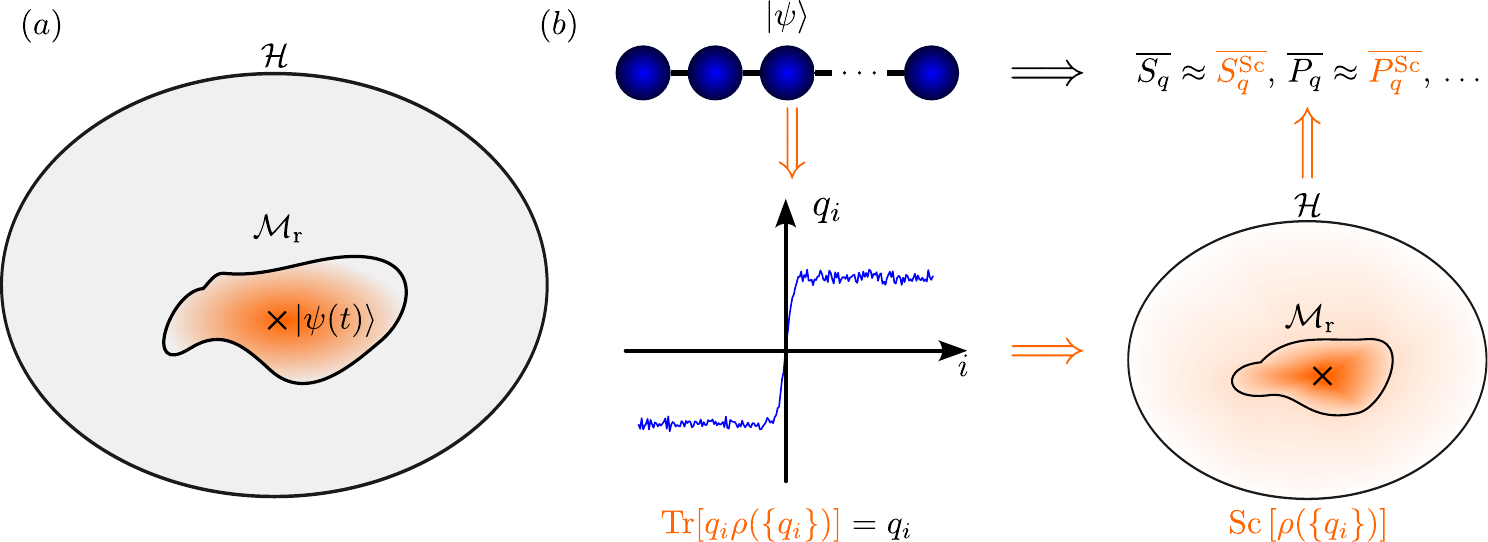}
    \caption{\textbf{Global restricted typicality and the application of the Scrooge ensemble framework to hydrodynamics.} (\textbf{a}) Conceptual representation of global restricted typicality within the full many-body Hilbert space $\mathcal{H}$. The exact time-evolved pure state $|\psi(t)\rangle$ (black cross) can be effectively described as a typical member of a dynamically restricted manifold $\mathcal{M}_{\rm r}$ (orange), defined by the effective macroscopic constraints obeyed by $|\psi(t)\rangle$. The global properties of $|\psi(t)\rangle$ correspond to a weighted average (indicated by color intensity) over this typical set. (\textbf{b}) Operational pipeline (orange arrows) for utilizing the Scrooge ensemble to capture global restricted typicality and reconstruct higher-order quantum information directly from macroscopic hydrodynamics. Starting from an exact microscopic state $|\psi(t)\rangle$, one extracts the instantaneous hydrodynamic profile $\langle \hat q_i \rangle$ of the slow modes, such as local magnetization or energy density. This macroscopic profile serves as an effective dynamical constraint, which is encoded into a density matrix $\rho(\{q_i\})$ via the maximum entropy principle~\cite{Jaynes57} and used to construct the associated Scrooge ensemble $\mathrm{Sc}[\rho(\{q_i\})]$. While defining a continuous global probability distribution across the entire $\mathcal{H}$ (orange gradient), this ensemble peaks sharply around the restricted manifold $\mathcal{M}_{\rm r}$. Ultimately, global, higher-order quantum properties—such as the R\'enyi entanglement ($S_q$) and participation ($P_q$) entropies—are obtained by evaluating their statistical averages over the Scrooge ensemble (with overlines $\overline{(\cdot)}$ indicating potential additional averaging over disorder or initial states). As indicated by the black arrow, these ensemble predictions faithfully reproduce the exact microscopic calculations.}
    \label{fig:cartoon}
\end{figure*}

\section{Scrooge ensemble\label{sec:scrooge}}
\subsection{Background}
The Scrooge ensemble, denoted as $\mathrm{Sc}[\rho]$, uniquely maximizes the accessible information entropy among all pure-state ensembles restricted to a fixed first statistical moment, $\mathbb{E}(\op{\Phi}{\Phi}) = \rho$~\cite{Jozsa94, Mark24}. Operationally, it is constructed by sampling pure states of the form $\ket{\Phi} = \frac{\sqrt{\rho}\ket{\phi}}{\norm{\sqrt{\rho}\ket{\phi}}}$, where $\ket{\phi}$ is a randomly drawn pure state with probability density $\mathcal{D} \braket{\phi|\rho|\phi}$ with respect to the Haar measure, where $\mathcal{D}$ is the Hilbert space dimension. Recent studies reveal that when $\rho$ encodes the exact dynamical constraints of a time-evolved state, $\mathrm{Sc}[\rho]$ provides a highly accurate statistical description of the system~\cite{Mark24}. This framework was explicitly demonstrated for the diagonal ensemble $\rho_{\mathrm d}$, which governs equilibration by retaining the time-invariant energy eigenbasis probabilities $\left|\left\langle E|\psi(0)\right\rangle\right|^2$; in this context, $\mathrm{Sc}[\rho_\mathrm{d}]$ predicts the properties of states within the saturation regime. Furthermore, under specific measurement basis assumptions, the statistical moments of the projected ensemble have been shown to converge identically to those of the Scrooge measure defined on the corresponding reduced subsystem, $\mathrm{Sc}[\rho_B]$.

Constructing a macroscopic density matrix $\rho$ via Jaynes' maximum entropy principle~\cite{Jaynes57} is ubiquitous in physics, most notably in the study of quantum thermalization~\cite{Deutsch91, Srednicki94,Rigol07,Rigol08,Linden09,Gogolin11,Cassidy11,Calabrese11,Polkovnikov11,Sirker14, Gogolin16}. For generic, chaotic systems, this principle naturally yields the standard canonical Gibbs density matrix, $\rho= e^{-\beta \hat H}/Z$~\cite{Deutsch91, Srednicki94}, which describes the effective thermalized state by constraining only total energy. For integrable models, this formulation is extended to the Generalized Gibbs Ensemble (GGE) to account for an extensive set of local conserved charges~\cite{Rigol07,Rigol08, Cramer08, Cassidy11, Gring12, Caux13, Mussardo13, Brockmann14, Sirker14, Ilievski15,Langen15, Vidmar16, Essler16}. These descriptions are fundamentally "effective" because they deliberately discard the exponentially large number of exact microscopic invariants, including the energy subspace projectors $\op{E}{E}$, retaining only the constraints strictly necessary to reproduce local late-time observables~\cite{Sirker14}. Crucially, elevating these effective thermal density matrices to the pure-state Scrooge measure, $\mathrm{Sc}[\rho]$, yields remarkably accurate predictions for late-time physics~\cite{Goldstein06S}, successfully capturing not only linear observables but also complex, non-linear global properties such as entanglement~\cite{Xu26}.

\subsection{The Scrooge framework for dynamically bottlenecked systems \label{sec:framework}}

To extend the statistical description to generic quantum many-body systems via the Scrooge ensemble beyond equilibrium, we examine systems characterized by a separation of timescales. When an exact global conservation law, $[\hat H,\hat Q]=0$, corresponds to an extensive charge that can be decomposed into local densities, $\hat Q = \sum_i \hat q_i$, the associated continuity equations significantly slow down the evolution of these local densities relative to other observables. Once the fast degrees of freedom effectively scramble~\cite{Sekino08, Crossley17,Kukuljan17, Glorioso18, Nahum18, Mi21, Ho22, Hahn24}, the instantaneous expectation values $q_i\equiv\bra{\psi(t)}\hat q_i \ket{\psi(t)}$ of these slow densities emerge as the primary dynamical constraints governing the time-evolved state $\ket{\psi(t)}$. In the following, we argue that this natural timescale separation can be employed to accurately describe the state $ \ket{\psi(t)}$ by a Scrooge ensemble constructed specifically from this set of bottlenecked slow modes. This effective approach stands in contrast to previous studies of the Scrooge ensemble~\cite{Jozsa94,Goldstein06S, Reimann08,Goldstein16,Liu24,Mark24,Teufel25,Manna25, Chang25, McGinley25, Mok26,Liu26,Wu26, Liu26nonlocal}, where mathematically exact constraints were used instead. For instance, obtaining the temporal ensemble requires fixing all $\mathcal{D}$ energy eigenstate populations of the initial state.

The application of this framework to quantum hydrodynamics is summarized in Fig.~\ref{fig:cartoon}. Given a single time-evolved state $\ket{\psi(t)}$, one first extracts the instantaneous expectation values $q_i$ of the slow modes, either via exact numerical simulation or by solving the underlying hydrodynamic equations. Following Jaynes' standard maximum entropy procedure~\cite{Jaynes57}, we require the constraint density matrix $\rho$ to exactly reproduce these values, $\Tr(\rho \hat q_i) = q_i$. Maximizing the von Neumann entropy, $-\Tr(\rho \ln \rho)$, subject to this requirement yields:

\begin{equation}
    \rho(\{ q_i \}) = e^{-\sum_{j}\beta_j\hat q_j}/Z,\label{eq:scrooge_rho}
\end{equation}
where $\beta_j$ are the instantaneous Lagrange multipliers enforcing the dynamical constraints and $Z=\Tr \left(e^{-\sum_{j}\beta_j\hat q_j}\right)$ is the partition function.
Hereafter, we denote the specific density matrix obtained via this maximum entropy procedure by $\rho(\{q_i\})$, reserving the symbol $\rho$ for general statements independent of this construction.
The state $\ket{\psi(t)}$ is then statistically modeled using the corresponding Scrooge ensemble, $\mathrm{Sc}[\rho(\{q_i\})]$. Finally, utilizing analytical formulas for the annealed averages within this ensemble, which we derive in Sec.~\ref{sec:analytic_main}, with more details provided in Appendix~\ref{sec:analytic_appendix}, one obtains the restricted typical values of non-linear quantum correlations, such as the entanglement entropy $S^{\rm Sc}$ or participation entropy $P^{\rm Sc}$, which can be directly benchmarked against exact microscopic calculations.

Traditionally, macroscopic hydrodynamical descriptions involve spatially coarse-graining this density matrix $\rho(\{q_i\})$ from Eq.~\eqref{eq:scrooge_rho}, operating under the assumption that local thermal equilibrium is established within mesoscopic fluid cells~\cite{Kadanoff63, Pines66, Spohn91, Doyon20}. In this work, we deliberately bypass this coarse-graining step for two critical reasons. First, retaining the exact microscopic operators $\hat q_i$ allows us to test the applicability of the Scrooge framework in complex systems where exact macroscopic hydrodynamical equations of motion are unknown~\cite{Kim13,Znidaric16,Luitz17b,Michailidis24}. Second, avoiding phenomenological fluid approximations allows us to cleanly isolate any errors originating purely from the Scrooge ensemble ansatz from the truncation errors that would otherwise be introduced by spatial coarse-graining.

While exact conservation laws serve as a prominent example of this mechanism, our statistical framework extends well beyond standard hydrodynamics to encompass any quantum system governed by bottlenecked relaxation. We conjecture that in the thermodynamic limit this effective operational procedure yields a statistical description of the pure state equivalent to imposing exact microscopic constraints, provided the following three conditions are met:
\begin{enumerate}
    \item The effective dimension of the restricted typical manifold $\mathcal{M}_{\rm r}$, defined as $\mathcal{D}_{r}=1/\Tr(\rho(\{q_i\})^2)$~\cite{Linden09}, is exponentially large in the system size.
    \item The set of slow observables encoded in $\rho(\{q_i\})$ is complete, ensuring that all relevant local observables are correctly predicted.
    \item All degrees of freedom dynamically independent of the slow modes relax and scramble on a timescale significantly shorter than the evolution of the bottlenecked modes.
\end{enumerate}

\section{Application to quench dynamics in ergodic systems \label{sec:numerics}}

We test our framework in ergodic systems whose dynamics are bottlenecked by a global symmetry, comparing the analytical predictions of the Scrooge ensemble, with $\rho(\{q_i\})$ given by Eq.~\eqref{eq:scrooge_rho},  against exact time-evolution dynamics obtained via an efficient Chebyshev polynomial expansion of the evolution operator~\cite{TalEzer84, Weisse06, Bera17, Sierant22}.

\subsection{Our numerical approach\label{sec:numerical_approach}}

To construct the density matrix $\rho(\{q_i\})$ in Eq.~\eqref{eq:scrooge_rho}, we first numerically simulate the expectation values $q_i(t)=\bra{\psi(t)}\hat q_i \ket{\psi(t)}$. Using these values, for each considered time step $t$, we solve a set of equations to determine the Lagrange multipliers $\beta_j(t)$:

\begin{equation}
    \frac{1}{Z}\Tr\left(e^{-\sum_{j}\beta_j(t)\hat q_j}\hat q_i\right) = q_i(t).
\end{equation}
In the simplest scenario, where the operators $\hat q_i$ act non-trivially on only a single site, the density matrix $\rho(\{q_i\})$ factorizes into a tensor product, reducing the problem to independently computing the single-site traces $\Tr(e^{-\beta_j\hat q_i}\hat q_i)$. However, in the general case, the densities $\hat q_i$ need not commute. This renders the system highly non-linear and couples the local constraints, necessitating a computationally inexpensive initial guess for the Lagrange multipliers $\beta_i(t)$. In such scenarios, we initialize the root-finding solver using a cluster Gutzwiller approximation~\cite{Buosante04, Yamamoto09, McIntosh12, Luehmann13} (detailed in Appendix~\ref{sec:gutzwiller}). Specifically, we select a spatial cluster of size $k$ amenable to exact diagonalization and introduce a mean-field approximation for the operator $\sum_j \beta_j \hat q_j$ within this subsystem, allowing us to rapidly estimate the local multipliers. By iteratively sweeping the position of this cluster across the system, we converge upon a global set of initial guesses $\beta_j^{(0)}(t)$.  We then refine these Lagrange multipliers using non-linear optimization schemes, as explained in detail in Appendix~\ref{sec:gutzwiller}.

After obtaining $\rho(\{q_i\})$, we evaluate the expectation values of interest within the associated Scrooge ensemble by adopting two complementary approaches. Whenever the structure of $\rho(\{q_i\})$ permits, we compute the averages analytically, as discussed in Sec.~\ref{sec:analytic_main}, with complete derivations presented in Appendix~\ref{sec:analytic_appendix}. However, evaluating these analytical formulas can become computationally expensive, as they sometimes require diagonalizing spatial blocks of the density matrix $\rho(\{q_i\})$. This limitation motivates a numerical approach based on sampling random states from the Scrooge ensemble, detailed in Appendix~\ref{sec:sampling}, which we employ for larger system sizes in this study. In this approach, to obtain a state $\ket{\Phi}$ distributed according to the probability density $P(\Phi)$ of the Scrooge ensemble, we first draw a Haar-random state $\ket{\phi}$ and assign it a weight $\bra{\phi}\rho(\{q_i\})\ket{\phi}$. We then act on this state with the operator $e^{-\sum_j\beta_j \hat q_j/2}\propto\sqrt{\rho(\{q_i\})}$, realizing its application via an imaginary-time variant of the Chebyshev polynomial expansion~\cite{Fehske08}. We omit the explicit partition function in this step, as the resulting expectation values do not depend on it due to the normalization of the sampled Scrooge states. Finally, the ensemble expectation value of interest is obtained by computing a weighted average across these sampled states.

This operational procedure yields the Scrooge ensemble prediction for any non-linear function of a single time-evolved state $\ket{\psi(t)}$. In the presence of disorder, we apply this method independently to each realization and perform an additional disorder average, denoted by $\overline{(\cdot)}$.

\subsection{Analytical annealed averages in the Scrooge ensemble \label{sec:analytic_main}}

Our analytical results are based on the replica formalism~\cite{Calabrese04, Daley12, Islam15, Brydges19}, in which quantities extending beyond the linear expectation values of observables are encoded within a replicated Hilbert space, $\mathcal{H}^{\otimes q}$. In the simplest case of linear quantities ($q=1$), we evaluate the first statistical moment of the ensemble, which is by definition equivalent to
the density matrix $\rho$:

\begin{equation}
    \mathbb{E}^{\rm Sc}(\op{\Phi}{\Phi})=\int\mathrm{d}\Phi\, P(\Phi)\op{\Phi}{\Phi}\equiv \rho,
\end{equation}
where $\mathbb{E}^{\rm Sc}$ denotes the average over the Scrooge ensemble, which is parameterized by a uniquely defined probability density $P(\Phi)$. For an observable $O$ acting on $\mathcal{H}$, we can use this first moment to express its expectation value within the Scrooge ensemble:

\begin{equation}
    O^{\rm Sc}=\int\mathrm{d}\Phi\, P(\Phi)\Tr(\op{\Phi}{\Phi}O)=\Tr(\rho O),\label{eq:expectation_q1}
\end{equation}
where we have introduced the shorthand notation $O^{\rm Sc} \equiv \mathbb{E}^{\rm Sc}(O)$. Generalizing this logic to an arbitrary integer $q$, we denote the first moment as $\rho^{(1)}$ and define the higher-order statistical moments as:

\begin{equation}
    \rho^{(q)}=\mathbb{E}^{\rm Sc}\left(\op{\Phi}{\Phi}^{\otimes q}\right).
\end{equation}
This definition allows us to directly extend Eq.~\eqref{eq:expectation_q1} to any operator $O^{(q)}$ acting on the replicated Hilbert space $\mathcal{H}^{\otimes q}$:

\begin{equation}
    \mathbb{E}^{\rm Sc}(O^{(q)})=\Tr(\rho^{(q)}O^{(q)}).
\end{equation}
To obtain analytical predictions, we utilize an approximate formula for the moments of the Scrooge ensemble~\cite{Mark24,McGinley25}:

\begin{equation}
    \rho^{(q)}\approx\rho^{\otimes q}\sum_{\pi\in\mathrm{Sym}(q)}T_\pi,\label{eq:scrooge_moments_main}
\end{equation}
where the sum runs over permutations $\pi$ of the symmetric group $\mathrm{Sym}(q)$, and $T_{\pi}$ is a representation of the permutation $\pi$ in the replicated Hilbert space $\mathcal{H}^{\otimes q}$, acting on tensor products as $T_{\pi}\bigotimes_i\ket{\psi_i}=\bigotimes_i\ket{\psi_{\pi^{-1}(i)}}$. The relative error introduced by this approximation is bounded from above~\cite{McGinley25} by $\mathcal{O}\left[q^2\Tr(\rho^2)^{1/4}\right]$. Because the purity $\Tr(\rho^2)$ must vanish exponentially with system size in order to invoke the typicality of the Scrooge ensemble~\cite{Teufel25}, this bound ensures that the moment formula remains highly accurate as long as typicality holds.

We focus on two non-linear quantities: the R\'enyi participation entropy $P_q$ in the computational basis $\ket{\sigma}$, and the half-chain R\'enyi entanglement entropy $S_q$. These are defined as:

\begin{equation}
    P_q=\frac{1}{1-q}\ln\left[\sum_\sigma\left|\braket{\sigma|\Phi}\right|^{2q}\right]=\frac{1}{1-q}\ln I_q,\label{eq:participation_entropy}
\end{equation}
and
\begin{equation}
    S_q=\frac{1}{1-q}\ln\Tr\left[(\Tr_B\op{\Phi}{\Phi})^q\right]=\frac{1}{1-q}\ln s_q,\label{eq:entanglement_entropy}
\end{equation}
where we have introduced the inverse participation ratio $I_q=\sum_\sigma\left|\braket{\sigma|\Phi}\right|^{2q}$ and the subsystem purity $s_q=\Tr\left[(\Tr_{B}\op{\Phi}{\Phi})^q\right]$ for a spatial bipartition of the system into subsystems $A$ and $B$, corresponding here to the left and right halves of the system.

Using the replica formalism, we can write the inverse participation ratio as:

\begin{equation}
    I_q=\Tr\left[ \op{\Phi}{\Phi}^{\otimes q}\mathcal{B}_q\right],
\end{equation}
where $\mathcal{B}_q=\sum_{\sigma}\op{\sigma}{\sigma}^{\otimes q}$ is the corresponding boundary operator~\cite{Stephan10, Turkeshi23measuring, Sierant22book}. Because $\mathcal{B}_q$ is independent of the specific state $\ket{\Phi}$, its Scrooge ensemble average evaluates directly to:

\begin{equation}
    I_q^{\rm Sc}=\Tr\left[\rho^{(q)}\mathcal{B}_q\right].\label{eq:ipr_moment_boundary}
\end{equation}
Substituting the moment approximation from Eq.~\eqref{eq:scrooge_moments_main}, we note that the operator $\mathcal{B}_q$ is invariant under replica permutations, as it is composed of tensor products of identical states across all $q$ replicas. This yields $T_{\pi}\mathcal{B}_q=\mathcal{B}_q$ for any permutation $\pi\in\mathrm{Sym}(q)$, which simplifies the trace to:

\begin{equation}
    I_q^{\rm Sc}\approx q! \Tr\left[\rho^{\otimes q}\mathcal{B}_q\right]=q!\sum_{\sigma}\bra{\sigma}\rho\ket{\sigma}^q.\label{eq:scrooge_ipr}
\end{equation}
Crucially, this average depends only on the diagonal elements of the density matrix in the computational basis $\ket{\sigma}$.

Equation~\eqref{eq:scrooge_ipr} directly enables the computation of the annealed average participation entropy, where the ensemble averaging is formally performed before evaluating the logarithm in Eq.~\eqref{eq:participation_entropy}.
Denoting the diagonal probabilities as $p_\sigma=\bra{\sigma}\rho\ket{\sigma}$, we  obtain:

\begin{equation}
    P_q^{\rm Sc} \approx \frac{1}{1-q}\ln\left(\sum_{\sigma}p_\sigma^q\right) + \frac{\ln q!}{1-q}. \label{eq:scrooge_P_general_main}
\end{equation}

The results for the entanglement entropy follow analogously. We express the subsystem purity using the corresponding boundary operator $\mathcal{A}_q$:

\begin{equation}
    s_q=\Tr\left[\op{\Phi}{\Phi}^{\otimes q}\mathcal{A}_q\right],\label{eq:subsystem_purity_boundary_operator}
\end{equation}
where $\mathcal{A}_q=T_{C_q}^A\otimes T_{\mathds{1}}^B$. The representations $T^A_{C_q}$ and $T^B_{\mathds{1}}$ act independently on the subspaces $\mathcal{H}_A$ and $\mathcal{H}_B$ associated with the spatial bipartition $\mathcal{H}=\mathcal{H}_A\otimes \mathcal{H}_B$. The permutations $C_q,\mathds{1}\in\mathrm{Sym}(q)$ correspond to a forward cyclic shift and the identity operation, respectively. The Scrooge ensemble average of Eq.~\eqref{eq:subsystem_purity_boundary_operator} then reads:

\begin{equation}
    s_q^{\rm Sc}=\Tr\left[ \rho^{(q)}\mathcal{A}_q \right].
\end{equation}
Upon substituting the moment approximation from Eq.~\eqref{eq:scrooge_moments_main}, evaluating this trace for arbitrary $q$ is more involved than for the inverse participation ratio in Eq.~\eqref{eq:ipr_moment_boundary}, as $\mathcal{A}_q$ explicitly relies on the group structure of $\mathrm{Sym}(q)$. In Appendix~\ref{sec:analytic_appendix}, we use the representation theory of $\mathrm{Sym}(q)$ to derive a computationally practical analytical formula for arbitrary $q$, which is exact for separable $\rho$ and serves as an accurate approximation for density matrices with clustering of correlations. Here, however, we focus on the simplest non-trivial case, $q=2$, with formulas valid for a general $\rho$. For $q=2$, the cyclic permutation reduces to a simple swap operator between the two replicas, $C_2\equiv \mathbb{S}$, which yields:

\begin{align}
    s_2^{\rm Sc} & =\Tr\left[ \rho^{\otimes2}(T_{\mathds{1}}+T_{\mathbb{S}})\left(T^A_{\mathbb{S}}\otimes T^B_{\mathds{1}}\right)\right] \nonumber                              \\
                 & =\Tr\left[\rho^{\otimes2}\left(T^A_{\mathbb{S}}\otimes T^B_{\mathds{1}}\right) +\rho^{\otimes2}\left(T^A_{\mathds{1}}\otimes T^B_{\mathbb{S}}\right) \right] \\
                 & =\Tr(\rho_A^2)+\Tr(\rho_B^2). \nonumber
\end{align}
While in principle for $q>1$ this annealed average provides only a lower bound for the exact quenched average, $S_q^{\rm Sc}\geq \frac{1}{1-q}\ln s_q^{\rm Sc}$, the difference between the two quantities is exponentially small in the system size due to typicality~\cite{Teufel25}. Therefore, the exact quenched average of the entanglement entropy can be accurately approximated by its annealed counterpart:

\begin{equation}
    S_2^{\rm Sc} \approx -\ln\left[ \Tr(\rho_A^2)+\Tr(\rho_B^2) \right].\label{eq:scrooge_S_general_main}
\end{equation}

While the results presented apply strictly to integer values of $q\geq2$, it is possible to extend them to arbitrary $q$ via analytic continuation. Because the formula for the average participation entropy, Eq.~\eqref{eq:scrooge_P_general_main}, is already analytic in $q$, we can readily evaluate the $q\to1$ limit using the standard replica trick:

\begin{equation}
    P_1^{\rm Sc} = -\lim_{q\to1}\frac{\partial\ln I_q^{\rm Sc}}{\partial q}=-\sum_{\sigma} p_\sigma \ln p_\sigma + \gamma - 1,\label{eq:scrooge_P1_general_main}
\end{equation}
where $\gamma$ is the Euler-Mascheroni constant. An analogous generalization is possible for the entanglement entropy; however, due to the combinatorial nature of its boundary operator, this analytic continuation is rigorously achievable only in the thermodynamic limit. We discuss this limiting behavior in Sec.~\ref{sec:integrable_main} and provide a complete proof of the extension in Appendix~\ref{sec:integrable_appendix}.

\subsection{Numerical results in $\mathrm{U}(1)$ symmetric periodically driven system\label{sec:floq}}

To study the accuracy of the Scrooge ensemble in describing nonlinear quantities characterizing nonclassicality of the time-evolved states, we numerically test our framework on the spin-$1/2$ Floquet XXZ chain of size $L$, where the conservation of the total spin projection $\hat S^z_{\rm tot} = \sum_{i}\hat S_i^z$ bottlenecks the macroscopic evolution. We consider a disordered variant of this model, inspired by Ref.~\cite{Michailidis24}, with the Floquet evolution operator $U$ defined as:

\begin{equation}
    \begin{gathered}
        \ \ U=U_e U_o,\quad U_e=\prod_{i=1}^{L/2}U_{2i,2i-1},\quad U_o=\prod_{i=1}^{L/2}U_{2i-1,2i-2},\\
        U_{i,i+1}=e^{-i\left[ J(\hat S_{i}^+ \hat S_{i+1}^- + \mathrm{h.c.})+2\Delta \hat S_{i}^z \hat S_{i+1}^z+h_i \hat S_i^z+h_{i+1} \hat S_{i+1}^z \right]}.
        \label{eq:floq_XXZ}
    \end{gathered}
\end{equation}
Here, the periodic driving consists of alternating interactions on even and odd bonds during each half-period. Throughout this section, we impose periodic boundary conditions. We set the interaction parameters to $J=\pi/4$ and $\Delta=J-0.2$, with the on-site fields $h_i$ drawn uniformly from the interval $[-2,2]$. The local charge densities in this model correspond directly to the spin-$z$ projection operators, $\hat q_j=\hat S_j^z$. In the absence of the disorder fields $h_i$, this chain is integrable~\cite{Gritsev17} and displays ballistic transport, whereas the introduction of a staggered magnetic field is known to yield diffusive behavior with associated hydrodynamic corrections~\cite{Michailidis24}.

We first consider a quantum Riemann problem~\cite{Antal99, Gobert05, Bertini16, Hauschild16, Misguich17, Ljubotina17}, i.e., a quench protocol in which the initial state is partitioned into two distinct macroscopic stationary states joined by a sharp spatial discontinuity. Specifically, we investigate the time evolution of a domain-wall initial state, $\ket{\psi_0}= \ket{\uparrow\cdots\uparrow\downarrow\cdots\downarrow}$, evaluating both the second R\'enyi entanglement entropy $S_2$ and the participation entropy $P_2$ in the computational spin basis. We simulate $100$ disorder realizations and present the numerical results obtained from a quenched average over the disorder, denoted as $\overline{S_2}$ and $\overline{P_2}$. In this scenario, the density matrix from Eq.~\eqref{eq:scrooge_rho} assumes a simple product form, allowing us to explicitly determine the Lagrange multipliers as $\beta_i(t)=-2\mathrm{arctanh}[{2}q_i(t)]$, where $q_i(t)=\bra{\psi(t)}\hat S_i^z\ket{\psi(t)}$. This enables the direct application of the analytical Scrooge ensemble formulas in Eqs.~\eqref{eq:scrooge_S_general_main} and \eqref{eq:scrooge_P_general_main}. We note that in this section, we do not utilize the full unconstrained density matrix; because the initial state resides entirely within the $S_{\rm tot}^z=0$ symmetry sector (for even system sizes $L$), we restrict our annealed average evaluations strictly to this single sector. While evaluating the full matrix $\rho(\{q_i\})$ would, in principle, yield slightly different numerical predictions for the Scrooge ensemble, we argue in Sec.~\ref{sec:typical_manifold} — with a complete proof provided in Appendix~\ref{sec:can_appendix} — that for our physical setup these corrections are strictly subleading in the thermodynamic limit.

\begin{figure}[ht]
    \centering
    \includegraphics[width=\linewidth]{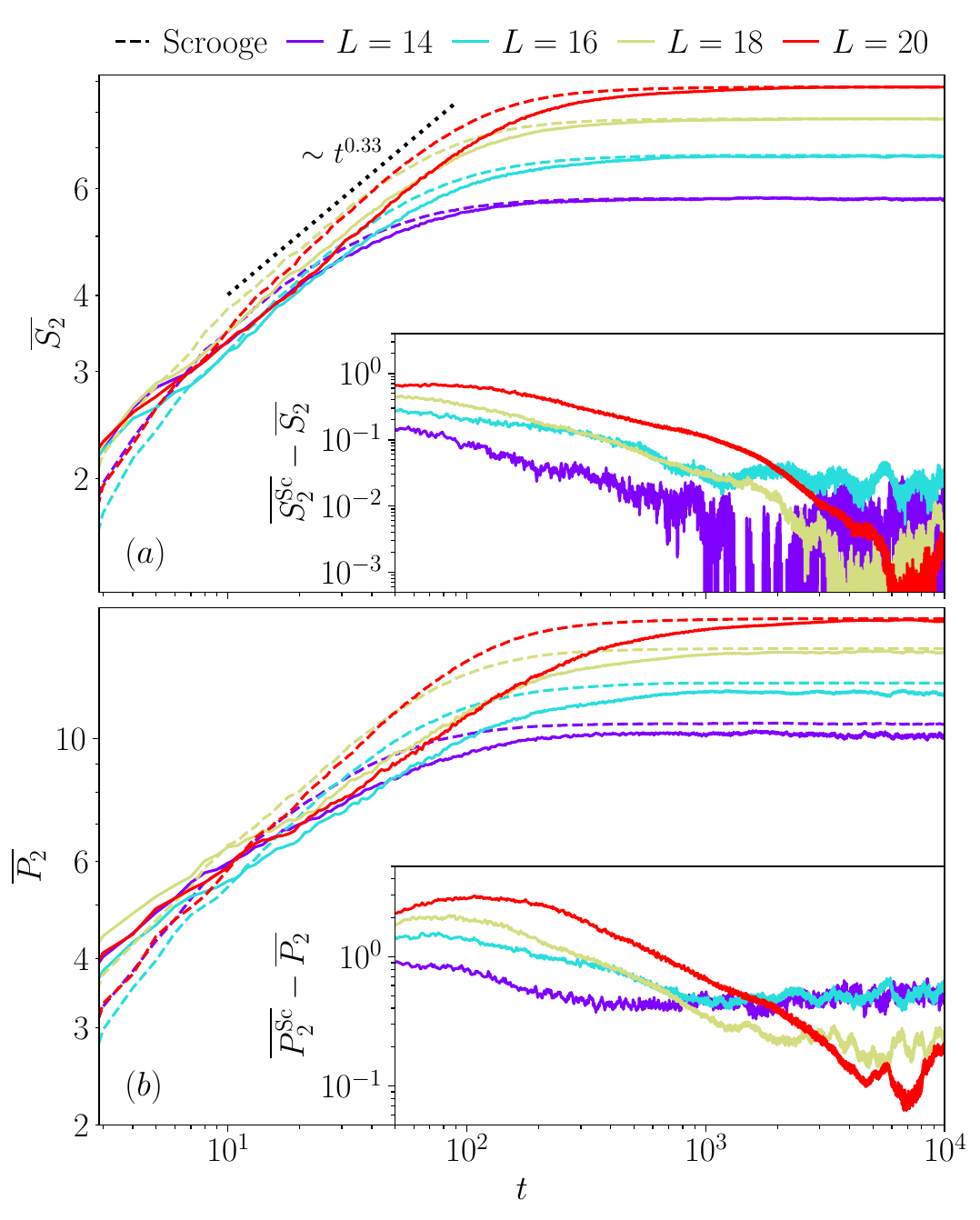}
    \caption{\textbf{Hydrodynamic constraints on nonlinear quantum information.} Time evolution of the initial domain-wall state $\ket{\psi_0}=\ket{\uparrow\cdots\uparrow\downarrow\cdots\downarrow}$ governed by the Floquet XXZ model [Eq.~\eqref{eq:floq_XXZ}] for various system sizes $L$, plotted on a log-log scale. The panels show (\textbf{a}) the second R\'enyi entanglement entropy $\overline{S_2}$ and (\textbf{b}) the participation entropy $\overline{P_2}$. Solid lines denote the exact unitary dynamics, and dashed lines represent the analytical Scrooge ensemble predictions constructed from instantaneous local profiles [Eqs.~\eqref{eq:scrooge_S_general_main} and \eqref{eq:scrooge_P_general_main}]. Panel (\textbf{a}) includes a dotted power-law guide to the eye, fitted for $L=20$ over the time window $t\in[10,90]$ and horizontally offset for improved readability. The insets share the main $x$-axis for $t \ge 50$ and display the absolute difference between the exact evolution and the Scrooge ensemble predictions, demonstrating a clear power-law decay. This illustrates the late-time convergence of microscopic complexity onto macroscopic hydrodynamic constraints.}
    \label{fig:domain_wall_floq}
\end{figure}

The results are depicted in Fig.~\ref{fig:domain_wall_floq}, where the exact disorder-averaged entropies $\overline{S_2}$ and $\overline{P_2}$ are compared against the Scrooge ensemble predictions, $\overline{S_2^{\rm Sc}}$ and $\overline{P_2^{\rm Sc}}$. Both entropies exhibit power-law growth at early to intermediate times. Remarkably, as shown in Fig.~\ref{fig:domain_wall_floq}(a), the exponent of this growth for the entanglement entropy is precisely reproduced by the Scrooge ensemble predictions relying solely on the local charge expectation values. The participation entropy in Fig.~\ref{fig:domain_wall_floq}(b) also demonstrates relatively good agreement between $\overline{P_2}$ and  $\overline{P_2^{\rm Sc}}$. As the system approaches equilibrium, we observe that the error $\overline{S_2^{\rm Sc}}-\overline{S_2}$ exhibits a power-law decay with a universal exponent, which is depicted in the insets of Fig.~\ref{fig:domain_wall_floq}. The late-time saturation values are also accurately captured, with absolute errors on the order of $10^{-3}$ for the entanglement entropy and $10^{-1}$ for the participation entropy. Finally, we note that for the system sizes considered, the time regime where the Scrooge ensemble predictions are most accurate corresponds exactly to the emergence of spatially uniform Lagrange multipliers, $\beta_i(t)\approx\beta(t)$. This precisely coincides with the regime where standard hydrodynamic approximations are expected to become increasingly accurate.

\begin{figure}[ht]
    \centering
    \includegraphics[width=\linewidth]{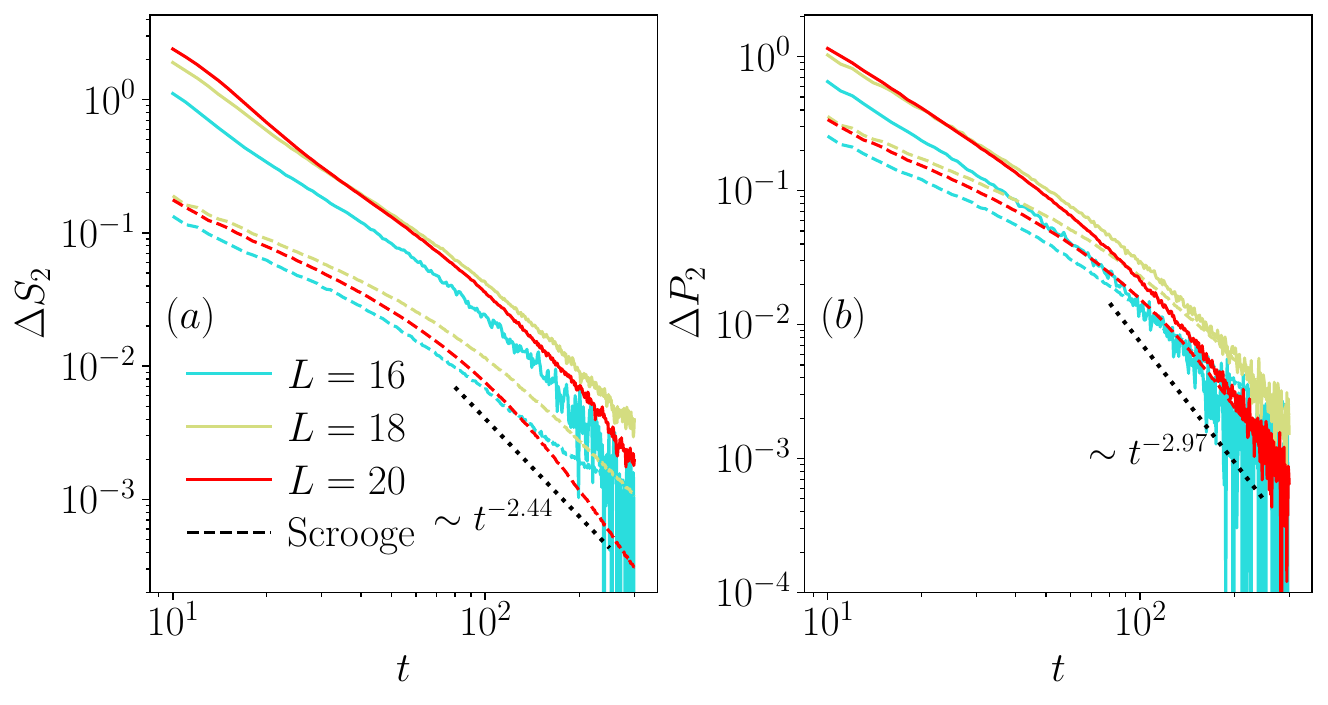}
    \caption{\textbf{Late-time hydrodynamic relaxation.} Time evolution of the initial N\'eel state $\ket{\psi_0}=\ket{\uparrow\downarrow\cdots}$ under the dynamics of the Floquet XXZ model [Eq.~\eqref{eq:floq_XXZ}] for system sizes $L \ge 16$. The panels show (\textbf{a}) the distance from the infinite-time saturation value of the second R\'enyi entanglement entropy, $\Delta \overline{S_2}$, and (\textbf{b}) the corresponding distance for the participation entropy, $\Delta \overline{P_2}$. Solid lines denote the exact microscopic unitary dynamics, and dashed lines represent the analytical Scrooge ensemble predictions constructed from instantaneous local profiles [Eqs.~\eqref{eq:scrooge_S_general_main} and \eqref{eq:scrooge_P_general_main}]. Dotted lines indicate power-law fits for $L=20$ over the time interval $t\in[80,250]$, which are horizontally offset for clarity. Upon the emergence of universal hydrodynamic behavior, both quantities exhibit the same power-law decay toward their respective saturation limits.}
    \label{fig:neel_floq}
\end{figure}

We next consider another inhomogeneous quench, initiated from the N\'eel state $\ket{\psi_0}=\ket{\uparrow\downarrow\cdots}$. In this case, we find that at early times, the Scrooge ensemble substantially overestimates both entropies. Under the evolution operator defined in Eq.~\eqref{eq:floq_XXZ}, the spatial charge profile $q_i$ of this specific initial state rapidly relaxes close to its stationary distribution. However, the entanglement does not have sufficient time to build up correspondingly, as the growth of non-local correlations remains causally bottlenecked. Because the Scrooge ensemble constructed via the density matrix $\rho(\{q_i\})$ in Eq.~\eqref{eq:scrooge_rho} represents a maximally agnostic distribution over the Hilbert space constrained solely by the instantaneous local densities $\hat q_i$, it retains no memory of the system's prior history. Consequently, a typical state drawn from the Scrooge ensemble exhibits significantly larger entropies than those physically attainable during the early stages of the evolution. In this transient regime, the validity conditions established in Sec.~\ref{sec:framework} are not fully satisfied. Improving the accuracy of the Scrooge ensemble prediction here would require incorporating a constraint into $\rho(\{q_i\})$ that explicitly captures the causal connection to the initial state. This physical intuition explains the comparatively high accuracy of the Scrooge predictions for the domain-wall quench in Fig.~\ref{fig:domain_wall_floq}, where the evolving macroscopic density profile inherently encodes the causal lightcone. We formalize this intuition for integrable systems in Sec.~\ref{sec:integrable_main}, demonstrating that in the Euler scaling limit of generalized hydrodynamics, the Scrooge ensemble, supplemented with causal information, exactly recovers the formula for the von Neumann entanglement entropy originally conjectured for homogeneous quenches by Alba and Calabrese~\cite{Alba17}, and later extended to inhomogeneous quenches~\cite{Alba18, Bertini18i, Alba19}. Additional results in support of this correspondence are provided in Appendix~\ref{sec:integrable_appendix}.

At late times following the N\'eel state quench, where causal constraints become less prominent, we analyze the approach of both R\'enyi entropies to their numerically extracted saturation values, defining $\Delta \overline{S_2}(t)=\overline{S_2}(\infty)-\overline{S_2}(t)$ and $\Delta \overline{P_2}(t)=\overline{P_2}(\infty)-\overline{P_2}(t)$. The late-time relaxation of the entanglement entropy, $\Delta S_2$, was previously studied in Ref.~\cite{Rakovszky19}, which demonstrated that the approach to saturation is governed by a single dominant hydrodynamic exponent dependent on the initial state. Our results in Fig.~\ref{fig:neel_floq} reveal analogous behavior for both quantities considered here. We observe that at intermediate times, these power-law exponents do not perfectly match between the exact numerics and the Scrooge ensemble predictions; we interpret this discrepancy as a manifestation of non-equilibrated dynamical modes. However, following this intermediate transient period, the Scrooge ensemble exactly reproduces the dominant late-time exponent for both $\Delta \overline{S_2}$ and $\Delta \overline{P_2}$, confirming that the relaxation at these timescales is dictated entirely by the dominant hydrodynamic modes. Even at late times, the absence of causal information continues to slightly affect the Scrooge ensemble, causing it to systematically predict an entanglement entropy closer to its saturation value. In contrast, the participation entropy remains highly robust against this effect. This systematic offset, however, poses no obstacle to accurately predicting the leading hydrodynamic exponents for either quantity.

\subsection{Numerical results in generic Hamiltonian system}

Having established the regime of accuracy for our Scrooge ensemble framework, alongside its causality-induced limitations within the $\mathrm{U}(1)$-symmetric Floquet model [Eq.~\eqref{eq:floq_XXZ}], we now investigate energy transport. In many respects, the previously considered $\mathrm{U}(1)$ symmetry constitutes the simplest possible scenario; the local charge densities can be chosen as strictly single-site operators, which reduces the underlying constrained density matrix $\rho(\{q_i\})$ to a trivial tensor product form. For generic Hamiltonian systems, however, local energy densities inherently involve multi-site interaction terms, breaking this simple product structure. This complexity limits the range of system sizes for which the exact analytical formulas from Sec.~\ref{sec:analytic_main} are practical to compute. To evaluate the Scrooge ensemble predictions for larger systems, we instead rely on direct numerical sampling and the principles of quantum typicality.

To systematically improve the short-time description and address the causal limitations of density-only constraints, we extend our Scrooge ensemble by incorporating local currents $\hat{j}_i$, defined via the lattice continuity equation:

\begin{equation}
    \partial_t \hat{q}_i = \hat{j}_{i-1} - \hat{j}_{i}.
\end{equation}
By defining the generalized density matrix as
\begin{equation}
    \rho(\{q_i,j_i\}) = \frac{1}{Z} \exp\left(-\sum_i \beta_i \hat{q}_i - \sum_i \gamma_i \hat{j}_i\right),
\end{equation}
where the Lagrange multipliers $\beta_i$ and $\gamma_i$ are fixed by the conditions $\Tr(\rho(\{q_i,j_i\})\hat q_i)=q_i$ and $\Tr(\rho(\{q_i,j_i\}) \hat j_i)=j_i\equiv\bra{\psi(t)}\hat j_i\ket{\psi(t)}$, we effectively constrain both the instantaneous densities and their exact first time derivatives, thereby partially capturing the dynamical flow of conserved quantities. We note that incorporating the full hierarchy of time derivatives at time $t$ would be equivalent to specifying the complete history of the densities from the initial state, which would fully restore the causality of local transport. This current-based extension therefore aims to enhance the fidelity of reconstructed non-linear quantities specifically during the early-time transient regime. At late times, as the currents decay, this extended ensemble reduces back to the density-only description. Nevertheless, it is important to emphasize that even if all local time derivatives were included, the physical description would remain fundamentally incomplete; capturing the exact causal front of the entanglement spreading inherently requires tracking the non-local correlations of the charges as well.

As a concrete model exhibiting energy transport, we consider the paradigmatic Ising model~\cite{Kim13}:
\begin{equation}
    \hat{H} = g \sum_{i=1}^L \hat{\sigma}_i^x + h \sum_{i=2}^{L-1} \hat{\sigma}_i^z + (h-J)(\hat{\sigma}_1^z + \hat{\sigma}_L^z) + J \sum_{i=1}^{L-1} \hat{\sigma}_i^z \hat{\sigma}_{i+1}^z,
\end{equation}
with the standard ergodic choice of the longitudinal field $h=(\sqrt{5}+1)/4$, the transverse field $g=(\sqrt{5}+5)/8$, and the interaction $J=1$. The longitudinal field $h$ is modified at the boundaries to reduce finite-size effects by keeping the spin-flip energy cost identical between the bulk and boundaries. Because the conserved charge $\hat{Q}$ does not uniquely define local densities $\hat{q}_i$ subject to the global constraint $\hat{Q} = \sum_i\hat{q}_i$ and locality, we partition the energy into bond-centered densities. Specifically, we combine the interaction on the bond $(i, i+1)$ with a weighted sum of the on-site energies from its adjacent sites:
\begin{equation}
    \hat{q}_i = J \hat{\sigma}_i^z \hat{\sigma}_{i+1}^z + w_i(g \hat{\sigma}_i^x + h_i \hat{\sigma}_i^z) + w_{i+1}(g \hat{\sigma}_{i+1}^x + h_{i+1} \hat{\sigma}_{i+1}^z).
\end{equation}
Here, the effective longitudinal field is $h_i = h$ in the bulk, and $h_1 = h_L = h - J$ at the open boundaries. To ensure that the sum of all local densities exactly recovers the global Hamiltonian, the spatial weights are chosen as $w_k = 1/2$ in the bulk and $w_1 = w_L = 1$ at the chain ends. Defining the density symmetrically in the bulk ensures that the local energy current $\hat{j}_i$, extracted via the lattice continuity equation, takes a remarkably simple form. For the bulk sites $i=2,\dots,L-1$, this site-centered current operator reduces exactly to:
\begin{equation}
    \hat{j}_i = J g \hat{\sigma}_i^y (\hat{\sigma}_{i+1}^z - \hat{\sigma}_{i-1}^z).
\end{equation}

\begin{figure}[ht]
    \centering
    \includegraphics[width=\linewidth]{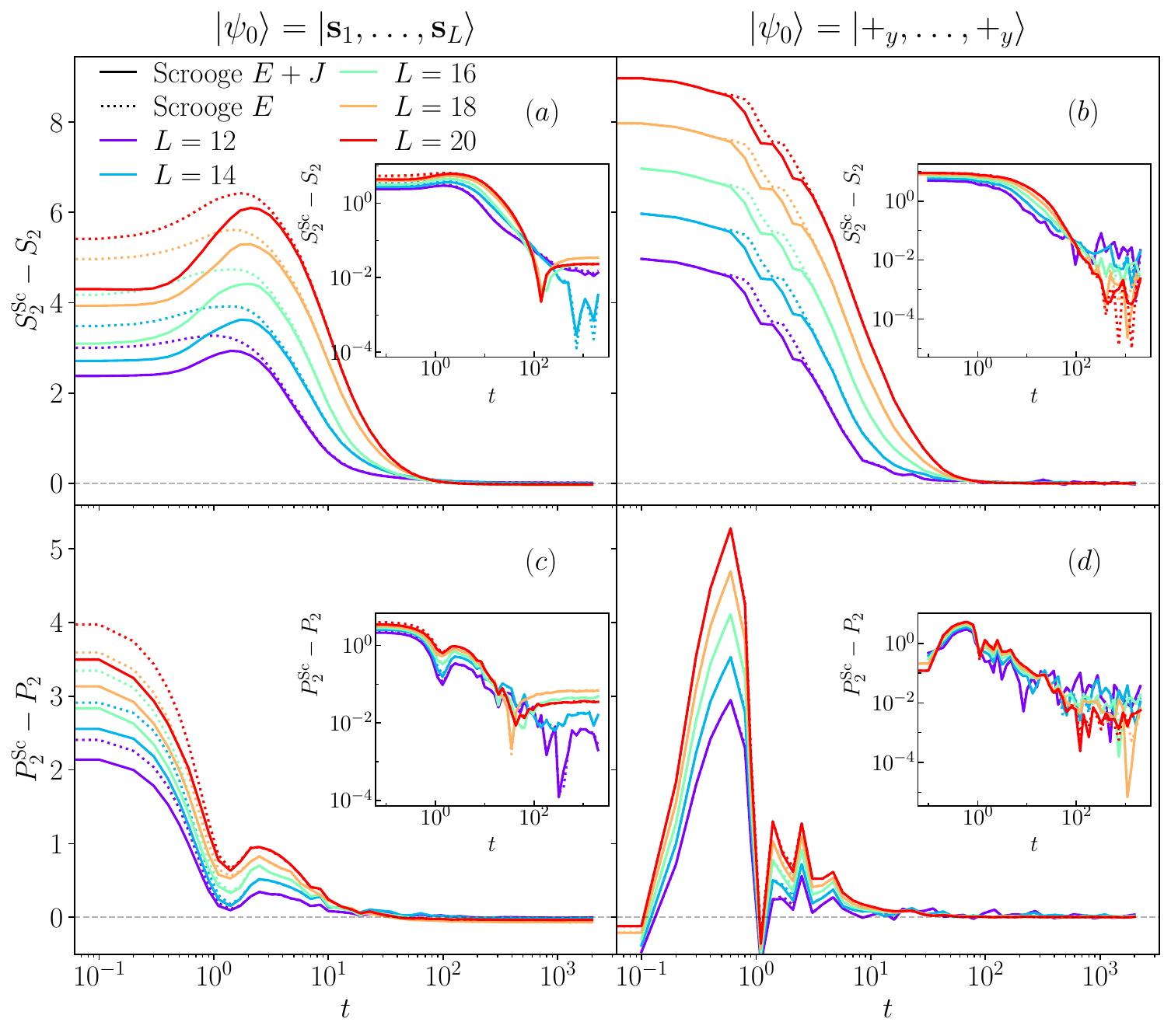}
    \caption{\textbf{Robustness of the late-time predictions of the Scrooge ensemble against minimal-information initial states.} Time evolution of the deviations between the Scrooge ensemble predictions and exact dynamics for (\textbf{a, b}) the second R\'enyi entanglement entropy $S_2$ and (\textbf{c, d}) the participation entropy $P_2$. The transverse-field Ising model (TFIM) is initialized in states that provide minimal local information to the ensemble: a set of random product states $|\psi_0\rangle = \ket{\mathbf{s}_1, \dots, \mathbf{s}_L}$, where the plotted quantities represent the statistical average of the entropies computed for each individual pure-state realization (\textbf{a, c}), and a uniformly polarized state $|\psi_0\rangle = \ket{+_y, \dots, +_y}$ (\textbf{b, d}). Despite this lack of initial information, the exact dynamics successfully converge to the Scrooge predictions following a period of non-universal, early-time evolution. Incorporating higher-order hydrodynamic information improves the description of these initial states; accordingly, solid lines denote the Scrooge ensemble constrained by both energy and current (Scrooge $E+J$), while dotted lines represent constraints from energy alone (Scrooge $E$). Colors indicate system sizes ranging from $L=12$ to $L=20$. Insets show the power-law decay of the deviations from the main panels, reaching a small saturation mismatch on the order of $10^{-2}$.}
    \label{fig:ising}
\end{figure}

To probe the out-of-equilibrium dynamics, we initialize the system in a generic random product state, $\ket{\psi_0} = \ket{\mathbf{s}_1, \dots, \mathbf{s}_L}$, where $\ket{\mathbf{s}_i} = \cos(\theta_i/2)\ket{\uparrow}_i + e^{i\phi_i}\sin(\theta_i/2)\ket{\downarrow}_i$. For each site $i$, the polar and azimuthal angles $\theta_i$ and $\phi_i$ are drawn independently and uniformly from the intervals $[0, \pi]$ and $[0, 2\pi]$, respectively, yielding a completely unentangled yet spatially inhomogeneous initial profile. Measured quantities are averaged over this set of initial states, yielding $\overline{S_2}$ and $\overline{P_2}$. We consider this a stringent test for our Scrooge ensemble framework, as these initial states exhibit ballistic entropy growth while the spread of energy remains purely diffusive. Nonetheless, as seen from the error of the Scrooge framework in Figs.~\ref{fig:ising}(a) and \ref{fig:ising}(c), both quantities exhibit significant deviations at early times. This early-time error is noticeably reduced by incorporating the local currents $\hat{j}_i$ into the framework. At later times, however, the error rapidly decreases—decaying as a power law for the participation entropy and exponentially for the entanglement entropy—reaching $10^{-2}$ in the saturation regime, as shown in the insets of the corresponding figures. In this late-time regime, we observe no difference between the descriptions with and without the currents, as the macroscopic local currents have already decayed.

Finally, we consider a homogeneous quench from the initial state $|\psi_0\rangle = \ket{+_y, \dots, +_y}$, where $\hat{\sigma}_i^y\ket{+_y}=\ket{+_y}$. This specific state is completely orthogonal to both the local densities $\hat{q}_i$ and the currents $\hat{j}_i$, such that their initial expectation values strictly vanish: $\bra{\psi_0}\hat{q}_i\ket{\psi_0}=\bra{\psi_0}\hat{j}_i\ket{\psi_0}=0$. Consequently, the constrained density matrix $\rho(\{q_i,j_i\})$ at $t=0$ reduces to the maximally mixed state, $\mathds{1}/\mathcal{D}$, meaning the Scrooge ensemble trivially predicts the Haar averages for both studied entropies. In this extreme case, local transport does not govern the early-time dynamics of higher-order quantum information, and capturing non-local correlations would be fundamentally necessary to improve the approximation~\cite{Nava26}. Figures~\ref{fig:ising}(b) and \ref{fig:ising}(d) make it clear that the inclusion of local currents offers no substantial improvement, aside from a minimal correction in a narrow time window around $t=1$. Yet, even for this featureless initial profile, the Scrooge description becomes increasingly accurate at late times; the error exhibits the same qualitative decay as observed for the random product states, ultimately yielding predictions for the saturation values with an accuracy on the order of $10^{-2}$.

This remarkable late-time precision stands in stark contrast to standard approaches that approximate the saturation of these non-linear measures using uniform Haar-random ensembles~\cite{Vidmar17, Roberts17, Cotler17, Lu19, Bianchi22, Tirrito25}. It is well established that eigenstates of ergodic Hamiltonians are not featureless random vectors. Instead, the intrinsic spectral properties of time-independent Hamiltonians~\cite{Roberts17, Cotler17}, alongside strict global conservation laws~\cite{Vidmar17, Lu19, Bianchi22}, fundamentally constrain the dynamically explored Hilbert space. Because of these spectral and dynamical constraints, non-linear quantities fail to reach the maximal Haar prediction~\cite{Vidmar17, Lu19, Tirrito25}. Crucially, the discrepancy between the true saturation value and the Haar approximation is not merely a finite-size effect, but a macroscopic error that diverges as the system size increases~\cite{Vidmar17, Lu19}. In the absence of additional symmetries, this error is strictly extensive in the system size $L$~\cite{Tirrito25}. For instance, for the system sizes considered in this study, this error in the participation entropy is already of order $1$. Our results demonstrate that these extensive discrepancies can be successfully captured without resorting to full spectral information; they are largely explained simply by the non-trivial, stationary spatial distribution of the conserved charge after the system reaches equilibrium.

\section{Restricted typicality for integrable systems\label{sec:integrable_main}}

The application of the Scrooge ensemble framework to generic systems with global conservation laws, where hydrodynamical behavior emerges at late times, paints a clear physical picture: the non-linear properties of the state are described increasingly well by the local conserved densities as time progresses. However, during the early stages of evolution, this description misses crucial causal information. Here, we utilize the generalized hydrodynamics (GHD) formalism~\cite{Castro-Alvaredo16, Bertini16, Doyon17, Bulchandani18, DeNardis18, Doyon20, Essler23, Doyon25} for integrable models to formalize this intuition. We analytically demonstrate that the formula for the entanglement entropy following an inhomogeneous quench of an integrable system~\cite{Alba17,Alba18,Bertini18i,Alba19},

\begin{equation}
    S_A(t) = \int \mathrm{d}\lambda \int_A \mathrm{d}x \, s_{YY}(x, t, \lambda)\Big[1 - \chi_{A}\big(x_{-\lambda}(x, t)\big)\Big],\label{eq:Alba-Calabrese}
\end{equation}
hereafter referred to as the Alba-Calabrese formula, is perfectly reproduced once we complement the Scrooge ensemble framework with the semiclassical quasiparticle picture~\cite{Calabrese05}. In Eq.~\eqref{eq:Alba-Calabrese}, $s_{YY}(x, t, \lambda)$ represents the instantaneous Yang-Yang entropy density~\cite{Yang69} and $\chi_A$ is the characteristic function of subsystem $A$. Here, we assume a single quasiparticle species without losing generality. In the quasiparticle picture~\cite{Calabrese05}, the initial state acts as a source of excitations that travel ballistically through the system after the quench. Pairs of excitations emitted from the same spatial point at $t=0$ are mutually entangled; bipartite entanglement between $A$ and its complement is generated whenever one member of the pair resides inside $A$ while its partner resides outside. The above formula assumes that quasiparticle excitations in the initial state are produced strictly in pairs, with possible extensions to more general quenches~\cite{Bertini18p,Bastianello18}. The coordinate $x_{-\lambda}(x, t)$ denotes the absolute spatial position at time $t$ of the entangled partner carrying quasimomentum $-\lambda$, which originated at $t=0$ from the same spatial point as the excitation with quasimomentum $\lambda$ currently observed at position $x$. In noninteracting systems where modes propagate at constant bare velocities $v(\lambda)$, this partner coordinate reduces to the simple linear shift $x_{-\lambda}(x, t) = x - 2v(\lambda)t$. In contrast, for interacting integrable models, determining this position requires dynamically tracing curved characteristic trajectories via the GHD equations~\cite{Alba19}.

Because the Alba-Calabrese formula describes the von Neumann entanglement entropy, we must consider the analytic continuation $q\to 1$ of the Scrooge ensemble entanglement entropy in the thermodynamic limit. We base this on the formula for $q=2$ [Eq.~\eqref{eq:scrooge_S_general_main}] and its generalization to any integer $q\geq 2$, detailed in Eqs.~\eqref{eq:subsystem_purity_q} and \eqref{eq:scrooge_S} of Appendix~\ref{sec:analytic_appendix}. A rigorous proof of this limiting behavior is provided in Appendix~\ref{sec:integrable_appendix}. The fundamental building blocks of the average entanglement in the Scrooge ensemble are the subsystem purities $\Tr(\rho_A^l)$ and $\Tr(\rho_B^l)$ with $1\leq l \leq q$, which decay exponentially with the system size $L$. Each contributing term in the average relates to a permutation $\pi$ in the formula~\eqref{eq:scrooge_moments_main} and takes the form $\prod_l\left(\Tr\rho_X^{l}\right)^{k_l}$, subject to the constraint $\sum_{l} l k_l=q$ for $X\in\{A,B\}$. Here, $l$ is the length of a cycle in the decomposition, while $k_l$ is its multiplicity in permutation $\pi$.  As demonstrated in Appendix~\ref{sec:integrable_appendix}, permutations other than $\mathds{1}$ and $C_q^{-1}$ are exponentially penalized, leaving $\Tr(\rho_A^q)$ and $\Tr(\rho_B^q)$ as the dominant terms in the thermodynamic limit. If these two terms share identical leading exponential scaling with $L$, the R\'enyi entropy reduces to $S_q^{\rm Sc}=\frac{1}{1-q} \ln \Tr(\rho_A^q)+\mathcal{O}(1)$, where the correction is strictly subleading in $L$. Alternatively, if their leading scalings differ, the formula takes the same form but without the $\mathcal{O}(1)$ correction, with $A$ denoting the subsystem whose contribution survives. This structure enables a direct analytic continuation of the leading-order contribution to $S_q^{\rm Sc}$ as $q\to 1$:

\begin{equation}
    S_1^{\rm Sc}=-\Tr(\rho_A \ln \rho_A),\label{eq:scrooge_S1_thermodynamic_limit}
\end{equation}
which yields the desired von Neumann entanglement entropy formula in the thermodynamic limit, up to subleading corrections.

Integrable systems possess an infinite tower of macroscopic local conserved charges $\hat Q_{n}=\int \mathrm{d}x\, \hat q_n(x)$, where $\hat q_{n}(x)$ corresponds to the local charge densities. The Jaynes' maximum entropy state corresponding to this charge distribution is
\begin{equation}
    \rho(\{q_n(x)\})=\frac{1}{Z}\exp\left(-\sum_n\int\mathrm{d}x\,\beta_n(x)\hat q_n(x)\right),\label{eq:ghd_rho}
\end{equation}
with spatially dependent Lagrange multipliers $\beta_n(x)$. We proceed in the Euler scaling limit, $x,t\to\infty$ as $x/t=\mathrm{const}$, which is also the regime where the Alba-Calabrese formula~\eqref{eq:Alba-Calabrese} is obtained. This limit naturally justifies the local density approximation, allowing us to partition the system into macroscopic fluid cells within which $\beta_n(x)$ is approximately constant. To explicitly evaluate the Scrooge von Neumann entropy~\eqref{eq:scrooge_S1_thermodynamic_limit} for this density matrix within a given fluid cell, we rely on the Bethe ansatz formalism~\cite{Takahashi99thermo,Mussardo10, Franchini17, Doyon20}, which serves as the microscopic foundation of GHD. Considering the eigenstates $\ket{\lambda_1,\cdots,\lambda_L}$, we define the single-particle eigenvalues $h_n(\lambda_k)$ of the conserved charge $\hat Q_n$:
\begin{equation}
    \hat Q_n\ket{\lambda_1,\cdots,\lambda_L}=\sum_{k}h_n(\lambda_k)\ket{\lambda_1,\cdots,\lambda_L}.
\end{equation}
Due to the Bethe equations, only discrete quasimomenta $\lambda$ are allowed. This motivates introducing the local quasimomentum density of particles $\rho_{\rm p}(x,\lambda)$, which leads directly to the formula for the local densities:
\begin{equation}
    q_n(x)= \int\mathrm{d}\lambda\,h_n(\lambda)\rho_{\rm p}(x,\lambda).
\end{equation}
The explicit spatial dependence in $\rho_{\rm p}(x,\lambda)$ indicates that this density may differ between fluid cells. In a generic state of the quantum system, not all available quasimomenta are occupied, motivating the introduction of the local total density of states $\rho_{\rm t}(x,\lambda)$:
\begin{equation}
    2\pi \rho_{\rm t}(x,\lambda)=p'(\lambda)+\int \mathrm{d}\lambda'\,\rho_{\rm p}(x,\lambda') \varphi(\lambda-\lambda'),
\end{equation}
where $\varphi(\lambda-\lambda')$ denotes the two-body scattering amplitude, and $p'(\lambda)$ is the derivative of the bare momentum with respect to the quasimomentum. This allows us to naturally define the local hole density $\rho_{\rm h}(x,\lambda)\equiv \rho_{\rm t}(x,\lambda)-\rho_{\rm p}(x,\lambda)$.

Under the local density approximation, we consider a single fluid cell $X$ of spatial length $\ell_X$ in the Euler scaling limit with the density matrix $\rho(X)=\exp\left[-\sum_n\int_X \mathrm{d}x\, \beta_n(x)\hat q_n(x)\right]/Z(X)$. The thermodynamic entropy of this cell is defined as $S(X)=-\Tr[\rho(X) \ln \rho(X)]$, which expands to:
\begin{equation}
    S(X)=\ln Z(X)+\sum_n \beta_n(X)\Tr[\rho(X) \hat Q_n(X)],\label{eq:cell_entropy}
\end{equation}
where $\hat Q_n(X)=\int_X\mathrm{d}x\, \hat q_n(x)$ denotes the total charge contained within the fluid cell. The partition function $Z(X)$ can be evaluated as a functional integral over all possible macroscopic quasiparticle distributions $\rho_{\rm p}'$. This formulation requires incorporating a combinatorial weight that accounts for the number of microscopic realizations of each macrostate $\rho_{\rm p}'$—this is precisely the Yang-Yang entropy $s_{YY}(\rho_{\rm p}')$. Upon including this statistical weight, the partition function takes the form:
\begin{equation}
    Z(X)=\int \mathrm{d}[\rho_{\rm p}']\, e^{\ell_X\int\mathrm{d}\lambda\,\left(s_{YY}(\rho_{\rm p}')-\sum_n \beta_n(X) h_n(\lambda)\rho_{\rm p}'(X,\lambda)\right)},\label{eq:cell_partition_function}
\end{equation}
with the Yang-Yang entropy density~\cite{Yang69} defined as:
\begin{equation}
    s_{YY}(\rho_{\rm p})= \rho_{\rm t}\ln\rho_{\rm t} - \rho_{\rm p}\ln\rho_{\rm p}-\rho_{\rm h}\ln \rho_{\rm h}.
\end{equation}
Evaluating the integral~\eqref{eq:cell_partition_function} via the saddle-point method isolates a unique dominant quasiparticle density $\rho_{\rm p}(\lambda)$. Consequently, to leading order, the thermodynamic entropy~\eqref{eq:cell_entropy} reduces to:
\begin{equation}
    S(X)=\ell_X \int\mathrm{d}\lambda\, s_{YY}(\rho_{\rm p}).\label{eq:cell_entropy_result}
\end{equation}
The Yang-Yang entropy density naturally inherits spatial and temporal dependence from the instantaneous global hydrodynamic density matrix $ \rho(\{q_n(x)\})$.

The crucial observation is that, in contrast to the exact state of the system $\ket{\psi(t)}$, the density matrix $ \rho(\{q_n(x)\})$ from Eq.~\eqref{eq:ghd_rho}, for a system with a valid GGE description, exhibits a clustering of correlations~\cite{Doyon17w}.  This is evident when we note that the construction of the density matrix~\eqref{eq:ghd_rho} uses local operators $\hat{q}_n$; hence, it fundamentally cannot encode all nonlocal correlations of the pure state $\ket{\psi(t)}$. As detailed in Appendix~\ref{sec:integrable_appendix}, this clustering ensures that the mutual information between adjacent macroscopic fluid cells is subleading~\cite{Wolf08} with respect to the thermodynamic entropy of the cell~\eqref{eq:cell_entropy_result}, thereby allowing us to treat different fluid cells independently in the Euler scaling limit. Consequently, the leading contribution to the Scrooge entanglement entropy,~\eqref{eq:scrooge_S1_thermodynamic_limit}, simplifies to a spatial integral of the thermodynamic entropy~\eqref{eq:cell_entropy_result} over the subsystem $A$:

\begin{equation}
    S_1^{\rm Sc}(t)=\int\mathrm{d}\lambda\int_A \mathrm{d}x\, s_{YY}[\rho_{\rm p}(x,t,\lambda)].\label{eq:Scrooge-Bertini}
\end{equation}
The Yang-Yang entropy density depends on space and time strictly through the quasiparticle density $\rho_{\rm p}(x,t,\lambda)$. We emphasize that Eq.~\eqref{eq:Scrooge-Bertini} is not equivalent to the Alba-Calabrese formula~\eqref{eq:Alba-Calabrese}, as it lacks causal information regarding the finite propagation velocity of excitations. Because Eq.~\eqref{eq:Scrooge-Bertini} is derived solely from the instantaneous local charge densities $\bra{\psi(t)}\hat q_n(x)\ket{\psi(t)}$ via the maximum-entropy principle based on the Scrooge ensemble, it represents the average entanglement of a typical pure state sharing that exact macroscopic charge profile. For a general initial state, this static average is blind to the light cones of correlations between excitations and can systematically violate the Lieb-Robinson bound~\cite{Lieb72}. For instance, in a homogeneous quench, Eq.~\eqref{eq:Scrooge-Bertini} predicts that the entanglement entropy immediately jumps to its saturation value—identical to the asymptotic limit of Eq.~\eqref{eq:Alba-Calabrese}—entirely bypassing the linear transient growth.

There is an important class of initial configurations where the local charge densities fully encode the underlying light cones: the junction of two different stationary states. More generally, one can consider quasistationary states—namely, those for which the reduced density matrix of an arbitrary subsystem commutes with the Hamiltonian up to boundary terms. For this class, one can demonstrate that the entanglement entropy of $\ket{\psi(t)}$ is strictly additive~\cite{Bertini18i}, thus leading precisely to Eq.~\eqref{eq:Scrooge-Bertini}. While Ref.~\cite{Bertini18i} originally established this for noninteracting systems, the additivity result generalizes to interacting models based on later findings~\cite{Alba18,Alba19}. This also provides an explanation for the relatively good Scrooge ensemble predictions at all times for the domain-wall initial state in Sec.~\ref{sec:floq}, as it falls exactly into this class.

In contrast, for generic initial states, the instantaneous spatial profile $q_n(x)$ does not encode the causal propagation history of excitations. Consequently, Eq.~\eqref{eq:Scrooge-Bertini} integrates over all excitations present inside $A$. Without the relevant causal restrictions, typical random states in the Scrooge ensemble overwhelmingly place the entangled partner outside $A$, treating the pair as mutually entangled with the complement regardless of how much time has elapsed. This reflects a fundamental mathematical feature of the Scrooge ensemble: even for a strictly short-range density matrix $\rho$, the sampled pure states inherently exhibit long-range correlations between causally disconnected regions~\cite{Liu26nonlocal}. To strictly respect causality, one must complement the instantaneous Scrooge ensemble result with the semiclassical quasiparticle picture by demanding that an entangled partner has actually had sufficient time to physically propagate out of subsystem $A$. This kinematic requirement is achieved precisely by introducing the geometric kernel $\Big[1 - \chi_{A}\big(x_{-\lambda}(x, t)\big)\Big]$, which systematically subtracts trajectories where both entangled partners remain trapped inside $A$. Remarkably, this minimal, causally motivated modification naturally recovers exactly the Alba-Calabrese formula in Eq.~\eqref{eq:Alba-Calabrese}. While it is generally difficult to encode dynamic causality directly into a single static density operator, we conjecture that this result corresponds precisely to evaluating the Scrooge ensemble restricted strictly to causally accessible states at time $t$. Recent results seem to support this intuition, demonstrating that entanglement dynamics after a quench can be captured using a transport framework for specific non-local charges, instead of standard local integrals of motion~\cite{Travaglino26}.

This provides a direct physical demonstration of how macroscopic, globally restricted typicality
emerges in generic Bethe ansatz integrable quantum systems. Because unitary quantum evolution is constrained by the transport of an infinite tower of conserved charges $\hat Q_n$, the state cannot explore the entire Hilbert space freely. Instead, through the Scrooge ensemble framework—which maximizes entropy subject to these macroscopic conservation laws—we demonstrate that, from the perspective of subsystem entanglement, the time-evolved pure state $\ket{\psi(t)}$ behaves indistinguishably from a typical state drawn from this constrained Hilbert space manifold, provided one explicitly accounts for the causal light cones governing the spread of quantum information.

\section{Structure of the typical manifold\label{sec:typical_manifold}}

The standard Scrooge ensemble provides a powerful framework for encoding macroscopic physical restrictions, such as emergent hydrodynamic profiles, directly into the background density matrix $\rho$. However, this formulation inherently acts as a soft statistical restriction: for a strictly positive $\rho$, every state in the Hilbert space acquires a nonzero probability density. This presents a fundamental conceptual mismatch for isolated quantum systems, where exact unitary time evolution is permanently confined by strict kinematic boundaries. A closed system cannot violate exact global symmetries, nor can it instantaneously develop macroscopic spatial charge fluctuations that circumvent local causality and transport constraints. Instead, as conceptually illustrated in Fig.~\ref{fig:cartoon}(a), the exact pure state remains rigidly trapped on a dynamically restricted submanifold of the Hilbert space, $\mathcal{M}_{\rm r}$. This physical confinement is the foundation of \textit{global restricted typicality}---the principle saying that a dynamically bottlenecked system behaves as a typical, random pure state drawn exclusively from this exact macroscopic constraint surface. To formally establish this phenomenon and justify the operational use of the unconstrained Scrooge ensemble, we  rigorously demonstrate that projecting the statistical measure onto these exact dynamical manifolds leaves the leading-order predictions for nonlinear functions of the state studied in this work intact.

To do so, we introduce generalizations of the Scrooge ensemble that enforce these macroscopic constraints strictly. By direct analogy to the standard ensembles of equilibrium statistical mechanics, we term these the canonical and microcanonical Scrooge ensembles. The detailed mathematical proofs establishing their geometric properties and scaling limits are relegated to Appendices~\ref{sec:can_appendix} and \ref{sec:microcan_appendix}. Here, we focus on the main physical results and their implications for the structure of the typical manifold.

\subsection{The canonical Scrooge ensemble\label{sec:can_main}}

We first address the role of global symmetries by defining the \textit{canonical Scrooge ensemble}. As argued in Sec.~\ref{sec:floq}, in models governed by a continuous $U(1)$ symmetry generated by $\hat{M}$, the standard Scrooge ensemble accurately captures the spatial profile of the local charges. Nevertheless, the standard Scrooge ensemble, analogously to the grand canonical ensemble of classical statistical mechanics, allows the total global charge to fluctuate. By rigidly projecting the probability distribution into the exact global symmetry sector (e.g., a fixed total magnetization $M$), we restrict the ensemble to the physically accessible Hilbert space.
Importantly, because the global charge commutes with the density matrix defining the Scrooge ensemble, $[\rho, \hat{M}] = 0$, this strict projection amounts to a uniform rescaling of the probabilities within the target symmetry sector: $p_\sigma^{\rm can} = (Z^{\rm Sc}/Z^{\rm can}) p_\sigma^{\rm Sc}$, where $Z^{\rm Sc}=\Tr(e^{-\sum_j\beta_j\hat q_j})$ is the unconstrained partition function and $Z^{\rm can}$ is its restriction to the target symmetry sector. This simple proportionality guarantees that macroscopic corrections to non-linear functionals, such as participation and entanglement entropies, can be systematically extracted by evaluating the corresponding restricted partition functions.

Evaluating this ratio of partition functions in the thermodynamic limit (see Appendix~\ref{sec:can_appendix} for a rigorous derivation) yields:
\begin{equation}
    \frac{Z^{\rm Sc}}{Z^{\rm can}} \approx \sqrt{2\pi \Sigma^2},\label{eq:can_ratio}
\end{equation}
where $\Sigma^2 \equiv \mathbb{E}^{\rm Sc}(\hat{M}^2) - \mathbb{E}^{\rm Sc}(\hat{M})^2$ is the total macroscopic charge variance. To demonstrate how this relation extracts canonical corrections for ensemble-averaged non-linear functions of the state, we compute those corrections to the participation entropy. The canonical inverse participation ratio is given by $I_q^{\rm can}\approx q!Z^{\rm can}_{q\beta}/(Z^{\rm can})^q$, where $Z^{\rm can}_{q\beta}$ is evaluated with Lagrange multipliers rescaled as $\beta_j\to q\beta_j$. For this modified ensemble, the expected total charge shifts to $\tilde M=\Tr(e^{-\sum_j q\beta_j\hat q_j}\hat M)/Z^{\rm Sc}_{q\beta}$. Consequently, the saddle-point integration yielding Eq.~\eqref{eq:can_ratio} must now accommodate this macroscopic shift, introducing an additional exponential penalty:

\begin{equation}
    \frac{Z^{\rm Sc}_{q\beta}}{Z^{\rm can}_{q\beta}} \approx \sqrt{2\pi\Sigma^2_{q\beta}}\exp\left(\frac{(M-\tilde M)^2}{2\Sigma^2_{q\beta}}\right).
\end{equation}

Substituting these relations into the Scrooge average of the R\'enyi participation entropy~\eqref{eq:scrooge_P_general_main} yields the canonical correction:
\begin{equation}
    P_q^{\rm can} = P_q^{\rm Sc} - \frac{(M - \tilde{M})^2}{2(1-q)\Sigma_{q\beta}^2} + \mathcal{O}(\ln \Sigma^2).
\end{equation}
The term $\frac{(M - \tilde{M})^2}{2(1-q)\Sigma_{q\beta}^2}$ may be physically interpreted as the missing relevant contribution from basis elements outside the target symmetry sector $M$. Because this correction is, in general, extensive, it highlights that higher-order R\'enyi participation entropies experience a macroscopic dependence on the chosen symmetry sector, closely mirroring the analogous behavior of symmetry-resolved entanglement entropies~\cite{Piroli22}. This establishes that typicality does not apply to all global non-linear quantities. A prominent example is the probability $\left|\braket{n|\Phi}\right|^2$. For states $\ket{\Phi}$ drawn from the Scrooge ensemble, these probabilities exhibit a Porter-Thomas distribution~\cite{Mark24} with a relative variance that remains non-vanishing regardless of the Hilbert space dimension.

Remarkably, however, this extensive correction completely vanishes in the physical regimes most relevant to this work. For instance, for antisymmetric macroscopic density profiles, such as the N\'eel and domain-wall states explored in Sec.~\ref{sec:floq}, one obtains $\tilde{M} = M$, canceling the extensive correction. Furthermore, for arbitrary spatial profiles, this extensive penalty systematically vanishes in the analytic continuation to the von Neumann limit $q \to 1$. In these regimes, the symmetry constraint leaves only a strictly subleading, variance-driven $\mathcal{O}(\ln L)$ correction, rendering the unconstrained and canonical Scrooge ensembles equivalent.

\subsection{Microcanonical Scrooge ensemble}

While the canonical projection successfully resolves the unphysical mixing of global symmetry sectors, individual states in this ensemble can still exhibit anomalous macroscopic spatial configurations that deviate radically from the exact time-evolved state $\ket{\psi(t)}$. When unitary evolution is strictly bottlenecked by local transport, these broad spatial fluctuations are dynamically inaccessible. The exact pure state is therefore confined to a strictly smaller, idealized hydrodynamic subspace $\mathcal{M}_{\rm r}$ that evolves together with the state.

As the final step in formalizing the phenomenon of global restricted typicality, we demonstrate (compare Appendix~\ref{sec:microcan_appendix}) that the Scrooge ensemble defined strictly on this manifold — which we term the \textit{microcanonical Scrooge ensemble} — yields identical results to its unrestricted variant. This ensemble is defined by rigidly enforcing on states of the ensemble $\ket{\Phi}$ an extensive number of continuous local amplitude constraints, $\langle \Phi | \hat{q}_j | \Phi \rangle = q_j$, corresponding to the exact local conserved densities. We emphasize that while standard ensemble equivalence in statistical mechanics is conventionally established only locally for reduced density matrices (via partial traces over a bath), our geometric proof establishes a significantly stronger result. Despite the severity of this geometric restriction, we demonstrate the exact measure concentration of the probability density $P(\Phi)$ onto the microcanonical manifold. The mechanism driving this concentration is that the generalized Gibbs state $ \rho(\{q_n(x)\})$ associated with the hydrodynamic profile possesses an extensive thermodynamic entropy, which forces the state purity $\Tr[ \rho(\{q_n(x)\})^2]$ to decay exponentially with the system size $L$, as we demonstrate in Appendix~\ref{sec:microcan_appendix}. Consequently, the statistical variance of the local charges is rigorously bounded by this purity,
\begin{equation}
    \mathrm{Var}^{\rm Sc}(\hat q_j) \leq ||\hat{q}_j||^2 \mathrm{Tr}(\rho^2),
\end{equation}
and hence it also vanishes in the thermodynamic limit.

By leveraging this vanishing variance via the Chebyshev inequality~\cite{Billingsley95}, alongside the Lyusternik-Graves theorem~\cite{Ioffe15} to establish the metric regularity of this microcanonical manifold, we  prove that the expectation values of physical observables and functionals of the state  concentrate precisely on this restricted subspace. Specifically, for any Lipschitz-continuous functional $g$ whose Lipschitz constant grows at most polynomially with the system size, the deviation between the standard Scrooge ensemble prediction and the exact microcanonical average is exponentially suppressed in $L$:
\begin{equation}
    \mathbb{E}^{\rm Sc}[g] \approx \mathbb{E}^{\rm Sc}_{\rm mc}[g].
\end{equation}
This equivalence also applies to nonlinear functionals of the state, such as the von Neumann entanglement and participation entropies, as well as the R\'enyi entanglement entropy. The latter is demonstrated to concentrate under the condition $S_{\infty}(\rho)>\frac{2(q-1)}{q}S_1(\rho_A)$, which formalizes the intuition emphasized throughout this work: the state $\ket{\psi}$ used to construct $\rho(\{q_i\})$ must not be almost fully determined by the local charge expectations $q_i$ alone.

Remarkably, this measure concentration can be mapped directly to projective geometry. Let $\ket{n}$ denote the eigenstates of the density matrix $\rho(\{q_i\})$ with corresponding eigenvalues $e^{-E_n}/Z$. We can arrange the target microcanonical densities $q_j$ into a vector $\mathbf{q}$, and define the diagonal charge expectation values for each eigenstate as analogous vectors $\mathbf{v}^{(n)}$ with components $v_j^{(n)} = \bra{n}\hat{q}_j\ket{n}$.  As derived in Appendix~\ref{sec:microcan_appendix}, the microcanonical partition function evaluates  to the first weight-moment of a multivariate rational simplex spline~\cite{Micchelli79,Farin93,Prautzsch02}:
\begin{equation}
    Z_{\rm mc} = \mathcal{D}! \, M^{(1)}_{\rm rat}\Big( \mathbf{q} \;\Big|\; \mathbf{v}^{(1)},\dots, \mathbf{v}^{(\mathcal{D})}; e^{-E_{1}},\dots,e^{-E_\mathcal{D}} \Big).
\end{equation}
In this geometric representation, the microcanonical manifold $\mathcal{M}_r$ determines the coordinate of the spline $\mathbf{q}$, while the expectation values $\mathbf{v}^{(n)}$ dictate the spatial vertices of a high-dimensional probability simplex. Simultaneously, the exponential Boltzmann weights $e^{-E_n}$ induce a rational projective transformation that shifts the center of the probability distribution precisely to the microcanonical manifold $\mathcal{M}_r$ determined by $\mathbf{q}$. Physically, this geometric transformation explains the emergence of restricted typicality: it exponentially suppresses the probability density of atypical spatial configurations, forcing the overwhelming majority of the statistical weight to concentrate exactly onto the microcanonical manifold $\mathcal{M}_r$.

Ultimately, these results validate the physical robustness of the standard Scrooge ensemble. Because for a wide class of functionals of the state both the global symmetry projections and the continuous local spatial restrictions yield only  subleading corrections in the thermodynamic limit, the analytically tractable ``grand canonical'' Scrooge formulation perfectly captures the exact leading-order dynamics of the locally constrained quantum state.

\section{Discussion}

In this work, we introduced a framework to describe the time evolution of quantum systems restricted by a few relevant macroscopic constraints using a maximally agnostic ensemble of pure states: the Scrooge ensemble. We have confirmed its accuracy and gauged its fundamental limitations, which stem primarily from the omission of causal information, through both analytical and numerical analysis. We also note that for finite system sizes and evolution times, the theoretical criteria for accuracy established in Sec.~\ref{sec:framework}, especially criterion 3, are only approximately fulfilled. Conceptually, our framework serves as a direct demonstration that the maximum entropy principle of Ref.~\cite{Mark24} succeeds not only under mathematically exact constraints—which restrict the available Hilbert space but are typically inaccessible in physical systems—but also under effective dynamical restrictions imposed by macroscopic bottlenecks such as local transport. Furthermore, we provided a rigorous geometric understanding of this phenomenon, formally demonstrating that such macroscopic constraints force the exact pure state to undergo measure concentration onto a dynamically restricted typical manifold.

By utilizing physical transport bottlenecks as effective dynamical constraints, our results provide a natural non-equilibrium counterpart to the Eigenstate Thermalization Hypothesis (ETH)~\cite{Deutsch91,Srednicki94,Rigol08,DAlessio16}. While standard ETH guarantees that deterministic pure states locally mimic thermal ensembles at late times, the maximum entropy principle underlying our approach reveals that this statistical equivalence is remarkably resilient.
We demonstrate that our generalized statistical ensemble successfully captures global, nonlinear quantum properties even far from equilibrium, provided the macroscopic transport restrictions are properly accounted for. In contrast, the standard ETH inherently leaves these properties unaddressed.

The most fundamental open question is whether an accurate early-time description can be achieved by implementing causal restrictions for instance through an appropriate construction of the density matrix $\rho$ within our framework, or via the broader maximum entropy approach of Ref.~\cite{Mark24}. For integrable systems, this causal information might be embedded directly into $ \rho(\{q_n(x)\})$ using two-point quasiparticle correlations, which we conjecture would recover the Alba-Calabrese formula~\cite{Alba17,Alba18,Bertini18i,Alba19} in its most general form. In contrast, generic ergodic systems currently lack an analogous microscopic description capable of fully capturing these dynamic constraints.

The framework we introduced may be directly generalized beyond exact conservation laws to a broad class of models featuring dynamical bottlenecks. For instance, in systems exhibiting weak Hilbert space fragmentation~\cite{Sala20, Khemani20, Guardado20, Moudgalya22, Kohlert23, Wang25}, our approach can be applied by projecting onto an exact Krylov sector with bottlenecked modes corresponding to a conserved charge and a dipole moment. Physical setups with clear timescale separations—which are particularly suitable for our framework—have been the focus of state-of-the-art experimental studies, including transport bottlenecks between two ergodic reservoirs~\cite{Brantut12,Brantut13,Husmann15,Krinner15,Hausler17,Lebrat18}, one-dimensional superfluids in adjustable double-well potentials~\cite{Gring12,Langen15,Schweigler17,Pigneur18}, spin transport in strongly interacting Fermi gases~\cite{Sommer11f,Luciuk17}, charge and spin transport in uniform box potentials~\cite{Nichols19,Brown19}, and one-dimensional Bose gases constrained by weak integrability breaking~\cite{Kinoshita06,Mazets08,Tan10}. Furthermore, recent advancements in quantum measurement techniques~\cite{Islam15, Kaufman16, Elben18, Brydges19, Elben23} now allow direct access to the relevant nonlinear quantities, such as the entanglement and participation entropies.

Finally, we highlight the computational and predictive utility of our framework. To rigorously validate our theoretical predictions throughout this work, we relied on exact numerical simulations of the quantum dynamics to evaluate the instantaneous expectation values of the conserved densities and to subsequently determine the corresponding Lagrange multipliers. While this exact approach is strictly limited to small system sizes due to the exponential growth of the Hilbert space, the true power of the Scrooge ensemble lies in its ability to bypass this computational bottleneck entirely. Because the framework requires only the macroscopic spatial profiles of the slow modes as inputs, these profiles can be efficiently obtained by solving classical hydrodynamic equations—such as GHD for integrable models or macroscopic diffusion equations for generic chaotic systems. By mapping these classically evolved charge densities to the background Lagrange multipliers, our framework provides a highly efficient semi-analytical shortcut. It enables the prediction of complex, nonlinear quantum properties, such as the entanglement and participation entropies, at length scales and late times that remain fundamentally out of reach for state-of-the-art tensor network algorithms and exact quantum dynamics simulations.

\acknowledgements
The work of K.P. and J.Z. was funded by the National Science Centre, Poland, project  2021/43/I/ST3/01142 -- OPUS call within the WEAVE programme.
P.S. acknowledges fellowship within the “Generación D” initiative, Red.es, Ministerio para la Transformación Digital y de la Función Pública, for talent attraction (C005/24-ED CV1), funded by the European Union NextGenerationEU funds through PRTR.
We gratefully acknowledge the Polish high-performance computing infrastructure PLGrid (HPC Centers: ACK Cyfronet AGH) for providing computer facilities and support within the computational grant no. PLG/2026/019648.

\begin{appendices}

    \section{Cluster Gutzwiller approximation and further refinement of the maximum entropy state \label{sec:gutzwiller}}

    Constructing the Scrooge ensemble requires determining the density matrix $\rho$, which in our framework is parameterized by the instantaneous expectation values $q_j(t)$ of dynamically bottlenecked modes $\hat{q}_j$, yielding $\rho(\{q_i\}) = e^{-\sum_j \beta_j \hat{q}_j}/Z$ as described in Eq.~\eqref{eq:scrooge_rho}. Finding the exact set of instantaneous Lagrange multipliers $\beta_j(t)$ requires strictly solving the system of equations $\Tr[\rho(\{q_i\}) \hat{q}_j] = q_j(t)$. Because the observables $\hat{q}_j$—which in our applications correspond to local charge densities and currents—are generally non-commuting operators with overlapping spatial supports, the density matrix $ \rho(\{q_i\})$ does not factorize. Thus, the system of equations that determines $ \rho(\{q_i\})$ is generally highly non-linear. Directly employing a global root-finding algorithm to determine the Lagrange multipliers $\beta_j$ is computationally prohibitive for large system sizes $L$, as each step of such an algorithm requires computationally expensive evaluations of matrix exponentials across the full Hilbert space.

    To make this tractable, we employ a hybrid numerical optimization scheme. This consists of generating a high-quality initial guess based on a cluster Gutzwiller approximation~\cite{Buosante04, Yamamoto09, McIntosh12, Luehmann13}, where we find exact solutions for the Lagrange multipliers on a local cluster of $k$ spins and combine those partial solutions into a single global guess $\beta_j^{(0)}$, followed by a global optimization step.

    \subsection{Construction of cluster Gutzwiller initial guess}
    To obtain the initial estimate for the Lagrange multipliers, we decompose the system into small, tractable spatial clusters of size $k$ (typically $k \sim 3-5$). For each such cluster, parameterized by its position $c$, we truncate the operator $\sum_j\beta_j\hat q_j$ to a reduced Hilbert space of dimension $2^k$ for spin $1/2$ degrees of freedom. This is achieved by incorporating only those operators $\hat q_j$ fully contained within the cluster, alongside a mean-field approximation for terms with support across the cluster boundary. Specifically, for operators $\hat q_j$ spanning the boundary, we implement this mean-field approximation by replacing the degrees of freedom external to the cluster with their exact instantaneous expectation values computed in the pure state of the system $\ket{\psi(t)}$.

    For systems governed by strict global symmetries, such as $\mathrm{U}(1)$ charge conservation, this formulation can be extended to explicitly enforce the global constraint. To achieve this, each eigenstate of the local cluster needs to be reweighted by the exact combinatorial probability that the external mean-field bath supplies the specific complementary charge required to preserve the total global charge. Because the mean-field bath treats the external sites as independent variables with known target densities, this probability for the case of $\mathrm{U}(1)$ symmetry can be evaluated efficiently using the Poisson binomial distribution.

    For each cluster position $c$, we solve the reduced system of equations using a Levenberg-Marquardt root-finding algorithm~\cite{Marquardt63}, which utilizes exact diagonalization on the small $2^k$-dimensional subspace to find the local Lagrange multipliers $\tilde{\beta}_{j,c}$ that satisfy the target constraints within that window. Because the $k$-site clusters overlap, a single physical bond or site $j$ is covered by multiple different windows, yielding several independent local estimates for the same Lagrange multiplier.

    To construct a single global guess, we aggregate these local estimates using a spatially weighted average. To naturally minimize the error introduced by the mean-field approximation, the weight assigned to a local estimate is chosen to be proportional to its spatial distance from the nearest edge of the cluster, meaning estimates originating from the bulk of a cluster contribute the most to our final guess. A small regularization constant is added to prevent strictly zero weights, and crucially, if a cluster boundary coincides with the physical edge of the entire system, the distance penalty for that edge is removed, as no mean-field approximation occurs there. Using these weights, the global initial guess is constructed as:
    \begin{equation}
        {\beta}_{j}^{(0)} = \sum_{c} w_{j,c} \tilde{\beta}_{j,c},
    \end{equation}
    where $w_{j,c}$ denotes the normalized distance-dependent weight of the estimate for position $j$ originating from cluster $c$ (such that $\sum_c w_{j,c} = 1$). This provides a highly accurate, physically motivated starting point for the subsequent global refinement.

    \subsection{Global refinement of the initial guess}
    While the cluster Gutzwiller approximation captures the density matrix surprisingly well, the mean-field truncation leaves small residual errors in the expectation values predicted using the initial multipliers $\beta_j^{(0)}$. To eliminate these discrepancies, we perform a global refinement over the entire $L$-site Hilbert space. The target multipliers correspond precisely to the global minimum of the function $\ln Z+\sum_j\beta_j q_j$, whose exact gradient with respect to $\beta_j$ is the residual error:
    \begin{equation}
        \frac{\partial \ln Z}{\partial \beta_j} + q_j(t) = q_j(t) - \Tr[\rho(\{q_i\}) \hat{q}_j].
    \end{equation}

    Although the hybrid scheme reduces the global optimization to just a few iterations, evaluating this exact gradient traditionally relies on matrix exponentiation to compute the expectation values of local observables, limiting the practically accessible system sizes to $L\approx12$, possibly extending to $L \approx 16$ in the presence of $\mathrm{U}(1)$ symmetry. To break this bottleneck, we can combine quantum typicality with a Chebyshev polynomial expansion.

    Instead of computing exact traces over the full Hilbert space, we approximate the trace of an observable by taking its expectation value with respect to a small set of $N$ unnormalized Gaussian random vectors $\ket{\tilde\phi_m}$, allowing for stochastic trace evaluation~\cite{Weisse06}:
    \begin{equation}
        \Tr(\rho O) \approx  \frac{1}{N} \sum_{m=1}^N \braket{\tilde\phi_m|\sqrt{\rho} O \sqrt{\rho}|\tilde\phi_m}.
    \end{equation}
    Because evaluating the exact partition function $Z$ required to normalize $\rho$ is also computationally prohibitive, we approximate both the numerator and the denominator stochastically. By evaluating both traces using the same set of random vectors, the $1/N$ normalization factors cancel, yielding a ratio of empirical averages:
    \begin{equation}
        \Tr(\rho O) \approx \frac{\sum_{m=1}^N\braket{\tilde\phi_m|e^{-\frac{1}{2}\sum_j \beta_j \hat{q}_j} O e^{-\frac{1}{2}\sum_j \beta_j \hat{q}_j}|\tilde\phi_m}}{\sum_{m=1}^N\braket{\tilde\phi_m|e^{-\sum_j \beta_j \hat{q}_j}|\tilde\phi_m}}.
    \end{equation}
    This formula requires evaluating the action of the unnormalized density matrix on an arbitrary state vector, $\ket{\tilde \Phi_m} = e^{-\frac{1}{2}\sum_j \beta_j \hat{q}_j}\ket{\tilde \phi_m}$. To compute this exponential action without relying on diagonalization, we shift and scale its eigenspectrum to the domain $[-1, 1]$. By dynamically calculating the extremal eigenvalues of $\sum_j\beta_j\hat q_j$ via Lanczos iteration~\cite{Lanczos50}, we define the scaled operator $\tilde{H} = (\sum_j \beta_j \hat{q}_j - E_{\rm mid}\mathds{1})/\Delta E$, where $E_{\rm mid}$ and $\Delta E$ are the center and half-width of the spectral range of $\sum_j\beta_j\hat q_j$, and $\mathds{1}$ is the identity matrix. The state vector is then obtained through the expansion~\cite{TalEzer84, Weisse06, Bera17, Sierant22}:
    \begin{multline}
        e^{-\frac{1}{2}\sum_j \beta_j \hat{q}_j}|\tilde\phi_m\rangle \approx \\
        e^{-E_{\rm mid}/2} \left[ I_0(\Delta E/2) |t_0\rangle + 2 \sum_{k=1}^M (-1)^k I_k(\Delta E/2) |t_k\rangle \right].
    \end{multline}
    Here, $I_k$ are the modified Bessel functions of the first kind, and $|t_k\rangle = T_k(\tilde{H})|\tilde\phi_m\rangle$ are the Chebyshev vectors generated iteratively via sparse matrix-vector multiplications, initialized with $|t_0\rangle = |\tilde \phi_m\rangle$. Because the modified Bessel coefficients decay super-exponentially for $k > \Delta E$, truncating the series at a finite order $M$ achieves machine-precision accuracy for the required observables. Furthermore, we enforce numerical stability during deep expansions using backward Clenshaw recurrence~\cite{Clenshaw55}.

    By circumventing the need for full diagonalization, this stochastic trace evaluation allows the Jacobian-free Krylov subspace routine to efficiently converge upon the refined Lagrange multipliers. Crucially, to preserve the continuity of the gradient and ensure the stability of the Krylov solver, the set of random vectors $\{\ket{\tilde\phi_m}\}$ is sampled exactly once at the initialization of the global refinement procedure and kept fixed across all optimization steps. This robust approach extends the accessible system sizes up to $L \approx 24$ in the presence of $\mathrm{U}(1)$ symmetry.

    \section{Analytical computation of annealed averages in the Scrooge ensemble\label{sec:analytic_appendix}}

    \subsection{Derivation of boundary operator for inverse participation ratio}

    The inverse participation ratio (IPR) of a pure state $\ket{\Phi}$ is given as
    \begin{equation}
        I_2 = \sum_{\sigma} |\braket{\sigma| \Phi}|^{4} = \sum_{\sigma}\braket{\sigma |\rho_\Phi |\sigma }^2,
    \end{equation}
    with $\rho_\Phi=\op{\Phi}{\Phi}$ being the density matrix representing $\ket{\Phi}$. Equivalently:

    \begin{align}
        I_2 & =\sum_{\sigma}\braket{\sigma |\rho_\Phi |\sigma } \braket{\sigma |\rho_\Phi |\sigma }
        = \sum_\sigma \Tr\left( \op{\sigma}{\sigma} \rho_\Phi \right)^2 \nonumber                                                                                            \\
            & =\sum_{\sigma}\Tr \left[ ( \op{\sigma}{\sigma} \rho_\Phi) ^{\otimes2} \right]=
        \sum_{\sigma}\Tr \left[ \op{\sigma}{\sigma}^{\otimes2} \rho_\Phi^{\otimes2} \right] \nonumber                                                                        \\
            & =\Tr \left[ \left(\sum_{\sigma}\op{\sigma}{\sigma}^{\otimes2} \right) \rho_\Phi^{\otimes2} \right] = \Tr \left[ \mathcal{B}_2 \, \rho_\Phi^{\otimes2} \right],
    \end{align}
    where we introduce the boundary operator $\mathcal{B}_2$:

    \begin{equation}
        \mathcal{B}_2 = \sum_{\sigma}\op{\sigma}{\sigma}^{\otimes2}
        = (\op{\uparrow}{\uparrow}^{\otimes2} + \op{\downarrow}{\downarrow}^{\otimes2})^{\otimes L} = \Lambda^{\otimes L},
    \end{equation}
    where $\Lambda = (\op{\uparrow}{\uparrow}^{\otimes2} + \op{\downarrow}{\downarrow}^{\otimes2})$ appears if the $\sigma$ sum runs over all computational basis states $\{\ket{\uparrow},\ket{\downarrow}\}^{\otimes L}$. Let us note here, that if the summation over $\sigma$ is restricted to a subset of all computational basis states, e.g. a specific symmetry sector, there are corrections to the following formulas. Those corrections are the main topic of the Appendix~\ref{sec:can_appendix}.

    The derivation of the higher participation ratio $I_q=\sum_{\sigma} |\braket{\sigma| \Phi}|^{2q}$ is a simple generalization of the $q=2$ case, the boundary operator is $\mathcal{B}_q=(\op{\uparrow}{\uparrow}^{\otimes q} + \op{\downarrow}{\downarrow}^{\otimes q})^{\otimes L} = \Lambda_q^{\otimes L}$, with $\Lambda_q = (\op{\uparrow}{\uparrow}^{\otimes q} + \op{\downarrow}{\downarrow}^{\otimes q})$.

    \subsection{Annealed average of the inverse participation ratio}

    Let us denote the exact average of a quantity within the Scrooge ensemble by the superscript $\rm Sc$, and the averaging operation itself as $\mathbb{E}^{\rm Sc}$. The formula for the average IPR reads:
    \begin{equation}
        I_q^{\rm Sc} =
        \mathbb{E}^{\rm Sc} \left(\Tr \left[ \mathcal{B}_q \, \rho_\Phi^{\otimes q} \right] \right) = \Tr \left[ \mathcal{B}_q \, \mathbb{E}^{\rm Sc}( \rho_\Phi^{\otimes q} ) \right],
    \end{equation}
    where $\mathbb{E}^{\rm Sc}( \rho_\Phi^{\otimes q} ) $ is the $q$-th moment of the Scrooge ensemble, which can be obtained from Schur-Weyl duality~\cite{Fulton04,Mark24,McGinley25}:
    \begin{equation}
        \mathbb{E}^{\rm Sc}( \rho_\Phi^{\otimes q} ) \approx \rho^{\otimes q}\sum_{\pi\in \mathrm{Sym}(q)}T_\pi,\label{eq:scrooge_moment_supmat}
    \end{equation}
    where $T_\pi$ are representation operators of permutations $\pi$ in the symmetric group $\mathrm{Sym}(q)$ acting on the $q$-fold replicated Hilbert space~\cite{Calabrese04, Daley12, Islam15, Brydges19}. The generalized boundary operator is $\mathcal{B}_q=\sum_{\sigma}(\op{\sigma}{\sigma})^{\otimes q}$. Because $\mathcal{B}_q$ is constructed from tensor products of identical states across all $q$ replicas, it is strictly invariant under any replica permutation, meaning $T_\pi \mathcal{B}_q = \mathcal{B}_q$. Consequently, the permutations have a trivial effect inside the trace, yielding a factor of $q!$:
    \begin{equation}
        I_q^{\rm Sc} \approx q! \Tr\left[ \mathcal{B}_q \rho^{\otimes q} \right] = q! \sum_{\sigma} \braket{\sigma | \rho | \sigma}^q.
    \end{equation}
    Denoting the diagonal elements of the density matrix in the computational basis as $p_\sigma = \braket{\sigma | \rho | \sigma}$, the annealed average of the participation entropy $P_q^{\rm Sc} = \frac{1}{1-q}\ln I_q^{\rm Sc}$ thus simplifies to:
    \begin{equation}
        P_q^{\rm Sc} \approx \frac{1}{1-q}\ln\left(\sum_{\sigma} p_\sigma^q\right) + \frac{\ln q!}{1-q}. \label{eq:scrooge_P_general}
    \end{equation}
    As the obtained formula is analytic in $q$, it allows computing the von Neumann participation entropy using the replica trick, giving:
    \begin{equation}
        P_1^{\rm Sc} \approx -\lim_{q\to1}\frac{\partial\ln I_q^{\rm Sc}}{\partial q}=-\sum_{\sigma} p_\sigma \ln p_\sigma + \gamma - 1,
    \end{equation}
    where $\gamma$ is the Euler-Mascheroni constant.

    Note that in theory, this annealed calculation represents a lower bound on the quenched average for $q>1$ (and an upper bound for $q<1$), namely $\mathbb{E}^{\rm Sc}(P_q) \geq \frac{1}{1-q}\ln \mathbb{E}^{\rm Sc} (I_q)$. For quantities that can be computed by tracing out at least $\mathcal{O}(\ln L)$ sites of the system, the difference between them is exponentially small in the system size due to the measure concentration in the Scrooge ensemble~\cite{Teufel25,McGinley25}.

    In the specific case where $\rho$ is a separable product state, $\rho = \bigotimes_{i=1}^L \rho_i$, the probabilities factorize and the participation entropy is strictly additive. Parameterizing the local probabilities via the local magnetization $p_{i,\uparrow} = \frac{1}{2} + m_i$, we immediately obtain:
    \begin{equation}
        P_q^{\rm Sc} \approx \frac{1}{1-q}\sum_{i=1}^{L}\ln\left[\left(\frac{1}{2}-m_i\right)^q+\left(\frac{1}{2}+m_i\right)^q\right]+\frac{\ln q!}{1-q}.\label{eq:scrooge_P_separable}
    \end{equation}

    \subsection{Derivation of boundary operator for subsystem purity}

    Following an analogous calculation, we now derive the subsystem purity $s_q=\Tr_{A}(\rho^{q}_{\Phi,A})$, where the reduced density matrix is obtained by tracing out the subsystem $B$, namely $\rho_{\Phi,A}=\Tr_B(\rho_{\Phi})$. This quantity can be used to approximate the quenched average of the entanglement entropy. While we present the proof for $q=2$, it easily generalizes to any arbitrary integer $q\geq2$. Using the identity $\Tr(X^2) = \Tr(T_{\mathbb{S}}X^{\otimes 2})$, we obtain:

    \begin{align}
        s_2 & =\Tr_{A}(\rho^{2}_{\Phi,A})=\Tr_{A}\left(T^A_{\mathbb{S}} \rho^{\otimes 2}_{\Phi,A}\right) \nonumber                                              \\
            & =\Tr_{A}\left[T^A_{\mathbb{S}}\left(\Tr_{B}\rho_\Phi\right)^{\otimes2}\right] =\Tr_{A}\left[T^A_{\mathbb{S}}\Tr_{B}(\rho_\Phi^{\otimes2})\right].
    \end{align}
    Since the swap operator acts only on the replicated copies of subsystem $A$, we can combine the partial traces into a full trace over the replicated Hilbert space by introducing the identity operator $T^B_{\mathds{1}}$ on subsystem $B$, yielding:
    \begin{equation}
        s_2=\Tr\left[\left(T^A_{\mathbb{S}}\otimes T^B_{\mathds{1}}\right) \rho_\Phi^{\otimes2}\right].
    \end{equation}

    An analogous calculation for higher-order subsystem purities $s_q=\Tr_{A}(\rho_{\Phi,A}^q)$ leads to the boundary operator $\mathcal{A}_q=T^A_{C_q}\otimes T^B_{\mathds{1}}$, where $C_q$ is the cyclic permutation:
    \begin{equation}
        T_{C_q}\ket{\sigma_1\sigma_2\ldots\sigma_{q-1}\sigma_q}=\ket{\sigma_q\sigma_1\ldots\sigma_{q-2}\sigma_{q-1}}.
    \end{equation}

    \subsection{Annealed average of the subsystem purity}

    After the ensemble average~\eqref{eq:scrooge_moment_supmat}, this simplifies to:
    \begin{align}
        s_2^{\rm Sc} & \approx\Tr\left[\left(T^A_{\mathbb{S}}\otimes T^B_{\mathds{1}}\right) \rho^{\otimes2}(T_{\mathbb{S}}+T_{\mathds{1}})\right] \nonumber                        \\
                     & =\Tr\left[\left(T^A_{\mathbb{S}}\otimes T^B_{\mathds{1}}\right) \rho^{\otimes2}+\left(T^A_{\mathds{1}}\otimes T^B_{\mathbb{S}}\right) \rho^{\otimes2}\right] \\
                     & =\Tr(\rho_A^2)+\Tr(\rho_B^2), \nonumber
    \end{align}
    which gives the annealed entanglement entropy:
    \begin{equation}
        S_2^{\rm Sc}\approx-\ln(s_2^{\rm Sc})=-\ln\left[\Tr(\rho_A^2)+\Tr(\rho_B^2)\right].
    \end{equation}
    For the special case of separable $\rho$ corresponding to $\hat q_i=\hat S_i^z$ and a symmetric or antisymmetric spin profile, the half-chain entanglement reduces to:
    \begin{equation}
        m^2_i=m^2_{L+1-i}\implies S_2^{\rm Sc}=\frac{1}{2}P_2^{\rm Sc}-\frac{1}{2}\ln 2,
    \end{equation}
    which indicates that both quantities exhibit the same qualitative behavior in this case.

    For higher-order subsystem purities ($q \ge 3$), the exact joint trace over the replicas yields complex cross-boundary correlation terms (e.g., $\Tr[\rho(\rho_A \otimes \rho_B)]$) unless $\rho$ is a separable state across the subsystem cut, namely $\rho = \rho_A \otimes \rho_B$. Let us note that the state $\rho$ resulting from our construction possesses a finite correlation length (see Appendix~\ref{sec:integrable_appendix}); hence, formulas assuming this factorized structure can be treated as a reasonable approximation of the exact annealed averages even in the general case. Assuming this separability condition holds, which in our framework is strictly true for purely local one-site constraints, the trace factorizes into a product of local traces, yielding:
    \begin{equation}
        s_q^{\rm Sc}\approx\sum_{\pi\in \mathrm{Sym}(q)}\Tr_{A}\left(T^A_{\pi C_q}\rho_A^{\otimes q}\right)\Tr_B\left(T^B_{\pi}\rho_{B}^{\otimes q}\right).
    \end{equation}
    To evaluate this expression, we can use knowledge of representation theory of the symmetric group $\mathrm{Sym}(q)$. First, let us note that if we have a permutation consisting of a single cycle $\pi_l$ of length $l$, then we get:
    \begin{equation}
        \Tr\rho^{l}_{A}=\Tr_{A}(T^A_{\pi_l} \rho^{\otimes l}_{A}).
    \end{equation}
    We used an analogous formula for the cyclic permutation $C_q$ to arrive at the boundary operator for subsystem purity. Now let us use the decomposition into cycles of the permutation $\pi C_q$:
    \begin{equation}
        \Tr_{A}(T_{\pi C_q}^A \rho_A^{\otimes q})=\prod_{l=1}^{q}(\Tr \rho_A^{l})^{k_l},
    \end{equation}
    where $l$ is a length of the cycle and $k_l$ is the number of repetitions of this cycle length in the decomposition of $\pi C_q$. The value of $\Tr_B(T^B_{\pi}\rho_{B}^{\otimes q})$ is computed analogously. For brevity, let us label the decomposition of an arbitrary permutation $\pi$ into cycles with a tuple $\nu_{\pi}=(1^{k_1},2^{k_2},\ldots,q^{k_q})$, where each cycle length is repeated as many times as in the decomposition, so $k_l$ times. Obviously, $\sum_l l k_l=q$. Due to the invariance of cycle structure under conjugation, and its uniqueness for each conjugacy class, $\nu$ labels conjugacy classes of permutations. We define a moment $M_\nu(\rho)$:
    \begin{equation}
        M_\nu(\rho) = \prod_{l}\left(\Tr\rho^{l}\right)^{k_l}.\label{eq:permutation_moment}
    \end{equation}
    Now the computation of subsystem purity $s_q^{\rm Sc}$ reduces to summing over all integer partitions $\nu$ of $q$, which we denote as $\nu\vdash q$, and then counting how many permutations of each label $\nu$ contribute to the sum. This can be summarized as:

    \begin{equation}
        s_q^{\rm Sc}\approx\sum_{\nu_{1}\vdash q}\sum_{\nu_2 \vdash q}N_{\nu_1,\nu_2}M_{\nu_1}(\rho_A)M_{\nu_2}(\rho_B),\label{eq:subsystem_purity_q}
    \end{equation}
    where $N_{\nu_1,\nu_2}$ counts the number of pairs $(\pi_1,\pi_2)$ such that $\nu_1=\nu_{\pi_1},\ \nu_2=\nu_{\pi_2}, \pi_1 = \pi_2 C_q$, which can be efficiently computed using results from representation theory of the symmetric group:
    \begin{equation}
        N_{\nu_1,\nu_2}=\mathcal{N}_{\nu_1}\mathcal{N}_{\nu_2}\sum_{r=0}^{q-1}\frac{(-1)^r}{q! \binom{q-1}{r}}
        \chi_{\nu_1}^{(q-r,1^r)}\chi_{\nu_2}^{(q-r,1^r)},\label{eq:count_frobenius}
    \end{equation}
    where $\mathcal{N}_{\nu}$ is the number of permutations within conjugacy class $\nu$:
    \begin{equation}
        \mathcal{N}_\nu=\frac{q!}{\prod_{j}(j^{k_j}k_j!)},
    \end{equation}
    with $k_j$ being the number of repetitions of $j$ in $\nu$, and $\chi_{\nu}^{(q-r,1^r)}$ is the character of the irreducible representation of the symmetric group $\mathrm{Sym}(q)$, associated with the hook partition $(q-r,1^r)$, evaluated on the conjugacy class $\nu$. Those characters are efficiently computed using the Murnaghan-Nakayama rule~\cite{Fulton04}. Finally, this gives:
    \begin{equation}
        S_q^{\rm Sc}\approx\frac{1}{1-q}\ln s_q^{\rm Sc}.\label{eq:scrooge_S}
    \end{equation}

    Let us briefly prove the formula~\eqref{eq:count_frobenius}. We start from the Frobenius formula~\cite{Lando04}, which states that for group $G$, the number of solutions to $\pi_1\cdots \pi_k=\mathds{1}$, where $\pi_i$ is in the conjugacy class $\nu_i$, is:
    \begin{equation}
        \tilde N_{\nu_1,\ldots,\nu_k}=\frac{\mathcal{N}_{\nu_1}\cdots\mathcal{N}_{\nu_k}}{\mathcal{N}_{G}}\sum_{\nu}\frac{\chi^\nu_{\nu_1}\cdots\chi_{\nu_k}^\nu}{(\chi_{\mathds{1}}^\nu)^{k-2}},
    \end{equation}
    where the sum runs over the irreducible representations of $G$. Let us apply it to our scenario by noticing that $N_{\nu_1,\nu_2}$ is the number of solutions to $\pi_1=\pi_2C_q$, where $\nu_{\pi_1}=\nu_1$ and $\nu_{\pi_2}=\nu_2$. Since $C_q$ is fixed, let us denote its conjugacy class by $\nu_C$. Let us note that $\pi$ and $\pi^{-1}$ share the same conjugacy class. Thus, as $C_q$ is fixed, we can obtain the desired number of pairs as:
    \begin{equation}
        N_{\nu_1,\nu_2}=\tilde N_{\nu_1,\nu_2,\nu_C}/\mathcal{N}_{\nu_C}.
    \end{equation}
    We then substitute $\mathcal{N}_{\mathrm{Sym}(q)}=q!$, and notice that following the Murnaghan-Nakayama rule, the only representations of $\mathrm{Sym}(q)$ that give non-zero $\chi^\nu_{\nu_C}$ are hook representations, associated with the partition $(q-r,1^r)$. For those, we have $\chi^{(q-r,1^{r})}_{\nu_C}=(-1)^r$. We finalize the proof by noting that the dimension of these representations is $\chi_{\mathds{1}}^{(q-r,1^r)}=\binom{q-1}{r}$.

    \section{Sampling from the Scrooge ensemble \label{sec:sampling}}

    \subsection{Exact sampling}

    Sampling from the Scrooge ensemble can be efficiently realized in the eigenbasis of the background density matrix $\rho\ket{n}=p_n\ket{n}$ by choosing one eigenvector $\ket{k}$ with probability given by eigenvalue $p_k$, and then generating an unnormalized vector $\ket{\tilde \phi}$:

    \begin{equation}
        \ket{\tilde \phi}=\sum_{n}c_n\ket{n},
    \end{equation}
    where $c_n$'s for $n\neq k$ are drawn from the standard complex (as they mean to emulate behavior of time evolved states) Gaussian distribution with probability density $p(c_n)=\frac{1}{\pi}e^{-|c_n|^2}$ independently, while $|c_k|^2$ is generated from $\Gamma(2,1)$ distribution,
    with uniformly random phase, giving the probability density $p(c_k)=\frac{1}{\pi}|c_k|^2e^{-|c_k|^2}$. As we demonstrate later, such a choice of the distribution $p(c_k)$ leads to the joint probability distribution $P(\tilde \phi)$ that factorizes into a norm dependent part and the desired $\bra{\phi}\rho \ket{\phi}$. Such $c_k$ can be numerically obtained~\cite{Devroye86} from three random numbers $u_1,u_2,u_3$ drawn uniformly from $(0,1]$:
    \begin{equation}
        c_k=\sqrt{-\ln (u_1 u_2)}e^{i2\pi u_3}.
    \end{equation}
    This procedure generates states with probability density:
    \begin{equation}
        P(\tilde\phi)\propto\sum_k p_k |c_k|^2\prod_{n}e^{-|c_n|^2}=\norm{\tilde \phi}^2 e^{-\norm{\tilde \phi}^2} \braket{\phi|\rho| \phi},\label{eq:scrooge_exact_sampling}
    \end{equation}
    where we use $|c_k|^2=\braket{\tilde \phi | k}\braket{k|\tilde \phi}$ and the spectral decomposition of $\rho$. After normalization of $\ket{\tilde \phi}$ to obtain $\ket{\phi}$, the part dependent on the norm contributes only to the normalization factor of $P(\phi)$, thus this procedure generates states with probability $P(\phi)\propto \braket{ \phi|\rho| \phi}$. Therefore, the sampled Scrooge state is:

    \begin{equation}
        \ket{\Phi}=\frac{\sqrt{\rho}\ket{\tilde \phi}}{\sqrt{\braket{\tilde \phi|\rho|\tilde \phi}}}=\frac{\sqrt{\rho}\ket{\phi}}{\sqrt{\braket{ \phi|\rho|\phi}}}.
    \end{equation}

    \subsection{Stochastic sampling}

    The analytical formulas for the participation and entanglement entropies, such as Eqs.~\eqref{eq:scrooge_P_general_main}, \eqref{eq:scrooge_S_general_main} and~\eqref{eq:subsystem_purity_q} provide practical approximations to annealed averages. When practically accessible, these algebraic evaluations are computationally optimal: the participation entropy requires only the diagonal elements of $\rho$ in the computational basis, while the entanglement entropy $S_q^{\rm Sc}$ can be efficiently computed by extracting the spectra of the reduced density matrices $\rho_A$ and $\rho_B$ to evaluate the required subsystem purities, such as $\Tr(\rho_A^q)$ and $\Tr(\rho_B^{q})$.

    However, for a general non-commuting set of constraints, $\rho$ does not factorize and becomes a dense matrix. In such cases, the exact algebraic evaluation of these traces and spectra scales prohibitively with system size. To circumvent this, we evaluate the entropies stochastically by sampling a finite set of typical pure states directly from the Scrooge ensemble.

    For each sample $m$, we construct the Scrooge state using an unnormalized complex Gaussian random vector $\ket{\tilde{\phi}_m}$ drawn from the isotropic Gaussian distribution $P(\tilde{\phi}) \propto e^{-\braket{\tilde{\phi}|\tilde{\phi}}}$. We then filter this state through the dynamical constraints to obtain the unnormalized Scrooge state, $\ket{\tilde{\Phi}_m} = e^{-\frac{1}{2}\sum_j \beta_j \hat{q}_j}\ket{\tilde{\phi}_m}$. As described in Appendix~\ref{sec:gutzwiller}, this imaginary time evolution is executed efficiently using a dynamically scaled Chebyshev polynomial expansion, fully bypassing exact diagonalization of $\sum_j\beta_j \hat q_j$.

    Weighting each sample by its exact squared norm, $Z_m = \braket{\tilde{\Phi}_m | \tilde{\Phi}_m}$, inherently enforces the correct physical measure. Because the normalized initial vectors $\ket{\phi_m} = \ket{\tilde{\phi}_m}/\norm{\tilde{\phi}_m}$ are uniformly distributed across the Hilbert space (Haar measure), weighting the statistical average by $Z_m \propto \braket{\phi_m | e^{-\sum_j \beta_j \hat{q}_j} | \phi_m}$ mathematically absorbs the radial integration of the Gaussian measure and correctly biases the uniform angular distribution. This importance sampling guarantees that the normalized states, $\ket{\Phi_m} = \ket{\tilde{\Phi}_m}/\sqrt{Z_m}$, recover expectation values over the exact Scrooge probability density $P(\Phi) \propto \braket{\phi | \rho | \phi}$.

    In practice, to avoid numerical underflow associated with exponentially small normalization constants in large systems, we evaluate the quantum correlations directly from the unnormalized state $\ket{\tilde{\Phi}_m}$. Taking the R\'enyi-$q$ entanglement entropy as a concrete example, we first compute the unnormalized subsystem purity, $\tilde{s}_{q,m} = \Tr[(\Tr_B \op{\tilde{\Phi}_m}{\tilde{\Phi}_m})^q]$. Because the exact normalized subsystem purity scales as $s_{q,m} = \tilde{s}_{q,m} / Z_m^q$, the normalized pure-state entanglement entropy $S_{q,m}$ for each sample is robustly recovered via a logarithmic shift:
    \begin{equation}
        S_{q,m} = \frac{1}{1-q}\Big[ \ln \tilde{s}_{q,m} - q \ln Z_m \Big].
    \end{equation}
    This procedure naturally extends to other non-linear quantities. For instance, to evaluate the participation entropy $P_{q,m}$, one simply replaces the unnormalized subsystem purity with the unnormalized inverse participation ratio, $\tilde{I}_{q,m} = \sum_\sigma |\braket{\sigma|\tilde{\Phi}_m}|^{2q}$.

    Finally, the Scrooge ensemble average is recovered by performing a weighted quenched average over the $M$ samples:
    \begin{equation}
        \mathbb{E}^{\rm Sc}(S_q) \approx \frac{\sum_{m=1}^M Z_m S_{q,m}}{\sum_{m=1}^M Z_m}.
    \end{equation}
    Crucially, the vector norm $\norm{\tilde{\phi}_m}$ is a statistically independent variable from the normalized direction $\ket{\phi_m}$. Consequently, the dependence on the radial integration fully cancels out from the ratio of these averages, guaranteeing that this unnormalized procedure generates results identical to those derived using exactly normalized Haar states.

    \section{Proofs for generalized hydrodynamics\label{sec:integrable_appendix}}

    In this appendix, we provide the theoretical derivations supporting the application of the Scrooge ensemble to generalized hydrodynamics (GHD) discussed in Sec.~\ref{sec:integrable_main}. First, we physically justify reducing the subsystem von Neumann entropy to a spatial integral over independent macroscopic fluid cells, relying on the clustering of correlations of the global maximum-entropy GHD density matrix $\rho(\{q_n(x)\})$, Eq.~\eqref{eq:ghd_rho}, in the Euler scaling limit. Subsequently, we demonstrate the validity of the analytic continuation $q \to 1$ for the Scrooge ensemble entanglement entropy in the thermodynamic limit using the properties of the symmetric group.

    \subsection{Entanglement additivity in the Euler scaling limit}

    To justify the reduction of the Scrooge entanglement entropy [Eq.~\eqref{eq:scrooge_S1_thermodynamic_limit}] to a spatial integral over independent fluid cells, we rely on the fact that the global GHD density matrix $\rho(\{q_n(x)\})$ exhibits clustering of correlations~\cite{Araki69, Doyon17w}. This means that for any two bounded operators $O_A$ and $O_B$ with spatial supports $A$ and $B$ separated by a spatial distance $d(A,B)$, their connected correlation function is bounded:
    \begin{multline}
        \left|\Tr(\rho O_A O_B)-\Tr(\rho O_A)\Tr(\rho O_B)\right|\leq \\
        C\norm{O_A}\norm{O_B}f(d(A,B)),
    \end{multline}
    where $C>0$ is a constant dependent only on the spatial supports of the operators, and $f(d)$ is a strictly decaying function of distance. Generally, this clustering function $f(d)$ decays either algebraically or exponentially. We note that at sufficiently high temperatures, thermal states possess a finite correlation length, leading to an exponential clustering of correlations~\cite{Kliesch14}.

    This clustering of correlations physically constrains the scaling of the mutual information $I(A:B)$. Denoting the von Neumann entropy as $S_A=-\Tr(\rho_A\ln \rho_A)$, the mutual information is defined as:
    \begin{equation}
        I(A:B)=S_{A}+S_B-S_{A\cup B}.
    \end{equation}
    Splitting $A$ into $N$ fluid cells $X_k$ and iteratively applying this formula, we obtain:
    \begin{equation}
        S_A=\sum_{k=1}^N S_{X_k}-\sum_{k=1}^{N-1}I(X_1\cup\cdots\cup X_k:X_{k+1}).
    \end{equation}
    The macroscopic nature of the Euler scaling limit combined with clustering of correlations physically ensures that the mutual information between distinct fluid cells scales sub-extensively. By analogy, in local one-dimensional Hamiltonians with algebraic clustering of correlations, the mutual information is known to scale at most logarithmically with the subsystem size~\cite{Wolf08}, which would make the sum of these boundary terms at most of order $\mathcal{O}(\frac{\ell_A}{\ell_X}\ln \ell_X)$. This boundary contribution is strictly subleading compared to the extensive $\mathcal{O}(\ell_A)$ bulk entropy. Thus, the leading extensive contribution to the Scrooge entanglement entropy is exactly the integrated bulk entropy, justifying the result presented in the main text:
    \begin{equation}
        S_1^{\rm Sc}(t)=\int\mathrm{d}\lambda\int_A \mathrm{d}x\, s_{YY}[\rho_{\rm p}(x,t,\lambda)].\label{eq:Scrooge-Bertini_appendix}
    \end{equation}

    \subsection{Analytic continuation in the thermodynamic limit}

    We demonstrate the analytic continuation $q\to 1$ for the annealed average entanglement entropy of the Scrooge ensemble [Eq.~\eqref{eq:scrooge_S}] in the thermodynamic limit. We note that in this limit, the annealed average is identical to the quenched average~\cite{Teufel25, McGinley25} for any state $\rho$ with vanishing purity, which is strictly the case for the macroscopic GHD density matrix $\rho(\{q_n(x)\})$ in Eq.~\eqref{eq:ghd_rho}. The subsystem purities scale as $\Tr(\rho_A^l)\sim e^{-(l-1)a_l^A L_A}$, where $a_l^A$ represents the spatially averaged $l$-th R\'enyi entropy density over subsystem $A$. As we established in the previous subsection, density matrices $\rho$ of interest exhibit clustering of correlations, indicating that in the Euler scaling limit we can utilize formulas assuming separable $\rho$ with strictly subleading corrections in the thermodynamic limit. The scaling for the products of moments $M_\nu(\rho)$ defined in Eq.~\eqref{eq:permutation_moment} reads:
    \begin{multline}
        M_{\nu_1}(\rho_A)M_{\nu_2}(\rho_B)\sim \\
        e^{-\sum_{l}k_l^{(1)}(l-1)a_l^A L_A-\sum_{m}k_m^{(2)}(m-1)a_m^B L_B}\equiv e^{-E(\pi_1,\pi_2)},\label{eq:combinatorial_moment_scaling}
    \end{multline}
    where $E(\pi_1,\pi_2)$ defines an effective scaling exponent for permutations $\pi_1$ and $\pi_2$ belonging to the conjugacy classes $\nu_1$ and $\nu_2$, respectively. We will demonstrate that this exponent is uniquely minimized when $\pi_1=C_q$ and $\pi_2=\mathds{1}$.

    Because the cycle lengths of any permutation in the symmetric group $\mathrm{Sym}(q)$ are bounded by $1 \le l \le q$, the monotonicity of R\'enyi entropies guarantees $a_l^A\geq a_q^A$ and $a_l^B\geq a_q^B$. Thus, we can bound the effective exponent from below:
    \begin{equation}
        E(\pi_1,\pi_2)\geq a_q^A L_A \sum_{l}k_l^{(1)}(l-1)+a_q^B L_B\sum_{m}k_m^{(2)}(m-1).\label{eq:moment_effective_exponent}
    \end{equation}
    The remaining combinatorial sums $T(\mathds{1},\pi_i)=\sum_{l}k_l^{(i)}(l-1)$ correspond exactly to the minimum number of transpositions required to connect the permutation $\pi_i$ to the identity $\mathds{1}$. The number of transpositions $T(\pi_1,\pi_2)$ needed to transform $\pi_1$ into $\pi_2$ defines a valid metric (the Cayley distance) on the space of permutations~\cite{Deza16}. This allows us to invoke the triangle inequality:
    \begin{equation}
        T(\mathds{1},C_q)\leq T(\mathds{1},\pi_2^{-1})+T(\pi_2^{-1},C_q)=T(\mathds{1},\pi_2)+T(\mathds{1},\pi_1),
    \end{equation}
    where the last equality follows from the constraint $\pi_1=\pi_2C_q$ imposed on the contributing pairs of permutations in the Scrooge-averaged subsystem purity. Because the distance between the full cyclic permutation $C_q$ and the identity is exactly $T(\mathds{1},C_q)=q-1$, we obtain the lower bound:
    \begin{equation}
        T(\mathds{1},\pi_1)\geq q-1-T(\mathds{1},\pi_2).
    \end{equation}
    Substituting this bound into Eq.~\eqref{eq:moment_effective_exponent} yields:
    \begin{equation}
        E(\pi_1,\pi_2)\geq (q-1)a_q^A L_A + T(\mathds{1},\pi_2)(a_q^B L_B - a_q^A L_A).
    \end{equation}

    Without loss of generality, we can assume $a_q^B L_B \geq a_q^A L_A$ (otherwise, we simply exchange the subsystem labels $A$ and $B$). Considering the strict inequality first ($a_q^B L_B > a_q^A L_A$), we observe that any permutation $\pi_2 \neq \mathds{1}$ introduces a strictly positive, extensive penalty to $E(\pi_1,\pi_2)$. This translates to a severe exponential suppression of those moments in Eq.~\eqref{eq:combinatorial_moment_scaling} in the thermodynamic limit. Consequently, the only surviving contribution arises from the pair $\pi_1=C_q$ and $\pi_2=\mathds{1}$. For these corresponding conjugacy classes, the multiplicity is exactly $N_{\nu_1,\nu_2}=1$, yielding the Scrooge R\'enyi entanglement entropy in the thermodynamic limit:
    \begin{equation}
        S_q^{\rm Sc}\approx\frac{1}{1-q}\ln \Tr(\rho_A^q).   \label{eq:scrooge_Sq_thermodynamic_limit}
    \end{equation}

    We note that the leading term of the subsystem purity, even for complex $q$ with $\mathrm{Re}(q)\geq 1$, satisfies $|\Tr(\rho_A^q)|\leq 1$. Furthermore, the full Scrooge subsystem purity also shares this global bound $|s_q^{\rm Sc}|\leq 1$. Thus, both functions trivially satisfy the growth bounds of Carlson's~\cite{Boas54} theorem, guaranteeing a unique analytic continuation in $q$ and allowing us to rigorously compute the von Neumann entropy limit $q\to 1$ in the leading order:
    \begin{equation}
        S_1^{\rm Sc}\approx-\Tr(\rho_A \ln \rho_A).\label{eq:scrooge_S1_thermodynamic_limit_appendix}
    \end{equation}
    Furthermore, because the extensive penalty is strictly positive for all integers $q \ge 2$, the difference between the full purity and the leading term is exponentially suppressed on these discrete values. Since their analytic continuations are globally bounded, this suppression rigorously extends to the complex plane, ensuring that all other combinatorial contributions vanish uniformly with $L$ in a neighborhood of $q=1$. By the Moore-Osgood theorem~\cite{Taylor65}, this uniform convergence rigorously guarantees that the thermodynamic limit and the analytic continuation $q\to 1$ safely commute up to subleading corrections.

    The critical edge case where $a_q^B L_B$ and $a_q^A L_A$ do not differ by an extensive amount requires a slightly modified approach. In this scenario, we assume that the macroscopic equality of the exponents in the leading order, $a_q^B L_B = a_q^A L_A$, holds for all integer indices $q \geq 2$. Assuming strict inequalities between the lower-order R\'enyi entropies (which physically corresponds to a non-flat entanglement spectrum), we have $a_l^A > a_q^A$ and $a_l^B > a_q^B$ for $l < q$. Under this assumption, we can similarly demonstrate the severe exponential suppression of all permutation cross-terms, leaving only $\Tr(\rho_A^q)$ and $\Tr(\rho_B^q)$ as the dominant contributions to the sum. Because the extensive exponents match for all integers $q \geq 2$, the subsystem purities agree at leading exponential order, $\Tr(\rho_A^q) \approx \Tr(\rho_B^q)$. By standard analytic continuation, this macroscopic equivalence extends smoothly to the $q \to 1$ limit. Consequently, the limit $q\to 1$ of the Scrooge R\'enyi entanglement entropy, Eq.~\eqref{eq:scrooge_Sq_thermodynamic_limit}, yields the leading contribution to the von Neumann entanglement entropy, Eq.~\eqref{eq:scrooge_S1_thermodynamic_limit_appendix}.

    \section{Canonical Scrooge ensemble}\label{sec:can_appendix}

    The standard Scrooge ensemble $\mathrm{Sc}[\rho]$ provides a powerful framework for encoding macroscopic physical restrictions, such as those coming from hydrodynamics, directly into the background density matrix $\rho$. However, this formulation inherently acts as a soft restriction: for a strictly positive $\rho$, every state $\ket{\Phi}$ in the Hilbert space acquires a non-zero probability density. This presents a fundamental mismatch for strictly isolated quantum quenches governed by a global $U(1)$ symmetry, where the exact time evolution is permanently confined to a fixed particle-number subspace. This dichotomy closely mirrors the relation between the grand canonical and canonical ensembles in equilibrium statistical mechanics. To bridge this gap, we introduce the canonical Scrooge ensemble: while the standard Scrooge ensemble allows encoding the spatial profile of the $U(1)$ charge, our canonical extension strictly projects this underlying distribution into the exact global symmetry sector. In this section, we provide a rigorous analytical proof for equivalence of those ensembles for non-linear functionals of the quantum state of importance to this work. We demonstrate that this strict constraint preserves the leading-order scaling of the von Neumann participation entropy, contributing only a universal, subleading correction in the thermodynamic limit. While higher R\'enyi entropies can generically acquire an extensive large-deviation correction~\cite{Piroli22}, we show that this penalty can be circumvented for a subclass of density matrices $\rho$ investigated in this work.

    We begin by restricting the ensemble to a specific symmetry sector defined by a fixed total magnetization $M$. Operating in the computational basis $|\sigma\rangle = |\sigma_1, \ldots, \sigma_L\rangle$ where $\sigma_i = \pm 1/2$, we strictly filter the Hilbert space to include only those basis states that satisfy the global constraint:
    \begin{equation}
        \sum_{i=1}^{L} \sigma_i=M.
    \end{equation}
    By the principle of maximum entropy, enforcing the local hydrodynamic profile via expectation values (analogous to the grand canonical ensemble) alongside this strict global conservation law yields the canonical probability distribution:

    \begin{equation}
        p_\sigma^{\rm can}=\frac{Z^{\rm Sc}}{Z^{\rm can}} p_\sigma^{\rm Sc},\label{eq:U1_rho_eigval}
    \end{equation}
    where the canonical partition function $Z^{\rm can}$ sums exclusively over the basis states within the restricted symmetry sector. By the standard ensemble equivalence, the local expectation values of this projected distribution converge to their unrestricted expectations in the thermodynamic limit, including the charge densities $q_i=\Tr(\rho\hat q_i)$. Importantly, in the case when the global $U(1)$ charge commutes with the density matrix $\rho$, this simple scalar proportionality extends beyond the diagonal probabilities to the off-diagonal quantum coherences within the allowed symmetry sector. Consequently, the restricted density matrix is governed entirely by this uniform scaling, which guarantees that corrections to macroscopic quantum properties can be systematically extracted by evaluating the corresponding restricted partition functions. For instance, provided the density matrix is diagonal in the chosen participation basis, as is the case for local $\mathrm{U}(1)$ charge densities $\hat{q}_i = \hat{S}_i^z$, the average participation entropy in the Scrooge ensemble may be directly related to the partition function, with the inverse participation ratio:

    \begin{equation}
        I^{\rm Sc}_q\approx q!\frac{Z^{\rm Sc}_{q\beta}}{\left(Z^{\rm Sc}\right)^q}.\label{eq:I_gc}
    \end{equation}
    Here, $Z_{q\beta}$ represents a variant of the partition function where all local Lagrange multipliers are scaled by $q$. In the remainder of this section we shall denote quantities computed for such rescaled multipliers with the subscript $q\beta$.

    \subsection{Saddle-point evaluation of the canonical partition function}

    Now let us consider the evaluation of the symmetry-restricted partition function $Z^{\rm can}$, which can be extracted by introducing a complex fugacity parameter $z$ into a generating function $G(z)$~\cite{Beale11}:
    \begin{equation}
        G(z) = \sum_{\sigma} \left( Z^{\rm Sc} p_\sigma^{\rm Sc} \right) z^{\sum_{i=1}^L \sigma_i} = \sum_{M'} Z^{\rm can}_{M'} z^{M'}.
    \end{equation}
    The expansion into symmetry sectors $Z^{\rm can}_{M'}$ follows when we notice that evaluating $G(1)$ perfectly recovers the full unconstrained partition function $Z^{\rm Sc}$, while the power of $z$ tracks the total global charge $M'$ of each basis state $\ket{\sigma}$. Therefore, our target canonical partition function $Z^{\rm can}$ is exactly the coefficient of $z^M$ in the expansion of $G(z)$. Let us note that for a pure state $\ket{\psi(t)}$ trapped within the symmetry sector $M$, the Scrooge density matrix $\rho(\{q_i\})$ following the construction based on charge densities we provide in Sec.~\ref{sec:numerical_approach} matches the expectation of the symmetry generator $\hat M$ with this sector, $\Tr(\rho(\{q_i\})\hat M)=M$. Thus, for brevity, we denote the canonical partition function of this distinguished sector simply as $Z^{\rm can}\equiv Z^{\rm can}_{M}$, explicitly reserving the subscript $M'$ for arbitrary sectors. We can extract this coefficient formally using Cauchy's integral formula:
    \begin{equation}
        Z^{\rm can} = \frac{1}{2\pi i} \oint_{\mathcal{C}} \frac{G(z)}{z^{M+1}} \mathrm{d}z,
    \end{equation}
    where the contour $\mathcal{C}$ encloses the origin. In the thermodynamic limit, this integral can be evaluated using the saddle-point approximation. To do so, we deform $\mathcal{C}$ to the unit circle and map the fugacity $z$ to a purely imaginary chemical potential via the substitution $z = e^\mu$ (with $\mu \in [-i\pi, i\pi]$), yielding:
    \begin{equation}
        Z^{\rm can} = \frac{1}{2\pi i} \int_{-i\pi}^{i\pi} G(e^\mu) e^{-M\mu} \mathrm{d}\mu.
    \end{equation}
    Rewriting the integrand as $e^{-F(\mu)}$ with $ F(\mu) = -\ln G(e^\mu) + M\mu$, the integral is dominated by the saddle point $\mu^*$ satisfying $ F'(\mu^*) = 0$:
    \begin{equation}
        \left. \frac{\partial}{\partial \mu} \ln G(e^\mu) \right|_{\mu^*} = M.
    \end{equation}

    This condition is immediately satisfied by our earlier observation on construction of density matrices of interest $\rho(\{q_i\})$. Since the logarithmic derivative of $G$ with respect to $\mu$ mathematically generates the expected total charge, evaluating it at the point $\mu=0$ perfectly recovers the built-in macroscopic constraint discussed above:
    \begin{equation}
        \left. \frac{\partial}{\partial \mu} \ln G(e^\mu) \right|_{\mu=0} = \mathbb{E}^{\rm Sc}\left(\hat M\right)=\langle \hat M\rangle= M.
    \end{equation}
    This establishes $\mu^*=0$ (corresponding to $z^*=1$) as the exact saddle point, universally valid for any background density matrix $\rho$ of interest.

    With the saddle point $\mu^*=0$ established, we evaluate the partition function by expanding the exact phase function $F(\mu)$ to second order. Parameterizing the imaginary integration contour near the saddle point as $\mu = iy$, we note that in the thermodynamic limit, the extensive scaling of the total charge variance causes the integrand to concentrate around $y=0$, allowing us to extend the integration bounds to infinity, yielding:

    \begin{equation}
        Z^{\rm can} \approx \frac{1}{2\pi i} \int_{-\infty}^\infty e^{-F(0) + y^2  F''(0)/2} (i\mathrm{d}y) = \frac{e^{-F(0)}}{\sqrt{2\pi (-F''(0))}}.
    \end{equation}
    The term $e^{-F(0)}$ trivially evaluates to $G(1) = Z^{\rm Sc}$. The denominator is determined by the second derivative evaluated at the saddle point. By substituting the explicit form of the generating function, $G(e^\mu) = Z^{\rm Sc}\mathbb{E}^{\rm Sc}( e^{\mu \hat{M}} )$, this second logarithmic derivative formally isolates the cumulant, perfectly yielding the total variance of the U(1) charge in the standard Scrooge ensemble:
    \begin{equation}
        -F''(0) = \mathbb{E}^{\rm Sc}\left(\hat{M}^2\right)  - \mathbb{E}^{\rm Sc}\left( \hat{M} \right)^2 \equiv \Sigma^2.
    \end{equation}
    This total variance $\Sigma^2$ incorporates all local fluctuations and spatial connected correlators of the charge densities. Substituting these evaluations back into the saddle-point approximation, we obtain the universal scaling relation:
    \begin{equation}
        Z^{\rm can} \approx \frac{Z^{\rm Sc}}{\sqrt{2\pi \Sigma^2}}. \label{eq:U1_Z}
    \end{equation}

    To understand this variance in the Scrooge ensemble, we first consider the fundamental scenario where the only restriction on the state comes from the U(1) symmetry $\hat M=\sum_{i}\hat S_i^z$, which enforces strictly local charge profile $m_i \equiv \langle \hat{S}_i^z \rangle=\mathbb{E}^{\rm Sc}(\hat{S}_i^z)$. Under this condition, the resulting maximum-entropy state factorizes into a direct product of local density matrices, $\rho = \bigotimes_{i=1}^L \rho_i$. Consequently, all spatial connected correlators identically vanish:
    \begin{equation}
        \mathbb{E}^{\rm Sc}\left(\hat{S}_i^z \hat{S}_j^z\right) - \mathbb{E}^{\rm Sc}\left(\hat{S}_i^z\right)\mathbb{E}^{\rm Sc}\left(\hat{S}_j^z\right) = 0 \quad \text{for } i \neq j.
    \end{equation}
    The total macroscopic variance $\Sigma^2$ therefore reduces exactly to the sum of decoupled local on-site variances:
    \begin{equation}
        \Sigma^2 = \sum_{i=1}^L \left[ \mathbb{E}^{\rm Sc}\left( (\hat{S}_i^z)^2 \right) - \mathbb{E}^{\rm Sc}\left( \hat{S}_i^z \right)^2 \right].
    \end{equation}
    For spin-$1/2$ degrees of freedom, the on-site square operator is strictly proportional to the identity, yielding $\mathbb{E}^{\rm Sc}((\hat{S}_i^z)^2) = 1/4$. The global charge variance thus evaluates to a simple spatial sum over the hydrodynamic profile:
    \begin{equation}
        \Sigma^2 = \sum_{i=1}^L \left( \frac{1}{4} - m_i^2 \right).
    \end{equation}
    This provides a clear physical picture: local regions that are fully polarized ($m_i = \pm 1/2$) are perfectly "frozen" and contribute zero variance, whereas locally unpolarized regions ($m_i \approx 0$) act as infinite-temperature thermal sources that drive the extensive charge fluctuations of the unconstrained ensemble. In more general scenarios where the unconstrained Scrooge ensemble is built from non-local macroscopic observables—such as the energy density of an interacting Hamiltonian—the state ceases to be strictly separable. The total variance $\Sigma^2$ is subsequently dressed by non-vanishing spatial connected correlators.

    \subsection{Symmetry-induced corrections to ensemble averages}

    With the universal scaling of the restricted partition function established, we can now analytically evaluate the symmetry-induced corrections to the participation entropies using the canonical variant of Eq.~\eqref{eq:I_gc}:

    \begin{equation}
        I_q^{\rm can}\approx q!\frac{Z^{\rm can}_{q\beta}}{(Z^{\rm can})^q}.\label{eq:I_can}
    \end{equation}
    Notably, the partition function $Z^{\rm can}_{q\beta}$ does not generally follow the simple scaling of Eq.~\eqref{eq:U1_Z}. Rescaling the Lagrange multipliers to $q\beta$ physically corresponds to changing the density profile, which, in the simplest case of a $\mathrm{U}(1)$ symmetry with local densities $\hat q_i=\hat S_i^z$, becomes:
    \begin{equation}
        \tilde{m}_i = \frac{\left(\frac{1}{2}+m_i\right)^q}{\left(\frac{1}{2}+m_i\right)^q + \left(\frac{1}{2}-m_i\right)^q} - \frac{1}{2}.
    \end{equation}
    Consequently, the unrestricted ensemble governed by the rescaled multipliers naturally peaks at a new expectation of the total charge, $\tilde{M} = \sum_i \tilde{m}_i$, which generally differs from the physical target sector $M$. Because the canonical projection must still strictly evaluate the partition function at the original sector $M$, the saddle point $\mu^*$ shifts away from zero to account for this macroscopic mismatch. Expanding the integral around this shifted saddle point introduces an exponential penalty:

    \begin{equation}
        Z^{\rm can}_{q\beta} \approx \frac{Z_{q\beta}^{\rm Sc}}{\sqrt{2\pi \Sigma_{q\beta}^2}}\exp\left( - \frac{(M - \tilde{M})^2}{2\Sigma_{q\beta}^2} \right).
    \end{equation}
    We note that for macroscopic shifts where $M - \tilde{M} = \mathcal{O}(L)$, the exact large-deviation penalty is formally governed by the full Cram\'er rate function~\cite{Dembo10}. However, because our physical argument relies strictly on establishing the generic extensive $\mathcal{O}(L)$ scaling of this penalty and its exact cancellation when $M = \tilde{M}$, the Gaussian approximation presented here, valid for moderate deviations, perfectly suffices to capture the relevant macroscopic behavior.

    Substituting this relation into Eq.~\eqref{eq:I_can} yields the canonical correction to the R\'enyi participation entropy:

    \begin{multline}
        P_q^{\rm can} \approx P_q^{\rm Sc} + \frac{q}{2(1-q)}\ln(2\pi\Sigma^2)\\ - \frac{1}{2(1-q)}\ln(2\pi \Sigma_{q\beta}^2) - \frac{(M - \tilde{M})^2}{2(1-q)\Sigma_{q\beta}^2}. \label{eq:U1_scrooge_P}
    \end{multline}
    As discussed in Sec.~\ref{sec:can_main}, the final term, $\frac{(M - \tilde{M})^2}{2(1-q)\Sigma_{q\beta}^2}$, constitutes an extensive penalty that introduces a macroscopic dependence on the chosen symmetry sector for higher-order R\'enyi participation entropies, demonstrating that typicality does not universally apply to all global non-linear quantities.

    Global restricted typicality can, however, be fully restored in specific physical regimes. For example, antisymmetric macroscopic density profiles can force $\tilde M=M$. Such configurations are of profound physical relevance, encompassing both the N\'eel and domain-wall initial states analyzed in Sec.~\ref{sec:floq}. In these protected cases, the extensive penalty explicitly vanishes, and the remaining variance-driven corrections are strictly subleading, scaling as $\mathcal{O}(\ln L)$. This restores the equivalence between the unrestricted and canonical Scrooge ensembles for a much broader class of global functionals of the quantum state.

    Furthermore, some global quantities remain unconditionally equivalent between the unrestricted and canonical ensembles. By taking the analytic continuation of the participation entropy to the von Neumann limit ($q\to 1$), the extensive penalty systematically vanishes for an arbitrary density profile, yielding:

    \begin{equation}
        P_1^{\rm can} \approx P_1^{\rm Sc} - \frac{1}{2}\ln(2\pi\Sigma^2) + \frac{1}{2\Sigma^2} \left. \frac{\partial \Sigma_{q\beta}^2}{\partial q} \right|_{q=1},
    \end{equation}
    where the final $\mathcal{O}(1)$ term measures the spatial asymmetry of the charge profile. This cancellation again perfectly parallels the established behavior of von Neumann entanglement entropies in $\mathrm{U}(1)$-symmetric systems, where the extensive symmetry-sector dependence inherent to $q>1$ R\'enyi entropies completely drops out in the $q\to 1$ limit~\cite{Piroli22}.

    Evaluating the subsystem purity for the canonical R\'enyi entanglement entropy requires explicitly accounting for subsystem charge fluctuations. Because the strict global constraint fixes the total charge to $M_A+M_B=M$, tracing over the environment $B$ couples the subsystems, forcing the reduced density matrix to become a classical mixture of subsystem charge sectors $M'_A$. Consequently, the trace of the $q$-th power of the reduced density matrix of the Scrooge ensemble evaluates to a coupled sum over the restricted subsystem partition functions:

    \begin{equation}
        \Tr[(\rho_A^{\rm can})^q] = \frac{1}{(Z^{\rm can})^q} \sum_{M'_A} Z^{\rm can}_{A, M'_A,q\beta} \left( Z^{\rm can}_{B, M - M'_A} \right)^q,\label{eq:U1_scrooge_S}
    \end{equation}
    where $Z^{\rm can}_{A,M'_A}$ denotes the partition function restricted to basis states from the sector $M'_A$ in the subsystem $A$, and $Z^{\rm can}_{A, M'_A,q\beta}$ denotes its variant evaluated with rescaled Lagrange multipliers. The full Scrooge-averaged canonical subsystem purity $s_q^{\rm can}$ can then be computed by combining these subsystem traces over all relevant replica permutations. While the exact algebraic form of the resulting entanglement entropy $S_q^{\rm can}$ differs from the participation entropy due to the subsystem variances and these replica sums, the universal scaling behavior is identical. In the thermodynamic limit, evaluating these subsystem traces yields a strict, extensive $\mathcal{O}(L)$ volume-law penalty for generic states at $q \neq 1$. Analogously to the participation entropy, this extensive penalty is eliminated for antisymmetric configurations, and it vanishes in the von Neumann $q\to 1$ limit unconditionally, leaving only a subleading correction $\mathcal{O}(\ln L)$.

    \section{Microcanonical Scrooge ensemble\label{sec:microcan_appendix}}

    While the canonical projection successfully resolves the nonphysical mixing of global symmetry sectors, it leaves a critical dynamical loophole: individual states within the canonical Scrooge ensemble can still exhibit macroscopic spatial charge distributions that radically deviate from the exact time-evolved state $\ket{\psi(t)}$ used to construct this ensemble. Because isolated unitary evolution is strictly bottlenecked by local transport, such anomalous spatial configurations are dynamically inaccessible. As conceptually illustrated in Fig.~\ref{fig:cartoon}(a), the exact pure state of a closed system is therefore permanently confined to a strictly smaller, idealized subspace $\mathcal{M}_{\rm r}$ defined by these exact spatial constraints. This confinement underpins a profound notion of restricted typicality: the system behaves locally as a typical state drawn exclusively from this restricted manifold. To bridge this fundamental gap, we introduce the microcanonical Scrooge ensemble. By rigidly projecting the Scrooge ensemble onto this exact hydrodynamic manifold, we establish a precise statistical framework for global restricted typicality. We demonstrate that this microcanonical projection imposes only an exponentially vanishing correction to ensemble-averaged macroscopic observables.

    \subsection{Local charge fluctuations in the Scrooge ensemble}

    We define the microcanonical Scrooge ensemble by rigorously enforcing a set of continuous local constraints, $\langle \Phi | \hat{q}_j | \Phi \rangle = q_j$, for all local charges $\hat{q}_j$ that parameterize the standard hydrodynamic Scrooge ensemble. Unlike a canonical projection onto a discrete global superselection sector, this imposes an extensive number of continuous amplitude constraints. We now demonstrate that this severe restriction introduces only an exponentially small correction. We begin by evaluating the statistical variance of the local charges within the unconstrained Scrooge ensemble:
    \begin{equation}
        \mathrm{Var}^{\rm Sc}(\hat q_j) = \mathbb{E}^{\rm Sc}\left[\langle \Phi | \hat{q}_j | \Phi \rangle^2\right] - \mathbb{E}^{\rm Sc}[\langle \Phi | \hat{q}_j | \Phi \rangle]^2.
    \end{equation}
    By definition, the ensemble average perfectly recovers the target macroscopic profile of the hydrodynamic density matrix $\rho(\{q_i\})$, such that $\mathbb{E}^{\rm Sc}[\langle \Phi | \hat{q}_j | \Phi \rangle]=\mathrm{Tr}(\rho(\{q_i\}) \hat q_j)=q_j$. The second moment can be expressed in terms of Haar-random pure states $\ket{\phi}$ as:
    \begin{multline}
        \mathbb{E}^{\rm Sc}\left[\langle \Phi | \hat{q}_j | \Phi \rangle^2\right]=\\
        \int\mathrm{d}\phi P_{\rm Haar}(\phi) \,
        \mathcal{D}\bra{\phi}\rho\ket{\phi}
        \left(\frac{\bra{\phi}\sqrt{\rho}\hat q_j \sqrt{\rho}\ket{\phi}}{\bra{\phi}\rho\ket{\phi}}\right)^2.
    \end{multline}
    It can be evaluated efficiently by expressing it as a trace over a doubled Hilbert space: $\mathbb{E}^{\rm Sc}[\langle \Phi | \hat{q}_j | \Phi \rangle^2] = \mathrm{Tr}((\hat{q}_j \otimes \hat{q}_j) \rho^{(2)})$. Substituting the second moment of the Scrooge ensemble~\cite{McGinley25}, $\mathbb{E}^{\rm Sc}[\op{\Phi}{\Phi}^{\otimes 2}] \approx \rho^{\otimes 2}(T_{\mathds 1} + T_{\mathbb{S}})$, splits the trace into two components. The relative error of this approximation scales as $\Tr(\rho^2)^{1/4}$, vanishing in the thermodynamic limit as we will argue later. The identity term yields the disconnected expectation values, $\mathrm{Tr}((\hat{q}_j \rho \otimes \hat{q}_j \rho)T_{\mathds 1}) = \mathrm{Tr}(\hat{q}_j \rho)^2 = q_j^2$, which perfectly cancels the squared mean in the variance. The swap term simplifies to:

    \begin{equation}
        \mathrm{Tr}((\hat{q}_j \rho \otimes \hat{q}_j \rho)T_{\mathbb{S}}) = \mathrm{Tr}(\hat{q}_j \rho \hat{q}_j \rho).
    \end{equation}
    The variance therefore evaluates to
    \begin{equation}
        \mathrm{Var}^{\rm Sc}(\hat q_j) = \mathrm{Tr}(\hat{q}_j \rho \hat{q}_j \rho).
    \end{equation}
    This trace is bounded by $\mathrm{Tr}(\hat{q}_j \rho \hat{q}_j \rho)\leq||\hat{q}_j||^2 \mathrm{Tr}(\rho^2)$, which implies $\mathrm{Var}^{\rm Sc}(\hat q_j) \leq \mathcal{O}(\mathrm{Tr}(\rho^2))$.

    \subsection{Exponential decay of purity for the density matrix constructed from local densities}

    For any state $\ket{\psi}$ that is not uniquely determined by its local charges, the corresponding generalized Gibbs state $\rho(\{q_i\})$ is strictly mixed. Provided the macroscopic profile avoids frozen zero-entropy boundary macrostates possessing finite Lagrange multipliers, this implies an extensive, volume-law thermodynamic R\'enyi entropy, which translates directly to an exponentially decaying purity, $\mathrm{Tr}(\rho^2)=e^{-a_2L}$, with a strictly positive rate $a_2>0$. This rate can be connected to thermodynamics of the system when we express the state purity as a ratio of partition functions:
    \begin{equation}
        \mathrm{Tr}(\rho^2)=\frac{Z(2\beta)}{Z(\beta)^2},
    \end{equation}
    where $Z(2\beta)=\mathrm{Tr}(e^{-2\sum_j \beta_j\hat q_j})$. Each partition function scales with the extensive free energy $F(\beta)$ as $Z(\beta)=e^{- F(\beta)}=e^{- Lf(\beta)}$, governed by an intensive free energy density $f(\beta)\sim\mathcal{O}(1)$. Substituting this scaling form directly into the partition function ratio yields $\mathrm{Tr}(\rho^2) = e^{-L[f(2\beta) - 2f(\beta)]} \equiv e^{-a_2L}$. The rate of decay of this purity can be related to the intensive free energy density as $a_2 = f(2\beta) - 2f(\beta)$. This confirms the exponential suppression of the purity. Consequently, the statistical variance $\mathrm{Var}^{\rm Sc}(\hat q_j)$ strictly vanishes in the thermodynamic limit, establishing that the local charge profiles of individual pure states within the unconstrained Scrooge ensemble undergo exact, deterministic concentration onto the microcanonical manifold, as demonstrated later.

    \subsection{Equivalence of Scrooge and microcanonical Scrooge averages}

    This vanishing variance allows us to rigorously bound the probability of violating any single local constraint via Chebyshev inequality~\cite{Billingsley95}:
    \begin{equation}
        P_{\rm Sc}\left(\left|\bra{\Phi}\hat q_j\ket{\Phi}-q_j\right|\geq \epsilon\right) \leq \frac{\mathrm{Var}^{\rm Sc}(\hat q_j)}{\epsilon^2}.
    \end{equation}
    Applying the union bound across all $L$ spatial constraints, the total probability $P_{\rm viol}$ that a randomly sampled state falls outside the typical hydrodynamic manifold is exponentially suppressed:
    \begin{equation}
        P_{\rm viol} \leq L \frac{e^{-a_2L}}{\epsilon^2}.
    \end{equation}
    Consider any bounded functional $g(\Phi)$ defined on the Hilbert space, such as the participation entropy $P_q(\Phi)$ or the entanglement entropy $S_q(\Phi)$. Assuming a global bound $|g(\Phi)| \leq g_{\max}$, we partition the full ensemble average into contributions from the restricted hydrodynamic manifold $\mathcal{M}_{\rm r,\epsilon}$, where each constraint is satisfied with a tolerance $\epsilon$, and the complement over the Hilbert space $\mathcal{M}_{\rm r,\epsilon}^{\rm c}$:

    \begin{equation}
        \mathbb{E}^{\rm Sc}[g]=\int_{\mathcal{M}_{\rm r,\epsilon}}\mathrm{d}\Phi P_{\rm Sc}(\Phi)g(\Phi)+\int_{\mathcal{M}_{\rm r,\epsilon}^{\rm c}}\mathrm{d}\Phi P_{\rm Sc}(\Phi)g(\Phi).
    \end{equation}
    The contribution from the remaining subspace is strictly bounded by $P_{\rm viol}$:
    \begin{equation}
        \left|\int_{\mathcal{M}_{\rm r,\epsilon}^{\rm c}}\mathrm{d}\Phi P_{\rm Sc}(\Phi)g(\Phi)\right|\leq g_{\max} P_{\rm viol}.
    \end{equation}
    For physical quantities where $g_{\max}$ grows at most polynomially with system size $L$, this atypical contribution vanishes asymptotically.

    Let us now demonstrate that for a Lipschitz-continuous functional $g(\Phi)$ satisfying
    \begin{equation}
        |g(\Phi_1)-g(\Phi_2)|\leq \eta(L)\| \ket{\Phi_1}-\ket{\Phi_2} \|,
    \end{equation}
    the remaining integral collapses exactly onto the microcanonical expectation value $\mathbb{E}^{\rm Sc}_{\rm mc}[g]$, provided that the Lipschitz constant $\eta(L)$ grows at most polynomially with $L$. We start by defining an auxiliary constraint mapping $\mathbf{V}(\Phi)$, whose components $V_i(\Phi) = \bra{\Phi}\hat q_i\ket{\Phi}-q_i$ measure the violation of the microcanonical restriction. Treating the state space as a projective manifold to properly account for normalization and global phase, we evaluate the tangent gradient of this mapping. This calculation yields the $L\times L$ symmetric Gram matrix $\nabla V_i\cdot \nabla V_j= 2\bra{\Phi}\{\hat q_i, \hat q_j\}\ket{\Phi} - 4\bra{\Phi}\hat q_i\ket{\Phi}\bra{\Phi}\hat q_j\ket{\Phi}$. A proof analogous to the one presented previously for the single-point expectation value demonstrates that this Gram matrix tightly concentrates around the expectation value in the Scrooge ensemble:
    \begin{equation}
        \nabla V_i\cdot \nabla V_j= 2\mathbb{E}^{\rm Sc}[\bra{\Phi}\{\hat q_i, \hat q_j\}\ket{\Phi}]-4q_iq_j+\mathcal{O}(e^{-a_2L}).
    \end{equation}
    The singular values of the Jacobian are strictly positive as long as this Gram matrix is positive-definite. This positivity condition is physically guaranteed, failing only at extreme geometric boundaries of the microcanonical manifold where the target charges $q_i$ are able to fully determine the state. Denoting the smallest singular value of the Jacobian as $\sigma_{\min}>0$, we note that the Gram matrix corresponds to the spatial correlation matrix of local charges in the macroscopic state $\rho$. Because the considered class of states exhibits a clustering of correlations, the off-diagonal matrix elements of the Gram matrix  decay with spatial distance. Assuming no exact degeneracies, this spatial decay mathematically prevents the singular values from vanishing exponentially with $L$, guaranteeing that $\sigma_{\min}$ decays at worst polynomially with system size, $\sigma_{\min} \ge \mathcal{O}(L^{-k})$. Furthermore, for systems exhibiting exponential clustering of correlations, such as when $\rho$ is a thermal state of a local Hamiltonian~\cite{Araki69}, this scaling typically saturates at $\sigma_{\min} = \mathcal{O}(1)$. Because the Gram matrix is invertible, the auxiliary mapping $\mathbf{V}(\Phi)$ satisfies the strict surjectivity assumption of the Lyusternik-Graves theorem~\cite{Ioffe15}. This proves that the mapping is metrically regular on the exact microcanonical manifold $\mathcal{M}_r$. To extend this local theorem to a uniform global bound over the shell $\mathcal{M}_{\rm r,\epsilon}$, we note that $\mathbf{V}(\Phi)$ is a quadratic mapping in the state amplitudes, ensuring its Jacobian is Lipschitz continuous with a constant growing at most polynomially with $L$. This ensures that the locally valid neighborhood guaranteed by the theorem shrinks at worst polynomially with system size. Because we dynamically scale the constraint tolerance $\epsilon$ to shrink exponentially with $L$, the typical shell $\mathcal{M}_{\rm r,\epsilon}$ asymptotically resides entirely within this uniform regularity neighborhood, establishing the global bound:
    \begin{equation}
        \min\limits_{\Phi_{\rm mc}\in\mathcal{M}_r}
        \| \ket{\Phi}-\ket{\Phi_{\mathrm{mc}}} \| \leq \frac{1}{\sigma_{\min}} \| \mathbf{V}(\Phi) \|.
    \end{equation}
    Because the typical shell $\mathcal{M}_{\rm r,\epsilon}$ is defined by the bounded violation $|V_i(\Phi)|\leq\epsilon$, the Euclidean norm of the constraint vector is strictly bounded by $\| \mathbf{V}(\Phi) \| \leq \sqrt{L}\epsilon$. Denoting a state $\ket{\Phi_{\rm mc}}$ in the microcanonical manifold $\mathcal{M}_r$ that realizes this minimum distance to $\ket{\Phi}$, and substituting this geometric bound directly into the Lipschitz condition limits the maximum variation of the quantity inside the typical shell:
    \begin{equation}
        |g(\Phi)-g(\Phi_{\rm mc})| \leq \frac{\eta(L)}{\sigma_{\min}} \sqrt{L}\epsilon.
    \end{equation}
    Consequently, replacing $g(\Phi)$ with $g(\Phi_{\rm mc})$ inside the shell integral incurs an error strictly bounded by $\mathcal{O}(\eta(L)\sqrt{L}\epsilon/\sigma_{\rm min})$. By dynamically scaling the constraint tolerance as e.g. $\epsilon \sim e^{-a_2L/3}$, this exponential suppression ensures that the error vanishes. In the thermodynamic limit, the integral over the $\epsilon$-thick shell converges to the integral over the exact manifold. Thus, we can define the exact microcanonical average $\mathbb{E}_{\rm mc}^{\rm Sc}$ as the limit of finite-$\epsilon$ averages when $\epsilon\to0$. Combining this with the vanishing atypical tail $P_{\rm viol}$, we recover exact ensemble equivalence up to an exponentially vanishing finite-size correction:
    \begin{equation}
        \mathbb{E}^{\rm Sc}[g] = \mathbb{E}^{\rm Sc}_{\rm mc}[g] + \mathcal{O}\left(\frac{\eta(L)}{\sigma_{\rm min}}\sqrt{L}\epsilon\right)+\mathcal{O}(g_{\rm max}P_{\rm  viol}),
    \end{equation}
    where  $\epsilon$ decreasing exponentially with $L$ suppresses both errors as long as $\lim_{L\to\infty}L\frac{e^{-a_2L}}{\epsilon^2}=0$.

    This geometric proof natively extends to non-linear quantities like von Neumann entanglement ($S_1$) and participation ($P_1$) entropies by applying the Fannes-Audenaert inequality~\cite{Audenaert07},
    \begin{equation}
        |S_1(\Phi_1) - S_1(\Phi_2)| \leq T \ln(d-1) - T \ln T-(1-T)\ln (1-T),
    \end{equation}
    where the pure-state trace distance is bounded by $T = \sqrt{1 - |\langle \Phi_1 | \Phi_2 \rangle|^2} \leq \| \ket{\Phi_1} - \ket{\Phi_2} \|$, and $d$ is the dimension of the relevant state space (e.g., $d=2^{L_A}$ for subsystem entanglement, or $d=2^L$ for global participation). Substituting the metrically regular displacement bound $T \leq \mathcal{O}(\sqrt{L}\epsilon/\sigma_{\rm min})$, and noting that both the spatial dimension ($\ln d \propto L$) and the logarithmic singularity ($-\ln(\sqrt{L}\epsilon) \sim \mathcal{O}(L)$) introduce only linear penalties, the integral error over the typical shell is strictly bounded by $\mathcal{O}(L^{3/2}\epsilon/\sigma_{\rm min})$. Because this polynomial penalty is trivially overcome by the exponentially vanishing tolerance $\epsilon$, ensemble equivalence is perfectly preserved in the thermodynamic limit:
    \begin{equation}
        \mathbb{E}^{\rm Sc}[S_1] = \mathbb{E}^{\rm Sc}_{\rm mc}[S_1] + \mathcal{O}\left(\frac{1}{\sigma_{\rm min}}L^{3/2}\epsilon\right) + \mathcal{O}(L P_{\rm viol}).
    \end{equation}

    To extend this proof to the R\'enyi entropy $S_q$ for $q>1$, we must overcome the lack of global Lipschitz continuity. Because of this, providing a global upper bound for $|S_q(\Phi)-S_q(\Phi_{\rm mc})|$ must rely on other methods, the most direct of which is integrating $\nabla S_q$ over a path connecting $\ket{\Phi}$ and $\ket{\Phi_{\rm mc}}$ and utilizing the fundamental theorem of calculus together with the Cauchy-Schwarz inequality. We stress that the local charge constraints do not prevent the reduced state on $A$ from being nearly maximally mixed. For such states, the gradient norm $\| \nabla S_q \| = \frac{2q}{q-1}\sqrt{s_{2q-1}}/s_q \leq \frac{2q}{q-1} s_q^{-1/(2q)}$ rapidly diverges, which causes the strategy of integrating $\nabla S_q$ to yield an excessively loose bound based on the worst-case limits for $s_q$. To bypass this problem, we introduce the truncated entropy $\tilde S_q(\Phi) = \min\big(S_q(\Phi), 2S_1(\rho_A)\big)$. By the properties of the minimum function, capping the maximum entropy bounds the minimum purity from below, rendering the truncated functional globally Lipschitz continuous:
    \begin{equation}
        \| \nabla \tilde{S}_q \| \leq \tilde{\eta}_q := \frac{2q}{q-1}e^{\frac{q-1}{q}S_1(\rho_A)}.
    \end{equation}

    Because $\tilde{S}_q$ is globally Lipschitz, its statistical fluctuations within the unrestricted Scrooge ensemble are governed by the generalized L\'evy lemma for Scrooge measures~\cite{Teufel25}. To apply this, we first note the exact bound $\mathbb{E}^{\rm Sc}[\tilde{S}_q] \leq \mathbb{E}^{\rm Sc}[S_q] \leq S_1(\rho_A)$, where the second inequality follows from the monotonicity of R\'enyi entropies and the concavity of the von Neumann entropy together with $\mathbb{E}^{\rm Sc}[\Tr_B\op{\Phi}{\Phi}]=\rho_A$. Because the mean is bounded by $S_1(\rho_A)$, the probability that the entropy $S_q$ exceeds the truncation threshold $2S_1(\rho_A)$ implies a macroscopic deviation of at least $S_1(\rho_A)$ from the mean. This allows us to strictly bound the atypical tail:
    $P_{\rm Sc}\big(S_q > 2S_1(\rho_A)\big) \leq 6\exp\big(-\tilde C S_1(\rho_A)^2/\tilde{\eta}_q^2\|\rho\|\big)$. Evaluating this exponent yields:
    \begin{multline}
        P_{\rm Sc}\Big(S_q(\Phi) > 2S_1(\rho_A)\Big) \leq \\
        6\exp\Big(-\frac{\tilde C(q-1)^2 S_1(\rho_A)^2}{4q^2}\, e^{S_\infty(\rho)-\frac{2(q-1)}{q}S_1(\rho_A)}\Big).\label{eq:teufel_tail}
    \end{multline}
    Provided the state $\rho$ satisfies the condition $S_\infty(\rho) > \frac{2(q-1)}{q}S_1(\rho_A)$, this atypical tail is double-exponentially suppressed. Because $S_q$ is bounded by $L_A \ln d$, the truncation incurs a negligible error:

    \begin{equation}
        \mathbb{E}^{\rm Sc}[S_q] = \mathbb{E}^{\rm Sc}[\tilde{S}_q] + \mathcal{O}\big(L_A P_{\rm atyp}\big),
    \end{equation}
    where we introduce the probability of the atypical state $P_{\rm atyp}\equiv P_{\rm Sc}(S_q > 2S_1(\rho_A))$.

    Having established that $\tilde{S}_q$ faithfully reproduces the averages of the Scrooge ensemble, we evaluate its concentration onto the microcanonical manifold via the Lyusternik-Graves theorem. Let us consider a path $\gamma(t)$ realizing the shortest distance between a state $\gamma(0)=\ket{\Phi} \in \mathcal{M}_{\rm r,\epsilon}$ and its exact microcanonical projection $\gamma(1)=\ket{\Phi_{\rm mc}}\in\mathcal{M}_r$. Applying the fundamental theorem of calculus and the Cauchy-Schwarz inequality yields:
    \begin{equation}
        |\tilde{S}_q(\Phi)-\tilde{S}_q(\Phi_{\rm mc})| \leq \int_0^1 \norm{\nabla \tilde{S}_q(\gamma(t))} \norm{\dot \gamma(t)} dt.\label{eq:sq_path_integration}
    \end{equation}
    After bounding the gradient with its global supremum $\tilde{\eta}_q$ and recognizing the remaining integral as arc length, we obtain an inequality valid uniformly across the typical shell:
    \begin{equation}
        |\tilde{S}_q(\Phi)-\tilde{S}_q(\Phi_{\rm mc})| \leq \tilde{\eta}_q \| \ket{\Phi}-\ket{\Phi_{\rm mc}} \|.
    \end{equation}
    Substituting the bound resulting from metric regularity, $\| \ket{\Phi}-\ket{\Phi_{\rm mc}} \| \leq \frac{\sqrt{L}}{\sigma_{\min}}\epsilon$, guarantees that the maximum geometric deviation is strictly bounded by $\frac{\sqrt{L}}{\sigma_{\min}}\epsilon \tilde{\eta}_q$.

    Setting the constraint tolerance dynamically as $\epsilon = e^{-bL}$, this geometric deviation vanishes exponentially for any scaling rate $b > \frac{q-1}{q}\frac{S_1(\rho_A)}{L}$. Since ensuring $P_{\rm viol} \to 0$ simultaneously requires $b < a_2 / 2$, a valid tolerance strictly exists whenever $a_2 L > 2\frac{q-1}{q}S_1(\rho_A)$. Let us note that for a state satisfying $S_\infty(\rho) > \frac{2(q-1)}{q}S_1(\rho_A)$ both bounds can be satisfied at once, as a direct consequence of the inequality $a_2 L = S_2(\rho) \geq S_\infty(\rho)$.

    Finally, the microcanonical expectation value is equivalent to the unrestricted Scrooge average conditioned on the shell $\mathcal{M}_{\rm r,\epsilon}$. By conditional probability, the fraction of atypical states exceeding the truncation threshold strictly within the shell is bounded by $P_{\rm atyp} / (1 - P_{\rm viol})$. Because this ratio vanishes in the thermodynamic limit, the expectation value of the truncated entropy $\tilde S_q$ matches that of the exact entropy $S_q$ on the microcanonical manifold up to an exponentially suppressed error:
    \begin{equation}
        \mathbb{E}^{\rm Sc}_{\rm mc}[S_q] = \mathbb{E}^{\rm Sc}_{\rm mc}[\tilde{S}_q] + \mathcal{O}(L_A P_{\rm atyp}).
    \end{equation}
    Combining these components establishes the restricted typicality of the R\'enyi entropy:
    \begin{equation}
        \mathbb{E}^{\rm Sc}[S_q] = \mathbb{E}^{\rm Sc}_{\rm mc}[S_q] + \mathcal{O}\left(\frac{\sqrt{L}}{\sigma_{\min}}\epsilon \tilde{\eta}_q\right) + \mathcal{O}\big[L_A (P_{\rm viol}+P_{\rm atyp})\big].
    \end{equation}

    \subsection{Geometric structure of microcanonical partition function}

    Having established the exact concentration of physical quantities up to exponentially suppressed corrections, we now turn to the geometric characterization of the microcanonical manifold itself. To fully understand the structure of this constrained Hilbert space, it is necessary to rigorously quantify the effective state-space statistical weight of the typical subspace $\mathcal{M}_{\rm r}$. Within the framework of statistical mechanics, this density is captured by the microcanonical partition function, $Z_{\rm mc}$. In what follows, we perform an exact analytical evaluation of $Z_{\rm mc}$ to explicitly demonstrate how the continuous local amplitude constraints restrict the available Hilbert space.

    We begin by defining the unnormalized Scrooge density matrix $\tilde{\rho} = e^{-\sum_{j}\beta_j \hat{q}_j}$ and expressing the unrestricted partition function as an exact integral over auxiliary Haar-random pure states $\ket{\phi}$:

    \begin{equation}
        Z = \Tr \left(\tilde{\rho}\right) = \mathcal{D}\int\mathrm{d}\phi \, P_{\rm Haar}(\phi)\bra{\phi} \tilde{\rho} \ket{\phi}.
    \end{equation}

    To rigorously evaluate the microcanonical partition function, the microcanonical constraints must be imposed directly onto the states of the Scrooge ensemble $\ket{\Phi}=\sqrt{\tilde{\rho}}\ket{\phi}/\sqrt{\bra{\phi}\tilde{\rho}\ket{\phi}}$, which yields the restriction $\bra{\Phi}\hat q_j\ket{\Phi}=q_j$ as:
    \begin{equation}
        \frac{\bra{\phi} e^{-\frac{1}{2}\sum_k \beta_k \hat{q}_k} \hat{q}_j e^{-\frac{1}{2}\sum_k \beta_k \hat{q}_k} \ket{\phi}}{\bra{\phi} e^{-\sum_k \beta_k \hat{q}_k} \ket{\phi}} = q_j.
    \end{equation}
    By embedding these conditions via Dirac delta functions, applying the scaling identity $\prod_j \delta(X_j/Y - q_j) = Y^L \prod_j \delta(X_j - q_j Y)$ to clear the denominator, and utilizing auxiliary variables $z_j$ for their integral representation, we obtain:

    \begin{multline}
        Z_{\rm mc} = \mathcal{D}\int\left[\prod_j \frac{\mathrm{d} z_j}{2\pi}\right]\mathrm{d}\phi\, P_{\rm Haar}(\phi) \bra{\phi} \tilde \rho \ket{\phi}^{L+1} \\
        \times \exp\left[ i\sum_j z_j \bra{\phi} \left(\sqrt{\tilde \rho}\hat{q}_j \sqrt{\tilde\rho} - q_j \tilde\rho \right) \ket{\phi} \right].
    \end{multline}
    We can demonstrate that the relation between both partition functions is:

    \begin{equation}
        Z_{\rm mc} = Z\int\left[\prod_j \frac{\mathrm{d} z_j}{2\pi}\right] e^{-i\sum_j z_j q_j} \mathbb{E}^{\rm Sc}\left[  e^{i\sum_j z_j \bra{\Phi}\hat q_j \ket{\Phi}} \right].
    \end{equation}

    Remarkably, the microcanonical partition function $Z_{\rm mc}$ can be evaluated analytically under minimal approximations. We begin the derivation by expressing the Haar measure, relating it to the volume of the Hilbert space:

    \begin{equation}
        P_{\rm Haar}(\phi)\mathrm{d}\phi = \frac{\Gamma(\mathcal{D})}{\pi^\mathcal{D}} \, \delta(\langle \phi | \phi \rangle - 1)\mathrm{d}\phi.
    \end{equation}
    Enforcing this norm constraint via the Fourier representation of the Dirac Delta function introduces a new auxiliary variable $\nu$, yielding:

    \begin{multline}
        Z_{\rm mc} = \frac{\mathcal{D}!}{\pi^\mathcal{D}}\int\mathrm{d}\phi \frac{\mathrm{d}\nu}{2\pi} \left[ \prod_{j} \frac{\mathrm{d}z_j}{2\pi} \right] \left( \bra{\phi} \tilde{\rho} \ket{\phi} \right)^{L+1} e^{i\nu \left( \langle\phi|\phi\rangle - 1 \right)} \\
        \times \exp\left[ i\sum_j z_j \bra{\phi} \left( \sqrt{\tilde{\rho}}\hat{q}_j\sqrt{\tilde{\rho}} - q_j \tilde{\rho} \right) \ket{\phi} \right].
    \end{multline}

    To evaluate this integral, we transform to the most natural basis: the eigenbasis of the unnormalized Scrooge density matrix $\tilde{\rho}$. Defining $\tilde{\rho}=e^{-\sum_j \beta_j \hat q_j}$ yields the eigenvalue equation $\tilde{\rho}\ket{n} = e^{-E_n}\ket{n}$, with $\sum_{j}\beta_j\hat q_j\ket{n}=E_n\ket{n}$. We expand the auxiliary Haar-random state using real non-negative amplitudes $A_n$ and phases $\theta_n$: $\ket{\phi} = \sum_n \sqrt{A_n} e^{i\theta_n}\ket{n}$. The diagonal inner product governing the geometric scaling reads:

    \begin{equation}
        \bra{\phi} \tilde{\rho} \ket{\phi} = \sum_n A_n e^{-E_n}.
    \end{equation}
    However, because the local charges $\hat{q}_i$ do not generally commute with $\sum_{j}\beta_j\hat q_j$, the constraint term develops off-diagonal, phase-dependent cross-terms:

    \begin{multline}
        \bra{\phi} \sqrt{\tilde{\rho}} \hat{q}_j \sqrt{\tilde{\rho}} \ket{\phi} = \\
        \sum_{n, m} \sqrt{A_n A_m} e^{i(\theta_m - \theta_n)} e^{-\frac{1}{2}(E_n + E_m)} \bra{n}\hat{q}_j\ket{m}.
    \end{multline}

    Expressing the flat integration measure over the complex plane in terms of these amplitude and phase variables, $\mathrm{d}\phi_n \mathrm{d}\phi_n^* =\frac{1}{2} \mathrm{d}A_n \mathrm{d}\theta_n$, we substitute these expressions back into the microcanonical partition function. The full integral takes the exact form:

    \begin{multline}
        Z_{\rm mc} = \frac{\mathcal{D}!}{(2\pi)^\mathcal{D}}\int\left( \prod_l \mathrm{d}A_l \mathrm{d}\theta_l \right) \frac{\mathrm{d}\nu}{2\pi} \left[ \prod_{j} \frac{\mathrm{d}z_j}{2\pi} \right]\\
        \times \left( \sum_n A_n e^{-E_n} \right)^{L+1} e^{i\nu\left(\sum_n A_n - 1\right)} \\
        \exp\Bigg[ i \sum_j z_j \Bigg(\sum_{n, m} \sqrt{A_n A_m} e^{i(\theta_m - \theta_n)} e^{-\frac{1}{2}(E_n + E_m)} \bra{n} \hat{q}_j \ket{m} \\
            - q_j \sum_n A_n e^{-E_n} \Bigg) \Bigg].
    \end{multline}
    To evaluate the multi-dimensional phase integral over $\{\theta_n\}$, we invoke the random-phase approximation, assuming that the oscillatory, off-diagonal cross-terms self-average to zero upon integration. For generic non-commuting constraints, this step constitutes an approximation. However, for effective Hamiltonians $\sum_k \beta_k \hat{q}_k$ satisfying the Eigenstate Thermalization Hypothesis (ETH), the omitted terms are strictly subleading for typical macroscopic constraints $\mathbf{q}$. This approximation only breaks down for highly atypical constraints $\mathbf{q}$ located near the boundary of the allowed configuration space, where the off-diagonal contributions may become necessary to satisfy the target constraint. We emphasize that this phase integration is mathematically exact if the effective Hamiltonian and all local charges mutually commute. Performing the remaining integrations over the decoupled phases yields an overall prefactor of $(2\pi)^{\mathcal{D}}$. The microcanonical partition function thus reduces to a real amplitude integral depending exclusively on the diagonal matrix elements:

    \begin{multline}
        Z_{\rm mc} = \mathcal{D}! \int\left( \prod_n \mathrm{d}A_n \right) \frac{\mathrm{d}\nu}{2\pi} \left[ \prod_{j} \frac{\mathrm{d}z_j}{2\pi} \right] \\
        \times\left( \sum_n A_n e^{-E_n} \right)^{L+1}e^{i\nu\left(\sum_n A_n - 1\right)}  \\
        \times \exp\left[ i \sum_j z_j \sum_{n} A_n e^{-E_n} \left( \bra{n} \hat{q}_j \ket{n} - q_j \right) \right].
    \end{multline}
    Defining the diagonal charge expectation values $v_j^{(n)}=\bra{n} \hat{q}_j \ket{n}$, we first absorb the polynomial prefactor by introducing a continuous auxiliary source field $s$, which allows us to express the term as a derivative: $\left( \sum_n A_n e^{-E_n} \right)^{L+1} = \left. \frac{\partial^{L+1}}{\partial s^{L+1}} \exp\left( s \sum_n A_n e^{-E_n} \right) \right|_{s=0}$. Pulling this derivative outside the integral, all remaining terms in the integrand become strictly exponential and linear in the amplitudes $A_n$.

    To simplify the notation, we define the Boltzmann-weighted local charges $\tilde{v}_j^{(n)} = e^{-E_n} v_j^{(n)}$ and their corresponding weighted targets $\tilde{q}_j^{(n)} = e^{-E_n} q_j$. Collecting all terms within the exponent into the form $e^{-\sum_n w_n(s) A_n}$, we identify the resulting poles:
    \begin{equation}
        w_n(s) = -i\nu - s e^{-E_n} - i\sum_j z_j \left( \tilde{v}_j^{(n)} - \tilde{q}_j^{(n)} \right).
    \end{equation}

    Because the physical amplitudes are strictly non-negative, $A_n \in [0, \infty)$, integration along the real axes for $\nu$ and $z_j$ yields a real part $\text{Re}(w_n(s)) = -s e^{-E_n}$. For the derivative at $s=0$ to be mathematically well-defined, the amplitude integrals must be absolutely convergent in an open neighborhood around $s=0$; however, any small $s > 0$ triggers an immediate divergence along the real axis. To guarantee absolute convergence and establish a stable domain for the source field, we invoke Cauchy's integral theorem to analytically continue the $\nu$ contour into the upper half of the complex plane, namely $\nu \to \nu + i\epsilon$, with $\epsilon > 0$. By deforming the multi-dimensional contour such that $\text{Re}(w_n) = \epsilon - s e^{-E_n} > 0$ for all $n$, the integrals over $A_n$ become strictly convergent Laplace transforms.

    Evaluating these exact integrals over $A_n$ yields:
    \begin{multline}
        Z_{\mathrm{mc}} = \mathcal{D}! \left. \frac{\partial^{L+1}}{\partial s^{L+1}} \int_{-\infty+i\epsilon}^{\infty+i\epsilon} \frac{\mathrm{d}\nu}{2\pi} \right. \\
        \left. \int_{\mathbb{R}^L}\left[ \prod_{j} \frac{\mathrm{d}z_j}{2\pi} \right] e^{-i\nu}     \prod_n \frac{1}{w_n(s)} \right|_{s=0}.
    \end{multline}

    To evaluate this expression, it is natural to treat the complex $\nu$-integral as an inverse Laplace transform. Introducing the auxiliary complex variable $\tilde{\nu} = -i\nu$, the integration maps exactly to a vertical Bromwich contour extending from $\epsilon - i\infty$ to $\epsilon + i\infty$:

    \begin{multline}
        I(s, \mathbf{z}) =\\ \int_{\epsilon - i\infty}^{\epsilon + i\infty} \frac{\mathrm{d}\tilde{\nu}}{2\pi i} \frac{e^{\tilde{\nu}}}{\prod_n \left( \tilde{\nu} - \left[s e^{-E_n} + i \mathbf{z} \cdot (\tilde{\mathbf{v}}^{(n)} - \tilde{\mathbf{q}}^{(n)})\right] \right)}.
    \end{multline}
    This is precisely the inverse Laplace transform of a product of simple poles, evaluated at the Laplace parameter $t=1$. Applying Cauchy's Residue Theorem to this contour integral yields the discrete divided difference of the exponential function across the set of poles $X_n = s e^{-E_n} + i \mathbf{z} \cdot (\tilde{\mathbf{v}}^{(n)} - \tilde{\mathbf{q}}^{(n)})$:

    \begin{equation}
        I(s, \mathbf{z}) = \sum_{n} \frac{e^{X_n}}{\prod_{m \neq n} (X_n - X_m)}.
    \end{equation}

    We can map this discrete sum to a continuous geometry using the Hermite-Genocchi formula~\cite{Micchelli79}. This theorem rigorously transforms the divided difference into a bounded integral over the standard $(\mathcal{D}-1)$-dimensional probability simplex $\Delta^{\mathcal{D}-1}$, defined by $t_n \geq 0$ and $\sum_n t_n = 1$:

    \begin{multline}
        I(s, \mathbf{z}) = \int_{\Delta^{\mathcal{D}-1}} \mathrm{d}^{\mathcal{D}-1}\mathbf{t} \, \exp\left( s \sum_n t_n e^{-E_n} \right)\\ \times \exp\left( i \mathbf{z} \cdot \sum_n t_n (\tilde{\mathbf{v}}^{(n)} - \tilde{\mathbf{q}}^{(n)}) \right).
    \end{multline}
    Substituting this evaluated characteristic function back into the formula for $Z_{\rm mc}$, we can now safely perform the remaining $L$-dimensional Fourier transform over the real spatial variables $\mathbf{z}$. Because $\mathbf{z}$ appears only in the oscillatory exponent, the integral trivially evaluates to an $L$-dimensional Dirac delta function, enforcing the microcanonical constraints directly on the simplex:

    \begin{multline}
        Z_{\rm mc} = \mathcal{D}! \left. \frac{\partial^{L+1}}{\partial s^{L+1}} \int_{\Delta^{\mathcal{D}-1}} \mathrm{d}^{\mathcal{D}-1}\mathbf{t} \, \exp\left( s \sum_n t_n e^{-E_n} \right) \right. \\
        \times \left. \delta\left( \sum_n t_n (\tilde{\mathbf{v}}^{(n)} - \tilde{\mathbf{q}}^{(n)}) \right) \right|_{s=0}.
    \end{multline}
    Applying the auxiliary derivative at $s=0$ and expanding the rescaled targets as $\tilde{\mathbf{q}}^{(n)} = e^{-E_n}\mathbf{q}$, we factor the argument of the Dirac delta function into a normalized geometric form:

    \begin{multline}
        Z_{\rm mc} = \mathcal{D}! \int_{\Delta^{\mathcal{D}-1}} \mathrm{d}^{\mathcal{D}-1}\mathbf{t} \left( \sum_n t_n e^{-E_n} \right)^{L+1}\\
        \times \delta\left( \left[\sum_n t_n e^{-E_n}\right] \left[ \frac{\sum_n t_n \tilde{\mathbf{v}}^{(n)}}{\sum_n t_n e^{-E_n}} - \mathbf{q} \right] \right).
    \end{multline}
    By invoking the fundamental scaling property of the multidimensional Dirac delta function, $\delta(Y \mathbf{x}) = Y^{-L} \delta(\mathbf{x})$ for an $L$-dimensional vector $\mathbf{x}$ and scalar $Y>0$, and expanding $\tilde{\mathbf{v}}^{(n)} = e^{-E_n}\mathbf{v}^{(n)}$, the microcanonical partition function reduces to a geometric projection over the simplex:

    \begin{multline}
        Z_{\rm mc}= \mathcal{D}! \int_{\Delta^{\mathcal{D}-1}} \mathrm{d}^{\mathcal{D}-1}\mathbf{t} \left( \sum_n t_n e^{-E_n} \right) \\
        \times \delta\left( \frac{\sum_n t_n e^{-E_n} \mathbf{v}^{(n)}}{\sum_n t_n e^{-E_n}} - \mathbf{q} \right).\label{eq:Z_mc_first_moment}
    \end{multline}

    Equation~\eqref{eq:Z_mc_first_moment} reveals a profound geometric structure: the integral corresponds exactly to the first weight-moment of a multivariate rational simplex spline. To formalize this identification, we recall that the canonical, multivariate B-spline $M(\mathbf{q}|V)$ is geometrically defined~\cite{Micchelli79,Prautzsch02} as the cross-sectional volume of a uniform probability simplex spanning vertices $V$, projected onto a specified coordinate $\mathbf{q}$ in a lower-dimensional affine space. This is expressed analytically as:

    \begin{multline}
        M\Big( \mathbf{q} \;\Big|\; \mathbf{v}^{(1)},\dots, \mathbf{v}^{(\mathcal{D})} \Big)=\\
        \int_{\Delta^{\mathcal{D}-1}} \mathrm{d}^{\mathcal{D}-1}\mathbf{t}
        \delta\left( {\sum_n t_n  \mathbf{v}^{(n)}} - \mathbf{q} \right).
    \end{multline}
    The rational generalization of this spline~\cite{Farin93} is obtained by assigning a strictly positive weight, $e^{-E_n}$, to each vertex. Geometrically, this weighting is achieved by embedding the simplex vertices into an $(L+1)$-dimensional space with the following transformation:

    \begin{equation}
        \hat{\mathbf{v}}^{(n)}=
        \begin{pmatrix}
            e^{-E_n}\mathbf{v}^{(n)} \\
            e^{-E_n}
        \end{pmatrix}.
    \end{equation}
    By evaluating the uniform spline in this elevated space along a ray parameterized by $u$, and projecting onto the hyperplane where the introduced weight coordinate is normalized to unity, the rational simplex spline is defined as:

    \begin{multline}
        M_{\rm rat}\Big( \mathbf{q} \;\Big|\; \mathbf{v}^{(1)},\dots, \mathbf{v}^{(\mathcal{D})}; e^{-E_{1}},\dots,e^{-E_\mathcal{D}} \Big)= \\
        \int_0^\infty \mathrm{d} u \,u^{L}M\Big( u\mathbf{q},u \;\Big|\; \hat{\mathbf{v}}^{(1)},\dots, \hat{\mathbf{v}}^{(\mathcal{D})} \Big).
    \end{multline}
    Applying this framework to Eq.~\eqref{eq:Z_mc_first_moment}, we find that the microcanonical partition function evaluates exactly to the first moment of this rational spline:

    \begin{multline}
        M_{\rm rat}^{(1)}\Big( \mathbf{q} \;\Big|\; \mathbf{v}^{(1)},\dots, \mathbf{v}^{(\mathcal{D})}; e^{-E_{1}},\dots,e^{-E_\mathcal{D}} \Big)= \\
        \int_0^\infty \mathrm{d} u \,u^{L+1}M\Big( u\mathbf{q},u \;\Big|\; \hat{\mathbf{v}}^{(1)},\dots, \hat{\mathbf{v}}^{(\mathcal{D})} \Big),
    \end{multline}
    leading to:
    \begin{equation}
        Z_{\rm mc} = \mathcal{D}! \, M^{(1)}_{\rm rat}\Big( \mathbf{q} \;\Big|\; \mathbf{v}^{(1)},\dots, \mathbf{v}^{(\mathcal{D})}; e^{-E_{1}},\dots,e^{-E_\mathcal{D}} \Big).\label{eq:Z_mc_exact}
    \end{equation}
    While the standard rational spline evaluates the normalized geometric probability density of projecting onto the constraint $\mathbf{q}$, its first moment scales this density by the total statistical weight contained within that constrained volume. This implies that $M^{(1)}_{\rm rat}$ directly measures density of the Scrooge statistical weight of the microcanonical manifold. In particular, if we fix the vertices and weights and consider $\mathbf{q}$ as the only free variable, we obtain $\int\mathrm{d}^{L}\mathbf{q}\,Z_{\rm mc}(\mathbf{q})=Z$, recovering the full partition function from the microcanonical slices.

    This geometric description provides yet another justification for the restricted typicality, with a dominant probabilistic weight being assigned to the microcanonical manifold $\mathcal{M}_r$. To see that, we note that $Z_{\rm mc}(\mathbf{q}')/Z$ represents the probability density of sampling a state with macroscopic charge profile $\mathbf{q}'$ from the unconstrained Scrooge ensemble. Using Eq.~\eqref{eq:Z_mc_first_moment}, the expectation value of this charge profile evaluates exactly to the target constraint: $\mathbb{E}^{\rm Sc}[\mathbf{q}']=\int \mathrm{d}\mathbf{q}'\,\mathbf{q}' Z_{\rm mc}(\mathbf{q'})/Z=\mathbf{q}$. Meanwhile, the statistical variance of the charge is given by:
    \begin{equation}
        \Sigma^{\rm mc}_{ij}=\mathbb{E}^{\rm Sc}\left[ \bra{\Phi}\hat q_i \ket{\Phi}\bra{\Phi}\hat q_j \ket{\Phi}\right]-q_i q_j=\Tr (\rho \hat q_i \rho \hat q_j),
    \end{equation}
    which according to previous considerations vanishes exponentially with the system size, $\Sigma^{\rm mc}_{ij}\sim \Tr(\rho^2)=e^{-a_2L}$. This means that the statistical weight of every manifold corresponding to a $\mathbf{q}'$ that does not match the ensemble average $\mathbf{q}$ is exponentially suppressed.

    Ultimately, Eq.~\eqref{eq:Z_mc_exact} establishes a rigorous dictionary between the quantum statistical mechanics of the Scrooge ensemble and the mathematics of projective geometry. The conserved physical charges $\hat{q}_j$ dictate the spatial vertices of a high-dimensional probability simplex, while the weights $e^{-E_n}$ induce a rational projective transformation that distorts the integration measure. This duality serves as a powerful reminder that measure concentration in quantum systems is fundamentally a geometric phenomenon.

\end{appendices}


\begin{thebibliography}{219}%
\makeatletter
\providecommand \@ifxundefined [1]{%
 \@ifx{#1\undefined}
}%
\providecommand \@ifnum [1]{%
 \ifnum #1\expandafter \@firstoftwo
 \else \expandafter \@secondoftwo
 \fi
}%
\providecommand \@ifx [1]{%
 \ifx #1\expandafter \@firstoftwo
 \else \expandafter \@secondoftwo
 \fi
}%
\providecommand \natexlab [1]{#1}%
\providecommand \enquote  [1]{``#1''}%
\providecommand \bibnamefont  [1]{#1}%
\providecommand \bibfnamefont [1]{#1}%
\providecommand \citenamefont [1]{#1}%
\providecommand \href@noop [0]{\@secondoftwo}%
\providecommand \href [0]{\begingroup \@sanitize@url \@href}%
\providecommand \@href[1]{\@@startlink{#1}\@@href}%
\providecommand \@@href[1]{\endgroup#1\@@endlink}%
\providecommand \@sanitize@url [0]{\catcode `\\12\catcode `\$12\catcode
  `\&12\catcode `\#12\catcode `\^12\catcode `\_12\catcode `\%12\relax}%
\providecommand \@@startlink[1]{}%
\providecommand \@@endlink[0]{}%
\providecommand \url  [0]{\begingroup\@sanitize@url \@url }%
\providecommand \@url [1]{\endgroup\@href {#1}{\urlprefix }}%
\providecommand \urlprefix  [0]{URL }%
\providecommand \Eprint [0]{\href }%
\providecommand \doibase [0]{https://doi.org/}%
\providecommand \selectlanguage [0]{\@gobble}%
\providecommand \bibinfo  [0]{\@secondoftwo}%
\providecommand \bibfield  [0]{\@secondoftwo}%
\providecommand \translation [1]{[#1]}%
\providecommand \BibitemOpen [0]{}%
\providecommand \bibitemStop [0]{}%
\providecommand \bibitemNoStop [0]{.\EOS\space}%
\providecommand \EOS [0]{\spacefactor3000\relax}%
\providecommand \BibitemShut  [1]{\csname bibitem#1\endcsname}%
\let\auto@bib@innerbib\@empty
\bibitem [{\citenamefont {Schrödinger}(1927)}]{Schrodinger27}%
  \BibitemOpen
  \bibfield  {author} {\bibinfo {author} {\bibfnamefont {E.}~\bibnamefont
  {Schrödinger}},\ }\bibfield  {title} {\bibinfo {title} {Energieaustausch
  nach der wellenmechanik},\ }\href
  {https://doi.org/https://doi.org/10.1002/andp.19273881504} {\bibfield
  {journal} {\bibinfo  {journal} {Annalen der Physik}\ }\textbf {\bibinfo
  {volume} {388}},\ \bibinfo {pages} {956} (\bibinfo {year} {1927})},\ \Eprint
  {https://arxiv.org/abs/https://onlinelibrary.wiley.com/doi/pdf/10.1002/andp.19273881504}
  {https://onlinelibrary.wiley.com/doi/pdf/10.1002/andp.19273881504}
  \BibitemShut {NoStop}%
\bibitem [{\citenamefont {Neumann}(1929)}]{vonNeumann29}%
  \BibitemOpen
  \bibfield  {author} {\bibinfo {author} {\bibfnamefont {J.~v.}\ \bibnamefont
  {Neumann}},\ }\bibfield  {title} {\bibinfo {title} {Beweis des
  {Ergodensatzes} und {desH}-{Theorems} in der neuen {Mechanik}},\ }\href
  {https://doi.org/10.1007/BF01339852} {\bibfield  {journal} {\bibinfo
  {journal} {Zeitschrift für Physik}\ }\textbf {\bibinfo {volume} {57}},\
  \bibinfo {pages} {30} (\bibinfo {year} {1929})}\BibitemShut {NoStop}%
\bibitem [{\citenamefont {von Neumann}(2010)}]{vonNeumann10}%
  \BibitemOpen
  \bibfield  {author} {\bibinfo {author} {\bibfnamefont {J.}~\bibnamefont {von
  Neumann}},\ }\bibfield  {title} {\bibinfo {title} {Proof of the ergodic
  theorem and the h-theorem in quantum mechanics},\ }\href
  {https://doi.org/10.1140/epjh/e2010-00008-5} {\bibfield  {journal} {\bibinfo
  {journal} {The European Physical Journal H}\ }\textbf {\bibinfo {volume}
  {35}},\ \bibinfo {pages} {201} (\bibinfo {year} {2010})}\BibitemShut
  {NoStop}%
\bibitem [{\citenamefont {Popescu}\ \emph {et~al.}(2006)\citenamefont
  {Popescu}, \citenamefont {Short},\ and\ \citenamefont {Winter}}]{Popescu06}%
  \BibitemOpen
  \bibfield  {author} {\bibinfo {author} {\bibfnamefont {S.}~\bibnamefont
  {Popescu}}, \bibinfo {author} {\bibfnamefont {A.~J.}\ \bibnamefont {Short}},\
  and\ \bibinfo {author} {\bibfnamefont {A.}~\bibnamefont {Winter}},\
  }\bibfield  {title} {\bibinfo {title} {Entanglement and the foundations of
  statistical mechanics},\ }\href {https://doi.org/10.1038/nphys444} {\bibfield
   {journal} {\bibinfo  {journal} {Nature Physics}\ }\textbf {\bibinfo {volume}
  {2}},\ \bibinfo {pages} {754} (\bibinfo {year} {2006})}\BibitemShut {NoStop}%
\bibitem [{\citenamefont {Goldstein}\ \emph
  {et~al.}(2006{\natexlab{a}})\citenamefont {Goldstein}, \citenamefont
  {Lebowitz}, \citenamefont {Tumulka},\ and\ \citenamefont
  {Zangh\`{\i}}}]{Goldstein06}%
  \BibitemOpen
  \bibfield  {author} {\bibinfo {author} {\bibfnamefont {S.}~\bibnamefont
  {Goldstein}}, \bibinfo {author} {\bibfnamefont {J.~L.}\ \bibnamefont
  {Lebowitz}}, \bibinfo {author} {\bibfnamefont {R.}~\bibnamefont {Tumulka}},\
  and\ \bibinfo {author} {\bibfnamefont {N.}~\bibnamefont {Zangh\`{\i}}},\
  }\bibfield  {title} {\bibinfo {title} {Canonical typicality},\ }\href
  {https://doi.org/10.1103/PhysRevLett.96.050403} {\bibfield  {journal}
  {\bibinfo  {journal} {Phys. Rev. Lett.}\ }\textbf {\bibinfo {volume} {96}},\
  \bibinfo {pages} {050403} (\bibinfo {year} {2006}{\natexlab{a}})}\BibitemShut
  {NoStop}%
\bibitem [{\citenamefont {Goldstein}\ \emph {et~al.}(2010)\citenamefont
  {Goldstein}, \citenamefont {Lebowitz}, \citenamefont {Mastrodonato},
  \citenamefont {Tumulka},\ and\ \citenamefont {Zanghì}}]{Goldstein10N}%
  \BibitemOpen
  \bibfield  {author} {\bibinfo {author} {\bibfnamefont {S.}~\bibnamefont
  {Goldstein}}, \bibinfo {author} {\bibfnamefont {J.~L.}\ \bibnamefont
  {Lebowitz}}, \bibinfo {author} {\bibfnamefont {C.}~\bibnamefont
  {Mastrodonato}}, \bibinfo {author} {\bibfnamefont {R.}~\bibnamefont
  {Tumulka}},\ and\ \bibinfo {author} {\bibfnamefont {N.}~\bibnamefont
  {Zanghì}},\ }\bibfield  {title} {\bibinfo {title} {Normal typicality and von
  neumann’s quantum ergodic theorem},\ }\href
  {https://doi.org/10.1098/rspa.2009.0635} {\bibfield  {journal} {\bibinfo
  {journal} {Proceedings of the Royal Society A: Mathematical, Physical and
  Engineering Sciences}\ }\textbf {\bibinfo {volume} {466}},\ \bibinfo {pages}
  {3203} (\bibinfo {year} {2010})},\ \Eprint
  {https://arxiv.org/abs/https://royalsocietypublishing.org/rspa/article-pdf/466/2123/3203/785392/rspa.2009.0635.pdf}
  {https://royalsocietypublishing.org/rspa/article-pdf/466/2123/3203/785392/rspa.2009.0635.pdf}
  \BibitemShut {NoStop}%
\bibitem [{\citenamefont {Gemmer}\ \emph {et~al.}(2009)\citenamefont {Gemmer},
  \citenamefont {Michel},\ and\ \citenamefont {Mahler}}]{Gemmer09}%
  \BibitemOpen
  \bibfield  {author} {\bibinfo {author} {\bibfnamefont {J.}~\bibnamefont
  {Gemmer}}, \bibinfo {author} {\bibfnamefont {M.}~\bibnamefont {Michel}},\
  and\ \bibinfo {author} {\bibfnamefont {G.}~\bibnamefont {Mahler}},\
  }\href@noop {} {\emph {\bibinfo {title} {Quantum thermodynamics: Emergence of
  thermodynamic behavior within composite quantum systems}}},\ Vol.\ \bibinfo
  {volume} {784}\ (\bibinfo  {publisher} {Springer Science \& Business Media},\
  \bibinfo {year} {2009})\BibitemShut {NoStop}%
\bibitem [{\citenamefont {Lloyd}(2013)}]{Lloyd13}%
  \BibitemOpen
  \bibfield  {author} {\bibinfo {author} {\bibfnamefont {S.}~\bibnamefont
  {Lloyd}},\ }\href {https://arxiv.org/abs/1307.0378} {\bibinfo {title} {Pure
  state quantum statistical mechanics and black holes}} (\bibinfo {year}
  {2013}),\ \Eprint {https://arxiv.org/abs/1307.0378} {arXiv:1307.0378
  [quant-ph]} \BibitemShut {NoStop}%
\bibitem [{\citenamefont {Brody}\ and\ \citenamefont
  {Hughston}(1998)}]{Brody98}%
  \BibitemOpen
  \bibfield  {author} {\bibinfo {author} {\bibfnamefont {D.~C.}\ \bibnamefont
  {Brody}}\ and\ \bibinfo {author} {\bibfnamefont {L.~P.}\ \bibnamefont
  {Hughston}},\ }\bibfield  {title} {\bibinfo {title} {The quantum canonical
  ensemble},\ }\href {https://doi.org/10.1063/1.532661} {\bibfield  {journal}
  {\bibinfo  {journal} {Journal of Mathematical Physics}\ }\textbf {\bibinfo
  {volume} {39}},\ \bibinfo {pages} {6502} (\bibinfo {year} {1998})},\ \Eprint
  {https://arxiv.org/abs/https://pubs.aip.org/aip/jmp/article-pdf/39/12/6502/19291502/6502\_1\_online.pdf}
  {https://pubs.aip.org/aip/jmp/article-pdf/39/12/6502/19291502/6502\_1\_online.pdf}
  \BibitemShut {NoStop}%
\bibitem [{\citenamefont {Bender}\ and\ \citenamefont
  {Monou}(2005)}]{Bender05}%
  \BibitemOpen
  \bibfield  {author} {\bibinfo {author} {\bibfnamefont {C.~M.}\ \bibnamefont
  {Bender}}\ and\ \bibinfo {author} {\bibfnamefont {M.}~\bibnamefont {Monou}},\
  }\bibfield  {title} {\bibinfo {title} {New quasi-exactly solvable sextic
  polynomial potentials},\ }\href {https://doi.org/10.1088/0305-4470/38/10/009}
  {\bibfield  {journal} {\bibinfo  {journal} {Journal of Physics A:
  Mathematical and General}\ }\textbf {\bibinfo {volume} {38}},\ \bibinfo
  {pages} {2179} (\bibinfo {year} {2005})}\BibitemShut {NoStop}%
\bibitem [{\citenamefont {Brody}\ \emph {et~al.}(2007)\citenamefont {Brody},
  \citenamefont {Hook},\ and\ \citenamefont {Hughston}}]{Brody07}%
  \BibitemOpen
  \bibfield  {author} {\bibinfo {author} {\bibfnamefont {D.~C.}\ \bibnamefont
  {Brody}}, \bibinfo {author} {\bibfnamefont {D.~W.}\ \bibnamefont {Hook}},\
  and\ \bibinfo {author} {\bibfnamefont {L.~P.}\ \bibnamefont {Hughston}},\
  }\bibfield  {title} {\bibinfo {title} {Quantum phase transitions without
  thermodynamic limits},\ }\href {https://doi.org/10.1098/rspa.2007.1865}
  {\bibfield  {journal} {\bibinfo  {journal} {Proceedings of the Royal Society
  A: Mathematical, Physical and Engineering Sciences}\ }\textbf {\bibinfo
  {volume} {463}},\ \bibinfo {pages} {2021} (\bibinfo {year} {2007})},\ \Eprint
  {https://arxiv.org/abs/https://royalsocietypublishing.org/rspa/article-pdf/463/2084/2021/704852/rspa.2007.1865.pdf}
  {https://royalsocietypublishing.org/rspa/article-pdf/463/2084/2021/704852/rspa.2007.1865.pdf}
  \BibitemShut {NoStop}%
\bibitem [{\citenamefont {Müller}\ \emph {et~al.}(2011)\citenamefont
  {Müller}, \citenamefont {Gross},\ and\ \citenamefont {Eisert}}]{Muller11}%
  \BibitemOpen
  \bibfield  {author} {\bibinfo {author} {\bibfnamefont {M.~P.}\ \bibnamefont
  {Müller}}, \bibinfo {author} {\bibfnamefont {D.}~\bibnamefont {Gross}},\
  and\ \bibinfo {author} {\bibfnamefont {J.}~\bibnamefont {Eisert}},\
  }\bibfield  {title} {\bibinfo {title} {Concentration of {Measure} for
  {Quantum} {States} with a {Fixed} {Expectation} {Value}},\ }\href
  {https://doi.org/10.1007/s00220-011-1205-1} {\bibfield  {journal} {\bibinfo
  {journal} {Communications in Mathematical Physics}\ }\textbf {\bibinfo
  {volume} {303}},\ \bibinfo {pages} {785} (\bibinfo {year}
  {2011})}\BibitemShut {NoStop}%
\bibitem [{\citenamefont {Ji}\ and\ \citenamefont {Fine}(2011)}]{Ji11}%
  \BibitemOpen
  \bibfield  {author} {\bibinfo {author} {\bibfnamefont {K.}~\bibnamefont
  {Ji}}\ and\ \bibinfo {author} {\bibfnamefont {B.~V.}\ \bibnamefont {Fine}},\
  }\bibfield  {title} {\bibinfo {title} {Nonthermal statistics in isolated
  quantum spin clusters after a series of perturbations},\ }\href
  {https://doi.org/10.1103/PhysRevLett.107.050401} {\bibfield  {journal}
  {\bibinfo  {journal} {Phys. Rev. Lett.}\ }\textbf {\bibinfo {volume} {107}},\
  \bibinfo {pages} {050401} (\bibinfo {year} {2011})}\BibitemShut {NoStop}%
\bibitem [{\citenamefont {Fresch}\ and\ \citenamefont {Moro}(2013)}]{Fresch13}%
  \BibitemOpen
  \bibfield  {author} {\bibinfo {author} {\bibfnamefont {B.}~\bibnamefont
  {Fresch}}\ and\ \bibinfo {author} {\bibfnamefont {G.~J.}\ \bibnamefont
  {Moro}},\ }\bibfield  {title} {\bibinfo {title} {Typical response of quantum
  pure states},\ }\href {https://doi.org/10.1140/epjb/e2013-40023-6} {\bibfield
   {journal} {\bibinfo  {journal} {The European Physical Journal B}\ }\textbf
  {\bibinfo {volume} {86}},\ \bibinfo {pages} {233} (\bibinfo {year}
  {2013})}\BibitemShut {NoStop}%
\bibitem [{\citenamefont {Reimann}(2007)}]{Reimann07}%
  \BibitemOpen
  \bibfield  {author} {\bibinfo {author} {\bibfnamefont {P.}~\bibnamefont
  {Reimann}},\ }\bibfield  {title} {\bibinfo {title} {Typicality for
  generalized microcanonical ensembles},\ }\href
  {https://doi.org/10.1103/PhysRevLett.99.160404} {\bibfield  {journal}
  {\bibinfo  {journal} {Phys. Rev. Lett.}\ }\textbf {\bibinfo {volume} {99}},\
  \bibinfo {pages} {160404} (\bibinfo {year} {2007})}\BibitemShut {NoStop}%
\bibitem [{\citenamefont {Mark}\ \emph {et~al.}(2024)\citenamefont {Mark},
  \citenamefont {Surace}, \citenamefont {Elben}, \citenamefont {Shaw},
  \citenamefont {Choi}, \citenamefont {Refael}, \citenamefont {Endres},\ and\
  \citenamefont {Choi}}]{Mark24}%
  \BibitemOpen
  \bibfield  {author} {\bibinfo {author} {\bibfnamefont {D.~K.}\ \bibnamefont
  {Mark}}, \bibinfo {author} {\bibfnamefont {F.}~\bibnamefont {Surace}},
  \bibinfo {author} {\bibfnamefont {A.}~\bibnamefont {Elben}}, \bibinfo
  {author} {\bibfnamefont {A.~L.}\ \bibnamefont {Shaw}}, \bibinfo {author}
  {\bibfnamefont {J.}~\bibnamefont {Choi}}, \bibinfo {author} {\bibfnamefont
  {G.}~\bibnamefont {Refael}}, \bibinfo {author} {\bibfnamefont
  {M.}~\bibnamefont {Endres}},\ and\ \bibinfo {author} {\bibfnamefont
  {S.}~\bibnamefont {Choi}},\ }\bibfield  {title} {\bibinfo {title} {Maximum
  {Entropy} {Principle} in {Deep} {Thermalization} and in {Hilbert}-{Space}
  {Ergodicity}},\ }\href {https://doi.org/10.1103/PhysRevX.14.041051}
  {\bibfield  {journal} {\bibinfo  {journal} {Physical Review X}\ }\textbf
  {\bibinfo {volume} {14}},\ \bibinfo {pages} {041051} (\bibinfo {year}
  {2024})}\BibitemShut {NoStop}%
\bibitem [{\citenamefont {Nakata}\ \emph {et~al.}(2012)\citenamefont {Nakata},
  \citenamefont {Turner},\ and\ \citenamefont {Murao}}]{Nakata12}%
  \BibitemOpen
  \bibfield  {author} {\bibinfo {author} {\bibfnamefont {Y.}~\bibnamefont
  {Nakata}}, \bibinfo {author} {\bibfnamefont {P.~S.}\ \bibnamefont {Turner}},\
  and\ \bibinfo {author} {\bibfnamefont {M.}~\bibnamefont {Murao}},\ }\bibfield
   {title} {\bibinfo {title} {Phase-random states: Ensembles of states with
  fixed amplitudes and uniformly distributed phases in a fixed basis},\ }\href
  {https://doi.org/10.1103/PhysRevA.86.012301} {\bibfield  {journal} {\bibinfo
  {journal} {Phys. Rev. A}\ }\textbf {\bibinfo {volume} {86}},\ \bibinfo
  {pages} {012301} (\bibinfo {year} {2012})}\BibitemShut {NoStop}%
\bibitem [{\citenamefont {Pilatowsky-Cameo}\ \emph {et~al.}(2023)\citenamefont
  {Pilatowsky-Cameo}, \citenamefont {Dag}, \citenamefont {Ho},\ and\
  \citenamefont {Choi}}]{Pilatowsky-Cameo23}%
  \BibitemOpen
  \bibfield  {author} {\bibinfo {author} {\bibfnamefont {S.}~\bibnamefont
  {Pilatowsky-Cameo}}, \bibinfo {author} {\bibfnamefont {C.~B.}\ \bibnamefont
  {Dag}}, \bibinfo {author} {\bibfnamefont {W.~W.}\ \bibnamefont {Ho}},\ and\
  \bibinfo {author} {\bibfnamefont {S.}~\bibnamefont {Choi}},\ }\bibfield
  {title} {\bibinfo {title} {Complete hilbert-space ergodicity in quantum
  dynamics of generalized fibonacci drives},\ }\href
  {https://doi.org/10.1103/PhysRevLett.131.250401} {\bibfield  {journal}
  {\bibinfo  {journal} {Phys. Rev. Lett.}\ }\textbf {\bibinfo {volume} {131}},\
  \bibinfo {pages} {250401} (\bibinfo {year} {2023})}\BibitemShut {NoStop}%
\bibitem [{\citenamefont {Pilatowsky-Cameo}\ \emph {et~al.}(2024)\citenamefont
  {Pilatowsky-Cameo}, \citenamefont {Marvian}, \citenamefont {Choi},\ and\
  \citenamefont {Ho}}]{Pilatowsky-Cameo24}%
  \BibitemOpen
  \bibfield  {author} {\bibinfo {author} {\bibfnamefont {S.}~\bibnamefont
  {Pilatowsky-Cameo}}, \bibinfo {author} {\bibfnamefont {I.}~\bibnamefont
  {Marvian}}, \bibinfo {author} {\bibfnamefont {S.}~\bibnamefont {Choi}},\ and\
  \bibinfo {author} {\bibfnamefont {W.~W.}\ \bibnamefont {Ho}},\ }\bibfield
  {title} {\bibinfo {title} {Hilbert-space ergodicity in driven quantum
  systems: Obstructions and designs},\ }\href
  {https://doi.org/10.1103/PhysRevX.14.041059} {\bibfield  {journal} {\bibinfo
  {journal} {Phys. Rev. X}\ }\textbf {\bibinfo {volume} {14}},\ \bibinfo
  {pages} {041059} (\bibinfo {year} {2024})}\BibitemShut {NoStop}%
\bibitem [{\citenamefont {Bartsch}\ and\ \citenamefont
  {Gemmer}(2009)}]{Bartsch09}%
  \BibitemOpen
  \bibfield  {author} {\bibinfo {author} {\bibfnamefont {C.}~\bibnamefont
  {Bartsch}}\ and\ \bibinfo {author} {\bibfnamefont {J.}~\bibnamefont
  {Gemmer}},\ }\bibfield  {title} {\bibinfo {title} {Dynamical typicality of
  quantum expectation values},\ }\href
  {https://doi.org/10.1103/PhysRevLett.102.110403} {\bibfield  {journal}
  {\bibinfo  {journal} {Phys. Rev. Lett.}\ }\textbf {\bibinfo {volume} {102}},\
  \bibinfo {pages} {110403} (\bibinfo {year} {2009})}\BibitemShut {NoStop}%
\bibitem [{\citenamefont {White}(2009)}]{White09me}%
  \BibitemOpen
  \bibfield  {author} {\bibinfo {author} {\bibfnamefont {S.~R.}\ \bibnamefont
  {White}},\ }\bibfield  {title} {\bibinfo {title} {Minimally entangled typical
  quantum states at finite temperature},\ }\href
  {https://doi.org/10.1103/PhysRevLett.102.190601} {\bibfield  {journal}
  {\bibinfo  {journal} {Phys. Rev. Lett.}\ }\textbf {\bibinfo {volume} {102}},\
  \bibinfo {pages} {190601} (\bibinfo {year} {2009})}\BibitemShut {NoStop}%
\bibitem [{\citenamefont {Sugiura}\ and\ \citenamefont
  {Shimizu}(2012)}]{Sugiura12}%
  \BibitemOpen
  \bibfield  {author} {\bibinfo {author} {\bibfnamefont {S.}~\bibnamefont
  {Sugiura}}\ and\ \bibinfo {author} {\bibfnamefont {A.}~\bibnamefont
  {Shimizu}},\ }\bibfield  {title} {\bibinfo {title} {Thermal pure quantum
  states at finite temperature},\ }\href
  {https://doi.org/10.1103/PhysRevLett.108.240401} {\bibfield  {journal}
  {\bibinfo  {journal} {Phys. Rev. Lett.}\ }\textbf {\bibinfo {volume} {108}},\
  \bibinfo {pages} {240401} (\bibinfo {year} {2012})}\BibitemShut {NoStop}%
\bibitem [{\citenamefont {Wietek}\ \emph {et~al.}(2021)\citenamefont {Wietek},
  \citenamefont {He}, \citenamefont {White}, \citenamefont {Georges},\ and\
  \citenamefont {Stoudenmire}}]{Wietek21}%
  \BibitemOpen
  \bibfield  {author} {\bibinfo {author} {\bibfnamefont {A.}~\bibnamefont
  {Wietek}}, \bibinfo {author} {\bibfnamefont {Y.-Y.}\ \bibnamefont {He}},
  \bibinfo {author} {\bibfnamefont {S.~R.}\ \bibnamefont {White}}, \bibinfo
  {author} {\bibfnamefont {A.}~\bibnamefont {Georges}},\ and\ \bibinfo {author}
  {\bibfnamefont {E.~M.}\ \bibnamefont {Stoudenmire}},\ }\bibfield  {title}
  {\bibinfo {title} {Stripes, antiferromagnetism, and the pseudogap in the
  doped hubbard model at finite temperature},\ }\href
  {https://doi.org/10.1103/PhysRevX.11.031007} {\bibfield  {journal} {\bibinfo
  {journal} {Phys. Rev. X}\ }\textbf {\bibinfo {volume} {11}},\ \bibinfo
  {pages} {031007} (\bibinfo {year} {2021})}\BibitemShut {NoStop}%
\bibitem [{\citenamefont {Calabrese}\ and\ \citenamefont
  {Cardy}(2004)}]{Calabrese04}%
  \BibitemOpen
  \bibfield  {author} {\bibinfo {author} {\bibfnamefont {P.}~\bibnamefont
  {Calabrese}}\ and\ \bibinfo {author} {\bibfnamefont {J.}~\bibnamefont
  {Cardy}},\ }\bibfield  {title} {\bibinfo {title} {Entanglement entropy and
  quantum field theory},\ }\href
  {https://doi.org/10.1088/1742-5468/2004/06/P06002} {\bibfield  {journal}
  {\bibinfo  {journal} {Journal of Statistical Mechanics: Theory and
  Experiment}\ }\textbf {\bibinfo {volume} {2004}},\ \bibinfo {pages} {P06002}
  (\bibinfo {year} {2004})}\BibitemShut {NoStop}%
\bibitem [{\citenamefont {Daley}\ \emph {et~al.}(2012)\citenamefont {Daley},
  \citenamefont {Pichler}, \citenamefont {Schachenmayer},\ and\ \citenamefont
  {Zoller}}]{Daley12}%
  \BibitemOpen
  \bibfield  {author} {\bibinfo {author} {\bibfnamefont {A.~J.}\ \bibnamefont
  {Daley}}, \bibinfo {author} {\bibfnamefont {H.}~\bibnamefont {Pichler}},
  \bibinfo {author} {\bibfnamefont {J.}~\bibnamefont {Schachenmayer}},\ and\
  \bibinfo {author} {\bibfnamefont {P.}~\bibnamefont {Zoller}},\ }\bibfield
  {title} {\bibinfo {title} {Measuring entanglement growth in quench dynamics
  of bosons in an optical lattice},\ }\href
  {https://doi.org/10.1103/PhysRevLett.109.020505} {\bibfield  {journal}
  {\bibinfo  {journal} {Phys. Rev. Lett.}\ }\textbf {\bibinfo {volume} {109}},\
  \bibinfo {pages} {020505} (\bibinfo {year} {2012})}\BibitemShut {NoStop}%
\bibitem [{\citenamefont {Islam}\ \emph {et~al.}(2015)\citenamefont {Islam},
  \citenamefont {Ma}, \citenamefont {Preiss}, \citenamefont {Eric~Tai},
  \citenamefont {Lukin}, \citenamefont {Rispoli},\ and\ \citenamefont
  {Greiner}}]{Islam15}%
  \BibitemOpen
  \bibfield  {author} {\bibinfo {author} {\bibfnamefont {R.}~\bibnamefont
  {Islam}}, \bibinfo {author} {\bibfnamefont {R.}~\bibnamefont {Ma}}, \bibinfo
  {author} {\bibfnamefont {P.~M.}\ \bibnamefont {Preiss}}, \bibinfo {author}
  {\bibfnamefont {M.}~\bibnamefont {Eric~Tai}}, \bibinfo {author}
  {\bibfnamefont {A.}~\bibnamefont {Lukin}}, \bibinfo {author} {\bibfnamefont
  {M.}~\bibnamefont {Rispoli}},\ and\ \bibinfo {author} {\bibfnamefont
  {M.}~\bibnamefont {Greiner}},\ }\bibfield  {title} {\bibinfo {title}
  {Measuring entanglement entropy in a quantum many-body system},\ }\href
  {https://doi.org/10.1038/nature15750} {\bibfield  {journal} {\bibinfo
  {journal} {Nature}\ }\textbf {\bibinfo {volume} {528}},\ \bibinfo {pages} {77
  EP } (\bibinfo {year} {2015})}\BibitemShut {NoStop}%
\bibitem [{\citenamefont {Brydges}\ \emph {et~al.}(2019)\citenamefont
  {Brydges}, \citenamefont {Elben}, \citenamefont {Jurcevic}, \citenamefont
  {Vermersch}, \citenamefont {Maier}, \citenamefont {Lanyon}, \citenamefont
  {Zoller}, \citenamefont {Blatt},\ and\ \citenamefont {Roos}}]{Brydges19}%
  \BibitemOpen
  \bibfield  {author} {\bibinfo {author} {\bibfnamefont {T.}~\bibnamefont
  {Brydges}}, \bibinfo {author} {\bibfnamefont {A.}~\bibnamefont {Elben}},
  \bibinfo {author} {\bibfnamefont {P.}~\bibnamefont {Jurcevic}}, \bibinfo
  {author} {\bibfnamefont {B.}~\bibnamefont {Vermersch}}, \bibinfo {author}
  {\bibfnamefont {C.}~\bibnamefont {Maier}}, \bibinfo {author} {\bibfnamefont
  {B.~P.}\ \bibnamefont {Lanyon}}, \bibinfo {author} {\bibfnamefont
  {P.}~\bibnamefont {Zoller}}, \bibinfo {author} {\bibfnamefont
  {R.}~\bibnamefont {Blatt}},\ and\ \bibinfo {author} {\bibfnamefont {C.~F.}\
  \bibnamefont {Roos}},\ }\bibfield  {title} {\bibinfo {title} {Probing rényi
  entanglement entropy via randomized measurements},\ }\href
  {https://doi.org/10.1126/science.aau4963} {\bibfield  {journal} {\bibinfo
  {journal} {Science}\ }\textbf {\bibinfo {volume} {364}},\ \bibinfo {pages}
  {260} (\bibinfo {year} {2019})},\ \Eprint
  {https://arxiv.org/abs/https://www.science.org/doi/pdf/10.1126/science.aau4963}
  {https://www.science.org/doi/pdf/10.1126/science.aau4963} \BibitemShut
  {NoStop}%
\bibitem [{\citenamefont {Bittel}\ \emph {et~al.}(2026)\citenamefont {Bittel},
  \citenamefont {Eisert}, \citenamefont {Leone}, \citenamefont {Mele},\ and\
  \citenamefont {Oliviero}}]{Bittel2026complete}%
  \BibitemOpen
  \bibfield  {author} {\bibinfo {author} {\bibfnamefont {L.}~\bibnamefont
  {Bittel}}, \bibinfo {author} {\bibfnamefont {J.}~\bibnamefont {Eisert}},
  \bibinfo {author} {\bibfnamefont {L.}~\bibnamefont {Leone}}, \bibinfo
  {author} {\bibfnamefont {A.~A.}\ \bibnamefont {Mele}},\ and\ \bibinfo
  {author} {\bibfnamefont {S.~F.}\ \bibnamefont {Oliviero}},\ }\bibfield
  {title} {\bibinfo {title} {A complete theory of the {C}lifford commutant},\
  }\href {https://doi.org/10.22331/q-2026-07-22-2171} {\bibfield  {journal}
  {\bibinfo  {journal} {{Quantum}}\ }\textbf {\bibinfo {volume} {10}},\
  \bibinfo {pages} {2171} (\bibinfo {year} {2026})}\BibitemShut {NoStop}%
\bibitem [{\citenamefont {Deneris}\ \emph {et~al.}(2026)\citenamefont
  {Deneris}, \citenamefont {Braccia}, \citenamefont {Bermejo}, \citenamefont
  {Diaz}, \citenamefont {Mele},\ and\ \citenamefont
  {Cerezo}}]{Deneris26comparing}%
  \BibitemOpen
  \bibfield  {author} {\bibinfo {author} {\bibfnamefont {A.~E.}\ \bibnamefont
  {Deneris}}, \bibinfo {author} {\bibfnamefont {P.}~\bibnamefont {Braccia}},
  \bibinfo {author} {\bibfnamefont {P.}~\bibnamefont {Bermejo}}, \bibinfo
  {author} {\bibfnamefont {N.~L.}\ \bibnamefont {Diaz}}, \bibinfo {author}
  {\bibfnamefont {A.~A.}\ \bibnamefont {Mele}},\ and\ \bibinfo {author}
  {\bibfnamefont {M.}~\bibnamefont {Cerezo}},\ }\bibfield  {title} {\bibinfo
  {title} {Analyzing the free states of one quantum resource theory as resource
  states of another},\ }\href
  {https://doi.org/https://doi.org/10.1002/qute.202500702} {\bibfield
  {journal} {\bibinfo  {journal} {Advanced Quantum Technologies}\ }\textbf
  {\bibinfo {volume} {9}},\ \bibinfo {pages} {e00702} (\bibinfo {year}
  {2026})}\BibitemShut {NoStop}%
\bibitem [{\citenamefont {Sierant}\ \emph
  {et~al.}(2026{\natexlab{a}})\citenamefont {Sierant}, \citenamefont
  {Turkeshi},\ and\ \citenamefont {Tarabunga}}]{Sierant26comm}%
  \BibitemOpen
  \bibfield  {author} {\bibinfo {author} {\bibfnamefont {P.}~\bibnamefont
  {Sierant}}, \bibinfo {author} {\bibfnamefont {X.}~\bibnamefont {Turkeshi}},\
  and\ \bibinfo {author} {\bibfnamefont {P.~S.}\ \bibnamefont {Tarabunga}},\
  }\href {https://arxiv.org/abs/2603.12392} {\bibinfo {title} {Theory of the
  matchgate commutant}} (\bibinfo {year} {2026}{\natexlab{a}}),\ \Eprint
  {https://arxiv.org/abs/2603.12392} {arXiv:2603.12392 [quant-ph]} \BibitemShut
  {NoStop}%
\bibitem [{\citenamefont {Tang}\ \emph {et~al.}(2026)\citenamefont {Tang},
  \citenamefont {Zhu}, \citenamefont {Zhang}, \citenamefont {Eisert},
  \citenamefont {Liu}, \citenamefont {Roth}, \citenamefont {Gühne},
  \citenamefont {Wang},\ and\ \citenamefont {Liu}}]{Tang26witness}%
  \BibitemOpen
  \bibfield  {author} {\bibinfo {author} {\bibfnamefont {Y.}~\bibnamefont
  {Tang}}, \bibinfo {author} {\bibfnamefont {C.}~\bibnamefont {Zhu}}, \bibinfo
  {author} {\bibfnamefont {Y.}~\bibnamefont {Zhang}}, \bibinfo {author}
  {\bibfnamefont {J.}~\bibnamefont {Eisert}}, \bibinfo {author} {\bibfnamefont
  {Z.-W.}\ \bibnamefont {Liu}}, \bibinfo {author} {\bibfnamefont
  {I.}~\bibnamefont {Roth}}, \bibinfo {author} {\bibfnamefont {O.}~\bibnamefont
  {Gühne}}, \bibinfo {author} {\bibfnamefont {X.}~\bibnamefont {Wang}},\ and\
  \bibinfo {author} {\bibfnamefont {Z.}~\bibnamefont {Liu}},\ }\href
  {https://arxiv.org/abs/2606.27105} {\bibinfo {title} {Witness expansion: A
  unified framework for analytical and measurable mixed-state resource
  detection}} (\bibinfo {year} {2026}),\ \Eprint
  {https://arxiv.org/abs/2606.27105} {arXiv:2606.27105 [quant-ph]} \BibitemShut
  {NoStop}%
\bibitem [{\citenamefont {Cotler}\ \emph {et~al.}(2023)\citenamefont {Cotler},
  \citenamefont {Mark}, \citenamefont {Huang}, \citenamefont {Hern\'andez},
  \citenamefont {Choi}, \citenamefont {Shaw}, \citenamefont {Endres},\ and\
  \citenamefont {Choi}}]{Cotler23}%
  \BibitemOpen
  \bibfield  {author} {\bibinfo {author} {\bibfnamefont {J.~S.}\ \bibnamefont
  {Cotler}}, \bibinfo {author} {\bibfnamefont {D.~K.}\ \bibnamefont {Mark}},
  \bibinfo {author} {\bibfnamefont {H.-Y.}\ \bibnamefont {Huang}}, \bibinfo
  {author} {\bibfnamefont {F.}~\bibnamefont {Hern\'andez}}, \bibinfo {author}
  {\bibfnamefont {J.}~\bibnamefont {Choi}}, \bibinfo {author} {\bibfnamefont
  {A.~L.}\ \bibnamefont {Shaw}}, \bibinfo {author} {\bibfnamefont
  {M.}~\bibnamefont {Endres}},\ and\ \bibinfo {author} {\bibfnamefont
  {S.}~\bibnamefont {Choi}},\ }\bibfield  {title} {\bibinfo {title} {Emergent
  quantum state designs from individual many-body wave functions},\ }\href
  {https://doi.org/10.1103/PRXQuantum.4.010311} {\bibfield  {journal} {\bibinfo
   {journal} {PRX Quantum}\ }\textbf {\bibinfo {volume} {4}},\ \bibinfo {pages}
  {010311} (\bibinfo {year} {2023})}\BibitemShut {NoStop}%
\bibitem [{\citenamefont {Claeys}\ and\ \citenamefont
  {Lamacraft}(2022)}]{Claeys22}%
  \BibitemOpen
  \bibfield  {author} {\bibinfo {author} {\bibfnamefont {P.~W.}\ \bibnamefont
  {Claeys}}\ and\ \bibinfo {author} {\bibfnamefont {A.}~\bibnamefont
  {Lamacraft}},\ }\bibfield  {title} {\bibinfo {title} {Emergent quantum state
  designs and biunitarity in dual-unitary circuit dynamics},\ }\href
  {https://doi.org/10.22331/q-2022-06-15-738} {\bibfield  {journal} {\bibinfo
  {journal} {Quantum}\ }\textbf {\bibinfo {volume} {6}},\ \bibinfo {pages}
  {738} (\bibinfo {year} {2022})}\BibitemShut {NoStop}%
\bibitem [{\citenamefont {Ippoliti}\ and\ \citenamefont
  {Ho}(2022)}]{Ippoliti22}%
  \BibitemOpen
  \bibfield  {author} {\bibinfo {author} {\bibfnamefont {M.}~\bibnamefont
  {Ippoliti}}\ and\ \bibinfo {author} {\bibfnamefont {W.~W.}\ \bibnamefont
  {Ho}},\ }\bibfield  {title} {\bibinfo {title} {Solvable model of deep
  thermalization with distinct design times},\ }\href
  {https://doi.org/10.22331/q-2022-12-29-886} {\bibfield  {journal} {\bibinfo
  {journal} {Quantum}\ }\textbf {\bibinfo {volume} {6}},\ \bibinfo {pages}
  {886} (\bibinfo {year} {2022})}\BibitemShut {NoStop}%
\bibitem [{\citenamefont {Choi}\ \emph {et~al.}(2023)\citenamefont {Choi},
  \citenamefont {Shaw}, \citenamefont {Madjarov}, \citenamefont {Xie},
  \citenamefont {Finkelstein}, \citenamefont {Covey}, \citenamefont {Cotler},
  \citenamefont {Mark}, \citenamefont {Huang}, \citenamefont {Kale},
  \citenamefont {Pichler}, \citenamefont {Brandão}, \citenamefont {Choi},\
  and\ \citenamefont {Endres}}]{Choi23}%
  \BibitemOpen
  \bibfield  {author} {\bibinfo {author} {\bibfnamefont {J.}~\bibnamefont
  {Choi}}, \bibinfo {author} {\bibfnamefont {A.~L.}\ \bibnamefont {Shaw}},
  \bibinfo {author} {\bibfnamefont {I.~S.}\ \bibnamefont {Madjarov}}, \bibinfo
  {author} {\bibfnamefont {X.}~\bibnamefont {Xie}}, \bibinfo {author}
  {\bibfnamefont {R.}~\bibnamefont {Finkelstein}}, \bibinfo {author}
  {\bibfnamefont {J.~P.}\ \bibnamefont {Covey}}, \bibinfo {author}
  {\bibfnamefont {J.~S.}\ \bibnamefont {Cotler}}, \bibinfo {author}
  {\bibfnamefont {D.~K.}\ \bibnamefont {Mark}}, \bibinfo {author}
  {\bibfnamefont {H.-Y.}\ \bibnamefont {Huang}}, \bibinfo {author}
  {\bibfnamefont {A.}~\bibnamefont {Kale}}, \bibinfo {author} {\bibfnamefont
  {H.}~\bibnamefont {Pichler}}, \bibinfo {author} {\bibfnamefont {F.~G. S.~L.}\
  \bibnamefont {Brandão}}, \bibinfo {author} {\bibfnamefont {S.}~\bibnamefont
  {Choi}},\ and\ \bibinfo {author} {\bibfnamefont {M.}~\bibnamefont {Endres}},\
  }\bibfield  {title} {\bibinfo {title} {Preparing random states and
  benchmarking with many-body quantum chaos},\ }\href
  {https://doi.org/10.1038/s41586-022-05442-1} {\bibfield  {journal} {\bibinfo
  {journal} {Nature}\ }\textbf {\bibinfo {volume} {613}},\ \bibinfo {pages}
  {468} (\bibinfo {year} {2023})}\BibitemShut {NoStop}%
\bibitem [{\citenamefont {Ippoliti}\ and\ \citenamefont
  {Ho}(2023)}]{Ippoliti23}%
  \BibitemOpen
  \bibfield  {author} {\bibinfo {author} {\bibfnamefont {M.}~\bibnamefont
  {Ippoliti}}\ and\ \bibinfo {author} {\bibfnamefont {W.~W.}\ \bibnamefont
  {Ho}},\ }\bibfield  {title} {\bibinfo {title} {Dynamical purification and the
  emergence of quantum state designs from the projected ensemble},\ }\href
  {https://doi.org/10.1103/PRXQuantum.4.030322} {\bibfield  {journal} {\bibinfo
   {journal} {PRX Quantum}\ }\textbf {\bibinfo {volume} {4}},\ \bibinfo {pages}
  {030322} (\bibinfo {year} {2023})}\BibitemShut {NoStop}%
\bibitem [{\citenamefont {Lucas}\ \emph {et~al.}(2023)\citenamefont {Lucas},
  \citenamefont {Piroli}, \citenamefont {De~Nardis},\ and\ \citenamefont
  {De~Luca}}]{Lucas23}%
  \BibitemOpen
  \bibfield  {author} {\bibinfo {author} {\bibfnamefont {M.}~\bibnamefont
  {Lucas}}, \bibinfo {author} {\bibfnamefont {L.}~\bibnamefont {Piroli}},
  \bibinfo {author} {\bibfnamefont {J.}~\bibnamefont {De~Nardis}},\ and\
  \bibinfo {author} {\bibfnamefont {A.}~\bibnamefont {De~Luca}},\ }\bibfield
  {title} {\bibinfo {title} {Generalized deep thermalization for free
  fermions},\ }\href {https://doi.org/10.1103/PhysRevA.107.032215} {\bibfield
  {journal} {\bibinfo  {journal} {Phys. Rev. A}\ }\textbf {\bibinfo {volume}
  {107}},\ \bibinfo {pages} {032215} (\bibinfo {year} {2023})}\BibitemShut
  {NoStop}%
\bibitem [{\citenamefont {Wilming}\ and\ \citenamefont
  {Roth}(2022)}]{Wilming22}%
  \BibitemOpen
  \bibfield  {author} {\bibinfo {author} {\bibfnamefont {H.}~\bibnamefont
  {Wilming}}\ and\ \bibinfo {author} {\bibfnamefont {I.}~\bibnamefont {Roth}},\
  }\href {https://doi.org/10.48550/arXiv.2202.01669} {\bibinfo {title}
  {High-temperature thermalization implies the emergence of quantum state
  designs}} (\bibinfo {year} {2022}),\ \bibinfo {note} {arXiv:2202.01669
  [quant-ph]}\BibitemShut {NoStop}%
\bibitem [{\citenamefont {Bhore}\ \emph {et~al.}(2023)\citenamefont {Bhore},
  \citenamefont {Desaules},\ and\ \citenamefont {Papi\ifmmode~\acute{c}\else
  \'{c}\fi{}}}]{Bhore23}%
  \BibitemOpen
  \bibfield  {author} {\bibinfo {author} {\bibfnamefont {T.}~\bibnamefont
  {Bhore}}, \bibinfo {author} {\bibfnamefont {J.-Y.}\ \bibnamefont
  {Desaules}},\ and\ \bibinfo {author} {\bibfnamefont {Z.}~\bibnamefont
  {Papi\ifmmode~\acute{c}\else \'{c}\fi{}}},\ }\bibfield  {title} {\bibinfo
  {title} {Deep thermalization in constrained quantum systems},\ }\href
  {https://doi.org/10.1103/PhysRevB.108.104317} {\bibfield  {journal} {\bibinfo
   {journal} {Phys. Rev. B}\ }\textbf {\bibinfo {volume} {108}},\ \bibinfo
  {pages} {104317} (\bibinfo {year} {2023})}\BibitemShut {NoStop}%
\bibitem [{\citenamefont {Chan}\ and\ \citenamefont {Luca}(2024)}]{Chan24}%
  \BibitemOpen
  \bibfield  {author} {\bibinfo {author} {\bibfnamefont {A.}~\bibnamefont
  {Chan}}\ and\ \bibinfo {author} {\bibfnamefont {A.~D.}\ \bibnamefont
  {Luca}},\ }\href {https://arxiv.org/abs/2402.16939} {\bibinfo {title}
  {Projected state ensemble of a generic model of many-body quantum chaos}}
  (\bibinfo {year} {2024}),\ \Eprint {https://arxiv.org/abs/2402.16939}
  {arXiv:2402.16939 [quant-ph]} \BibitemShut {NoStop}%
\bibitem [{\citenamefont {Varikuti}\ and\ \citenamefont
  {Bandyopadhyay}(2024)}]{Varikuti24}%
  \BibitemOpen
  \bibfield  {author} {\bibinfo {author} {\bibfnamefont {N.~D.}\ \bibnamefont
  {Varikuti}}\ and\ \bibinfo {author} {\bibfnamefont {S.}~\bibnamefont
  {Bandyopadhyay}},\ }\bibfield  {title} {\bibinfo {title} {Unraveling the
  emergence of quantum state designs in systems with symmetry},\ }\href
  {https://doi.org/10.22331/q-2024-08-29-1456} {\bibfield  {journal} {\bibinfo
  {journal} {Quantum}\ }\textbf {\bibinfo {volume} {8}},\ \bibinfo {pages}
  {1456} (\bibinfo {year} {2024})}\BibitemShut {NoStop}%
\bibitem [{\citenamefont {Vairogs}\ and\ \citenamefont
  {Yan}(2025)}]{Vairogs25}%
  \BibitemOpen
  \bibfield  {author} {\bibinfo {author} {\bibfnamefont {C.}~\bibnamefont
  {Vairogs}}\ and\ \bibinfo {author} {\bibfnamefont {B.}~\bibnamefont {Yan}},\
  }\bibfield  {title} {\bibinfo {title} {Extracting randomness from magic
  quantum states},\ }\href {https://doi.org/10.1103/3ttm-vhdt} {\bibfield
  {journal} {\bibinfo  {journal} {Phys. Rev. Res.}\ }\textbf {\bibinfo {volume}
  {7}},\ \bibinfo {pages} {L022069} (\bibinfo {year} {2025})}\BibitemShut
  {NoStop}%
\bibitem [{\citenamefont {Shrotriya}\ and\ \citenamefont
  {Ho}(2025)}]{Shrotriya25}%
  \BibitemOpen
  \bibfield  {author} {\bibinfo {author} {\bibfnamefont {H.}~\bibnamefont
  {Shrotriya}}\ and\ \bibinfo {author} {\bibfnamefont {W.~W.}\ \bibnamefont
  {Ho}},\ }\bibfield  {title} {\bibinfo {title} {Nonlocality of deep
  thermalization},\ }\href {https://doi.org/10.21468/SciPostPhys.18.3.107}
  {\bibfield  {journal} {\bibinfo  {journal} {SciPost Phys.}\ }\textbf
  {\bibinfo {volume} {18}},\ \bibinfo {pages} {107} (\bibinfo {year}
  {2025})}\BibitemShut {NoStop}%
\bibitem [{\citenamefont {Liu}\ and\ \citenamefont {Zhang}(2026)}]{Liu26}%
  \BibitemOpen
  \bibfield  {author} {\bibinfo {author} {\bibfnamefont {Z.}~\bibnamefont
  {Liu}}\ and\ \bibinfo {author} {\bibfnamefont {P.}~\bibnamefont {Zhang}},\
  }\href {https://arxiv.org/abs/2607.04864} {\bibinfo {title} {Emergence of the
  scrooge ensemble in the sachdev-ye-kitaev model}} (\bibinfo {year} {2026}),\
  \Eprint {https://arxiv.org/abs/2607.04864} {arXiv:2607.04864 [quant-ph]}
  \BibitemShut {NoStop}%
\bibitem [{\citenamefont {Anza}\ and\ \citenamefont {Hahn}(2026)}]{Anza26}%
  \BibitemOpen
  \bibfield  {author} {\bibinfo {author} {\bibfnamefont {F.}~\bibnamefont
  {Anza}}\ and\ \bibinfo {author} {\bibfnamefont {C.}~\bibnamefont {Hahn}},\
  }\href {https://arxiv.org/abs/2608.10110} {\bibinfo {title} {Non-equilibrium
  theory of projected ensembles}} (\bibinfo {year} {2026}),\ \Eprint
  {https://arxiv.org/abs/2608.10110} {arXiv:2608.10110 [quant-ph]} \BibitemShut
  {NoStop}%
\bibitem [{\citenamefont {Chitambar}\ and\ \citenamefont
  {Gour}(2019)}]{Chitambar19}%
  \BibitemOpen
  \bibfield  {author} {\bibinfo {author} {\bibfnamefont {E.}~\bibnamefont
  {Chitambar}}\ and\ \bibinfo {author} {\bibfnamefont {G.}~\bibnamefont
  {Gour}},\ }\bibfield  {title} {\bibinfo {title} {Quantum resource theories},\
  }\href {https://doi.org/10.1103/RevModPhys.91.025001} {\bibfield  {journal}
  {\bibinfo  {journal} {Rev. Mod. Phys.}\ }\textbf {\bibinfo {volume} {91}},\
  \bibinfo {pages} {025001} (\bibinfo {year} {2019})}\BibitemShut {NoStop}%
\bibitem [{\citenamefont {Bera}\ and\ \citenamefont
  {Schirò}(2025)}]{Bera25syk}%
  \BibitemOpen
  \bibfield  {author} {\bibinfo {author} {\bibfnamefont {S.}~\bibnamefont
  {Bera}}\ and\ \bibinfo {author} {\bibfnamefont {M.}~\bibnamefont {Schirò}},\
  }\bibfield  {title} {\bibinfo {title} {Non-stabilizerness of
  sachdev-ye-kitaev model},\ }\href
  {https://doi.org/10.21468/SciPostPhys.19.6.159} {\bibfield  {journal}
  {\bibinfo  {journal} {SciPost Phys.}\ }\textbf {\bibinfo {volume} {19}},\
  \bibinfo {pages} {159} (\bibinfo {year} {2025})}\BibitemShut {NoStop}%
\bibitem [{\citenamefont {Odavić}\ \emph {et~al.}(2023)\citenamefont
  {Odavić}, \citenamefont {Haug}, \citenamefont {Torre}, \citenamefont
  {Hamma}, \citenamefont {Franchini},\ and\ \citenamefont
  {Giampaolo}}]{Odavic24}%
  \BibitemOpen
  \bibfield  {author} {\bibinfo {author} {\bibfnamefont {J.}~\bibnamefont
  {Odavić}}, \bibinfo {author} {\bibfnamefont {T.}~\bibnamefont {Haug}},
  \bibinfo {author} {\bibfnamefont {G.}~\bibnamefont {Torre}}, \bibinfo
  {author} {\bibfnamefont {A.}~\bibnamefont {Hamma}}, \bibinfo {author}
  {\bibfnamefont {F.}~\bibnamefont {Franchini}},\ and\ \bibinfo {author}
  {\bibfnamefont {S.~M.}\ \bibnamefont {Giampaolo}},\ }\bibfield  {title}
  {\bibinfo {title} {Complexity of frustration: A new source of non-local
  non-stabilizerness},\ }\href {https://doi.org/10.21468/SciPostPhys.15.4.131}
  {\bibfield  {journal} {\bibinfo  {journal} {SciPost Phys.}\ }\textbf
  {\bibinfo {volume} {15}},\ \bibinfo {pages} {131} (\bibinfo {year}
  {2023})}\BibitemShut {NoStop}%
\bibitem [{\citenamefont {Fux}\ \emph {et~al.}(2024)\citenamefont {Fux},
  \citenamefont {Tirrito}, \citenamefont {Dalmonte},\ and\ \citenamefont
  {Fazio}}]{Fux24}%
  \BibitemOpen
  \bibfield  {author} {\bibinfo {author} {\bibfnamefont {G.~E.}\ \bibnamefont
  {Fux}}, \bibinfo {author} {\bibfnamefont {E.}~\bibnamefont {Tirrito}},
  \bibinfo {author} {\bibfnamefont {M.}~\bibnamefont {Dalmonte}},\ and\
  \bibinfo {author} {\bibfnamefont {R.}~\bibnamefont {Fazio}},\ }\bibfield
  {title} {\bibinfo {title} {Entanglement -- nonstabilizerness separation in
  hybrid quantum circuits},\ }\href
  {https://doi.org/10.1103/PhysRevResearch.6.L042030} {\bibfield  {journal}
  {\bibinfo  {journal} {Phys. Rev. Res.}\ }\textbf {\bibinfo {volume} {6}},\
  \bibinfo {pages} {L042030} (\bibinfo {year} {2024})}\BibitemShut {NoStop}%
\bibitem [{\citenamefont {Jasser}\ \emph {et~al.}(2025)\citenamefont {Jasser},
  \citenamefont {Odavi\ifmmode~\acute{c}\else \'{c}\fi{}},\ and\ \citenamefont
  {Hamma}}]{Jasser25}%
  \BibitemOpen
  \bibfield  {author} {\bibinfo {author} {\bibfnamefont {B.}~\bibnamefont
  {Jasser}}, \bibinfo {author} {\bibfnamefont {J.}~\bibnamefont
  {Odavi\ifmmode~\acute{c}\else \'{c}\fi{}}},\ and\ \bibinfo {author}
  {\bibfnamefont {A.}~\bibnamefont {Hamma}},\ }\bibfield  {title} {\bibinfo
  {title} {Stabilizer entropy and entanglement complexity in the
  sachdev-ye-kitaev model},\ }\href {https://doi.org/10.1103/rz86-47h3}
  {\bibfield  {journal} {\bibinfo  {journal} {Phys. Rev. B}\ }\textbf {\bibinfo
  {volume} {112}},\ \bibinfo {pages} {174204} (\bibinfo {year}
  {2025})}\BibitemShut {NoStop}%
\bibitem [{\citenamefont {Odavi\'{c}}\ \emph {et~al.}(2025)\citenamefont
  {Odavi\'{c}}, \citenamefont {Viscardi},\ and\ \citenamefont
  {Hamma}}]{Odavic25}%
  \BibitemOpen
  \bibfield  {author} {\bibinfo {author} {\bibfnamefont {J.}~\bibnamefont
  {Odavi\'{c}}}, \bibinfo {author} {\bibfnamefont {M.}~\bibnamefont
  {Viscardi}},\ and\ \bibinfo {author} {\bibfnamefont {A.}~\bibnamefont
  {Hamma}},\ }\bibfield  {title} {\bibinfo {title} {Stabilizer entropy in
  nonintegrable quantum evolutions},\ }\href
  {https://doi.org/10.1103/y9r6-dx7p} {\bibfield  {journal} {\bibinfo
  {journal} {Phys. Rev. B}\ }\textbf {\bibinfo {volume} {112}},\ \bibinfo
  {pages} {104301} (\bibinfo {year} {2025})}\BibitemShut {NoStop}%
\bibitem [{\citenamefont {Lami}\ \emph {et~al.}(2026)\citenamefont {Lami},
  \citenamefont {Luca}, \citenamefont {Turkeshi},\ and\ \citenamefont
  {Nardis}}]{Lami26}%
  \BibitemOpen
  \bibfield  {author} {\bibinfo {author} {\bibfnamefont {G.}~\bibnamefont
  {Lami}}, \bibinfo {author} {\bibfnamefont {A.~D.}\ \bibnamefont {Luca}},
  \bibinfo {author} {\bibfnamefont {X.}~\bibnamefont {Turkeshi}},\ and\
  \bibinfo {author} {\bibfnamefont {J.~D.}\ \bibnamefont {Nardis}},\ }\bibfield
   {title} {\bibinfo {title} {Quantum state design and emergent confinement
  mechanism in measured tensor network states},\ }\bibfield  {journal}
  {\bibinfo  {journal} {npj Quantum Information}\ }\href
  {https://doi.org/10.1038/s41534-026-01298-9} {10.1038/s41534-026-01298-9}
  (\bibinfo {year} {2026})\BibitemShut {NoStop}%
\bibitem [{\citenamefont {Magni}\ \emph {et~al.}(2025)\citenamefont {Magni},
  \citenamefont {Christopoulos}, \citenamefont {De~Luca},\ and\ \citenamefont
  {Turkeshi}}]{Magni25anti}%
  \BibitemOpen
  \bibfield  {author} {\bibinfo {author} {\bibfnamefont {B.}~\bibnamefont
  {Magni}}, \bibinfo {author} {\bibfnamefont {A.}~\bibnamefont
  {Christopoulos}}, \bibinfo {author} {\bibfnamefont {A.}~\bibnamefont
  {De~Luca}},\ and\ \bibinfo {author} {\bibfnamefont {X.}~\bibnamefont
  {Turkeshi}},\ }\bibfield  {title} {\bibinfo {title} {Anticoncentration in
  clifford circuits and beyond: From random tensor networks to pseudomagic
  states},\ }\href {https://doi.org/10.1103/p8dn-glcw} {\bibfield  {journal}
  {\bibinfo  {journal} {Phys. Rev. X}\ }\textbf {\bibinfo {volume} {15}},\
  \bibinfo {pages} {031071} (\bibinfo {year} {2025})}\BibitemShut {NoStop}%
\bibitem [{\citenamefont {Falc\~ao}\ \emph
  {et~al.}(2025{\natexlab{a}})\citenamefont {Falc\~ao}, \citenamefont
  {Tarabunga}, \citenamefont {Frau}, \citenamefont {Tirrito}, \citenamefont
  {Zakrzewski},\ and\ \citenamefont {Dalmonte}}]{Falcao25u1}%
  \BibitemOpen
  \bibfield  {author} {\bibinfo {author} {\bibfnamefont {P.~R.~N.}\
  \bibnamefont {Falc\~ao}}, \bibinfo {author} {\bibfnamefont {P.~S.}\
  \bibnamefont {Tarabunga}}, \bibinfo {author} {\bibfnamefont {M.}~\bibnamefont
  {Frau}}, \bibinfo {author} {\bibfnamefont {E.}~\bibnamefont {Tirrito}},
  \bibinfo {author} {\bibfnamefont {J.}~\bibnamefont {Zakrzewski}},\ and\
  \bibinfo {author} {\bibfnamefont {M.}~\bibnamefont {Dalmonte}},\ }\bibfield
  {title} {\bibinfo {title} {Nonstabilizerness in u(1) lattice gauge theory},\
  }\href {https://doi.org/10.1103/PhysRevB.111.L081102} {\bibfield  {journal}
  {\bibinfo  {journal} {Phys. Rev. B}\ }\textbf {\bibinfo {volume} {111}},\
  \bibinfo {pages} {L081102} (\bibinfo {year}
  {2025}{\natexlab{a}})}\BibitemShut {NoStop}%
\bibitem [{\citenamefont {Tirrito}\ \emph
  {et~al.}(2025{\natexlab{a}})\citenamefont {Tirrito}, \citenamefont
  {Tarabunga}, \citenamefont {Bhakuni}, \citenamefont {Dalmonte}, \citenamefont
  {Sierant},\ and\ \citenamefont {Turkeshi}}]{Tirrito2025universal}%
  \BibitemOpen
  \bibfield  {author} {\bibinfo {author} {\bibfnamefont {E.}~\bibnamefont
  {Tirrito}}, \bibinfo {author} {\bibfnamefont {P.~S.}\ \bibnamefont
  {Tarabunga}}, \bibinfo {author} {\bibfnamefont {D.~S.}\ \bibnamefont
  {Bhakuni}}, \bibinfo {author} {\bibfnamefont {M.}~\bibnamefont {Dalmonte}},
  \bibinfo {author} {\bibfnamefont {P.}~\bibnamefont {Sierant}},\ and\ \bibinfo
  {author} {\bibfnamefont {X.}~\bibnamefont {Turkeshi}},\ }\href
  {https://arxiv.org/abs/2506.12133} {\bibinfo {title} {Universal spreading of
  nonstabilizerness and quantum transport}} (\bibinfo {year}
  {2025}{\natexlab{a}}),\ \Eprint {https://arxiv.org/abs/2506.12133}
  {arXiv:2506.12133 [quant-ph]} \BibitemShut {NoStop}%
\bibitem [{\citenamefont {Lami}\ \emph {et~al.}(2025)\citenamefont {Lami},
  \citenamefont {De~Nardis},\ and\ \citenamefont {Turkeshi}}]{Lami25mps}%
  \BibitemOpen
  \bibfield  {author} {\bibinfo {author} {\bibfnamefont {G.}~\bibnamefont
  {Lami}}, \bibinfo {author} {\bibfnamefont {J.}~\bibnamefont {De~Nardis}},\
  and\ \bibinfo {author} {\bibfnamefont {X.}~\bibnamefont {Turkeshi}},\
  }\bibfield  {title} {\bibinfo {title} {Anticoncentration and state design of
  random tensor networks},\ }\href
  {https://doi.org/10.1103/PhysRevLett.134.010401} {\bibfield  {journal}
  {\bibinfo  {journal} {Phys. Rev. Lett.}\ }\textbf {\bibinfo {volume} {134}},\
  \bibinfo {pages} {010401} (\bibinfo {year} {2025})}\BibitemShut {NoStop}%
\bibitem [{\citenamefont {Aditya}\ \emph {et~al.}(2025)\citenamefont {Aditya},
  \citenamefont {Summer}, \citenamefont {Sierant},\ and\ \citenamefont
  {Turkeshi}}]{Aditya25mpemba}%
  \BibitemOpen
  \bibfield  {author} {\bibinfo {author} {\bibfnamefont {S.}~\bibnamefont
  {Aditya}}, \bibinfo {author} {\bibfnamefont {A.}~\bibnamefont {Summer}},
  \bibinfo {author} {\bibfnamefont {P.}~\bibnamefont {Sierant}},\ and\ \bibinfo
  {author} {\bibfnamefont {X.}~\bibnamefont {Turkeshi}},\ }\href
  {https://arxiv.org/abs/2509.22176} {\bibinfo {title} {Mpemba effects in
  quantum complexity}} (\bibinfo {year} {2025}),\ \Eprint
  {https://arxiv.org/abs/2509.22176} {arXiv:2509.22176 [quant-ph]} \BibitemShut
  {NoStop}%
\bibitem [{\citenamefont {Xiao}\ \emph {et~al.}(2026)\citenamefont {Xiao},
  \citenamefont {Zhang},\ and\ \citenamefont {Liu}}]{Xiao26non}%
  \BibitemOpen
  \bibfield  {author} {\bibinfo {author} {\bibfnamefont {Z.}~\bibnamefont
  {Xiao}}, \bibinfo {author} {\bibfnamefont {H.-K.}\ \bibnamefont {Zhang}},\
  and\ \bibinfo {author} {\bibfnamefont {S.}~\bibnamefont {Liu}},\ }\href
  {https://arxiv.org/abs/2605.04155} {\bibinfo {title} {Nonstabilizerness
  mpemba effects}} (\bibinfo {year} {2026}),\ \Eprint
  {https://arxiv.org/abs/2605.04155} {arXiv:2605.04155 [quant-ph]} \BibitemShut
  {NoStop}%
\bibitem [{\citenamefont {Falc\~ao}\ \emph
  {et~al.}(2025{\natexlab{b}})\citenamefont {Falc\~ao}, \citenamefont
  {Sierant}, \citenamefont {Zakrzewski},\ and\ \citenamefont
  {Tirrito}}]{Falcao25mbl}%
  \BibitemOpen
  \bibfield  {author} {\bibinfo {author} {\bibfnamefont {P.~R.~N.}\
  \bibnamefont {Falc\~ao}}, \bibinfo {author} {\bibfnamefont {P.}~\bibnamefont
  {Sierant}}, \bibinfo {author} {\bibfnamefont {J.}~\bibnamefont
  {Zakrzewski}},\ and\ \bibinfo {author} {\bibfnamefont {E.}~\bibnamefont
  {Tirrito}},\ }\bibfield  {title} {\bibinfo {title} {Nonstabilizerness
  dynamics in many-body localized systems},\ }\href
  {https://doi.org/10.1103/xfp5-hhs4} {\bibfield  {journal} {\bibinfo
  {journal} {Phys. Rev. Lett.}\ }\textbf {\bibinfo {volume} {135}},\ \bibinfo
  {pages} {240404} (\bibinfo {year} {2025}{\natexlab{b}})}\BibitemShut
  {NoStop}%
\bibitem [{\citenamefont {Turkeshi}\ \emph {et~al.}(2025)\citenamefont
  {Turkeshi}, \citenamefont {Tirrito},\ and\ \citenamefont
  {Sierant}}]{Turkeshi25spreading}%
  \BibitemOpen
  \bibfield  {author} {\bibinfo {author} {\bibfnamefont {X.}~\bibnamefont
  {Turkeshi}}, \bibinfo {author} {\bibfnamefont {E.}~\bibnamefont {Tirrito}},\
  and\ \bibinfo {author} {\bibfnamefont {P.}~\bibnamefont {Sierant}},\
  }\bibfield  {title} {\bibinfo {title} {Magic spreading in random quantum
  circuits},\ }\href {https://doi.org/10.1038/s41467-025-57704-x} {\bibfield
  {journal} {\bibinfo  {journal} {Nature Communications}\ }\textbf {\bibinfo
  {volume} {16}},\ \bibinfo {pages} {2575} (\bibinfo {year}
  {2025})}\BibitemShut {NoStop}%
\bibitem [{\citenamefont {Sierant}\ \emph
  {et~al.}(2026{\natexlab{b}})\citenamefont {Sierant}, \citenamefont
  {Stornati},\ and\ \citenamefont {Turkeshi}}]{Sierant26faf}%
  \BibitemOpen
  \bibfield  {author} {\bibinfo {author} {\bibfnamefont {P.}~\bibnamefont
  {Sierant}}, \bibinfo {author} {\bibfnamefont {P.}~\bibnamefont {Stornati}},\
  and\ \bibinfo {author} {\bibfnamefont {X.}~\bibnamefont {Turkeshi}},\
  }\bibfield  {title} {\bibinfo {title} {Fermionic magic resources of quantum
  many-body systems},\ }\href {https://doi.org/10.1103/3yx4-1j27} {\bibfield
  {journal} {\bibinfo  {journal} {PRX Quantum}\ }\textbf {\bibinfo {volume}
  {7}},\ \bibinfo {pages} {010302} (\bibinfo {year}
  {2026}{\natexlab{b}})}\BibitemShut {NoStop}%
\bibitem [{\citenamefont {Aditya}\ \emph
  {et~al.}(2026{\natexlab{a}})\citenamefont {Aditya}, \citenamefont
  {Turkeshi},\ and\ \citenamefont {Sierant}}]{Aditya26a}%
  \BibitemOpen
  \bibfield  {author} {\bibinfo {author} {\bibfnamefont {S.}~\bibnamefont
  {Aditya}}, \bibinfo {author} {\bibfnamefont {X.}~\bibnamefont {Turkeshi}},\
  and\ \bibinfo {author} {\bibfnamefont {P.}~\bibnamefont {Sierant}},\
  }\bibfield  {title} {\bibinfo {title} {Growth and spreading of quantum
  resources under random circuit dynamics},\ }\href
  {https://doi.org/10.1103/79vj-nx6r} {\bibfield  {journal} {\bibinfo
  {journal} {Phys. Rev. Res.}\ }\textbf {\bibinfo {volume} {8}},\ \bibinfo
  {pages} {033062} (\bibinfo {year} {2026}{\natexlab{a}})}\BibitemShut
  {NoStop}%
\bibitem [{\citenamefont {Trigueros}\ \emph {et~al.}(2026)\citenamefont
  {Trigueros}, \citenamefont {Sun}, \citenamefont {Turkeshi}, \citenamefont
  {Sierant},\ and\ \citenamefont {Tarabunga}}]{Trigueros26unitary}%
  \BibitemOpen
  \bibfield  {author} {\bibinfo {author} {\bibfnamefont {F.~B.}\ \bibnamefont
  {Trigueros}}, \bibinfo {author} {\bibfnamefont {Z.-H.}\ \bibnamefont {Sun}},
  \bibinfo {author} {\bibfnamefont {X.}~\bibnamefont {Turkeshi}}, \bibinfo
  {author} {\bibfnamefont {P.}~\bibnamefont {Sierant}},\ and\ \bibinfo {author}
  {\bibfnamefont {P.~S.}\ \bibnamefont {Tarabunga}},\ }\href
  {https://arxiv.org/abs/2606.23800} {\bibinfo {title} {Unitary designs from
  doped matchgate circuits}} (\bibinfo {year} {2026}),\ \Eprint
  {https://arxiv.org/abs/2606.23800} {arXiv:2606.23800 [quant-ph]} \BibitemShut
  {NoStop}%
\bibitem [{\citenamefont {Aditya}\ \emph
  {et~al.}(2026{\natexlab{b}})\citenamefont {Aditya}, \citenamefont {Tirrito},
  \citenamefont {Sierant},\ and\ \citenamefont {Turkeshi}}]{Aditya26coherence}%
  \BibitemOpen
  \bibfield  {author} {\bibinfo {author} {\bibfnamefont {S.}~\bibnamefont
  {Aditya}}, \bibinfo {author} {\bibfnamefont {E.}~\bibnamefont {Tirrito}},
  \bibinfo {author} {\bibfnamefont {P.}~\bibnamefont {Sierant}},\ and\ \bibinfo
  {author} {\bibfnamefont {X.}~\bibnamefont {Turkeshi}},\ }\href
  {https://arxiv.org/abs/2604.23192} {\bibinfo {title} {Coherence dynamics in
  quantum many-body systems with conservation laws}} (\bibinfo {year}
  {2026}{\natexlab{b}}),\ \Eprint {https://arxiv.org/abs/2604.23192}
  {arXiv:2604.23192 [quant-ph]} \BibitemShut {NoStop}%
\bibitem [{\citenamefont {Sierant}\ and\ \citenamefont
  {Turkeshi}(2026)}]{Sierant26magic}%
  \BibitemOpen
  \bibfield  {author} {\bibinfo {author} {\bibfnamefont {P.}~\bibnamefont
  {Sierant}}\ and\ \bibinfo {author} {\bibfnamefont {X.}~\bibnamefont
  {Turkeshi}},\ }\href {https://arxiv.org/abs/2603.00235} {\bibinfo {title}
  {Theory of magic phase transitions in encoding-decoding circuits}} (\bibinfo
  {year} {2026}),\ \Eprint {https://arxiv.org/abs/2603.00235} {arXiv:2603.00235
  [quant-ph]} \BibitemShut {NoStop}%
\bibitem [{\citenamefont {Falcão}\ \emph {et~al.}(2026)\citenamefont
  {Falcão}, \citenamefont {Zakrzewski},\ and\ \citenamefont
  {Sierant}}]{Falcao26faf}%
  \BibitemOpen
  \bibfield  {author} {\bibinfo {author} {\bibfnamefont {P.~R.~N.}\
  \bibnamefont {Falcão}}, \bibinfo {author} {\bibfnamefont {J.}~\bibnamefont
  {Zakrzewski}},\ and\ \bibinfo {author} {\bibfnamefont {P.}~\bibnamefont
  {Sierant}},\ }\href {https://arxiv.org/abs/2602.00245} {\bibinfo {title}
  {Fermionic magic resources in disordered quantum spin chains}} (\bibinfo
  {year} {2026}),\ \Eprint {https://arxiv.org/abs/2602.00245} {arXiv:2602.00245
  [quant-ph]} \BibitemShut {NoStop}%
\bibitem [{\citenamefont {Nava}\ \emph {et~al.}(2026)\citenamefont {Nava},
  \citenamefont {Artiaco}, \citenamefont {Gefen}, \citenamefont {Gornyi},
  \citenamefont {Tsitsishvili}, \citenamefont {Zazunov},\ and\ \citenamefont
  {Egger}}]{Nava26}%
  \BibitemOpen
  \bibfield  {author} {\bibinfo {author} {\bibfnamefont {A.}~\bibnamefont
  {Nava}}, \bibinfo {author} {\bibfnamefont {C.}~\bibnamefont {Artiaco}},
  \bibinfo {author} {\bibfnamefont {Y.}~\bibnamefont {Gefen}}, \bibinfo
  {author} {\bibfnamefont {I.}~\bibnamefont {Gornyi}}, \bibinfo {author}
  {\bibfnamefont {M.}~\bibnamefont {Tsitsishvili}}, \bibinfo {author}
  {\bibfnamefont {A.}~\bibnamefont {Zazunov}},\ and\ \bibinfo {author}
  {\bibfnamefont {R.}~\bibnamefont {Egger}},\ }\bibfield  {title} {\bibinfo
  {title} {Information transport and transport-induced entanglement in open
  fermion chains},\ }\href {https://doi.org/10.1103/5dks-395l} {\bibfield
  {journal} {\bibinfo  {journal} {Phys. Rev. Res.}\ } (\bibinfo {year}
  {2026})}\BibitemShut {NoStop}%
\bibitem [{\citenamefont {Aditya}\ \emph
  {et~al.}(2026{\natexlab{c}})\citenamefont {Aditya}, \citenamefont
  {Turkeshi},\ and\ \citenamefont {Sierant}}]{Aditya26equivalence}%
  \BibitemOpen
  \bibfield  {author} {\bibinfo {author} {\bibfnamefont {S.}~\bibnamefont
  {Aditya}}, \bibinfo {author} {\bibfnamefont {X.}~\bibnamefont {Turkeshi}},\
  and\ \bibinfo {author} {\bibfnamefont {P.}~\bibnamefont {Sierant}},\ }\href
  {https://arxiv.org/abs/2608.15197} {\bibinfo {title} {Equivalence of quantum
  resources under ergodic dynamics}} (\bibinfo {year} {2026}{\natexlab{c}}),\
  \Eprint {https://arxiv.org/abs/2608.15197} {arXiv:2608.15197 [quant-ph]}
  \BibitemShut {NoStop}%
\bibitem [{\citenamefont {Sekino}\ and\ \citenamefont
  {Susskind}(2008)}]{Sekino08}%
  \BibitemOpen
  \bibfield  {author} {\bibinfo {author} {\bibfnamefont {Y.}~\bibnamefont
  {Sekino}}\ and\ \bibinfo {author} {\bibfnamefont {L.}~\bibnamefont
  {Susskind}},\ }\bibfield  {title} {\bibinfo {title} {Fast scramblers},\
  }\href {https://doi.org/10.1088/1126-6708/2008/10/065} {\bibfield  {journal}
  {\bibinfo  {journal} {Journal of High Energy Physics}\ }\textbf {\bibinfo
  {volume} {2008}},\ \bibinfo {pages} {065} (\bibinfo {year}
  {2008})}\BibitemShut {NoStop}%
\bibitem [{\citenamefont {Kaufman}\ \emph {et~al.}(2016)\citenamefont
  {Kaufman}, \citenamefont {Tai}, \citenamefont {Lukin}, \citenamefont
  {Rispoli}, \citenamefont {Schittko}, \citenamefont {Preiss},\ and\
  \citenamefont {Greiner}}]{Kaufman16}%
  \BibitemOpen
  \bibfield  {author} {\bibinfo {author} {\bibfnamefont {A.~M.}\ \bibnamefont
  {Kaufman}}, \bibinfo {author} {\bibfnamefont {M.~E.}\ \bibnamefont {Tai}},
  \bibinfo {author} {\bibfnamefont {A.}~\bibnamefont {Lukin}}, \bibinfo
  {author} {\bibfnamefont {M.}~\bibnamefont {Rispoli}}, \bibinfo {author}
  {\bibfnamefont {R.}~\bibnamefont {Schittko}}, \bibinfo {author}
  {\bibfnamefont {P.~M.}\ \bibnamefont {Preiss}},\ and\ \bibinfo {author}
  {\bibfnamefont {M.}~\bibnamefont {Greiner}},\ }\bibfield  {title} {\bibinfo
  {title} {Quantum thermalization through entanglement in an isolated many-body
  system},\ }\href {https://doi.org/10.1126/science.aaf6725} {\bibfield
  {journal} {\bibinfo  {journal} {Science}\ }\textbf {\bibinfo {volume}
  {353}},\ \bibinfo {pages} {794} (\bibinfo {year} {2016})},\ \Eprint
  {https://arxiv.org/abs/https://www.science.org/doi/pdf/10.1126/science.aaf6725}
  {https://www.science.org/doi/pdf/10.1126/science.aaf6725} \BibitemShut
  {NoStop}%
\bibitem [{\citenamefont {Nahum}\ \emph {et~al.}(2018)\citenamefont {Nahum},
  \citenamefont {Vijay},\ and\ \citenamefont {Haah}}]{Nahum18}%
  \BibitemOpen
  \bibfield  {author} {\bibinfo {author} {\bibfnamefont {A.}~\bibnamefont
  {Nahum}}, \bibinfo {author} {\bibfnamefont {S.}~\bibnamefont {Vijay}},\ and\
  \bibinfo {author} {\bibfnamefont {J.}~\bibnamefont {Haah}},\ }\bibfield
  {title} {\bibinfo {title} {Operator spreading in random unitary circuits},\
  }\href {https://doi.org/10.1103/PhysRevX.8.021014} {\bibfield  {journal}
  {\bibinfo  {journal} {Phys. Rev. X}\ }\textbf {\bibinfo {volume} {8}},\
  \bibinfo {pages} {021014} (\bibinfo {year} {2018})}\BibitemShut {NoStop}%
\bibitem [{\citenamefont {Elben}\ \emph {et~al.}(2018)\citenamefont {Elben},
  \citenamefont {Vermersch}, \citenamefont {Dalmonte}, \citenamefont {Cirac},\
  and\ \citenamefont {Zoller}}]{Elben18}%
  \BibitemOpen
  \bibfield  {author} {\bibinfo {author} {\bibfnamefont {A.}~\bibnamefont
  {Elben}}, \bibinfo {author} {\bibfnamefont {B.}~\bibnamefont {Vermersch}},
  \bibinfo {author} {\bibfnamefont {M.}~\bibnamefont {Dalmonte}}, \bibinfo
  {author} {\bibfnamefont {J.~I.}\ \bibnamefont {Cirac}},\ and\ \bibinfo
  {author} {\bibfnamefont {P.}~\bibnamefont {Zoller}},\ }\bibfield  {title}
  {\bibinfo {title} {R\'enyi entropies from random quenches in atomic hubbard
  and spin models},\ }\href {https://doi.org/10.1103/PhysRevLett.120.050406}
  {\bibfield  {journal} {\bibinfo  {journal} {Phys. Rev. Lett.}\ }\textbf
  {\bibinfo {volume} {120}},\ \bibinfo {pages} {050406} (\bibinfo {year}
  {2018})}\BibitemShut {NoStop}%
\bibitem [{\citenamefont {Elben}\ \emph {et~al.}(2019)\citenamefont {Elben},
  \citenamefont {Vermersch}, \citenamefont {Roos},\ and\ \citenamefont
  {Zoller}}]{Elben19}%
  \BibitemOpen
  \bibfield  {author} {\bibinfo {author} {\bibfnamefont {A.}~\bibnamefont
  {Elben}}, \bibinfo {author} {\bibfnamefont {B.}~\bibnamefont {Vermersch}},
  \bibinfo {author} {\bibfnamefont {C.~F.}\ \bibnamefont {Roos}},\ and\
  \bibinfo {author} {\bibfnamefont {P.}~\bibnamefont {Zoller}},\ }\bibfield
  {title} {\bibinfo {title} {Statistical correlations between locally
  randomized measurements: A toolbox for probing entanglement in many-body
  quantum states},\ }\href {https://doi.org/10.1103/PhysRevA.99.052323}
  {\bibfield  {journal} {\bibinfo  {journal} {Phys. Rev. A}\ }\textbf {\bibinfo
  {volume} {99}},\ \bibinfo {pages} {052323} (\bibinfo {year}
  {2019})}\BibitemShut {NoStop}%
\bibitem [{\citenamefont {Huang}\ \emph {et~al.}(2020)\citenamefont {Huang},
  \citenamefont {Kueng},\ and\ \citenamefont {Preskill}}]{Huang20}%
  \BibitemOpen
  \bibfield  {author} {\bibinfo {author} {\bibfnamefont {H.-Y.}\ \bibnamefont
  {Huang}}, \bibinfo {author} {\bibfnamefont {R.}~\bibnamefont {Kueng}},\ and\
  \bibinfo {author} {\bibfnamefont {J.}~\bibnamefont {Preskill}},\ }\bibfield
  {title} {\bibinfo {title} {Predicting many properties of a quantum system
  from very few measurements},\ }\href
  {https://doi.org/10.1038/s41567-020-0932-7} {\bibfield  {journal} {\bibinfo
  {journal} {Nature Physics}\ }\textbf {\bibinfo {volume} {16}},\ \bibinfo
  {pages} {1050} (\bibinfo {year} {2020})}\BibitemShut {NoStop}%
\bibitem [{\citenamefont {Mi}\ \emph {et~al.}(2021)\citenamefont {Mi},
  \citenamefont {Roushan}, \citenamefont {Quintana}, \citenamefont {Mandrà},
  \citenamefont {Marshall}, \citenamefont {Neill}, \citenamefont {Arute},
  \citenamefont {Arya}, \citenamefont {Atalaya}, \citenamefont {Babbush},
  \citenamefont {Bardin}, \citenamefont {Barends}, \citenamefont {Basso},
  \citenamefont {Bengtsson}, \citenamefont {Boixo}, \citenamefont {Bourassa},
  \citenamefont {Broughton}, \citenamefont {Buckley}, \citenamefont {Buell},
  \citenamefont {Burkett}, \citenamefont {Bushnell}, \citenamefont {Chen},
  \citenamefont {Chiaro}, \citenamefont {Collins}, \citenamefont {Courtney},
  \citenamefont {Demura}, \citenamefont {Derk}, \citenamefont {Dunsworth},
  \citenamefont {Eppens}, \citenamefont {Erickson}, \citenamefont {Farhi},
  \citenamefont {Fowler}, \citenamefont {Foxen}, \citenamefont {Gidney},
  \citenamefont {Giustina}, \citenamefont {Gross}, \citenamefont {Harrigan},
  \citenamefont {Harrington}, \citenamefont {Hilton}, \citenamefont {Ho},
  \citenamefont {Hong}, \citenamefont {Huang}, \citenamefont {Huggins},
  \citenamefont {Ioffe}, \citenamefont {Isakov}, \citenamefont {Jeffrey},
  \citenamefont {Jiang}, \citenamefont {Jones}, \citenamefont {Kafri},
  \citenamefont {Kelly}, \citenamefont {Kim}, \citenamefont {Kitaev},
  \citenamefont {Klimov}, \citenamefont {Korotkov}, \citenamefont {Kostritsa},
  \citenamefont {Landhuis}, \citenamefont {Laptev}, \citenamefont {Lucero},
  \citenamefont {Martin}, \citenamefont {McClean}, \citenamefont {McCourt},
  \citenamefont {McEwen}, \citenamefont {Megrant}, \citenamefont {Miao},
  \citenamefont {Mohseni}, \citenamefont {Montazeri}, \citenamefont
  {Mruczkiewicz}, \citenamefont {Mutus}, \citenamefont {Naaman}, \citenamefont
  {Neeley}, \citenamefont {Newman}, \citenamefont {Niu}, \citenamefont
  {O’Brien}, \citenamefont {Opremcak}, \citenamefont {Ostby}, \citenamefont
  {Pato}, \citenamefont {Petukhov}, \citenamefont {Redd}, \citenamefont
  {Rubin}, \citenamefont {Sank}, \citenamefont {Satzinger}, \citenamefont
  {Shvarts}, \citenamefont {Strain}, \citenamefont {Szalay}, \citenamefont
  {Trevithick}, \citenamefont {Villalonga}, \citenamefont {White},
  \citenamefont {Yao}, \citenamefont {Yeh}, \citenamefont {Zalcman},
  \citenamefont {Neven}, \citenamefont {Aleiner}, \citenamefont {Kechedzhi},
  \citenamefont {Smelyanskiy},\ and\ \citenamefont {Chen}}]{Mi21}%
  \BibitemOpen
  \bibfield  {author} {\bibinfo {author} {\bibfnamefont {X.}~\bibnamefont
  {Mi}}, \bibinfo {author} {\bibfnamefont {P.}~\bibnamefont {Roushan}},
  \bibinfo {author} {\bibfnamefont {C.}~\bibnamefont {Quintana}}, \bibinfo
  {author} {\bibfnamefont {S.}~\bibnamefont {Mandrà}}, \bibinfo {author}
  {\bibfnamefont {J.}~\bibnamefont {Marshall}}, \bibinfo {author}
  {\bibfnamefont {C.}~\bibnamefont {Neill}}, \bibinfo {author} {\bibfnamefont
  {F.}~\bibnamefont {Arute}}, \bibinfo {author} {\bibfnamefont
  {K.}~\bibnamefont {Arya}}, \bibinfo {author} {\bibfnamefont {J.}~\bibnamefont
  {Atalaya}}, \bibinfo {author} {\bibfnamefont {R.}~\bibnamefont {Babbush}},
  \bibinfo {author} {\bibfnamefont {J.~C.}\ \bibnamefont {Bardin}}, \bibinfo
  {author} {\bibfnamefont {R.}~\bibnamefont {Barends}}, \bibinfo {author}
  {\bibfnamefont {J.}~\bibnamefont {Basso}}, \bibinfo {author} {\bibfnamefont
  {A.}~\bibnamefont {Bengtsson}}, \bibinfo {author} {\bibfnamefont
  {S.}~\bibnamefont {Boixo}}, \bibinfo {author} {\bibfnamefont
  {A.}~\bibnamefont {Bourassa}}, \bibinfo {author} {\bibfnamefont
  {M.}~\bibnamefont {Broughton}}, \bibinfo {author} {\bibfnamefont {B.~B.}\
  \bibnamefont {Buckley}}, \bibinfo {author} {\bibfnamefont {D.~A.}\
  \bibnamefont {Buell}}, \bibinfo {author} {\bibfnamefont {B.}~\bibnamefont
  {Burkett}}, \bibinfo {author} {\bibfnamefont {N.}~\bibnamefont {Bushnell}},
  \bibinfo {author} {\bibfnamefont {Z.}~\bibnamefont {Chen}}, \bibinfo {author}
  {\bibfnamefont {B.}~\bibnamefont {Chiaro}}, \bibinfo {author} {\bibfnamefont
  {R.}~\bibnamefont {Collins}}, \bibinfo {author} {\bibfnamefont
  {W.}~\bibnamefont {Courtney}}, \bibinfo {author} {\bibfnamefont
  {S.}~\bibnamefont {Demura}}, \bibinfo {author} {\bibfnamefont {A.~R.}\
  \bibnamefont {Derk}}, \bibinfo {author} {\bibfnamefont {A.}~\bibnamefont
  {Dunsworth}}, \bibinfo {author} {\bibfnamefont {D.}~\bibnamefont {Eppens}},
  \bibinfo {author} {\bibfnamefont {C.}~\bibnamefont {Erickson}}, \bibinfo
  {author} {\bibfnamefont {E.}~\bibnamefont {Farhi}}, \bibinfo {author}
  {\bibfnamefont {A.~G.}\ \bibnamefont {Fowler}}, \bibinfo {author}
  {\bibfnamefont {B.}~\bibnamefont {Foxen}}, \bibinfo {author} {\bibfnamefont
  {C.}~\bibnamefont {Gidney}}, \bibinfo {author} {\bibfnamefont
  {M.}~\bibnamefont {Giustina}}, \bibinfo {author} {\bibfnamefont {J.~A.}\
  \bibnamefont {Gross}}, \bibinfo {author} {\bibfnamefont {M.~P.}\ \bibnamefont
  {Harrigan}}, \bibinfo {author} {\bibfnamefont {S.~D.}\ \bibnamefont
  {Harrington}}, \bibinfo {author} {\bibfnamefont {J.}~\bibnamefont {Hilton}},
  \bibinfo {author} {\bibfnamefont {A.}~\bibnamefont {Ho}}, \bibinfo {author}
  {\bibfnamefont {S.}~\bibnamefont {Hong}}, \bibinfo {author} {\bibfnamefont
  {T.}~\bibnamefont {Huang}}, \bibinfo {author} {\bibfnamefont {W.~J.}\
  \bibnamefont {Huggins}}, \bibinfo {author} {\bibfnamefont {L.~B.}\
  \bibnamefont {Ioffe}}, \bibinfo {author} {\bibfnamefont {S.~V.}\ \bibnamefont
  {Isakov}}, \bibinfo {author} {\bibfnamefont {E.}~\bibnamefont {Jeffrey}},
  \bibinfo {author} {\bibfnamefont {Z.}~\bibnamefont {Jiang}}, \bibinfo
  {author} {\bibfnamefont {C.}~\bibnamefont {Jones}}, \bibinfo {author}
  {\bibfnamefont {D.}~\bibnamefont {Kafri}}, \bibinfo {author} {\bibfnamefont
  {J.}~\bibnamefont {Kelly}}, \bibinfo {author} {\bibfnamefont
  {S.}~\bibnamefont {Kim}}, \bibinfo {author} {\bibfnamefont {A.}~\bibnamefont
  {Kitaev}}, \bibinfo {author} {\bibfnamefont {P.~V.}\ \bibnamefont {Klimov}},
  \bibinfo {author} {\bibfnamefont {A.~N.}\ \bibnamefont {Korotkov}}, \bibinfo
  {author} {\bibfnamefont {F.}~\bibnamefont {Kostritsa}}, \bibinfo {author}
  {\bibfnamefont {D.}~\bibnamefont {Landhuis}}, \bibinfo {author}
  {\bibfnamefont {P.}~\bibnamefont {Laptev}}, \bibinfo {author} {\bibfnamefont
  {E.}~\bibnamefont {Lucero}}, \bibinfo {author} {\bibfnamefont
  {O.}~\bibnamefont {Martin}}, \bibinfo {author} {\bibfnamefont {J.~R.}\
  \bibnamefont {McClean}}, \bibinfo {author} {\bibfnamefont {T.}~\bibnamefont
  {McCourt}}, \bibinfo {author} {\bibfnamefont {M.}~\bibnamefont {McEwen}},
  \bibinfo {author} {\bibfnamefont {A.}~\bibnamefont {Megrant}}, \bibinfo
  {author} {\bibfnamefont {K.~C.}\ \bibnamefont {Miao}}, \bibinfo {author}
  {\bibfnamefont {M.}~\bibnamefont {Mohseni}}, \bibinfo {author} {\bibfnamefont
  {S.}~\bibnamefont {Montazeri}}, \bibinfo {author} {\bibfnamefont
  {W.}~\bibnamefont {Mruczkiewicz}}, \bibinfo {author} {\bibfnamefont
  {J.}~\bibnamefont {Mutus}}, \bibinfo {author} {\bibfnamefont
  {O.}~\bibnamefont {Naaman}}, \bibinfo {author} {\bibfnamefont
  {M.}~\bibnamefont {Neeley}}, \bibinfo {author} {\bibfnamefont
  {M.}~\bibnamefont {Newman}}, \bibinfo {author} {\bibfnamefont {M.~Y.}\
  \bibnamefont {Niu}}, \bibinfo {author} {\bibfnamefont {T.~E.}\ \bibnamefont
  {O’Brien}}, \bibinfo {author} {\bibfnamefont {A.}~\bibnamefont {Opremcak}},
  \bibinfo {author} {\bibfnamefont {E.}~\bibnamefont {Ostby}}, \bibinfo
  {author} {\bibfnamefont {B.}~\bibnamefont {Pato}}, \bibinfo {author}
  {\bibfnamefont {A.}~\bibnamefont {Petukhov}}, \bibinfo {author}
  {\bibfnamefont {N.}~\bibnamefont {Redd}}, \bibinfo {author} {\bibfnamefont
  {N.~C.}\ \bibnamefont {Rubin}}, \bibinfo {author} {\bibfnamefont
  {D.}~\bibnamefont {Sank}}, \bibinfo {author} {\bibfnamefont {K.~J.}\
  \bibnamefont {Satzinger}}, \bibinfo {author} {\bibfnamefont {V.}~\bibnamefont
  {Shvarts}}, \bibinfo {author} {\bibfnamefont {D.}~\bibnamefont {Strain}},
  \bibinfo {author} {\bibfnamefont {M.}~\bibnamefont {Szalay}}, \bibinfo
  {author} {\bibfnamefont {M.~D.}\ \bibnamefont {Trevithick}}, \bibinfo
  {author} {\bibfnamefont {B.}~\bibnamefont {Villalonga}}, \bibinfo {author}
  {\bibfnamefont {T.}~\bibnamefont {White}}, \bibinfo {author} {\bibfnamefont
  {Z.~J.}\ \bibnamefont {Yao}}, \bibinfo {author} {\bibfnamefont
  {P.}~\bibnamefont {Yeh}}, \bibinfo {author} {\bibfnamefont {A.}~\bibnamefont
  {Zalcman}}, \bibinfo {author} {\bibfnamefont {H.}~\bibnamefont {Neven}},
  \bibinfo {author} {\bibfnamefont {I.}~\bibnamefont {Aleiner}}, \bibinfo
  {author} {\bibfnamefont {K.}~\bibnamefont {Kechedzhi}}, \bibinfo {author}
  {\bibfnamefont {V.}~\bibnamefont {Smelyanskiy}},\ and\ \bibinfo {author}
  {\bibfnamefont {Y.}~\bibnamefont {Chen}},\ }\bibfield  {title} {\bibinfo
  {title} {Information scrambling in quantum circuits},\ }\href
  {https://doi.org/10.1126/science.abg5029} {\bibfield  {journal} {\bibinfo
  {journal} {Science}\ }\textbf {\bibinfo {volume} {374}},\ \bibinfo {pages}
  {1479} (\bibinfo {year} {2021})},\ \Eprint
  {https://arxiv.org/abs/https://www.science.org/doi/pdf/10.1126/science.abg5029}
  {https://www.science.org/doi/pdf/10.1126/science.abg5029} \BibitemShut
  {NoStop}%
\bibitem [{\citenamefont {Ho}\ and\ \citenamefont {Choi}(2022)}]{Ho22}%
  \BibitemOpen
  \bibfield  {author} {\bibinfo {author} {\bibfnamefont {W.~W.}\ \bibnamefont
  {Ho}}\ and\ \bibinfo {author} {\bibfnamefont {S.}~\bibnamefont {Choi}},\
  }\bibfield  {title} {\bibinfo {title} {Exact emergent quantum state designs
  from quantum chaotic dynamics},\ }\href
  {https://doi.org/10.1103/PhysRevLett.128.060601} {\bibfield  {journal}
  {\bibinfo  {journal} {Phys. Rev. Lett.}\ }\textbf {\bibinfo {volume} {128}},\
  \bibinfo {pages} {060601} (\bibinfo {year} {2022})}\BibitemShut {NoStop}%
\bibitem [{\citenamefont {Elben}\ \emph {et~al.}(2023)\citenamefont {Elben},
  \citenamefont {Flammia}, \citenamefont {Huang}, \citenamefont {Kueng},
  \citenamefont {Preskill}, \citenamefont {Vermersch},\ and\ \citenamefont
  {Zoller}}]{Elben23}%
  \BibitemOpen
  \bibfield  {author} {\bibinfo {author} {\bibfnamefont {A.}~\bibnamefont
  {Elben}}, \bibinfo {author} {\bibfnamefont {S.~T.}\ \bibnamefont {Flammia}},
  \bibinfo {author} {\bibfnamefont {H.-Y.}\ \bibnamefont {Huang}}, \bibinfo
  {author} {\bibfnamefont {R.}~\bibnamefont {Kueng}}, \bibinfo {author}
  {\bibfnamefont {J.}~\bibnamefont {Preskill}}, \bibinfo {author}
  {\bibfnamefont {B.}~\bibnamefont {Vermersch}},\ and\ \bibinfo {author}
  {\bibfnamefont {P.}~\bibnamefont {Zoller}},\ }\bibfield  {title} {\bibinfo
  {title} {The randomized measurement toolbox},\ }\href
  {https://doi.org/10.1038/s42254-022-00535-2} {\bibfield  {journal} {\bibinfo
  {journal} {Nature Reviews Physics}\ }\textbf {\bibinfo {volume} {5}},\
  \bibinfo {pages} {9} (\bibinfo {year} {2023})}\BibitemShut {NoStop}%
\bibitem [{\citenamefont {Rosenberg}\ \emph {et~al.}(2024)\citenamefont
  {Rosenberg}, \citenamefont {Andersen}, \citenamefont {Samajdar},
  \citenamefont {Petukhov}, \citenamefont {Hoke}, \citenamefont {Abanin},
  \citenamefont {Bengtsson}, \citenamefont {Drozdov}, \citenamefont {Erickson},
  \citenamefont {Klimov}, \citenamefont {Mi}, \citenamefont {Morvan},
  \citenamefont {Neeley}, \citenamefont {Neill}, \citenamefont {Acharya},
  \citenamefont {Allen}, \citenamefont {Anderson}, \citenamefont {Ansmann},
  \citenamefont {Arute}, \citenamefont {Arya}, \citenamefont {Asfaw},
  \citenamefont {Atalaya}, \citenamefont {Bardin}, \citenamefont {Bilmes},
  \citenamefont {Bortoli}, \citenamefont {Bourassa}, \citenamefont {Bovaird},
  \citenamefont {Brill}, \citenamefont {Broughton}, \citenamefont {Buckley},
  \citenamefont {Buell}, \citenamefont {Burger}, \citenamefont {Burkett},
  \citenamefont {Bushnell}, \citenamefont {Campero}, \citenamefont {Chang},
  \citenamefont {Chen}, \citenamefont {Chiaro}, \citenamefont {Chik},
  \citenamefont {Cogan}, \citenamefont {Collins}, \citenamefont {Conner},
  \citenamefont {Courtney}, \citenamefont {Crook}, \citenamefont {Curtin},
  \citenamefont {Debroy}, \citenamefont {Barba}, \citenamefont {Demura},
  \citenamefont {Paolo}, \citenamefont {Dunsworth}, \citenamefont {Earle},
  \citenamefont {Faoro}, \citenamefont {Farhi}, \citenamefont {Fatemi},
  \citenamefont {Ferreira}, \citenamefont {Burgos}, \citenamefont {Forati},
  \citenamefont {Fowler}, \citenamefont {Foxen}, \citenamefont {Garcia},
  \citenamefont {Genois}, \citenamefont {Giang}, \citenamefont {Gidney},
  \citenamefont {Gilboa}, \citenamefont {Giustina}, \citenamefont {Gosula},
  \citenamefont {Dau}, \citenamefont {Gross}, \citenamefont {Habegger},
  \citenamefont {Hamilton}, \citenamefont {Hansen}, \citenamefont {Harrigan},
  \citenamefont {Harrington}, \citenamefont {Heu}, \citenamefont {Hill},
  \citenamefont {Hoffmann}, \citenamefont {Hong}, \citenamefont {Huang},
  \citenamefont {Huff}, \citenamefont {Huggins}, \citenamefont {Ioffe},
  \citenamefont {Isakov}, \citenamefont {Iveland}, \citenamefont {Jeffrey},
  \citenamefont {Jiang}, \citenamefont {Jones}, \citenamefont {Juhas},
  \citenamefont {Kafri}, \citenamefont {Khattar}, \citenamefont {Khezri},
  \citenamefont {Kieferov{\'a}}, \citenamefont {Kim}, \citenamefont {Kitaev},
  \citenamefont {Klots}, \citenamefont {Korotkov}, \citenamefont {Kostritsa},
  \citenamefont {Kreikebaum}, \citenamefont {Landhuis}, \citenamefont {Laptev},
  \citenamefont {Lau}, \citenamefont {Laws}, \citenamefont {Lee}, \citenamefont
  {Lee}, \citenamefont {Lensky}, \citenamefont {Lester}, \citenamefont {Lill},
  \citenamefont {Liu}, \citenamefont {Locharla}, \citenamefont {Mandr{\`a}},
  \citenamefont {Martin}, \citenamefont {Martin}, \citenamefont {McClean},
  \citenamefont {McEwen}, \citenamefont {Meeks}, \citenamefont {Miao},
  \citenamefont {Mieszala}, \citenamefont {Montazeri}, \citenamefont
  {Movassagh}, \citenamefont {Mruczkiewicz}, \citenamefont {Nersisyan},
  \citenamefont {Newman}, \citenamefont {Ng}, \citenamefont {Nguyen},
  \citenamefont {Nguyen}, \citenamefont {Niu}, \citenamefont {O'Brien},
  \citenamefont {Omonije}, \citenamefont {Opremcak}, \citenamefont {Potter},
  \citenamefont {Pryadko}, \citenamefont {Quintana}, \citenamefont {Rhodes},
  \citenamefont {Rocque}, \citenamefont {Rubin}, \citenamefont {Saei},
  \citenamefont {Sank}, \citenamefont {Sankaragomathi}, \citenamefont
  {Satzinger}, \citenamefont {Schurkus}, \citenamefont {Schuster},
  \citenamefont {Shearn}, \citenamefont {Shorter}, \citenamefont {Shutty},
  \citenamefont {Shvarts}, \citenamefont {Sivak}, \citenamefont {Skruzny},
  \citenamefont {Smith}, \citenamefont {Somma}, \citenamefont {Sterling},
  \citenamefont {Strain}, \citenamefont {Szalay}, \citenamefont {Thor},
  \citenamefont {Torres}, \citenamefont {Vidal}, \citenamefont {Villalonga},
  \citenamefont {Heidweiller}, \citenamefont {White}, \citenamefont {Woo},
  \citenamefont {Xing}, \citenamefont {Yao}, \citenamefont {Yeh}, \citenamefont
  {Yoo}, \citenamefont {Young}, \citenamefont {Zalcman}, \citenamefont {Zhang},
  \citenamefont {Zhu}, \citenamefont {Zobrist}, \citenamefont {Neven},
  \citenamefont {Babbush}, \citenamefont {Bacon}, \citenamefont {Boixo},
  \citenamefont {Hilton}, \citenamefont {Lucero}, \citenamefont {Megrant},
  \citenamefont {Kelly}, \citenamefont {Chen}, \citenamefont {Smelyanskiy},
  \citenamefont {Khemani}, \citenamefont {Gopalakrishnan}, \citenamefont
  {Prosen},\ and\ \citenamefont {Roushan}}]{Rosenberg24}%
  \BibitemOpen
  \bibfield  {author} {\bibinfo {author} {\bibfnamefont {E.}~\bibnamefont
  {Rosenberg}}, \bibinfo {author} {\bibfnamefont {T.~I.}\ \bibnamefont
  {Andersen}}, \bibinfo {author} {\bibfnamefont {R.}~\bibnamefont {Samajdar}},
  \bibinfo {author} {\bibfnamefont {A.}~\bibnamefont {Petukhov}}, \bibinfo
  {author} {\bibfnamefont {J.~C.}\ \bibnamefont {Hoke}}, \bibinfo {author}
  {\bibfnamefont {D.}~\bibnamefont {Abanin}}, \bibinfo {author} {\bibfnamefont
  {A.}~\bibnamefont {Bengtsson}}, \bibinfo {author} {\bibfnamefont {I.~K.}\
  \bibnamefont {Drozdov}}, \bibinfo {author} {\bibfnamefont {C.}~\bibnamefont
  {Erickson}}, \bibinfo {author} {\bibfnamefont {P.~V.}\ \bibnamefont
  {Klimov}}, \bibinfo {author} {\bibfnamefont {X.}~\bibnamefont {Mi}}, \bibinfo
  {author} {\bibfnamefont {A.}~\bibnamefont {Morvan}}, \bibinfo {author}
  {\bibfnamefont {M.}~\bibnamefont {Neeley}}, \bibinfo {author} {\bibfnamefont
  {C.}~\bibnamefont {Neill}}, \bibinfo {author} {\bibfnamefont
  {R.}~\bibnamefont {Acharya}}, \bibinfo {author} {\bibfnamefont
  {R.}~\bibnamefont {Allen}}, \bibinfo {author} {\bibfnamefont
  {K.}~\bibnamefont {Anderson}}, \bibinfo {author} {\bibfnamefont
  {M.}~\bibnamefont {Ansmann}}, \bibinfo {author} {\bibfnamefont
  {F.}~\bibnamefont {Arute}}, \bibinfo {author} {\bibfnamefont
  {K.}~\bibnamefont {Arya}}, \bibinfo {author} {\bibfnamefont {A.}~\bibnamefont
  {Asfaw}}, \bibinfo {author} {\bibfnamefont {J.}~\bibnamefont {Atalaya}},
  \bibinfo {author} {\bibfnamefont {J.~C.}\ \bibnamefont {Bardin}}, \bibinfo
  {author} {\bibfnamefont {A.}~\bibnamefont {Bilmes}}, \bibinfo {author}
  {\bibfnamefont {G.}~\bibnamefont {Bortoli}}, \bibinfo {author} {\bibfnamefont
  {A.}~\bibnamefont {Bourassa}}, \bibinfo {author} {\bibfnamefont
  {J.}~\bibnamefont {Bovaird}}, \bibinfo {author} {\bibfnamefont
  {L.}~\bibnamefont {Brill}}, \bibinfo {author} {\bibfnamefont
  {M.}~\bibnamefont {Broughton}}, \bibinfo {author} {\bibfnamefont {B.~B.}\
  \bibnamefont {Buckley}}, \bibinfo {author} {\bibfnamefont {D.~A.}\
  \bibnamefont {Buell}}, \bibinfo {author} {\bibfnamefont {T.}~\bibnamefont
  {Burger}}, \bibinfo {author} {\bibfnamefont {B.}~\bibnamefont {Burkett}},
  \bibinfo {author} {\bibfnamefont {N.}~\bibnamefont {Bushnell}}, \bibinfo
  {author} {\bibfnamefont {J.}~\bibnamefont {Campero}}, \bibinfo {author}
  {\bibfnamefont {H.-S.}\ \bibnamefont {Chang}}, \bibinfo {author}
  {\bibfnamefont {Z.}~\bibnamefont {Chen}}, \bibinfo {author} {\bibfnamefont
  {B.}~\bibnamefont {Chiaro}}, \bibinfo {author} {\bibfnamefont
  {D.}~\bibnamefont {Chik}}, \bibinfo {author} {\bibfnamefont {J.}~\bibnamefont
  {Cogan}}, \bibinfo {author} {\bibfnamefont {R.}~\bibnamefont {Collins}},
  \bibinfo {author} {\bibfnamefont {P.}~\bibnamefont {Conner}}, \bibinfo
  {author} {\bibfnamefont {W.}~\bibnamefont {Courtney}}, \bibinfo {author}
  {\bibfnamefont {A.~L.}\ \bibnamefont {Crook}}, \bibinfo {author}
  {\bibfnamefont {B.}~\bibnamefont {Curtin}}, \bibinfo {author} {\bibfnamefont
  {D.~M.}\ \bibnamefont {Debroy}}, \bibinfo {author} {\bibfnamefont {A.~D.~T.}\
  \bibnamefont {Barba}}, \bibinfo {author} {\bibfnamefont {S.}~\bibnamefont
  {Demura}}, \bibinfo {author} {\bibfnamefont {A.~D.}\ \bibnamefont {Paolo}},
  \bibinfo {author} {\bibfnamefont {A.}~\bibnamefont {Dunsworth}}, \bibinfo
  {author} {\bibfnamefont {C.}~\bibnamefont {Earle}}, \bibinfo {author}
  {\bibfnamefont {L.}~\bibnamefont {Faoro}}, \bibinfo {author} {\bibfnamefont
  {E.}~\bibnamefont {Farhi}}, \bibinfo {author} {\bibfnamefont
  {R.}~\bibnamefont {Fatemi}}, \bibinfo {author} {\bibfnamefont {V.~S.}\
  \bibnamefont {Ferreira}}, \bibinfo {author} {\bibfnamefont {L.~F.}\
  \bibnamefont {Burgos}}, \bibinfo {author} {\bibfnamefont {E.}~\bibnamefont
  {Forati}}, \bibinfo {author} {\bibfnamefont {A.~G.}\ \bibnamefont {Fowler}},
  \bibinfo {author} {\bibfnamefont {B.}~\bibnamefont {Foxen}}, \bibinfo
  {author} {\bibfnamefont {G.}~\bibnamefont {Garcia}}, \bibinfo {author}
  {\bibfnamefont {{\'E}.}~\bibnamefont {Genois}}, \bibinfo {author}
  {\bibfnamefont {W.}~\bibnamefont {Giang}}, \bibinfo {author} {\bibfnamefont
  {C.}~\bibnamefont {Gidney}}, \bibinfo {author} {\bibfnamefont
  {D.}~\bibnamefont {Gilboa}}, \bibinfo {author} {\bibfnamefont
  {M.}~\bibnamefont {Giustina}}, \bibinfo {author} {\bibfnamefont
  {R.}~\bibnamefont {Gosula}}, \bibinfo {author} {\bibfnamefont {A.~G.}\
  \bibnamefont {Dau}}, \bibinfo {author} {\bibfnamefont {J.~A.}\ \bibnamefont
  {Gross}}, \bibinfo {author} {\bibfnamefont {S.}~\bibnamefont {Habegger}},
  \bibinfo {author} {\bibfnamefont {M.~C.}\ \bibnamefont {Hamilton}}, \bibinfo
  {author} {\bibfnamefont {M.}~\bibnamefont {Hansen}}, \bibinfo {author}
  {\bibfnamefont {M.~P.}\ \bibnamefont {Harrigan}}, \bibinfo {author}
  {\bibfnamefont {S.~D.}\ \bibnamefont {Harrington}}, \bibinfo {author}
  {\bibfnamefont {P.}~\bibnamefont {Heu}}, \bibinfo {author} {\bibfnamefont
  {G.}~\bibnamefont {Hill}}, \bibinfo {author} {\bibfnamefont {M.~R.}\
  \bibnamefont {Hoffmann}}, \bibinfo {author} {\bibfnamefont {S.}~\bibnamefont
  {Hong}}, \bibinfo {author} {\bibfnamefont {T.}~\bibnamefont {Huang}},
  \bibinfo {author} {\bibfnamefont {A.}~\bibnamefont {Huff}}, \bibinfo {author}
  {\bibfnamefont {W.~J.}\ \bibnamefont {Huggins}}, \bibinfo {author}
  {\bibfnamefont {L.~B.}\ \bibnamefont {Ioffe}}, \bibinfo {author}
  {\bibfnamefont {S.~V.}\ \bibnamefont {Isakov}}, \bibinfo {author}
  {\bibfnamefont {J.}~\bibnamefont {Iveland}}, \bibinfo {author} {\bibfnamefont
  {E.}~\bibnamefont {Jeffrey}}, \bibinfo {author} {\bibfnamefont
  {Z.}~\bibnamefont {Jiang}}, \bibinfo {author} {\bibfnamefont
  {C.}~\bibnamefont {Jones}}, \bibinfo {author} {\bibfnamefont
  {P.}~\bibnamefont {Juhas}}, \bibinfo {author} {\bibfnamefont
  {D.}~\bibnamefont {Kafri}}, \bibinfo {author} {\bibfnamefont
  {T.}~\bibnamefont {Khattar}}, \bibinfo {author} {\bibfnamefont
  {M.}~\bibnamefont {Khezri}}, \bibinfo {author} {\bibfnamefont
  {M.}~\bibnamefont {Kieferov{\'a}}}, \bibinfo {author} {\bibfnamefont
  {S.}~\bibnamefont {Kim}}, \bibinfo {author} {\bibfnamefont {A.}~\bibnamefont
  {Kitaev}}, \bibinfo {author} {\bibfnamefont {A.~R.}\ \bibnamefont {Klots}},
  \bibinfo {author} {\bibfnamefont {A.~N.}\ \bibnamefont {Korotkov}}, \bibinfo
  {author} {\bibfnamefont {F.}~\bibnamefont {Kostritsa}}, \bibinfo {author}
  {\bibfnamefont {J.~M.}\ \bibnamefont {Kreikebaum}}, \bibinfo {author}
  {\bibfnamefont {D.}~\bibnamefont {Landhuis}}, \bibinfo {author}
  {\bibfnamefont {P.}~\bibnamefont {Laptev}}, \bibinfo {author} {\bibfnamefont
  {K.-M.}\ \bibnamefont {Lau}}, \bibinfo {author} {\bibfnamefont
  {L.}~\bibnamefont {Laws}}, \bibinfo {author} {\bibfnamefont {J.}~\bibnamefont
  {Lee}}, \bibinfo {author} {\bibfnamefont {K.~W.}\ \bibnamefont {Lee}},
  \bibinfo {author} {\bibfnamefont {Y.~D.}\ \bibnamefont {Lensky}}, \bibinfo
  {author} {\bibfnamefont {B.~J.}\ \bibnamefont {Lester}}, \bibinfo {author}
  {\bibfnamefont {A.~T.}\ \bibnamefont {Lill}}, \bibinfo {author}
  {\bibfnamefont {W.}~\bibnamefont {Liu}}, \bibinfo {author} {\bibfnamefont
  {A.}~\bibnamefont {Locharla}}, \bibinfo {author} {\bibfnamefont
  {S.}~\bibnamefont {Mandr{\`a}}}, \bibinfo {author} {\bibfnamefont
  {O.}~\bibnamefont {Martin}}, \bibinfo {author} {\bibfnamefont
  {S.}~\bibnamefont {Martin}}, \bibinfo {author} {\bibfnamefont {J.~R.}\
  \bibnamefont {McClean}}, \bibinfo {author} {\bibfnamefont {M.}~\bibnamefont
  {McEwen}}, \bibinfo {author} {\bibfnamefont {S.}~\bibnamefont {Meeks}},
  \bibinfo {author} {\bibfnamefont {K.~C.}\ \bibnamefont {Miao}}, \bibinfo
  {author} {\bibfnamefont {A.}~\bibnamefont {Mieszala}}, \bibinfo {author}
  {\bibfnamefont {S.}~\bibnamefont {Montazeri}}, \bibinfo {author}
  {\bibfnamefont {R.}~\bibnamefont {Movassagh}}, \bibinfo {author}
  {\bibfnamefont {W.}~\bibnamefont {Mruczkiewicz}}, \bibinfo {author}
  {\bibfnamefont {A.}~\bibnamefont {Nersisyan}}, \bibinfo {author}
  {\bibfnamefont {M.}~\bibnamefont {Newman}}, \bibinfo {author} {\bibfnamefont
  {J.~H.}\ \bibnamefont {Ng}}, \bibinfo {author} {\bibfnamefont
  {A.}~\bibnamefont {Nguyen}}, \bibinfo {author} {\bibfnamefont
  {M.}~\bibnamefont {Nguyen}}, \bibinfo {author} {\bibfnamefont {M.~Y.}\
  \bibnamefont {Niu}}, \bibinfo {author} {\bibfnamefont {T.~E.}\ \bibnamefont
  {O'Brien}}, \bibinfo {author} {\bibfnamefont {S.}~\bibnamefont {Omonije}},
  \bibinfo {author} {\bibfnamefont {A.}~\bibnamefont {Opremcak}}, \bibinfo
  {author} {\bibfnamefont {R.}~\bibnamefont {Potter}}, \bibinfo {author}
  {\bibfnamefont {L.~P.}\ \bibnamefont {Pryadko}}, \bibinfo {author}
  {\bibfnamefont {C.}~\bibnamefont {Quintana}}, \bibinfo {author}
  {\bibfnamefont {D.~M.}\ \bibnamefont {Rhodes}}, \bibinfo {author}
  {\bibfnamefont {C.}~\bibnamefont {Rocque}}, \bibinfo {author} {\bibfnamefont
  {N.~C.}\ \bibnamefont {Rubin}}, \bibinfo {author} {\bibfnamefont
  {N.}~\bibnamefont {Saei}}, \bibinfo {author} {\bibfnamefont {D.}~\bibnamefont
  {Sank}}, \bibinfo {author} {\bibfnamefont {K.}~\bibnamefont
  {Sankaragomathi}}, \bibinfo {author} {\bibfnamefont {K.~J.}\ \bibnamefont
  {Satzinger}}, \bibinfo {author} {\bibfnamefont {H.~F.}\ \bibnamefont
  {Schurkus}}, \bibinfo {author} {\bibfnamefont {C.}~\bibnamefont {Schuster}},
  \bibinfo {author} {\bibfnamefont {M.~J.}\ \bibnamefont {Shearn}}, \bibinfo
  {author} {\bibfnamefont {A.}~\bibnamefont {Shorter}}, \bibinfo {author}
  {\bibfnamefont {N.}~\bibnamefont {Shutty}}, \bibinfo {author} {\bibfnamefont
  {V.}~\bibnamefont {Shvarts}}, \bibinfo {author} {\bibfnamefont
  {V.}~\bibnamefont {Sivak}}, \bibinfo {author} {\bibfnamefont
  {J.}~\bibnamefont {Skruzny}}, \bibinfo {author} {\bibfnamefont {W.~C.}\
  \bibnamefont {Smith}}, \bibinfo {author} {\bibfnamefont {R.~D.}\ \bibnamefont
  {Somma}}, \bibinfo {author} {\bibfnamefont {G.}~\bibnamefont {Sterling}},
  \bibinfo {author} {\bibfnamefont {D.}~\bibnamefont {Strain}}, \bibinfo
  {author} {\bibfnamefont {M.}~\bibnamefont {Szalay}}, \bibinfo {author}
  {\bibfnamefont {D.}~\bibnamefont {Thor}}, \bibinfo {author} {\bibfnamefont
  {A.}~\bibnamefont {Torres}}, \bibinfo {author} {\bibfnamefont
  {G.}~\bibnamefont {Vidal}}, \bibinfo {author} {\bibfnamefont
  {B.}~\bibnamefont {Villalonga}}, \bibinfo {author} {\bibfnamefont {C.~V.}\
  \bibnamefont {Heidweiller}}, \bibinfo {author} {\bibfnamefont
  {T.}~\bibnamefont {White}}, \bibinfo {author} {\bibfnamefont {B.~W.~K.}\
  \bibnamefont {Woo}}, \bibinfo {author} {\bibfnamefont {C.}~\bibnamefont
  {Xing}}, \bibinfo {author} {\bibfnamefont {Z.~J.}\ \bibnamefont {Yao}},
  \bibinfo {author} {\bibfnamefont {P.}~\bibnamefont {Yeh}}, \bibinfo {author}
  {\bibfnamefont {J.}~\bibnamefont {Yoo}}, \bibinfo {author} {\bibfnamefont
  {G.}~\bibnamefont {Young}}, \bibinfo {author} {\bibfnamefont
  {A.}~\bibnamefont {Zalcman}}, \bibinfo {author} {\bibfnamefont
  {Y.}~\bibnamefont {Zhang}}, \bibinfo {author} {\bibfnamefont
  {N.}~\bibnamefont {Zhu}}, \bibinfo {author} {\bibfnamefont {N.}~\bibnamefont
  {Zobrist}}, \bibinfo {author} {\bibfnamefont {H.}~\bibnamefont {Neven}},
  \bibinfo {author} {\bibfnamefont {R.}~\bibnamefont {Babbush}}, \bibinfo
  {author} {\bibfnamefont {D.}~\bibnamefont {Bacon}}, \bibinfo {author}
  {\bibfnamefont {S.}~\bibnamefont {Boixo}}, \bibinfo {author} {\bibfnamefont
  {J.}~\bibnamefont {Hilton}}, \bibinfo {author} {\bibfnamefont
  {E.}~\bibnamefont {Lucero}}, \bibinfo {author} {\bibfnamefont
  {A.}~\bibnamefont {Megrant}}, \bibinfo {author} {\bibfnamefont
  {J.}~\bibnamefont {Kelly}}, \bibinfo {author} {\bibfnamefont
  {Y.}~\bibnamefont {Chen}}, \bibinfo {author} {\bibfnamefont {V.}~\bibnamefont
  {Smelyanskiy}}, \bibinfo {author} {\bibfnamefont {V.}~\bibnamefont
  {Khemani}}, \bibinfo {author} {\bibfnamefont {S.}~\bibnamefont
  {Gopalakrishnan}}, \bibinfo {author} {\bibfnamefont {T.}~\bibnamefont
  {Prosen}},\ and\ \bibinfo {author} {\bibfnamefont {P.}~\bibnamefont
  {Roushan}},\ }\bibfield  {title} {\bibinfo {title} {Dynamics of magnetization
  at infinite temperature in a heisenberg spin chain},\ }\href
  {https://doi.org/10.1126/science.adi7877} {\bibfield  {journal} {\bibinfo
  {journal} {Science}\ }\textbf {\bibinfo {volume} {384}},\ \bibinfo {pages}
  {48} (\bibinfo {year} {2024})},\ \Eprint
  {https://arxiv.org/abs/https://www.science.org/doi/pdf/10.1126/science.adi7877}
  {https://www.science.org/doi/pdf/10.1126/science.adi7877} \BibitemShut
  {NoStop}%
\bibitem [{\citenamefont {Wienand}\ \emph {et~al.}(2024)\citenamefont
  {Wienand}, \citenamefont {Karch}, \citenamefont {Impertro}, \citenamefont
  {Schweizer}, \citenamefont {McCulloch}, \citenamefont {Vasseur},
  \citenamefont {Gopalakrishnan}, \citenamefont {Aidelsburger},\ and\
  \citenamefont {Bloch}}]{Wienand24}%
  \BibitemOpen
  \bibfield  {author} {\bibinfo {author} {\bibfnamefont {J.~F.}\ \bibnamefont
  {Wienand}}, \bibinfo {author} {\bibfnamefont {S.}~\bibnamefont {Karch}},
  \bibinfo {author} {\bibfnamefont {A.}~\bibnamefont {Impertro}}, \bibinfo
  {author} {\bibfnamefont {C.}~\bibnamefont {Schweizer}}, \bibinfo {author}
  {\bibfnamefont {E.}~\bibnamefont {McCulloch}}, \bibinfo {author}
  {\bibfnamefont {R.}~\bibnamefont {Vasseur}}, \bibinfo {author} {\bibfnamefont
  {S.}~\bibnamefont {Gopalakrishnan}}, \bibinfo {author} {\bibfnamefont
  {M.}~\bibnamefont {Aidelsburger}},\ and\ \bibinfo {author} {\bibfnamefont
  {I.}~\bibnamefont {Bloch}},\ }\bibfield  {title} {\bibinfo {title} {Emergence
  of fluctuating hydrodynamics in chaotic quantum systems},\ }\href
  {https://doi.org/10.1038/s41567-024-02611-z} {\bibfield  {journal} {\bibinfo
  {journal} {Nature Physics}\ }\textbf {\bibinfo {volume} {20}},\ \bibinfo
  {pages} {1732} (\bibinfo {year} {2024})}\BibitemShut {NoStop}%
\bibitem [{\citenamefont {Haar}(1933)}]{Haar33}%
  \BibitemOpen
  \bibfield  {author} {\bibinfo {author} {\bibfnamefont {A.}~\bibnamefont
  {Haar}},\ }\bibfield  {title} {\bibinfo {title} {Der massbegriff in der
  theorie der kontinuierlichen gruppen},\ }\href
  {http://www.jstor.org/stable/1968346} {\bibfield  {journal} {\bibinfo
  {journal} {Annals of Mathematics}\ }\textbf {\bibinfo {volume} {34}},\
  \bibinfo {pages} {147} (\bibinfo {year} {1933})}\BibitemShut {NoStop}%
\bibitem [{\citenamefont {Halmos}(1950)}]{Halmos50}%
  \BibitemOpen
  \bibfield  {author} {\bibinfo {author} {\bibfnamefont {P.~R.}\ \bibnamefont
  {Halmos}},\ }\bibinfo {title} {Haar measure},\ in\ \href
  {https://doi.org/10.1007/978-1-4684-9440-2_12} {\emph {\bibinfo {booktitle}
  {Measure Theory}}}\ (\bibinfo  {publisher} {Springer New York},\ \bibinfo
  {address} {New York, NY},\ \bibinfo {year} {1950})\ pp.\ \bibinfo {pages}
  {250--265}\BibitemShut {NoStop}%
\bibitem [{\citenamefont {Page}(1993)}]{Page93}%
  \BibitemOpen
  \bibfield  {author} {\bibinfo {author} {\bibfnamefont {D.~N.}\ \bibnamefont
  {Page}},\ }\bibfield  {title} {\bibinfo {title} {Average entropy of a
  subsystem},\ }\href {https://doi.org/10.1103/PhysRevLett.71.1291} {\bibfield
  {journal} {\bibinfo  {journal} {Phys. Rev. Lett.}\ }\textbf {\bibinfo
  {volume} {71}},\ \bibinfo {pages} {1291} (\bibinfo {year}
  {1993})}\BibitemShut {NoStop}%
\bibitem [{\citenamefont {Vidmar}\ and\ \citenamefont
  {Rigol}(2017)}]{Vidmar17}%
  \BibitemOpen
  \bibfield  {author} {\bibinfo {author} {\bibfnamefont {L.}~\bibnamefont
  {Vidmar}}\ and\ \bibinfo {author} {\bibfnamefont {M.}~\bibnamefont {Rigol}},\
  }\bibfield  {title} {\bibinfo {title} {Entanglement entropy of eigenstates of
  quantum chaotic hamiltonians},\ }\href
  {https://doi.org/10.1103/PhysRevLett.119.220603} {\bibfield  {journal}
  {\bibinfo  {journal} {Phys. Rev. Lett.}\ }\textbf {\bibinfo {volume} {119}},\
  \bibinfo {pages} {220603} (\bibinfo {year} {2017})}\BibitemShut {NoStop}%
\bibitem [{\citenamefont {Roberts}\ and\ \citenamefont
  {Yoshida}(2017)}]{Roberts17}%
  \BibitemOpen
  \bibfield  {author} {\bibinfo {author} {\bibfnamefont {D.~A.}\ \bibnamefont
  {Roberts}}\ and\ \bibinfo {author} {\bibfnamefont {B.}~\bibnamefont
  {Yoshida}},\ }\bibfield  {title} {\bibinfo {title} {Chaos and complexity by
  design},\ }\href {https://doi.org/10.1007/JHEP04(2017)121} {\bibfield
  {journal} {\bibinfo  {journal} {Journal of High Energy Physics}\ }\textbf
  {\bibinfo {volume} {2017}},\ \bibinfo {pages} {121} (\bibinfo {year}
  {2017})}\BibitemShut {NoStop}%
\bibitem [{\citenamefont {Cotler}\ \emph {et~al.}(2017)\citenamefont {Cotler},
  \citenamefont {Hunter-Jones}, \citenamefont {Liu},\ and\ \citenamefont
  {Yoshida}}]{Cotler17}%
  \BibitemOpen
  \bibfield  {author} {\bibinfo {author} {\bibfnamefont {J.}~\bibnamefont
  {Cotler}}, \bibinfo {author} {\bibfnamefont {N.}~\bibnamefont
  {Hunter-Jones}}, \bibinfo {author} {\bibfnamefont {J.}~\bibnamefont {Liu}},\
  and\ \bibinfo {author} {\bibfnamefont {B.}~\bibnamefont {Yoshida}},\
  }\bibfield  {title} {\bibinfo {title} {Chaos, complexity, and random
  matrices},\ }\href {https://doi.org/10.1007/JHEP11(2017)048} {\bibfield
  {journal} {\bibinfo  {journal} {Journal of High Energy Physics}\ }\textbf
  {\bibinfo {volume} {2017}},\ \bibinfo {pages} {48} (\bibinfo {year}
  {2017})}\BibitemShut {NoStop}%
\bibitem [{\citenamefont {Lu}\ and\ \citenamefont {Grover}(2019)}]{Lu19}%
  \BibitemOpen
  \bibfield  {author} {\bibinfo {author} {\bibfnamefont {T.-C.}\ \bibnamefont
  {Lu}}\ and\ \bibinfo {author} {\bibfnamefont {T.}~\bibnamefont {Grover}},\
  }\bibfield  {title} {\bibinfo {title} {Renyi entropy of chaotic
  eigenstates},\ }\href {https://doi.org/10.1103/PhysRevE.99.032111} {\bibfield
   {journal} {\bibinfo  {journal} {Phys. Rev. E}\ }\textbf {\bibinfo {volume}
  {99}},\ \bibinfo {pages} {032111} (\bibinfo {year} {2019})}\BibitemShut
  {NoStop}%
\bibitem [{\citenamefont {Bianchi}\ \emph {et~al.}(2022)\citenamefont
  {Bianchi}, \citenamefont {Hackl}, \citenamefont {Kieburg}, \citenamefont
  {Rigol},\ and\ \citenamefont {Vidmar}}]{Bianchi22}%
  \BibitemOpen
  \bibfield  {author} {\bibinfo {author} {\bibfnamefont {E.}~\bibnamefont
  {Bianchi}}, \bibinfo {author} {\bibfnamefont {L.}~\bibnamefont {Hackl}},
  \bibinfo {author} {\bibfnamefont {M.}~\bibnamefont {Kieburg}}, \bibinfo
  {author} {\bibfnamefont {M.}~\bibnamefont {Rigol}},\ and\ \bibinfo {author}
  {\bibfnamefont {L.}~\bibnamefont {Vidmar}},\ }\bibfield  {title} {\bibinfo
  {title} {{Volume-Law Entanglement Entropy of Typical Pure Quantum States}},\
  }\href {https://doi.org/10.1103/PRXQuantum.3.030201} {\bibfield  {journal}
  {\bibinfo  {journal} {PRX Quantum}\ }\textbf {\bibinfo {volume} {3}},\
  \bibinfo {pages} {030201} (\bibinfo {year} {2022})}\BibitemShut {NoStop}%
\bibitem [{\citenamefont {Tirrito}\ \emph
  {et~al.}(2025{\natexlab{b}})\citenamefont {Tirrito}, \citenamefont
  {Turkeshi},\ and\ \citenamefont {Sierant}}]{Tirrito25}%
  \BibitemOpen
  \bibfield  {author} {\bibinfo {author} {\bibfnamefont {E.}~\bibnamefont
  {Tirrito}}, \bibinfo {author} {\bibfnamefont {X.}~\bibnamefont {Turkeshi}},\
  and\ \bibinfo {author} {\bibfnamefont {P.}~\bibnamefont {Sierant}},\
  }\bibfield  {title} {\bibinfo {title} {Anticoncentration and
  nonstabilizerness spreading under ergodic quantum dynamics},\ }\href
  {https://doi.org/10.1103/1jzy-sk9r} {\bibfield  {journal} {\bibinfo
  {journal} {Phys. Rev. Lett.}\ }\textbf {\bibinfo {volume} {135}},\ \bibinfo
  {pages} {220401} (\bibinfo {year} {2025}{\natexlab{b}})}\BibitemShut
  {NoStop}%
\bibitem [{\citenamefont {Jozsa}\ \emph {et~al.}(1994)\citenamefont {Jozsa},
  \citenamefont {Robb},\ and\ \citenamefont {Wootters}}]{Jozsa94}%
  \BibitemOpen
  \bibfield  {author} {\bibinfo {author} {\bibfnamefont {R.}~\bibnamefont
  {Jozsa}}, \bibinfo {author} {\bibfnamefont {D.}~\bibnamefont {Robb}},\ and\
  \bibinfo {author} {\bibfnamefont {W.~K.}\ \bibnamefont {Wootters}},\
  }\bibfield  {title} {\bibinfo {title} {Lower bound for accessible information
  in quantum mechanics},\ }\href {https://doi.org/10.1103/PhysRevA.49.668}
  {\bibfield  {journal} {\bibinfo  {journal} {Phys. Rev. A}\ }\textbf {\bibinfo
  {volume} {49}},\ \bibinfo {pages} {668} (\bibinfo {year} {1994})}\BibitemShut
  {NoStop}%
\bibitem [{\citenamefont {Goldstein}\ \emph
  {et~al.}(2006{\natexlab{b}})\citenamefont {Goldstein}, \citenamefont
  {Lebowitz}, \citenamefont {Tumulka},\ and\ \citenamefont
  {Zanghì}}]{Goldstein06S}%
  \BibitemOpen
  \bibfield  {author} {\bibinfo {author} {\bibfnamefont {S.}~\bibnamefont
  {Goldstein}}, \bibinfo {author} {\bibfnamefont {J.~L.}\ \bibnamefont
  {Lebowitz}}, \bibinfo {author} {\bibfnamefont {R.}~\bibnamefont {Tumulka}},\
  and\ \bibinfo {author} {\bibfnamefont {N.}~\bibnamefont {Zanghì}},\
  }\bibfield  {title} {\bibinfo {title} {On the {Distribution} of the {Wave}
  {Function} for {Systems} in {Thermal} {Equilibrium}},\ }\href
  {https://doi.org/10.1007/s10955-006-9210-z} {\bibfield  {journal} {\bibinfo
  {journal} {Journal of Statistical Physics}\ }\textbf {\bibinfo {volume}
  {125}},\ \bibinfo {pages} {1193} (\bibinfo {year}
  {2006}{\natexlab{b}})}\BibitemShut {NoStop}%
\bibitem [{\citenamefont {Reimann}(2008)}]{Reimann08}%
  \BibitemOpen
  \bibfield  {author} {\bibinfo {author} {\bibfnamefont {P.}~\bibnamefont
  {Reimann}},\ }\bibfield  {title} {\bibinfo {title} {Typicality of {Pure}
  {States} {Randomly} {Sampled} {According} to the {Gaussian} {Adjusted}
  {Projected} {Measure}},\ }\href {https://doi.org/10.1007/s10955-008-9576-1}
  {\bibfield  {journal} {\bibinfo  {journal} {Journal of Statistical Physics}\
  }\textbf {\bibinfo {volume} {132}},\ \bibinfo {pages} {921} (\bibinfo {year}
  {2008})}\BibitemShut {NoStop}%
\bibitem [{\citenamefont {Goldstein}\ \emph {et~al.}(2016)\citenamefont
  {Goldstein}, \citenamefont {Lebowitz}, \citenamefont {Mastrodonato},
  \citenamefont {Tumulka},\ and\ \citenamefont {Zanghì}}]{Goldstein16}%
  \BibitemOpen
  \bibfield  {author} {\bibinfo {author} {\bibfnamefont {S.}~\bibnamefont
  {Goldstein}}, \bibinfo {author} {\bibfnamefont {J.~L.}\ \bibnamefont
  {Lebowitz}}, \bibinfo {author} {\bibfnamefont {C.}~\bibnamefont
  {Mastrodonato}}, \bibinfo {author} {\bibfnamefont {R.}~\bibnamefont
  {Tumulka}},\ and\ \bibinfo {author} {\bibfnamefont {N.}~\bibnamefont
  {Zanghì}},\ }\bibfield  {title} {\bibinfo {title} {Universal {Probability}
  {Distribution} for the {Wave} {Function} of a {Quantum} {System} {Entangled}
  with its {Environment}},\ }\href {https://doi.org/10.1007/s00220-015-2536-0}
  {\bibfield  {journal} {\bibinfo  {journal} {Communications in Mathematical
  Physics}\ }\textbf {\bibinfo {volume} {342}},\ \bibinfo {pages} {965}
  (\bibinfo {year} {2016})}\BibitemShut {NoStop}%
\bibitem [{\citenamefont {Liu}\ \emph {et~al.}(2024)\citenamefont {Liu},
  \citenamefont {Huang},\ and\ \citenamefont {Ho}}]{Liu24}%
  \BibitemOpen
  \bibfield  {author} {\bibinfo {author} {\bibfnamefont {C.}~\bibnamefont
  {Liu}}, \bibinfo {author} {\bibfnamefont {Q.~C.}\ \bibnamefont {Huang}},\
  and\ \bibinfo {author} {\bibfnamefont {W.~W.}\ \bibnamefont {Ho}},\
  }\bibfield  {title} {\bibinfo {title} {Deep thermalization in gaussian
  continuous-variable quantum systems},\ }\href
  {https://doi.org/10.1103/PhysRevLett.133.260401} {\bibfield  {journal}
  {\bibinfo  {journal} {Phys. Rev. Lett.}\ }\textbf {\bibinfo {volume} {133}},\
  \bibinfo {pages} {260401} (\bibinfo {year} {2024})}\BibitemShut {NoStop}%
\bibitem [{\citenamefont {Teufel}\ \emph {et~al.}(2025)\citenamefont {Teufel},
  \citenamefont {Tumulka},\ and\ \citenamefont {Vogel}}]{Teufel25}%
  \BibitemOpen
  \bibfield  {author} {\bibinfo {author} {\bibfnamefont {S.}~\bibnamefont
  {Teufel}}, \bibinfo {author} {\bibfnamefont {R.}~\bibnamefont {Tumulka}},\
  and\ \bibinfo {author} {\bibfnamefont {C.}~\bibnamefont {Vogel}},\ }\bibfield
   {title} {\bibinfo {title} {Canonical {Typicality} for {Other} {Ensembles}
  than {Micro}-canonical},\ }\href {https://doi.org/10.1007/s00023-024-01466-7}
  {\bibfield  {journal} {\bibinfo  {journal} {Annales Henri Poincaré}\
  }\textbf {\bibinfo {volume} {26}},\ \bibinfo {pages} {1477} (\bibinfo {year}
  {2025})}\BibitemShut {NoStop}%
\bibitem [{\citenamefont {Manna}\ \emph {et~al.}(2025)\citenamefont {Manna},
  \citenamefont {Roy},\ and\ \citenamefont {Sreejith}}]{Manna25}%
  \BibitemOpen
  \bibfield  {author} {\bibinfo {author} {\bibfnamefont {S.}~\bibnamefont
  {Manna}}, \bibinfo {author} {\bibfnamefont {S.}~\bibnamefont {Roy}},\ and\
  \bibinfo {author} {\bibfnamefont {G.~J.}\ \bibnamefont {Sreejith}},\
  }\bibfield  {title} {\bibinfo {title} {Projected ensemble in a system with
  locally supported conserved charges},\ }\href
  {https://doi.org/10.1103/PhysRevB.111.144302} {\bibfield  {journal} {\bibinfo
   {journal} {Phys. Rev. B}\ }\textbf {\bibinfo {volume} {111}},\ \bibinfo
  {pages} {144302} (\bibinfo {year} {2025})}\BibitemShut {NoStop}%
\bibitem [{\citenamefont {Chang}\ \emph {et~al.}(2025)\citenamefont {Chang},
  \citenamefont {Shrotriya}, \citenamefont {Ho},\ and\ \citenamefont
  {Ippoliti}}]{Chang25}%
  \BibitemOpen
  \bibfield  {author} {\bibinfo {author} {\bibfnamefont {R.-A.}\ \bibnamefont
  {Chang}}, \bibinfo {author} {\bibfnamefont {H.}~\bibnamefont {Shrotriya}},
  \bibinfo {author} {\bibfnamefont {W.~W.}\ \bibnamefont {Ho}},\ and\ \bibinfo
  {author} {\bibfnamefont {M.}~\bibnamefont {Ippoliti}},\ }\bibfield  {title}
  {\bibinfo {title} {Deep thermalization under charge-conserving quantum
  dynamics},\ }\href {https://doi.org/10.1103/PRXQuantum.6.020343} {\bibfield
  {journal} {\bibinfo  {journal} {PRX Quantum}\ }\textbf {\bibinfo {volume}
  {6}},\ \bibinfo {pages} {020343} (\bibinfo {year} {2025})}\BibitemShut
  {NoStop}%
\bibitem [{\citenamefont {McGinley}\ and\ \citenamefont
  {Schuster}(2025)}]{McGinley25}%
  \BibitemOpen
  \bibfield  {author} {\bibinfo {author} {\bibfnamefont {M.}~\bibnamefont
  {McGinley}}\ and\ \bibinfo {author} {\bibfnamefont {T.}~\bibnamefont
  {Schuster}},\ }\href {https://doi.org/10.48550/arXiv.2511.17172} {\bibinfo
  {title} {The {Scrooge} ensemble in many-body quantum systems}} (\bibinfo
  {year} {2025}),\ \bibinfo {note} {arXiv:2511.17172 [quant-ph] version:
  1}\BibitemShut {NoStop}%
\bibitem [{\citenamefont {Mok}\ \emph {et~al.}(2026)\citenamefont {Mok},
  \citenamefont {Haug}, \citenamefont {Ho},\ and\ \citenamefont
  {Preskill}}]{Mok26}%
  \BibitemOpen
  \bibfield  {author} {\bibinfo {author} {\bibfnamefont {W.-K.}\ \bibnamefont
  {Mok}}, \bibinfo {author} {\bibfnamefont {T.}~\bibnamefont {Haug}}, \bibinfo
  {author} {\bibfnamefont {W.~W.}\ \bibnamefont {Ho}},\ and\ \bibinfo {author}
  {\bibfnamefont {J.}~\bibnamefont {Preskill}},\ }\href
  {https://doi.org/10.48550/arXiv.2601.00266} {\bibinfo {title} {Nature is
  stingy: {Universality} of {Scrooge} ensembles in quantum many-body systems}}
  (\bibinfo {year} {2026}),\ \bibinfo {note} {arXiv:2601.00266
  [quant-ph]}\BibitemShut {NoStop}%
\bibitem [{\citenamefont {Xu}\ and\ \citenamefont {Zhang}(2026)}]{Xu26}%
  \BibitemOpen
  \bibfield  {author} {\bibinfo {author} {\bibfnamefont {C.}~\bibnamefont
  {Xu}}\ and\ \bibinfo {author} {\bibfnamefont {P.}~\bibnamefont {Zhang}},\
  }\href {https://doi.org/10.48550/arXiv.2605.20656} {\bibinfo {title}
  {Entanglement {Growth} from {Structured} {Initial} {States} in {Many}-{Body}
  {Localized} {Systems}}} (\bibinfo {year} {2026}),\ \bibinfo {note}
  {arXiv:2605.20656 [quant-ph]}\BibitemShut {NoStop}%
\bibitem [{\citenamefont {Wu}\ \emph {et~al.}(2026)\citenamefont {Wu},
  \citenamefont {Tong}, \citenamefont {Mao},\ and\ \citenamefont
  {Zhang}}]{Wu26}%
  \BibitemOpen
  \bibfield  {author} {\bibinfo {author} {\bibfnamefont {Y.}~\bibnamefont
  {Wu}}, \bibinfo {author} {\bibfnamefont {Y.}~\bibnamefont {Tong}}, \bibinfo
  {author} {\bibfnamefont {L.}~\bibnamefont {Mao}},\ and\ \bibinfo {author}
  {\bibfnamefont {P.}~\bibnamefont {Zhang}},\ }\href
  {https://arxiv.org/abs/2606.29042} {\bibinfo {title} {Exact hilbert-space
  ergodicity from continuous monitoring}} (\bibinfo {year} {2026}),\ \Eprint
  {https://arxiv.org/abs/2606.29042} {arXiv:2606.29042 [quant-ph]} \BibitemShut
  {NoStop}%
\bibitem [{\citenamefont {Igelspacher}\ \emph {et~al.}(2026)\citenamefont
  {Igelspacher}, \citenamefont {Tumulka},\ and\ \citenamefont
  {Vogel}}]{Igelspacher26}%
  \BibitemOpen
  \bibfield  {author} {\bibinfo {author} {\bibfnamefont {C.}~\bibnamefont
  {Igelspacher}}, \bibinfo {author} {\bibfnamefont {R.}~\bibnamefont
  {Tumulka}},\ and\ \bibinfo {author} {\bibfnamefont {C.}~\bibnamefont
  {Vogel}},\ }\href {https://arxiv.org/abs/2601.03253} {\bibinfo {title}
  {Grand-canonical typicality}} (\bibinfo {year} {2026}),\ \Eprint
  {https://arxiv.org/abs/2601.03253} {arXiv:2601.03253 [quant-ph]} \BibitemShut
  {NoStop}%
\bibitem [{\citenamefont {Liu}\ \emph {et~al.}(2026)\citenamefont {Liu},
  \citenamefont {McGinley}, \citenamefont {Schuster},\ and\ \citenamefont
  {Gosset}}]{Liu26nonlocal}%
  \BibitemOpen
  \bibfield  {author} {\bibinfo {author} {\bibfnamefont {Y.}~\bibnamefont
  {Liu}}, \bibinfo {author} {\bibfnamefont {M.}~\bibnamefont {McGinley}},
  \bibinfo {author} {\bibfnamefont {T.}~\bibnamefont {Schuster}},\ and\
  \bibinfo {author} {\bibfnamefont {D.}~\bibnamefont {Gosset}},\ }\href
  {https://arxiv.org/abs/2608.12255} {\bibinfo {title} {Conditional dependence
  and scrooge ensembles in shallow random quantum circuits}} (\bibinfo {year}
  {2026}),\ \Eprint {https://arxiv.org/abs/2608.12255} {arXiv:2608.12255
  [quant-ph]} \BibitemShut {NoStop}%
\bibitem [{\citenamefont {Sarma}\ \emph {et~al.}(2026)\citenamefont {Sarma},
  \citenamefont {Haug}, \citenamefont {Preskill},\ and\ \citenamefont
  {Mok}}]{Sarma26}%
  \BibitemOpen
  \bibfield  {author} {\bibinfo {author} {\bibfnamefont {S.}~\bibnamefont
  {Sarma}}, \bibinfo {author} {\bibfnamefont {T.}~\bibnamefont {Haug}},
  \bibinfo {author} {\bibfnamefont {J.}~\bibnamefont {Preskill}},\ and\
  \bibinfo {author} {\bibfnamefont {W.-K.}\ \bibnamefont {Mok}},\ }\href
  {https://arxiv.org/abs/2608.22939} {\bibinfo {title} {Universal equilibrium
  magic in quantum many-body systems}} (\bibinfo {year} {2026}),\ \Eprint
  {https://arxiv.org/abs/2608.22939} {arXiv:2608.22939 [quant-ph]} \BibitemShut
  {NoStop}%
\bibitem [{\citenamefont {Kadanoff}\ and\ \citenamefont
  {Martin}(1963)}]{Kadanoff63}%
  \BibitemOpen
  \bibfield  {author} {\bibinfo {author} {\bibfnamefont {L.~P.}\ \bibnamefont
  {Kadanoff}}\ and\ \bibinfo {author} {\bibfnamefont {P.~C.}\ \bibnamefont
  {Martin}},\ }\bibfield  {title} {\bibinfo {title} {Hydrodynamic equations and
  correlation functions},\ }\href
  {https://doi.org/https://doi.org/10.1016/0003-4916(63)90078-2} {\bibfield
  {journal} {\bibinfo  {journal} {Annals of Physics}\ }\textbf {\bibinfo
  {volume} {24}},\ \bibinfo {pages} {419} (\bibinfo {year} {1963})}\BibitemShut
  {NoStop}%
\bibitem [{\citenamefont {Pines}\ and\ \citenamefont
  {Nozi{\`e}res}(1966)}]{Pines66}%
  \BibitemOpen
  \bibfield  {author} {\bibinfo {author} {\bibfnamefont {D.}~\bibnamefont
  {Pines}}\ and\ \bibinfo {author} {\bibfnamefont {P.}~\bibnamefont
  {Nozi{\`e}res}},\ }\href@noop {} {\emph {\bibinfo {title} {The Theory of
  Quantum Liquids}}},\ Advanced book classics series\ (\bibinfo  {publisher}
  {W.A. Benjamin},\ \bibinfo {year} {1966})\BibitemShut {NoStop}%
\bibitem [{\citenamefont {Spohn}(1991)}]{Spohn91}%
  \BibitemOpen
  \bibfield  {author} {\bibinfo {author} {\bibfnamefont {H.}~\bibnamefont
  {Spohn}},\ }\href@noop {} {\emph {\bibinfo {title} {Large scale dynamics of
  interacting particles}}}\ (\bibinfo  {publisher} {Springer-Verlag,
  Heidelberg},\ \bibinfo {year} {1991})\BibitemShut {NoStop}%
\bibitem [{\citenamefont {Castro-Alvaredo}\ \emph {et~al.}(2016)\citenamefont
  {Castro-Alvaredo}, \citenamefont {Doyon},\ and\ \citenamefont
  {Yoshimura}}]{Castro-Alvaredo16}%
  \BibitemOpen
  \bibfield  {author} {\bibinfo {author} {\bibfnamefont {O.~A.}\ \bibnamefont
  {Castro-Alvaredo}}, \bibinfo {author} {\bibfnamefont {B.}~\bibnamefont
  {Doyon}},\ and\ \bibinfo {author} {\bibfnamefont {T.}~\bibnamefont
  {Yoshimura}},\ }\bibfield  {title} {\bibinfo {title} {Emergent hydrodynamics
  in integrable quantum systems out of equilibrium},\ }\href
  {https://doi.org/10.1103/PhysRevX.6.041065} {\bibfield  {journal} {\bibinfo
  {journal} {Phys. Rev. X}\ }\textbf {\bibinfo {volume} {6}},\ \bibinfo {pages}
  {041065} (\bibinfo {year} {2016})}\BibitemShut {NoStop}%
\bibitem [{\citenamefont {Bertini}\ \emph {et~al.}(2016)\citenamefont
  {Bertini}, \citenamefont {Collura}, \citenamefont {De~Nardis},\ and\
  \citenamefont {Fagotti}}]{Bertini16}%
  \BibitemOpen
  \bibfield  {author} {\bibinfo {author} {\bibfnamefont {B.}~\bibnamefont
  {Bertini}}, \bibinfo {author} {\bibfnamefont {M.}~\bibnamefont {Collura}},
  \bibinfo {author} {\bibfnamefont {J.}~\bibnamefont {De~Nardis}},\ and\
  \bibinfo {author} {\bibfnamefont {M.}~\bibnamefont {Fagotti}},\ }\bibfield
  {title} {\bibinfo {title} {Transport in out-of-equilibrium $xxz$ chains:
  Exact profiles of charges and currents},\ }\href
  {https://doi.org/10.1103/PhysRevLett.117.207201} {\bibfield  {journal}
  {\bibinfo  {journal} {Phys. Rev. Lett.}\ }\textbf {\bibinfo {volume} {117}},\
  \bibinfo {pages} {207201} (\bibinfo {year} {2016})}\BibitemShut {NoStop}%
\bibitem [{\citenamefont {Doyon}\ and\ \citenamefont
  {Yoshimura}(2017)}]{Doyon17}%
  \BibitemOpen
  \bibfield  {author} {\bibinfo {author} {\bibfnamefont {B.}~\bibnamefont
  {Doyon}}\ and\ \bibinfo {author} {\bibfnamefont {T.}~\bibnamefont
  {Yoshimura}},\ }\bibfield  {title} {\bibinfo {title} {A note on generalized
  hydrodynamics: inhomogeneous fields and other concepts},\ }\href
  {https://doi.org/10.21468/SciPostPhys.2.2.014} {\bibfield  {journal}
  {\bibinfo  {journal} {SciPost Phys.}\ }\textbf {\bibinfo {volume} {2}},\
  \bibinfo {pages} {014} (\bibinfo {year} {2017})}\BibitemShut {NoStop}%
\bibitem [{\citenamefont {Bulchandani}\ \emph {et~al.}(2018)\citenamefont
  {Bulchandani}, \citenamefont {Vasseur}, \citenamefont {Karrasch},\ and\
  \citenamefont {Moore}}]{Bulchandani18}%
  \BibitemOpen
  \bibfield  {author} {\bibinfo {author} {\bibfnamefont {V.~B.}\ \bibnamefont
  {Bulchandani}}, \bibinfo {author} {\bibfnamefont {R.}~\bibnamefont
  {Vasseur}}, \bibinfo {author} {\bibfnamefont {C.}~\bibnamefont {Karrasch}},\
  and\ \bibinfo {author} {\bibfnamefont {J.~E.}\ \bibnamefont {Moore}},\
  }\bibfield  {title} {\bibinfo {title} {Bethe-boltzmann hydrodynamics and spin
  transport in the xxz chain},\ }\href
  {https://doi.org/10.1103/PhysRevB.97.045407} {\bibfield  {journal} {\bibinfo
  {journal} {Phys. Rev. B}\ }\textbf {\bibinfo {volume} {97}},\ \bibinfo
  {pages} {045407} (\bibinfo {year} {2018})}\BibitemShut {NoStop}%
\bibitem [{\citenamefont {De~Nardis}\ \emph {et~al.}(2018)\citenamefont
  {De~Nardis}, \citenamefont {Bernard},\ and\ \citenamefont
  {Doyon}}]{DeNardis18}%
  \BibitemOpen
  \bibfield  {author} {\bibinfo {author} {\bibfnamefont {J.}~\bibnamefont
  {De~Nardis}}, \bibinfo {author} {\bibfnamefont {D.}~\bibnamefont {Bernard}},\
  and\ \bibinfo {author} {\bibfnamefont {B.}~\bibnamefont {Doyon}},\ }\bibfield
   {title} {\bibinfo {title} {Hydrodynamic diffusion in integrable systems},\
  }\href {https://doi.org/10.1103/PhysRevLett.121.160603} {\bibfield  {journal}
  {\bibinfo  {journal} {Phys. Rev. Lett.}\ }\textbf {\bibinfo {volume} {121}},\
  \bibinfo {pages} {160603} (\bibinfo {year} {2018})}\BibitemShut {NoStop}%
\bibitem [{\citenamefont {Doyon}(2020)}]{Doyon20}%
  \BibitemOpen
  \bibfield  {author} {\bibinfo {author} {\bibfnamefont {B.}~\bibnamefont
  {Doyon}},\ }\bibfield  {title} {\bibinfo {title} {Lecture notes on
  generalised hydrodynamics},\ }\href
  {https://doi.org/10.21468/SciPostPhysLectNotes.18} {\bibfield  {journal}
  {\bibinfo  {journal} {SciPost Phys. Lect. Notes}\ ,\ \bibinfo {pages} {18}}
  (\bibinfo {year} {2020})}\BibitemShut {NoStop}%
\bibitem [{\citenamefont {Essler}(2023)}]{Essler23}%
  \BibitemOpen
  \bibfield  {author} {\bibinfo {author} {\bibfnamefont {F.~H.}\ \bibnamefont
  {Essler}},\ }\bibfield  {title} {\bibinfo {title} {A short introduction to
  generalized hydrodynamics},\ }\href
  {https://doi.org/https://doi.org/10.1016/j.physa.2022.127572} {\bibfield
  {journal} {\bibinfo  {journal} {Physica A: Statistical Mechanics and its
  Applications}\ }\textbf {\bibinfo {volume} {631}},\ \bibinfo {pages} {127572}
  (\bibinfo {year} {2023})},\ \bibinfo {note} {lecture Notes of the 15th
  International Summer School of Fundamental Problems in Statistical
  Physics}\BibitemShut {NoStop}%
\bibitem [{\citenamefont {Doyon}\ \emph {et~al.}(2025)\citenamefont {Doyon},
  \citenamefont {Gopalakrishnan}, \citenamefont {M\o{}ller}, \citenamefont
  {Schmiedmayer},\ and\ \citenamefont {Vasseur}}]{Doyon25}%
  \BibitemOpen
  \bibfield  {author} {\bibinfo {author} {\bibfnamefont {B.}~\bibnamefont
  {Doyon}}, \bibinfo {author} {\bibfnamefont {S.}~\bibnamefont
  {Gopalakrishnan}}, \bibinfo {author} {\bibfnamefont {F.}~\bibnamefont
  {M\o{}ller}}, \bibinfo {author} {\bibfnamefont {J.}~\bibnamefont
  {Schmiedmayer}},\ and\ \bibinfo {author} {\bibfnamefont {R.}~\bibnamefont
  {Vasseur}},\ }\bibfield  {title} {\bibinfo {title} {Generalized
  hydrodynamics: A perspective},\ }\href
  {https://doi.org/10.1103/PhysRevX.15.010501} {\bibfield  {journal} {\bibinfo
  {journal} {Phys. Rev. X}\ }\textbf {\bibinfo {volume} {15}},\ \bibinfo
  {pages} {010501} (\bibinfo {year} {2025})}\BibitemShut {NoStop}%
\bibitem [{\citenamefont {Sala}\ \emph {et~al.}(2020)\citenamefont {Sala},
  \citenamefont {Rakovszky}, \citenamefont {Verresen}, \citenamefont {Knap},\
  and\ \citenamefont {Pollmann}}]{Sala20}%
  \BibitemOpen
  \bibfield  {author} {\bibinfo {author} {\bibfnamefont {P.}~\bibnamefont
  {Sala}}, \bibinfo {author} {\bibfnamefont {T.}~\bibnamefont {Rakovszky}},
  \bibinfo {author} {\bibfnamefont {R.}~\bibnamefont {Verresen}}, \bibinfo
  {author} {\bibfnamefont {M.}~\bibnamefont {Knap}},\ and\ \bibinfo {author}
  {\bibfnamefont {F.}~\bibnamefont {Pollmann}},\ }\bibfield  {title} {\bibinfo
  {title} {Ergodicity breaking arising from hilbert space fragmentation in
  dipole-conserving hamiltonians},\ }\href
  {https://doi.org/10.1103/physrevx.10.011047} {\bibfield  {journal} {\bibinfo
  {journal} {Physical Review X}\ }\textbf {\bibinfo {volume} {10}},\ \bibinfo
  {pages} {011047} (\bibinfo {year} {2020})}\BibitemShut {NoStop}%
\bibitem [{\citenamefont {Khemani}\ \emph {et~al.}(2020)\citenamefont
  {Khemani}, \citenamefont {Hermele},\ and\ \citenamefont
  {Nandkishore}}]{Khemani20}%
  \BibitemOpen
  \bibfield  {author} {\bibinfo {author} {\bibfnamefont {V.}~\bibnamefont
  {Khemani}}, \bibinfo {author} {\bibfnamefont {M.}~\bibnamefont {Hermele}},\
  and\ \bibinfo {author} {\bibfnamefont {R.}~\bibnamefont {Nandkishore}},\
  }\bibfield  {title} {\bibinfo {title} {{Localization from Hilbert space
  shattering: From theory to physical realizations}},\ }\href
  {https://doi.org/10.1103/PhysRevB.101.174204} {\bibfield  {journal} {\bibinfo
   {journal} {Phys. Rev. B}\ }\textbf {\bibinfo {volume} {101}},\ \bibinfo
  {pages} {174204} (\bibinfo {year} {2020})}\BibitemShut {NoStop}%
\bibitem [{\citenamefont {Guardado-Sanchez}\ \emph {et~al.}(2020)\citenamefont
  {Guardado-Sanchez}, \citenamefont {Morningstar}, \citenamefont {Spar},
  \citenamefont {Brown}, \citenamefont {Huse},\ and\ \citenamefont
  {Bakr}}]{Guardado20}%
  \BibitemOpen
  \bibfield  {author} {\bibinfo {author} {\bibfnamefont {E.}~\bibnamefont
  {Guardado-Sanchez}}, \bibinfo {author} {\bibfnamefont {A.}~\bibnamefont
  {Morningstar}}, \bibinfo {author} {\bibfnamefont {B.~M.}\ \bibnamefont
  {Spar}}, \bibinfo {author} {\bibfnamefont {P.~T.}\ \bibnamefont {Brown}},
  \bibinfo {author} {\bibfnamefont {D.~A.}\ \bibnamefont {Huse}},\ and\
  \bibinfo {author} {\bibfnamefont {W.~S.}\ \bibnamefont {Bakr}},\ }\bibfield
  {title} {\bibinfo {title} {{Subdiffusion and Heat Transport in a Tilted
  Two-Dimensional Fermi-Hubbard System}},\ }\href
  {https://doi.org/10.1103/PhysRevX.10.011042} {\bibfield  {journal} {\bibinfo
  {journal} {Phys. Rev. X}\ }\textbf {\bibinfo {volume} {10}},\ \bibinfo
  {pages} {011042} (\bibinfo {year} {2020})}\BibitemShut {NoStop}%
\bibitem [{\citenamefont {Moudgalya}\ and\ \citenamefont
  {Motrunich}(2022)}]{Moudgalya22}%
  \BibitemOpen
  \bibfield  {author} {\bibinfo {author} {\bibfnamefont {S.}~\bibnamefont
  {Moudgalya}}\ and\ \bibinfo {author} {\bibfnamefont {O.~I.}\ \bibnamefont
  {Motrunich}},\ }\bibfield  {title} {\bibinfo {title} {Hilbert space
  fragmentation and commutant algebras},\ }\href
  {https://doi.org/10.1103/PhysRevX.12.011050} {\bibfield  {journal} {\bibinfo
  {journal} {Phys. Rev. X}\ }\textbf {\bibinfo {volume} {12}},\ \bibinfo
  {pages} {011050} (\bibinfo {year} {2022})}\BibitemShut {NoStop}%
\bibitem [{\citenamefont {Kohlert}\ \emph {et~al.}(2023)\citenamefont
  {Kohlert}, \citenamefont {Scherg}, \citenamefont {Sala}, \citenamefont
  {Pollmann}, \citenamefont {Hebbe~Madhusudhana}, \citenamefont {Bloch},\ and\
  \citenamefont {Aidelsburger}}]{Kohlert23}%
  \BibitemOpen
  \bibfield  {author} {\bibinfo {author} {\bibfnamefont {T.}~\bibnamefont
  {Kohlert}}, \bibinfo {author} {\bibfnamefont {S.}~\bibnamefont {Scherg}},
  \bibinfo {author} {\bibfnamefont {P.}~\bibnamefont {Sala}}, \bibinfo {author}
  {\bibfnamefont {F.}~\bibnamefont {Pollmann}}, \bibinfo {author}
  {\bibfnamefont {B.}~\bibnamefont {Hebbe~Madhusudhana}}, \bibinfo {author}
  {\bibfnamefont {I.}~\bibnamefont {Bloch}},\ and\ \bibinfo {author}
  {\bibfnamefont {M.}~\bibnamefont {Aidelsburger}},\ }\bibfield  {title}
  {\bibinfo {title} {Exploring the regime of fragmentation in strongly tilted
  fermi-hubbard chains},\ }\href
  {https://doi.org/10.1103/PhysRevLett.130.010201} {\bibfield  {journal}
  {\bibinfo  {journal} {Phys. Rev. Lett.}\ }\textbf {\bibinfo {volume} {130}},\
  \bibinfo {pages} {010201} (\bibinfo {year} {2023})}\BibitemShut {NoStop}%
\bibitem [{\citenamefont {Wang}\ \emph {et~al.}(2025)\citenamefont {Wang},
  \citenamefont {Shi}, \citenamefont {Sun}, \citenamefont {Chen}, \citenamefont
  {Wang}, \citenamefont {Zhao}, \citenamefont {Liu}, \citenamefont {Ma},
  \citenamefont {Wang}, \citenamefont {Li}, \citenamefont {Zhang},
  \citenamefont {Liu}, \citenamefont {Deng}, \citenamefont {Li}, \citenamefont
  {He}, \citenamefont {Liu}, \citenamefont {Peng}, \citenamefont {Song},
  \citenamefont {Xue}, \citenamefont {Yu}, \citenamefont {Huang}, \citenamefont
  {Xiang}, \citenamefont {Zheng}, \citenamefont {Xu},\ and\ \citenamefont
  {Fan}}]{Wang25}%
  \BibitemOpen
  \bibfield  {author} {\bibinfo {author} {\bibfnamefont {Y.-Y.}\ \bibnamefont
  {Wang}}, \bibinfo {author} {\bibfnamefont {Y.-H.}\ \bibnamefont {Shi}},
  \bibinfo {author} {\bibfnamefont {Z.-H.}\ \bibnamefont {Sun}}, \bibinfo
  {author} {\bibfnamefont {C.-T.}\ \bibnamefont {Chen}}, \bibinfo {author}
  {\bibfnamefont {Z.-A.}\ \bibnamefont {Wang}}, \bibinfo {author}
  {\bibfnamefont {K.}~\bibnamefont {Zhao}}, \bibinfo {author} {\bibfnamefont
  {H.-T.}\ \bibnamefont {Liu}}, \bibinfo {author} {\bibfnamefont {W.-G.}\
  \bibnamefont {Ma}}, \bibinfo {author} {\bibfnamefont {Z.}~\bibnamefont
  {Wang}}, \bibinfo {author} {\bibfnamefont {H.}~\bibnamefont {Li}}, \bibinfo
  {author} {\bibfnamefont {J.-C.}\ \bibnamefont {Zhang}}, \bibinfo {author}
  {\bibfnamefont {Y.}~\bibnamefont {Liu}}, \bibinfo {author} {\bibfnamefont
  {C.-L.}\ \bibnamefont {Deng}}, \bibinfo {author} {\bibfnamefont {T.-M.}\
  \bibnamefont {Li}}, \bibinfo {author} {\bibfnamefont {Y.}~\bibnamefont {He}},
  \bibinfo {author} {\bibfnamefont {Z.-H.}\ \bibnamefont {Liu}}, \bibinfo
  {author} {\bibfnamefont {Z.-Y.}\ \bibnamefont {Peng}}, \bibinfo {author}
  {\bibfnamefont {X.}~\bibnamefont {Song}}, \bibinfo {author} {\bibfnamefont
  {G.}~\bibnamefont {Xue}}, \bibinfo {author} {\bibfnamefont {H.}~\bibnamefont
  {Yu}}, \bibinfo {author} {\bibfnamefont {K.}~\bibnamefont {Huang}}, \bibinfo
  {author} {\bibfnamefont {Z.}~\bibnamefont {Xiang}}, \bibinfo {author}
  {\bibfnamefont {D.}~\bibnamefont {Zheng}}, \bibinfo {author} {\bibfnamefont
  {K.}~\bibnamefont {Xu}},\ and\ \bibinfo {author} {\bibfnamefont
  {H.}~\bibnamefont {Fan}},\ }\bibfield  {title} {\bibinfo {title} {Exploring
  hilbert-space fragmentation on a superconducting processor},\ }\href
  {https://doi.org/10.1103/PRXQuantum.6.010325} {\bibfield  {journal} {\bibinfo
   {journal} {PRX Quantum}\ }\textbf {\bibinfo {volume} {6}},\ \bibinfo {pages}
  {010325} (\bibinfo {year} {2025})}\BibitemShut {NoStop}%
\bibitem [{\citenamefont {Jaynes}(1957)}]{Jaynes57}%
  \BibitemOpen
  \bibfield  {author} {\bibinfo {author} {\bibfnamefont {E.~T.}\ \bibnamefont
  {Jaynes}},\ }\bibfield  {title} {\bibinfo {title} {Information theory and
  statistical mechanics},\ }\href {https://doi.org/10.1103/PhysRev.106.620}
  {\bibfield  {journal} {\bibinfo  {journal} {Phys. Rev.}\ }\textbf {\bibinfo
  {volume} {106}},\ \bibinfo {pages} {620} (\bibinfo {year}
  {1957})}\BibitemShut {NoStop}%
\bibitem [{\citenamefont {Deutsch}(1991)}]{Deutsch91}%
  \BibitemOpen
  \bibfield  {author} {\bibinfo {author} {\bibfnamefont {J.~M.}\ \bibnamefont
  {Deutsch}},\ }\bibfield  {title} {\bibinfo {title} {Quantum statistical
  mechanics in a closed system},\ }\href
  {https://doi.org/10.1103/PhysRevA.43.2046} {\bibfield  {journal} {\bibinfo
  {journal} {Phys. Rev. A}\ }\textbf {\bibinfo {volume} {43}},\ \bibinfo
  {pages} {2046} (\bibinfo {year} {1991})}\BibitemShut {NoStop}%
\bibitem [{\citenamefont {Srednicki}(1994)}]{Srednicki94}%
  \BibitemOpen
  \bibfield  {author} {\bibinfo {author} {\bibfnamefont {M.}~\bibnamefont
  {Srednicki}},\ }\bibfield  {title} {\bibinfo {title} {Chaos and quantum
  thermalization},\ }\href {https://doi.org/10.1103/PhysRevE.50.888} {\bibfield
   {journal} {\bibinfo  {journal} {Phys. Rev. E}\ }\textbf {\bibinfo {volume}
  {50}},\ \bibinfo {pages} {888} (\bibinfo {year} {1994})}\BibitemShut
  {NoStop}%
\bibitem [{\citenamefont {Rigol}\ \emph {et~al.}(2007)\citenamefont {Rigol},
  \citenamefont {Dunjko}, \citenamefont {Yurovsky},\ and\ \citenamefont
  {Olshanii}}]{Rigol07}%
  \BibitemOpen
  \bibfield  {author} {\bibinfo {author} {\bibfnamefont {M.}~\bibnamefont
  {Rigol}}, \bibinfo {author} {\bibfnamefont {V.}~\bibnamefont {Dunjko}},
  \bibinfo {author} {\bibfnamefont {V.}~\bibnamefont {Yurovsky}},\ and\
  \bibinfo {author} {\bibfnamefont {M.}~\bibnamefont {Olshanii}},\ }\bibfield
  {title} {\bibinfo {title} {Relaxation in a completely integrable many-body
  quantum system: An ab initio study of the dynamics of the highly excited
  states of 1d lattice hard-core bosons},\ }\href
  {https://doi.org/10.1103/PhysRevLett.98.050405} {\bibfield  {journal}
  {\bibinfo  {journal} {Phys. Rev. Lett.}\ }\textbf {\bibinfo {volume} {98}},\
  \bibinfo {pages} {050405} (\bibinfo {year} {2007})}\BibitemShut {NoStop}%
\bibitem [{\citenamefont {Rigol}\ \emph {et~al.}(2008)\citenamefont {Rigol},
  \citenamefont {Dunjko},\ and\ \citenamefont {Olshanii}}]{Rigol08}%
  \BibitemOpen
  \bibfield  {author} {\bibinfo {author} {\bibfnamefont {M.}~\bibnamefont
  {Rigol}}, \bibinfo {author} {\bibfnamefont {V.}~\bibnamefont {Dunjko}},\ and\
  \bibinfo {author} {\bibfnamefont {M.}~\bibnamefont {Olshanii}},\ }\bibfield
  {title} {\bibinfo {title} {Thermalization and its mechanism for generic
  isolated quantum systems},\ }\href {https://doi.org/10.1038/nature06838}
  {\bibfield  {journal} {\bibinfo  {journal} {Nature}\ }\textbf {\bibinfo
  {volume} {452}},\ \bibinfo {pages} {854 EP } (\bibinfo {year}
  {2008})}\BibitemShut {NoStop}%
\bibitem [{\citenamefont {Linden}\ \emph {et~al.}(2009)\citenamefont {Linden},
  \citenamefont {Popescu}, \citenamefont {Short},\ and\ \citenamefont
  {Winter}}]{Linden09}%
  \BibitemOpen
  \bibfield  {author} {\bibinfo {author} {\bibfnamefont {N.}~\bibnamefont
  {Linden}}, \bibinfo {author} {\bibfnamefont {S.}~\bibnamefont {Popescu}},
  \bibinfo {author} {\bibfnamefont {A.~J.}\ \bibnamefont {Short}},\ and\
  \bibinfo {author} {\bibfnamefont {A.}~\bibnamefont {Winter}},\ }\bibfield
  {title} {\bibinfo {title} {Quantum mechanical evolution towards thermal
  equilibrium},\ }\href {https://doi.org/10.1103/PhysRevE.79.061103} {\bibfield
   {journal} {\bibinfo  {journal} {Phys. Rev. E}\ }\textbf {\bibinfo {volume}
  {79}},\ \bibinfo {pages} {061103} (\bibinfo {year} {2009})}\BibitemShut
  {NoStop}%
\bibitem [{\citenamefont {Gogolin}\ \emph {et~al.}(2011)\citenamefont
  {Gogolin}, \citenamefont {M\"uller},\ and\ \citenamefont
  {Eisert}}]{Gogolin11}%
  \BibitemOpen
  \bibfield  {author} {\bibinfo {author} {\bibfnamefont {C.}~\bibnamefont
  {Gogolin}}, \bibinfo {author} {\bibfnamefont {M.~P.}\ \bibnamefont
  {M\"uller}},\ and\ \bibinfo {author} {\bibfnamefont {J.}~\bibnamefont
  {Eisert}},\ }\bibfield  {title} {\bibinfo {title} {Absence of thermalization
  in nonintegrable systems},\ }\href
  {https://doi.org/10.1103/PhysRevLett.106.040401} {\bibfield  {journal}
  {\bibinfo  {journal} {Phys. Rev. Lett.}\ }\textbf {\bibinfo {volume} {106}},\
  \bibinfo {pages} {040401} (\bibinfo {year} {2011})}\BibitemShut {NoStop}%
\bibitem [{\citenamefont {Cassidy}\ \emph {et~al.}(2011)\citenamefont
  {Cassidy}, \citenamefont {Clark},\ and\ \citenamefont {Rigol}}]{Cassidy11}%
  \BibitemOpen
  \bibfield  {author} {\bibinfo {author} {\bibfnamefont {A.~C.}\ \bibnamefont
  {Cassidy}}, \bibinfo {author} {\bibfnamefont {C.~W.}\ \bibnamefont {Clark}},\
  and\ \bibinfo {author} {\bibfnamefont {M.}~\bibnamefont {Rigol}},\ }\bibfield
   {title} {\bibinfo {title} {Generalized thermalization in an integrable
  lattice system},\ }\href {https://doi.org/10.1103/PhysRevLett.106.140405}
  {\bibfield  {journal} {\bibinfo  {journal} {Phys. Rev. Lett.}\ }\textbf
  {\bibinfo {volume} {106}},\ \bibinfo {pages} {140405} (\bibinfo {year}
  {2011})}\BibitemShut {NoStop}%
\bibitem [{\citenamefont {Calabrese}\ \emph {et~al.}(2011)\citenamefont
  {Calabrese}, \citenamefont {Essler},\ and\ \citenamefont
  {Fagotti}}]{Calabrese11}%
  \BibitemOpen
  \bibfield  {author} {\bibinfo {author} {\bibfnamefont {P.}~\bibnamefont
  {Calabrese}}, \bibinfo {author} {\bibfnamefont {F.~H.~L.}\ \bibnamefont
  {Essler}},\ and\ \bibinfo {author} {\bibfnamefont {M.}~\bibnamefont
  {Fagotti}},\ }\bibfield  {title} {\bibinfo {title} {Quantum quench in the
  transverse-field ising chain},\ }\href
  {https://doi.org/10.1103/PhysRevLett.106.227203} {\bibfield  {journal}
  {\bibinfo  {journal} {Phys. Rev. Lett.}\ }\textbf {\bibinfo {volume} {106}},\
  \bibinfo {pages} {227203} (\bibinfo {year} {2011})}\BibitemShut {NoStop}%
\bibitem [{\citenamefont {Polkovnikov}\ \emph {et~al.}(2011)\citenamefont
  {Polkovnikov}, \citenamefont {Sengupta}, \citenamefont {Silva},\ and\
  \citenamefont {Vengalattore}}]{Polkovnikov11}%
  \BibitemOpen
  \bibfield  {author} {\bibinfo {author} {\bibfnamefont {A.}~\bibnamefont
  {Polkovnikov}}, \bibinfo {author} {\bibfnamefont {K.}~\bibnamefont
  {Sengupta}}, \bibinfo {author} {\bibfnamefont {A.}~\bibnamefont {Silva}},\
  and\ \bibinfo {author} {\bibfnamefont {M.}~\bibnamefont {Vengalattore}},\
  }\bibfield  {title} {\bibinfo {title} {Colloquium: Nonequilibrium dynamics of
  closed interacting quantum systems},\ }\href
  {https://doi.org/10.1103/RevModPhys.83.863} {\bibfield  {journal} {\bibinfo
  {journal} {Rev. Mod. Phys.}\ }\textbf {\bibinfo {volume} {83}},\ \bibinfo
  {pages} {863} (\bibinfo {year} {2011})}\BibitemShut {NoStop}%
\bibitem [{\citenamefont {Sirker}\ \emph {et~al.}(2014)\citenamefont {Sirker},
  \citenamefont {Konstantinidis}, \citenamefont {Andraschko},\ and\
  \citenamefont {Sedlmayr}}]{Sirker14}%
  \BibitemOpen
  \bibfield  {author} {\bibinfo {author} {\bibfnamefont {J.}~\bibnamefont
  {Sirker}}, \bibinfo {author} {\bibfnamefont {N.~P.}\ \bibnamefont
  {Konstantinidis}}, \bibinfo {author} {\bibfnamefont {F.}~\bibnamefont
  {Andraschko}},\ and\ \bibinfo {author} {\bibfnamefont {N.}~\bibnamefont
  {Sedlmayr}},\ }\bibfield  {title} {\bibinfo {title} {Locality and
  thermalization in closed quantum systems},\ }\href
  {https://doi.org/10.1103/PhysRevA.89.042104} {\bibfield  {journal} {\bibinfo
  {journal} {Phys. Rev. A}\ }\textbf {\bibinfo {volume} {89}},\ \bibinfo
  {pages} {042104} (\bibinfo {year} {2014})}\BibitemShut {NoStop}%
\bibitem [{\citenamefont {Gogolin}\ and\ \citenamefont
  {Eisert}(2016)}]{Gogolin16}%
  \BibitemOpen
  \bibfield  {author} {\bibinfo {author} {\bibfnamefont {C.}~\bibnamefont
  {Gogolin}}\ and\ \bibinfo {author} {\bibfnamefont {J.}~\bibnamefont
  {Eisert}},\ }\bibfield  {title} {\bibinfo {title} {Equilibration,
  thermalisation, and the emergence of statistical mechanics in closed quantum
  systems},\ }\href {https://doi.org/10.1088/0034-4885/79/5/056001} {\bibfield
  {journal} {\bibinfo  {journal} {Reports on Progress in Physics}\ }\textbf
  {\bibinfo {volume} {79}},\ \bibinfo {pages} {056001} (\bibinfo {year}
  {2016})}\BibitemShut {NoStop}%
\bibitem [{\citenamefont {Cramer}\ \emph {et~al.}(2008)\citenamefont {Cramer},
  \citenamefont {Dawson}, \citenamefont {Eisert},\ and\ \citenamefont
  {Osborne}}]{Cramer08}%
  \BibitemOpen
  \bibfield  {author} {\bibinfo {author} {\bibfnamefont {M.}~\bibnamefont
  {Cramer}}, \bibinfo {author} {\bibfnamefont {C.~M.}\ \bibnamefont {Dawson}},
  \bibinfo {author} {\bibfnamefont {J.}~\bibnamefont {Eisert}},\ and\ \bibinfo
  {author} {\bibfnamefont {T.~J.}\ \bibnamefont {Osborne}},\ }\bibfield
  {title} {\bibinfo {title} {Exact relaxation in a class of nonequilibrium
  quantum lattice systems},\ }\href
  {https://doi.org/10.1103/PhysRevLett.100.030602} {\bibfield  {journal}
  {\bibinfo  {journal} {Phys. Rev. Lett.}\ }\textbf {\bibinfo {volume} {100}},\
  \bibinfo {pages} {030602} (\bibinfo {year} {2008})}\BibitemShut {NoStop}%
\bibitem [{\citenamefont {Gring}\ \emph {et~al.}(2012)\citenamefont {Gring},
  \citenamefont {Kuhnert}, \citenamefont {Langen}, \citenamefont {Kitagawa},
  \citenamefont {Rauer}, \citenamefont {Schreitl}, \citenamefont {Mazets},
  \citenamefont {Smith}, \citenamefont {Demler},\ and\ \citenamefont
  {Schmiedmayer}}]{Gring12}%
  \BibitemOpen
  \bibfield  {author} {\bibinfo {author} {\bibfnamefont {M.}~\bibnamefont
  {Gring}}, \bibinfo {author} {\bibfnamefont {M.}~\bibnamefont {Kuhnert}},
  \bibinfo {author} {\bibfnamefont {T.}~\bibnamefont {Langen}}, \bibinfo
  {author} {\bibfnamefont {T.}~\bibnamefont {Kitagawa}}, \bibinfo {author}
  {\bibfnamefont {B.}~\bibnamefont {Rauer}}, \bibinfo {author} {\bibfnamefont
  {M.}~\bibnamefont {Schreitl}}, \bibinfo {author} {\bibfnamefont
  {I.}~\bibnamefont {Mazets}}, \bibinfo {author} {\bibfnamefont {D.~A.}\
  \bibnamefont {Smith}}, \bibinfo {author} {\bibfnamefont {E.}~\bibnamefont
  {Demler}},\ and\ \bibinfo {author} {\bibfnamefont {J.}~\bibnamefont
  {Schmiedmayer}},\ }\bibfield  {title} {\bibinfo {title} {Relaxation and
  prethermalization in an isolated quantum system},\ }\href
  {https://doi.org/10.1126/science.1224953} {\bibfield  {journal} {\bibinfo
  {journal} {Science}\ }\textbf {\bibinfo {volume} {337}},\ \bibinfo {pages}
  {1318} (\bibinfo {year} {2012})},\ \Eprint
  {https://arxiv.org/abs/https://www.science.org/doi/pdf/10.1126/science.1224953}
  {https://www.science.org/doi/pdf/10.1126/science.1224953} \BibitemShut
  {NoStop}%
\bibitem [{\citenamefont {Caux}\ and\ \citenamefont {Essler}(2013)}]{Caux13}%
  \BibitemOpen
  \bibfield  {author} {\bibinfo {author} {\bibfnamefont {J.-S.}\ \bibnamefont
  {Caux}}\ and\ \bibinfo {author} {\bibfnamefont {F.~H.~L.}\ \bibnamefont
  {Essler}},\ }\bibfield  {title} {\bibinfo {title} {Time evolution of local
  observables after quenching to an integrable model},\ }\href
  {https://doi.org/10.1103/PhysRevLett.110.257203} {\bibfield  {journal}
  {\bibinfo  {journal} {Phys. Rev. Lett.}\ }\textbf {\bibinfo {volume} {110}},\
  \bibinfo {pages} {257203} (\bibinfo {year} {2013})}\BibitemShut {NoStop}%
\bibitem [{\citenamefont {Mussardo}(2013)}]{Mussardo13}%
  \BibitemOpen
  \bibfield  {author} {\bibinfo {author} {\bibfnamefont {G.}~\bibnamefont
  {Mussardo}},\ }\bibfield  {title} {\bibinfo {title} {Infinite-time average of
  local fields in an integrable quantum field theory after a quantum quench},\
  }\href {https://doi.org/10.1103/PhysRevLett.111.100401} {\bibfield  {journal}
  {\bibinfo  {journal} {Phys. Rev. Lett.}\ }\textbf {\bibinfo {volume} {111}},\
  \bibinfo {pages} {100401} (\bibinfo {year} {2013})}\BibitemShut {NoStop}%
\bibitem [{\citenamefont {Brockmann}\ \emph {et~al.}(2014)\citenamefont
  {Brockmann}, \citenamefont {Wouters}, \citenamefont {Fioretto}, \citenamefont
  {Nardis}, \citenamefont {Vlijm},\ and\ \citenamefont {Caux}}]{Brockmann14}%
  \BibitemOpen
  \bibfield  {author} {\bibinfo {author} {\bibfnamefont {M.}~\bibnamefont
  {Brockmann}}, \bibinfo {author} {\bibfnamefont {B.}~\bibnamefont {Wouters}},
  \bibinfo {author} {\bibfnamefont {D.}~\bibnamefont {Fioretto}}, \bibinfo
  {author} {\bibfnamefont {J.~D.}\ \bibnamefont {Nardis}}, \bibinfo {author}
  {\bibfnamefont {R.}~\bibnamefont {Vlijm}},\ and\ \bibinfo {author}
  {\bibfnamefont {J.-S.}\ \bibnamefont {Caux}},\ }\bibfield  {title} {\bibinfo
  {title} {Quench action approach for releasing the néel state into the
  spin-1/2 xxz chain},\ }\href
  {https://doi.org/10.1088/1742-5468/2014/12/P12009} {\bibfield  {journal}
  {\bibinfo  {journal} {Journal of Statistical Mechanics: Theory and
  Experiment}\ }\textbf {\bibinfo {volume} {2014}},\ \bibinfo {pages} {P12009}
  (\bibinfo {year} {2014})}\BibitemShut {NoStop}%
\bibitem [{\citenamefont {Ilievski}\ \emph {et~al.}(2015)\citenamefont
  {Ilievski}, \citenamefont {De~Nardis}, \citenamefont {Wouters}, \citenamefont
  {Caux}, \citenamefont {Essler},\ and\ \citenamefont {Prosen}}]{Ilievski15}%
  \BibitemOpen
  \bibfield  {author} {\bibinfo {author} {\bibfnamefont {E.}~\bibnamefont
  {Ilievski}}, \bibinfo {author} {\bibfnamefont {J.}~\bibnamefont {De~Nardis}},
  \bibinfo {author} {\bibfnamefont {B.}~\bibnamefont {Wouters}}, \bibinfo
  {author} {\bibfnamefont {J.-S.}\ \bibnamefont {Caux}}, \bibinfo {author}
  {\bibfnamefont {F.~H.~L.}\ \bibnamefont {Essler}},\ and\ \bibinfo {author}
  {\bibfnamefont {T.}~\bibnamefont {Prosen}},\ }\bibfield  {title} {\bibinfo
  {title} {Complete generalized gibbs ensembles in an interacting theory},\
  }\href {https://doi.org/10.1103/PhysRevLett.115.157201} {\bibfield  {journal}
  {\bibinfo  {journal} {Phys. Rev. Lett.}\ }\textbf {\bibinfo {volume} {115}},\
  \bibinfo {pages} {157201} (\bibinfo {year} {2015})}\BibitemShut {NoStop}%
\bibitem [{\citenamefont {Langen}\ \emph {et~al.}(2015)\citenamefont {Langen},
  \citenamefont {Erne}, \citenamefont {Geiger}, \citenamefont {Rauer},
  \citenamefont {Schweigler}, \citenamefont {Kuhnert}, \citenamefont
  {Rohringer}, \citenamefont {Mazets}, \citenamefont {Gasenzer},\ and\
  \citenamefont {Schmiedmayer}}]{Langen15}%
  \BibitemOpen
  \bibfield  {author} {\bibinfo {author} {\bibfnamefont {T.}~\bibnamefont
  {Langen}}, \bibinfo {author} {\bibfnamefont {S.}~\bibnamefont {Erne}},
  \bibinfo {author} {\bibfnamefont {R.}~\bibnamefont {Geiger}}, \bibinfo
  {author} {\bibfnamefont {B.}~\bibnamefont {Rauer}}, \bibinfo {author}
  {\bibfnamefont {T.}~\bibnamefont {Schweigler}}, \bibinfo {author}
  {\bibfnamefont {M.}~\bibnamefont {Kuhnert}}, \bibinfo {author} {\bibfnamefont
  {W.}~\bibnamefont {Rohringer}}, \bibinfo {author} {\bibfnamefont {I.~E.}\
  \bibnamefont {Mazets}}, \bibinfo {author} {\bibfnamefont {T.}~\bibnamefont
  {Gasenzer}},\ and\ \bibinfo {author} {\bibfnamefont {J.}~\bibnamefont
  {Schmiedmayer}},\ }\bibfield  {title} {\bibinfo {title} {Experimental
  observation of a generalized gibbs ensemble},\ }\href
  {https://doi.org/10.1126/science.1257026} {\bibfield  {journal} {\bibinfo
  {journal} {Science}\ }\textbf {\bibinfo {volume} {348}},\ \bibinfo {pages}
  {207} (\bibinfo {year} {2015})},\ \Eprint
  {https://arxiv.org/abs/https://www.science.org/doi/pdf/10.1126/science.1257026}
  {https://www.science.org/doi/pdf/10.1126/science.1257026} \BibitemShut
  {NoStop}%
\bibitem [{\citenamefont {Vidmar}\ and\ \citenamefont
  {Rigol}(2016)}]{Vidmar16}%
  \BibitemOpen
  \bibfield  {author} {\bibinfo {author} {\bibfnamefont {L.}~\bibnamefont
  {Vidmar}}\ and\ \bibinfo {author} {\bibfnamefont {M.}~\bibnamefont {Rigol}},\
  }\bibfield  {title} {\bibinfo {title} {Generalized gibbs ensemble in
  integrable lattice models},\ }\href
  {https://doi.org/10.1088/1742-5468/2016/06/064007} {\bibfield  {journal}
  {\bibinfo  {journal} {Journal of Statistical Mechanics: Theory and
  Experiment}\ }\textbf {\bibinfo {volume} {2016}},\ \bibinfo {pages} {064007}
  (\bibinfo {year} {2016})}\BibitemShut {NoStop}%
\bibitem [{\citenamefont {Essler}\ and\ \citenamefont
  {Fagotti}(2016)}]{Essler16}%
  \BibitemOpen
  \bibfield  {author} {\bibinfo {author} {\bibfnamefont {F.~H.~L.}\
  \bibnamefont {Essler}}\ and\ \bibinfo {author} {\bibfnamefont
  {M.}~\bibnamefont {Fagotti}},\ }\bibfield  {title} {\bibinfo {title} {Quench
  dynamics and relaxation in isolated integrable quantum spin chains},\ }\href
  {https://doi.org/10.1088/1742-5468/2016/06/064002} {\bibfield  {journal}
  {\bibinfo  {journal} {Journal of Statistical Mechanics: Theory and
  Experiment}\ }\textbf {\bibinfo {volume} {2016}},\ \bibinfo {pages} {064002}
  (\bibinfo {year} {2016})}\BibitemShut {NoStop}%
\bibitem [{\citenamefont {Crossley}\ \emph {et~al.}(2017)\citenamefont
  {Crossley}, \citenamefont {Glorioso},\ and\ \citenamefont
  {Liu}}]{Crossley17}%
  \BibitemOpen
  \bibfield  {author} {\bibinfo {author} {\bibfnamefont {M.}~\bibnamefont
  {Crossley}}, \bibinfo {author} {\bibfnamefont {P.}~\bibnamefont {Glorioso}},\
  and\ \bibinfo {author} {\bibfnamefont {H.}~\bibnamefont {Liu}},\ }\bibfield
  {title} {\bibinfo {title} {Effective field theory of dissipative fluids},\
  }\href {https://doi.org/10.1007/JHEP09(2017)095} {\bibfield  {journal}
  {\bibinfo  {journal} {Journal of High Energy Physics}\ }\textbf {\bibinfo
  {volume} {2017}},\ \bibinfo {pages} {95} (\bibinfo {year}
  {2017})}\BibitemShut {NoStop}%
\bibitem [{\citenamefont {Kukuljan}\ \emph {et~al.}(2017)\citenamefont
  {Kukuljan}, \citenamefont {Grozdanov},\ and\ \citenamefont
  {Prosen}}]{Kukuljan17}%
  \BibitemOpen
  \bibfield  {author} {\bibinfo {author} {\bibfnamefont {I.}~\bibnamefont
  {Kukuljan}}, \bibinfo {author} {\bibfnamefont {S.~c.~v.}\ \bibnamefont
  {Grozdanov}},\ and\ \bibinfo {author} {\bibfnamefont {T.~c.~v.}\ \bibnamefont
  {Prosen}},\ }\bibfield  {title} {\bibinfo {title} {Weak quantum chaos},\
  }\href {https://doi.org/10.1103/PhysRevB.96.060301} {\bibfield  {journal}
  {\bibinfo  {journal} {Phys. Rev. B}\ }\textbf {\bibinfo {volume} {96}},\
  \bibinfo {pages} {060301(R)} (\bibinfo {year} {2017})}\BibitemShut {NoStop}%
\bibitem [{\citenamefont {Glorioso}\ and\ \citenamefont
  {Liu}(2018)}]{Glorioso18}%
  \BibitemOpen
  \bibfield  {author} {\bibinfo {author} {\bibfnamefont {P.}~\bibnamefont
  {Glorioso}}\ and\ \bibinfo {author} {\bibfnamefont {H.}~\bibnamefont {Liu}},\
  }\href {https://arxiv.org/abs/1805.09331} {\bibinfo {title} {Lectures on
  non-equilibrium effective field theories and fluctuating hydrodynamics}}
  (\bibinfo {year} {2018}),\ \Eprint {https://arxiv.org/abs/1805.09331}
  {arXiv:1805.09331 [hep-th]} \BibitemShut {NoStop}%
\bibitem [{\citenamefont {Hahn}\ \emph {et~al.}(2024)\citenamefont {Hahn},
  \citenamefont {Luitz},\ and\ \citenamefont {Chalker}}]{Hahn24}%
  \BibitemOpen
  \bibfield  {author} {\bibinfo {author} {\bibfnamefont {D.}~\bibnamefont
  {Hahn}}, \bibinfo {author} {\bibfnamefont {D.~J.}\ \bibnamefont {Luitz}},\
  and\ \bibinfo {author} {\bibfnamefont {J.~T.}\ \bibnamefont {Chalker}},\
  }\bibfield  {title} {\bibinfo {title} {Eigenstate correlations, the
  eigenstate thermalization hypothesis, and quantum information dynamics in
  chaotic many-body quantum systems},\ }\href
  {https://doi.org/10.1103/PhysRevX.14.031029} {\bibfield  {journal} {\bibinfo
  {journal} {Phys. Rev. X}\ }\textbf {\bibinfo {volume} {14}},\ \bibinfo
  {pages} {031029} (\bibinfo {year} {2024})}\BibitemShut {NoStop}%
\bibitem [{\citenamefont {Kim}\ and\ \citenamefont {Huse}(2013)}]{Kim13}%
  \BibitemOpen
  \bibfield  {author} {\bibinfo {author} {\bibfnamefont {H.}~\bibnamefont
  {Kim}}\ and\ \bibinfo {author} {\bibfnamefont {D.~A.}\ \bibnamefont {Huse}},\
  }\bibfield  {title} {\bibinfo {title} {Ballistic spreading of entanglement in
  a diffusive nonintegrable system},\ }\href
  {https://doi.org/10.1103/PhysRevLett.111.127205} {\bibfield  {journal}
  {\bibinfo  {journal} {Phys. Rev. Lett.}\ }\textbf {\bibinfo {volume} {111}},\
  \bibinfo {pages} {127205} (\bibinfo {year} {2013})}\BibitemShut {NoStop}%
\bibitem [{\citenamefont {\v{Z}nidari\v{c}}\ \emph {et~al.}(2016)\citenamefont
  {\v{Z}nidari\v{c}}, \citenamefont {Scardicchio},\ and\ \citenamefont
  {Varma}}]{Znidaric16}%
  \BibitemOpen
  \bibfield  {author} {\bibinfo {author} {\bibfnamefont {M.}~\bibnamefont
  {\v{Z}nidari\v{c}}}, \bibinfo {author} {\bibfnamefont {A.}~\bibnamefont
  {Scardicchio}},\ and\ \bibinfo {author} {\bibfnamefont {V.~K.}\ \bibnamefont
  {Varma}},\ }\bibfield  {title} {\bibinfo {title} {Diffusive and subdiffusive
  spin transport in the ergodic phase of a many-body localizable system},\
  }\href {https://doi.org/10.1103/PhysRevLett.117.040601} {\bibfield  {journal}
  {\bibinfo  {journal} {Phys. Rev. Lett.}\ }\textbf {\bibinfo {volume} {117}},\
  \bibinfo {pages} {040601} (\bibinfo {year} {2016})}\BibitemShut {NoStop}%
\bibitem [{\citenamefont {Luitz}\ and\ \citenamefont {Lev}(2017)}]{Luitz17b}%
  \BibitemOpen
  \bibfield  {author} {\bibinfo {author} {\bibfnamefont {D.~J.}\ \bibnamefont
  {Luitz}}\ and\ \bibinfo {author} {\bibfnamefont {Y.~B.}\ \bibnamefont
  {Lev}},\ }\bibfield  {title} {\bibinfo {title} {The ergodic side of the
  many-body localization transition},\ }\href
  {https://doi.org/10.1002/andp.201600350} {\bibfield  {journal} {\bibinfo
  {journal} {Annalen der Physik}\ }\textbf {\bibinfo {volume} {529}},\ \bibinfo
  {pages} {1600350} (\bibinfo {year} {2017})}\BibitemShut {NoStop}%
\bibitem [{\citenamefont {Michailidis}\ \emph {et~al.}(2024)\citenamefont
  {Michailidis}, \citenamefont {Abanin},\ and\ \citenamefont
  {Delacrétaz}}]{Michailidis24}%
  \BibitemOpen
  \bibfield  {author} {\bibinfo {author} {\bibfnamefont {A.~A.}\ \bibnamefont
  {Michailidis}}, \bibinfo {author} {\bibfnamefont {D.~A.}\ \bibnamefont
  {Abanin}},\ and\ \bibinfo {author} {\bibfnamefont {L.~V.}\ \bibnamefont
  {Delacrétaz}},\ }\bibfield  {title} {\bibinfo {title} {Corrections to
  {Diffusion} in {Interacting} {Quantum} {Systems}},\ }\href
  {https://doi.org/10.1103/PhysRevX.14.031020} {\bibfield  {journal} {\bibinfo
  {journal} {Physical Review X}\ }\textbf {\bibinfo {volume} {14}},\ \bibinfo
  {pages} {031020} (\bibinfo {year} {2024})}\BibitemShut {NoStop}%
\bibitem [{\citenamefont {Tal-Ezer}\ and\ \citenamefont
  {Kosloff}(1984)}]{TalEzer84}%
  \BibitemOpen
  \bibfield  {author} {\bibinfo {author} {\bibfnamefont {H.}~\bibnamefont
  {Tal-Ezer}}\ and\ \bibinfo {author} {\bibfnamefont {R.}~\bibnamefont
  {Kosloff}},\ }\bibfield  {title} {\bibinfo {title} {An accurate and efficient
  scheme for propagating the time dependent {Schr\"odinger} equation},\ }\href
  {https://doi.org/10.1063/1.448136} {\bibfield  {journal} {\bibinfo  {journal}
  {J. Chem. Phys.}\ }\textbf {\bibinfo {volume} {81}},\ \bibinfo {pages} {3967}
  (\bibinfo {year} {1984})},\ \Eprint
  {https://arxiv.org/abs/https://doi.org/10.1063/1.448136}
  {https://doi.org/10.1063/1.448136} \BibitemShut {NoStop}%
\bibitem [{\citenamefont {Wei\ss{}e}\ \emph {et~al.}(2006)\citenamefont
  {Wei\ss{}e}, \citenamefont {Wellein}, \citenamefont {Alvermann},\ and\
  \citenamefont {Fehske}}]{Weisse06}%
  \BibitemOpen
  \bibfield  {author} {\bibinfo {author} {\bibfnamefont {A.}~\bibnamefont
  {Wei\ss{}e}}, \bibinfo {author} {\bibfnamefont {G.}~\bibnamefont {Wellein}},
  \bibinfo {author} {\bibfnamefont {A.}~\bibnamefont {Alvermann}},\ and\
  \bibinfo {author} {\bibfnamefont {H.}~\bibnamefont {Fehske}},\ }\bibfield
  {title} {\bibinfo {title} {The kernel polynomial method},\ }\href
  {https://doi.org/10.1103/RevModPhys.78.275} {\bibfield  {journal} {\bibinfo
  {journal} {Rev. Mod. Phys.}\ }\textbf {\bibinfo {volume} {78}},\ \bibinfo
  {pages} {275} (\bibinfo {year} {2006})}\BibitemShut {NoStop}%
\bibitem [{\citenamefont {Bera}\ \emph {et~al.}(2017)\citenamefont {Bera},
  \citenamefont {De~Tomasi}, \citenamefont {Weiner},\ and\ \citenamefont
  {Evers}}]{Bera17}%
  \BibitemOpen
  \bibfield  {author} {\bibinfo {author} {\bibfnamefont {S.}~\bibnamefont
  {Bera}}, \bibinfo {author} {\bibfnamefont {G.}~\bibnamefont {De~Tomasi}},
  \bibinfo {author} {\bibfnamefont {F.}~\bibnamefont {Weiner}},\ and\ \bibinfo
  {author} {\bibfnamefont {F.}~\bibnamefont {Evers}},\ }\bibfield  {title}
  {\bibinfo {title} {Density propagator for many-body localization: Finite-size
  effects, transient subdiffusion, and exponential decay},\ }\href
  {https://doi.org/10.1103/PhysRevLett.118.196801} {\bibfield  {journal}
  {\bibinfo  {journal} {Phys. Rev. Lett.}\ }\textbf {\bibinfo {volume} {118}},\
  \bibinfo {pages} {196801} (\bibinfo {year} {2017})}\BibitemShut {NoStop}%
\bibitem [{\citenamefont {Sierant}\ and\ \citenamefont
  {Zakrzewski}(2022)}]{Sierant22}%
  \BibitemOpen
  \bibfield  {author} {\bibinfo {author} {\bibfnamefont {P.}~\bibnamefont
  {Sierant}}\ and\ \bibinfo {author} {\bibfnamefont {J.}~\bibnamefont
  {Zakrzewski}},\ }\bibfield  {title} {\bibinfo {title} {Challenges to
  observation of many-body localization},\ }\href
  {https://doi.org/10.1103/PhysRevB.105.224203} {\bibfield  {journal} {\bibinfo
   {journal} {Phys. Rev. B}\ }\textbf {\bibinfo {volume} {105}},\ \bibinfo
  {pages} {224203} (\bibinfo {year} {2022})}\BibitemShut {NoStop}%
\bibitem [{\citenamefont {Buonsante}\ \emph {et~al.}(2004)\citenamefont
  {Buonsante}, \citenamefont {Penna},\ and\ \citenamefont
  {Vezzani}}]{Buosante04}%
  \BibitemOpen
  \bibfield  {author} {\bibinfo {author} {\bibfnamefont {P.}~\bibnamefont
  {Buonsante}}, \bibinfo {author} {\bibfnamefont {V.}~\bibnamefont {Penna}},\
  and\ \bibinfo {author} {\bibfnamefont {A.}~\bibnamefont {Vezzani}},\
  }\bibfield  {title} {\bibinfo {title} {Fractional-filling loophole insulator
  domains for ultracold bosons in optical superlattices},\ }\href
  {https://doi.org/10.1103/PhysRevA.70.061603} {\bibfield  {journal} {\bibinfo
  {journal} {Phys. Rev. A}\ }\textbf {\bibinfo {volume} {70}},\ \bibinfo
  {pages} {061603(R)} (\bibinfo {year} {2004})}\BibitemShut {NoStop}%
\bibitem [{\citenamefont {Yamamoto}(2009)}]{Yamamoto09}%
  \BibitemOpen
  \bibfield  {author} {\bibinfo {author} {\bibfnamefont {D.}~\bibnamefont
  {Yamamoto}},\ }\bibfield  {title} {\bibinfo {title} {Correlated cluster
  mean-field theory for spin systems},\ }\href
  {https://doi.org/10.1103/PhysRevB.79.144427} {\bibfield  {journal} {\bibinfo
  {journal} {Phys. Rev. B}\ }\textbf {\bibinfo {volume} {79}},\ \bibinfo
  {pages} {144427} (\bibinfo {year} {2009})}\BibitemShut {NoStop}%
\bibitem [{\citenamefont {McIntosh}\ \emph {et~al.}(2012)\citenamefont
  {McIntosh}, \citenamefont {Pisarski}, \citenamefont {Gooding},\ and\
  \citenamefont {Zaremba}}]{McIntosh12}%
  \BibitemOpen
  \bibfield  {author} {\bibinfo {author} {\bibfnamefont {T.}~\bibnamefont
  {McIntosh}}, \bibinfo {author} {\bibfnamefont {P.}~\bibnamefont {Pisarski}},
  \bibinfo {author} {\bibfnamefont {R.~J.}\ \bibnamefont {Gooding}},\ and\
  \bibinfo {author} {\bibfnamefont {E.}~\bibnamefont {Zaremba}},\ }\bibfield
  {title} {\bibinfo {title} {Multisite mean-field theory for cold bosonic atoms
  in optical lattices},\ }\href {https://doi.org/10.1103/PhysRevA.86.013623}
  {\bibfield  {journal} {\bibinfo  {journal} {Phys. Rev. A}\ }\textbf {\bibinfo
  {volume} {86}},\ \bibinfo {pages} {013623} (\bibinfo {year}
  {2012})}\BibitemShut {NoStop}%
\bibitem [{\citenamefont {L\"uhmann}(2013)}]{Luehmann13}%
  \BibitemOpen
  \bibfield  {author} {\bibinfo {author} {\bibfnamefont {D.-S.}\ \bibnamefont
  {L\"uhmann}},\ }\bibfield  {title} {\bibinfo {title} {Cluster gutzwiller
  method for bosonic lattice systems},\ }\href
  {https://doi.org/10.1103/PhysRevA.87.043619} {\bibfield  {journal} {\bibinfo
  {journal} {Phys. Rev. A}\ }\textbf {\bibinfo {volume} {87}},\ \bibinfo
  {pages} {043619} (\bibinfo {year} {2013})}\BibitemShut {NoStop}%
\bibitem [{\citenamefont {Fehske}\ and\ \citenamefont
  {Schneider}(2008)}]{Fehske08}%
  \BibitemOpen
  \bibfield  {author} {\bibinfo {author} {\bibfnamefont {H.}~\bibnamefont
  {Fehske}}\ and\ \bibinfo {author} {\bibfnamefont {R.}~\bibnamefont
  {Schneider}},\ }\href {http://dx.doi.org/10.1007/978-3-540-74686-7} {\emph
  {\bibinfo {title} {Computational many-particle physics}}}\ (\bibinfo
  {publisher} {Springer},\ \bibinfo {address} {Germany},\ \bibinfo {year}
  {2008})\BibitemShut {NoStop}%
\bibitem [{\citenamefont {St\'ephan}\ \emph {et~al.}(2010)\citenamefont
  {St\'ephan}, \citenamefont {Misguich},\ and\ \citenamefont
  {Pasquier}}]{Stephan10}%
  \BibitemOpen
  \bibfield  {author} {\bibinfo {author} {\bibfnamefont {J.-M.}\ \bibnamefont
  {St\'ephan}}, \bibinfo {author} {\bibfnamefont {G.}~\bibnamefont
  {Misguich}},\ and\ \bibinfo {author} {\bibfnamefont {V.}~\bibnamefont
  {Pasquier}},\ }\bibfield  {title} {\bibinfo {title} {R\'enyi entropy of a
  line in two-dimensional ising models},\ }\href
  {https://doi.org/10.1103/PhysRevB.82.125455} {\bibfield  {journal} {\bibinfo
  {journal} {Phys. Rev. B}\ }\textbf {\bibinfo {volume} {82}},\ \bibinfo
  {pages} {125455} (\bibinfo {year} {2010})}\BibitemShut {NoStop}%
\bibitem [{\citenamefont {Turkeshi}\ \emph {et~al.}(2023)\citenamefont
  {Turkeshi}, \citenamefont {Schir\`o},\ and\ \citenamefont
  {Sierant}}]{Turkeshi23measuring}%
  \BibitemOpen
  \bibfield  {author} {\bibinfo {author} {\bibfnamefont {X.}~\bibnamefont
  {Turkeshi}}, \bibinfo {author} {\bibfnamefont {M.}~\bibnamefont {Schir\`o}},\
  and\ \bibinfo {author} {\bibfnamefont {P.}~\bibnamefont {Sierant}},\
  }\bibfield  {title} {\bibinfo {title} {Measuring nonstabilizerness via
  multifractal flatness},\ }\href {https://doi.org/10.1103/PhysRevA.108.042408}
  {\bibfield  {journal} {\bibinfo  {journal} {Phys. Rev. A}\ }\textbf {\bibinfo
  {volume} {108}},\ \bibinfo {pages} {042408} (\bibinfo {year}
  {2023})}\BibitemShut {NoStop}%
\bibitem [{\citenamefont {Sierant}\ and\ \citenamefont
  {Turkeshi}(2022)}]{Sierant22book}%
  \BibitemOpen
  \bibfield  {author} {\bibinfo {author} {\bibfnamefont {P.}~\bibnamefont
  {Sierant}}\ and\ \bibinfo {author} {\bibfnamefont {X.}~\bibnamefont
  {Turkeshi}},\ }\bibfield  {title} {\bibinfo {title} {Universal behavior
  beyond multifractality of wave functions at measurement-induced phase
  transitions},\ }\href {https://doi.org/10.1103/PhysRevLett.128.130605}
  {\bibfield  {journal} {\bibinfo  {journal} {Phys. Rev. Lett.}\ }\textbf
  {\bibinfo {volume} {128}},\ \bibinfo {pages} {130605} (\bibinfo {year}
  {2022})}\BibitemShut {NoStop}%
\bibitem [{\citenamefont {Gritsev}\ and\ \citenamefont
  {Polkovnikov}(2017)}]{Gritsev17}%
  \BibitemOpen
  \bibfield  {author} {\bibinfo {author} {\bibfnamefont {V.}~\bibnamefont
  {Gritsev}}\ and\ \bibinfo {author} {\bibfnamefont {A.}~\bibnamefont
  {Polkovnikov}},\ }\bibfield  {title} {\bibinfo {title} {Integrable floquet
  dynamics},\ }\href {https://doi.org/10.21468/SciPostPhys.2.3.021} {\bibfield
  {journal} {\bibinfo  {journal} {SciPost Phys.}\ }\textbf {\bibinfo {volume}
  {2}},\ \bibinfo {pages} {021} (\bibinfo {year} {2017})}\BibitemShut {NoStop}%
\bibitem [{\citenamefont {Antal}\ \emph {et~al.}(1999)\citenamefont {Antal},
  \citenamefont {R\'acz}, \citenamefont {R\'akos},\ and\ \citenamefont
  {Sch\"utz}}]{Antal99}%
  \BibitemOpen
  \bibfield  {author} {\bibinfo {author} {\bibfnamefont {T.}~\bibnamefont
  {Antal}}, \bibinfo {author} {\bibfnamefont {Z.}~\bibnamefont {R\'acz}},
  \bibinfo {author} {\bibfnamefont {A.}~\bibnamefont {R\'akos}},\ and\ \bibinfo
  {author} {\bibfnamefont {G.~M.}\ \bibnamefont {Sch\"utz}},\ }\bibfield
  {title} {\bibinfo {title} {Transport in the $\mathrm{XX}$ chain at zero
  temperature: Emergence of flat magnetization profiles},\ }\href
  {https://doi.org/10.1103/PhysRevE.59.4912} {\bibfield  {journal} {\bibinfo
  {journal} {Phys. Rev. E}\ }\textbf {\bibinfo {volume} {59}},\ \bibinfo
  {pages} {4912} (\bibinfo {year} {1999})}\BibitemShut {NoStop}%
\bibitem [{\citenamefont {Gobert}\ \emph {et~al.}(2005)\citenamefont {Gobert},
  \citenamefont {Kollath}, \citenamefont {Schollw\"ock},\ and\ \citenamefont
  {Sch\"utz}}]{Gobert05}%
  \BibitemOpen
  \bibfield  {author} {\bibinfo {author} {\bibfnamefont {D.}~\bibnamefont
  {Gobert}}, \bibinfo {author} {\bibfnamefont {C.}~\bibnamefont {Kollath}},
  \bibinfo {author} {\bibfnamefont {U.}~\bibnamefont {Schollw\"ock}},\ and\
  \bibinfo {author} {\bibfnamefont {G.}~\bibnamefont {Sch\"utz}},\ }\bibfield
  {title} {\bibinfo {title} {Real-time dynamics in spin-$\frac{1}{2}$ chains
  with adaptive time-dependent density matrix renormalization group},\ }\href
  {https://doi.org/10.1103/PhysRevE.71.036102} {\bibfield  {journal} {\bibinfo
  {journal} {Phys. Rev. E}\ }\textbf {\bibinfo {volume} {71}},\ \bibinfo
  {pages} {036102} (\bibinfo {year} {2005})}\BibitemShut {NoStop}%
\bibitem [{\citenamefont {Hauschild}\ \emph {et~al.}(2016)\citenamefont
  {Hauschild}, \citenamefont {Heidrich-Meisner},\ and\ \citenamefont
  {Pollmann}}]{Hauschild16}%
  \BibitemOpen
  \bibfield  {author} {\bibinfo {author} {\bibfnamefont {J.}~\bibnamefont
  {Hauschild}}, \bibinfo {author} {\bibfnamefont {F.}~\bibnamefont
  {Heidrich-Meisner}},\ and\ \bibinfo {author} {\bibfnamefont {F.}~\bibnamefont
  {Pollmann}},\ }\bibfield  {title} {\bibinfo {title} {Domain-wall melting as a
  probe of many-body localization},\ }\href
  {https://doi.org/10.1103/PhysRevB.94.161109} {\bibfield  {journal} {\bibinfo
  {journal} {Phys. Rev. B}\ }\textbf {\bibinfo {volume} {94}},\ \bibinfo
  {pages} {161109} (\bibinfo {year} {2016})}\BibitemShut {NoStop}%
\bibitem [{\citenamefont {Misguich}\ \emph {et~al.}(2017)\citenamefont
  {Misguich}, \citenamefont {Mallick},\ and\ \citenamefont
  {Krapivsky}}]{Misguich17}%
  \BibitemOpen
  \bibfield  {author} {\bibinfo {author} {\bibfnamefont {G.}~\bibnamefont
  {Misguich}}, \bibinfo {author} {\bibfnamefont {K.}~\bibnamefont {Mallick}},\
  and\ \bibinfo {author} {\bibfnamefont {P.~L.}\ \bibnamefont {Krapivsky}},\
  }\bibfield  {title} {\bibinfo {title} {Dynamics of the spin-$\frac{1}{2}$
  heisenberg chain initialized in a domain-wall state},\ }\href
  {https://doi.org/10.1103/PhysRevB.96.195151} {\bibfield  {journal} {\bibinfo
  {journal} {Phys. Rev. B}\ }\textbf {\bibinfo {volume} {96}},\ \bibinfo
  {pages} {195151} (\bibinfo {year} {2017})}\BibitemShut {NoStop}%
\bibitem [{\citenamefont {Ljubotina}\ \emph {et~al.}(2017)\citenamefont
  {Ljubotina}, \citenamefont {Žnidarič},\ and\ \citenamefont
  {Prosen}}]{Ljubotina17}%
  \BibitemOpen
  \bibfield  {author} {\bibinfo {author} {\bibfnamefont {M.}~\bibnamefont
  {Ljubotina}}, \bibinfo {author} {\bibfnamefont {M.}~\bibnamefont
  {Žnidarič}},\ and\ \bibinfo {author} {\bibfnamefont {T.}~\bibnamefont
  {Prosen}},\ }\bibfield  {title} {\bibinfo {title} {Spin diffusion from an
  inhomogeneous quench in an integrable system},\ }\href
  {https://doi.org/10.1038/ncomms16117} {\bibfield  {journal} {\bibinfo
  {journal} {Nature Communications}\ }\textbf {\bibinfo {volume} {8}},\
  \bibinfo {pages} {16117} (\bibinfo {year} {2017})}\BibitemShut {NoStop}%
\bibitem [{\citenamefont {Alba}\ and\ \citenamefont
  {Calabrese}(2017)}]{Alba17}%
  \BibitemOpen
  \bibfield  {author} {\bibinfo {author} {\bibfnamefont {V.}~\bibnamefont
  {Alba}}\ and\ \bibinfo {author} {\bibfnamefont {P.}~\bibnamefont
  {Calabrese}},\ }\bibfield  {title} {\bibinfo {title} {Entanglement and
  thermodynamics after a quantum quench in integrable systems},\ }\href
  {https://doi.org/10.1073/pnas.1703516114} {\bibfield  {journal} {\bibinfo
  {journal} {Proceedings of the National Academy of Sciences}\ }\textbf
  {\bibinfo {volume} {114}},\ \bibinfo {pages} {7947} (\bibinfo {year}
  {2017})},\ \Eprint
  {https://arxiv.org/abs/https://www.pnas.org/doi/pdf/10.1073/pnas.1703516114}
  {https://www.pnas.org/doi/pdf/10.1073/pnas.1703516114} \BibitemShut {NoStop}%
\bibitem [{\citenamefont {Alba}(2018)}]{Alba18}%
  \BibitemOpen
  \bibfield  {author} {\bibinfo {author} {\bibfnamefont {V.}~\bibnamefont
  {Alba}},\ }\bibfield  {title} {\bibinfo {title} {Entanglement and quantum
  transport in integrable systems},\ }\href
  {https://doi.org/10.1103/PhysRevB.97.245135} {\bibfield  {journal} {\bibinfo
  {journal} {Phys. Rev. B}\ }\textbf {\bibinfo {volume} {97}},\ \bibinfo
  {pages} {245135} (\bibinfo {year} {2018})}\BibitemShut {NoStop}%
\bibitem [{\citenamefont {Bertini}\ \emph
  {et~al.}(2018{\natexlab{a}})\citenamefont {Bertini}, \citenamefont {Fagotti},
  \citenamefont {Piroli},\ and\ \citenamefont {Calabrese}}]{Bertini18i}%
  \BibitemOpen
  \bibfield  {author} {\bibinfo {author} {\bibfnamefont {B.}~\bibnamefont
  {Bertini}}, \bibinfo {author} {\bibfnamefont {M.}~\bibnamefont {Fagotti}},
  \bibinfo {author} {\bibfnamefont {L.}~\bibnamefont {Piroli}},\ and\ \bibinfo
  {author} {\bibfnamefont {P.}~\bibnamefont {Calabrese}},\ }\bibfield  {title}
  {\bibinfo {title} {Entanglement evolution and generalised hydrodynamics:
  noninteracting systems},\ }\href {https://doi.org/10.1088/1751-8121/aad82e}
  {\bibfield  {journal} {\bibinfo  {journal} {Journal of Physics A:
  Mathematical and Theoretical}\ }\textbf {\bibinfo {volume} {51}},\ \bibinfo
  {pages} {39LT01} (\bibinfo {year} {2018}{\natexlab{a}})}\BibitemShut
  {NoStop}%
\bibitem [{\citenamefont {Alba}\ \emph {et~al.}(2019)\citenamefont {Alba},
  \citenamefont {Bertini},\ and\ \citenamefont {Fagotti}}]{Alba19}%
  \BibitemOpen
  \bibfield  {author} {\bibinfo {author} {\bibfnamefont {V.}~\bibnamefont
  {Alba}}, \bibinfo {author} {\bibfnamefont {B.}~\bibnamefont {Bertini}},\ and\
  \bibinfo {author} {\bibfnamefont {M.}~\bibnamefont {Fagotti}},\ }\bibfield
  {title} {\bibinfo {title} {Entanglement evolution and generalised
  hydrodynamics: interacting integrable systems},\ }\href
  {https://doi.org/10.21468/SciPostPhys.7.1.005} {\bibfield  {journal}
  {\bibinfo  {journal} {SciPost Phys.}\ }\textbf {\bibinfo {volume} {7}},\
  \bibinfo {pages} {005} (\bibinfo {year} {2019})}\BibitemShut {NoStop}%
\bibitem [{\citenamefont {Rakovszky}\ \emph {et~al.}(2019)\citenamefont
  {Rakovszky}, \citenamefont {Pollmann},\ and\ \citenamefont {von
  Keyserlingk}}]{Rakovszky19}%
  \BibitemOpen
  \bibfield  {author} {\bibinfo {author} {\bibfnamefont {T.}~\bibnamefont
  {Rakovszky}}, \bibinfo {author} {\bibfnamefont {F.}~\bibnamefont
  {Pollmann}},\ and\ \bibinfo {author} {\bibfnamefont {C.~W.}\ \bibnamefont
  {von Keyserlingk}},\ }\bibfield  {title} {\bibinfo {title} {Sub-ballistic
  growth of r\'enyi entropies due to diffusion},\ }\href
  {https://doi.org/10.1103/PhysRevLett.122.250602} {\bibfield  {journal}
  {\bibinfo  {journal} {Phys. Rev. Lett.}\ }\textbf {\bibinfo {volume} {122}},\
  \bibinfo {pages} {250602} (\bibinfo {year} {2019})}\BibitemShut {NoStop}%
\bibitem [{\citenamefont {Calabrese}\ and\ \citenamefont
  {Cardy}(2005)}]{Calabrese05}%
  \BibitemOpen
  \bibfield  {author} {\bibinfo {author} {\bibfnamefont {P.}~\bibnamefont
  {Calabrese}}\ and\ \bibinfo {author} {\bibfnamefont {J.}~\bibnamefont
  {Cardy}},\ }\bibfield  {title} {\bibinfo {title} {Evolution of entanglement
  entropy in one-dimensional systems},\ }\href
  {https://doi.org/10.1088/1742-5468/2005/04/P04010} {\bibfield  {journal}
  {\bibinfo  {journal} {Journal of Statistical Mechanics: Theory and
  Experiment}\ }\textbf {\bibinfo {volume} {2005}},\ \bibinfo {pages} {P04010}
  (\bibinfo {year} {2005})}\BibitemShut {NoStop}%
\bibitem [{\citenamefont {Yang}\ and\ \citenamefont {Yang}(1969)}]{Yang69}%
  \BibitemOpen
  \bibfield  {author} {\bibinfo {author} {\bibfnamefont {C.~N.}\ \bibnamefont
  {Yang}}\ and\ \bibinfo {author} {\bibfnamefont {C.~P.}\ \bibnamefont
  {Yang}},\ }\bibfield  {title} {\bibinfo {title} {Thermodynamics of a
  one‐dimensional system of bosons with repulsive delta‐function
  interaction},\ }\href {https://doi.org/10.1063/1.1664947} {\bibfield
  {journal} {\bibinfo  {journal} {Journal of Mathematical Physics}\ }\textbf
  {\bibinfo {volume} {10}},\ \bibinfo {pages} {1115} (\bibinfo {year}
  {1969})},\ \Eprint
  {https://arxiv.org/abs/https://pubs.aip.org/aip/jmp/article-pdf/10/7/1115/19101094/1115\_1\_online.pdf}
  {https://pubs.aip.org/aip/jmp/article-pdf/10/7/1115/19101094/1115\_1\_online.pdf}
  \BibitemShut {NoStop}%
\bibitem [{\citenamefont {Bertini}\ \emph
  {et~al.}(2018{\natexlab{b}})\citenamefont {Bertini}, \citenamefont
  {Tartaglia},\ and\ \citenamefont {Calabrese}}]{Bertini18p}%
  \BibitemOpen
  \bibfield  {author} {\bibinfo {author} {\bibfnamefont {B.}~\bibnamefont
  {Bertini}}, \bibinfo {author} {\bibfnamefont {E.}~\bibnamefont {Tartaglia}},\
  and\ \bibinfo {author} {\bibfnamefont {P.}~\bibnamefont {Calabrese}},\
  }\bibfield  {title} {\bibinfo {title} {Entanglement and diagonal entropies
  after a quench with no pair structure},\ }\href
  {https://doi.org/10.1088/1742-5468/aac73f} {\bibfield  {journal} {\bibinfo
  {journal} {Journal of Statistical Mechanics: Theory and Experiment}\ }\textbf
  {\bibinfo {volume} {2018}},\ \bibinfo {pages} {063104} (\bibinfo {year}
  {2018}{\natexlab{b}})}\BibitemShut {NoStop}%
\bibitem [{\citenamefont {Bastianello}\ and\ \citenamefont
  {Calabrese}(2018)}]{Bastianello18}%
  \BibitemOpen
  \bibfield  {author} {\bibinfo {author} {\bibfnamefont {A.}~\bibnamefont
  {Bastianello}}\ and\ \bibinfo {author} {\bibfnamefont {P.}~\bibnamefont
  {Calabrese}},\ }\bibfield  {title} {\bibinfo {title} {Spreading of
  entanglement and correlations after a quench with intertwined
  quasiparticles},\ }\href {https://doi.org/10.21468/SciPostPhys.5.4.033}
  {\bibfield  {journal} {\bibinfo  {journal} {SciPost Phys.}\ ,\ \bibinfo
  {pages} {033}} (\bibinfo {year} {2018})}\BibitemShut {NoStop}%
\bibitem [{\citenamefont {Takahashi}(1999)}]{Takahashi99thermo}%
  \BibitemOpen
  \bibfield  {author} {\bibinfo {author} {\bibfnamefont {M.}~\bibnamefont
  {Takahashi}},\ }\href
  {https://doi.org/https://doi.org/10.1017/CBO9780511524332} {\emph {\bibinfo
  {title} {Thermodynamics of one-dimensional solvable models}}}\ (\bibinfo
  {publisher} {Cambridge University Press, Cambridge},\ \bibinfo {year}
  {1999})\BibitemShut {NoStop}%
\bibitem [{\citenamefont {Mussardo}(2010)}]{Mussardo10}%
  \BibitemOpen
  \bibfield  {author} {\bibinfo {author} {\bibfnamefont {G.}~\bibnamefont
  {Mussardo}},\ }\href {https://books.google.pl/books?id=fakVDAAAQBAJ} {\emph
  {\bibinfo {title} {Statistical Field Theory: An Introduction to Exactly
  Solved Models in Statistical Physics}}},\ Oxford Graduate Texts\ (\bibinfo
  {publisher} {OUP Oxford},\ \bibinfo {year} {2010})\BibitemShut {NoStop}%
\bibitem [{\citenamefont {Franchini}(2017)}]{Franchini17}%
  \BibitemOpen
  \bibfield  {author} {\bibinfo {author} {\bibfnamefont {F.}~\bibnamefont
  {Franchini}},\ }\href {https://doi.org/10.1007/978-3-319-48487-7} {\emph
  {\bibinfo {title} {An {Introduction} to {Integrable} {Techniques} for
  {One}-{Dimensional} {Quantum} {Systems}}}},\ \bibinfo {series} {Lecture
  {Notes} in {Physics}}, Vol.\ \bibinfo {volume} {940}\ (\bibinfo  {publisher}
  {Springer International Publishing},\ \bibinfo {address} {Cham},\ \bibinfo
  {year} {2017})\BibitemShut {NoStop}%
\bibitem [{\citenamefont {Doyon}(2017)}]{Doyon17w}%
  \BibitemOpen
  \bibfield  {author} {\bibinfo {author} {\bibfnamefont {B.}~\bibnamefont
  {Doyon}},\ }\bibfield  {title} {\bibinfo {title} {Thermalization and
  {Pseudolocality} in {Extended} {Quantum} {Systems}},\ }\href
  {https://doi.org/10.1007/s00220-017-2836-7} {\bibfield  {journal} {\bibinfo
  {journal} {Communications in Mathematical Physics}\ }\textbf {\bibinfo
  {volume} {351}},\ \bibinfo {pages} {155} (\bibinfo {year}
  {2017})}\BibitemShut {NoStop}%
\bibitem [{\citenamefont {Wolf}\ \emph {et~al.}(2008)\citenamefont {Wolf},
  \citenamefont {Verstraete}, \citenamefont {Hastings},\ and\ \citenamefont
  {Cirac}}]{Wolf08}%
  \BibitemOpen
  \bibfield  {author} {\bibinfo {author} {\bibfnamefont {M.~M.}\ \bibnamefont
  {Wolf}}, \bibinfo {author} {\bibfnamefont {F.}~\bibnamefont {Verstraete}},
  \bibinfo {author} {\bibfnamefont {M.~B.}\ \bibnamefont {Hastings}},\ and\
  \bibinfo {author} {\bibfnamefont {J.~I.}\ \bibnamefont {Cirac}},\ }\bibfield
  {title} {\bibinfo {title} {Area laws in quantum systems: Mutual information
  and correlations},\ }\href {https://doi.org/10.1103/PhysRevLett.100.070502}
  {\bibfield  {journal} {\bibinfo  {journal} {Phys. Rev. Lett.}\ }\textbf
  {\bibinfo {volume} {100}},\ \bibinfo {pages} {070502} (\bibinfo {year}
  {2008})}\BibitemShut {NoStop}%
\bibitem [{\citenamefont {Lieb}\ and\ \citenamefont {Robinson}(1972)}]{Lieb72}%
  \BibitemOpen
  \bibfield  {author} {\bibinfo {author} {\bibfnamefont {E.~H.}\ \bibnamefont
  {Lieb}}\ and\ \bibinfo {author} {\bibfnamefont {D.~W.}\ \bibnamefont
  {Robinson}},\ }\bibfield  {title} {\bibinfo {title} {The finite group
  velocity of quantum spin systems},\ }\href
  {https://doi.org/10.1007/BF01645779} {\bibfield  {journal} {\bibinfo
  {journal} {Communications in Mathematical Physics}\ }\textbf {\bibinfo
  {volume} {28}},\ \bibinfo {pages} {251} (\bibinfo {year} {1972})}\BibitemShut
  {NoStop}%
\bibitem [{\citenamefont {Travaglino}\ and\ \citenamefont
  {Calabrese}(2026)}]{Travaglino26}%
  \BibitemOpen
  \bibfield  {author} {\bibinfo {author} {\bibfnamefont {R.}~\bibnamefont
  {Travaglino}}\ and\ \bibinfo {author} {\bibfnamefont {P.}~\bibnamefont
  {Calabrese}},\ }\href {https://arxiv.org/abs/2608.24613} {\bibinfo {title}
  {Transport interpretation of entanglement hamiltonian cumulants in integrable
  quantum quenches}} (\bibinfo {year} {2026}),\ \Eprint
  {https://arxiv.org/abs/2608.24613} {arXiv:2608.24613 [cond-mat.stat-mech]}
  \BibitemShut {NoStop}%
\bibitem [{\citenamefont {Piroli}\ \emph {et~al.}(2022)\citenamefont {Piroli},
  \citenamefont {Vernier}, \citenamefont {Collura},\ and\ \citenamefont
  {Calabrese}}]{Piroli22}%
  \BibitemOpen
  \bibfield  {author} {\bibinfo {author} {\bibfnamefont {L.}~\bibnamefont
  {Piroli}}, \bibinfo {author} {\bibfnamefont {E.}~\bibnamefont {Vernier}},
  \bibinfo {author} {\bibfnamefont {M.}~\bibnamefont {Collura}},\ and\ \bibinfo
  {author} {\bibfnamefont {P.}~\bibnamefont {Calabrese}},\ }\bibfield  {title}
  {\bibinfo {title} {Thermodynamic symmetry resolved entanglement entropies in
  integrable systems},\ }\href {https://doi.org/10.1088/1742-5468/ac7a2d}
  {\bibfield  {journal} {\bibinfo  {journal} {Journal of Statistical Mechanics:
  Theory and Experiment}\ }\textbf {\bibinfo {volume} {2022}},\ \bibinfo
  {pages} {073102} (\bibinfo {year} {2022})}\BibitemShut {NoStop}%
\bibitem [{\citenamefont {Billingsley}(1995)}]{Billingsley95}%
  \BibitemOpen
  \bibfield  {author} {\bibinfo {author} {\bibfnamefont {P.}~\bibnamefont
  {Billingsley}},\ }\href {https://books.google.pl/books?id=z39jQgAACAAJ}
  {\emph {\bibinfo {title} {Probability and Measure}}},\ Wiley Series in
  Probability and Statistics\ (\bibinfo  {publisher} {Wiley},\ \bibinfo {year}
  {1995})\BibitemShut {NoStop}%
\bibitem [{\citenamefont {Ioffe}(2015)}]{Ioffe15}%
  \BibitemOpen
  \bibfield  {author} {\bibinfo {author} {\bibfnamefont {A.~D.}\ \bibnamefont
  {Ioffe}},\ }\href {https://arxiv.org/abs/1505.07920} {\bibinfo {title}
  {Metric regularity. theory and applications - a survey}} (\bibinfo {year}
  {2015}),\ \Eprint {https://arxiv.org/abs/1505.07920} {arXiv:1505.07920
  [math.OC]} \BibitemShut {NoStop}%
\bibitem [{\citenamefont {Micchelli}(1979)}]{Micchelli79}%
  \BibitemOpen
  \bibfield  {author} {\bibinfo {author} {\bibfnamefont {C.~A.}\ \bibnamefont
  {Micchelli}},\ }\bibinfo {title} {On a numerically efficient method for
  computing multivariate b-splines},\ in\ \href
  {https://doi.org/10.1007/978-3-0348-6289-9_14} {\emph {\bibinfo {booktitle}
  {Multivariate Approximation Theory: Proceedings of the Conference held at the
  Mathematical Research Institute at Oberwolfach Black Forest, February 4--10,
  1979}}},\ \bibinfo {editor} {edited by\ \bibinfo {editor} {\bibfnamefont
  {W.}~\bibnamefont {Schempp}}\ and\ \bibinfo {editor} {\bibfnamefont
  {K.}~\bibnamefont {Zeller}}}\ (\bibinfo  {publisher} {Birkh{\"a}user Basel},\
  \bibinfo {address} {Basel},\ \bibinfo {year} {1979})\ pp.\ \bibinfo {pages}
  {211--248}\BibitemShut {NoStop}%
\bibitem [{\citenamefont {Farin}(1993)}]{Farin93}%
  \BibitemOpen
  \bibfield  {author} {\bibinfo {author} {\bibfnamefont {G.}~\bibnamefont
  {Farin}},\ }\bibfield  {title} {\bibinfo {title} {Chapter 15 - rational
  bézier and b-spline curves},\ }in\ \href
  {https://doi.org/https://doi.org/10.1016/B978-0-12-249052-1.50020-9} {\emph
  {\bibinfo {booktitle} {Curves and Surfaces for Computer-Aided Geometric
  Design (Third Edition)}}}\ (\bibinfo  {publisher} {Academic Press},\ \bibinfo
  {address} {Boston},\ \bibinfo {year} {1993})\ \bibinfo {edition} {third
  edition}\ ed.,\ pp.\ \bibinfo {pages} {251--266}\BibitemShut {NoStop}%
\bibitem [{\citenamefont {Prautzsch}\ \emph {et~al.}(2002)\citenamefont
  {Prautzsch}, \citenamefont {Boehm},\ and\ \citenamefont
  {Paluszny}}]{Prautzsch02}%
  \BibitemOpen
  \bibfield  {author} {\bibinfo {author} {\bibfnamefont {H.}~\bibnamefont
  {Prautzsch}}, \bibinfo {author} {\bibfnamefont {W.}~\bibnamefont {Boehm}},\
  and\ \bibinfo {author} {\bibfnamefont {M.}~\bibnamefont {Paluszny}},\
  }\bibinfo {title} {Simplex splines},\ in\ \href
  {https://doi.org/10.1007/978-3-662-04919-8_18} {\emph {\bibinfo {booktitle}
  {B{\'e}zier and B-Spline Techniques}}}\ (\bibinfo  {publisher} {Springer
  Berlin Heidelberg},\ \bibinfo {address} {Berlin, Heidelberg},\ \bibinfo
  {year} {2002})\ pp.\ \bibinfo {pages} {259--269}\BibitemShut {NoStop}%
\bibitem [{\citenamefont {D'Alessio}\ \emph {et~al.}(2016)\citenamefont
  {D'Alessio}, \citenamefont {Kafri}, \citenamefont {Polkovnikov},\ and\
  \citenamefont {Rigol}}]{DAlessio16}%
  \BibitemOpen
  \bibfield  {author} {\bibinfo {author} {\bibfnamefont {L.}~\bibnamefont
  {D'Alessio}}, \bibinfo {author} {\bibfnamefont {Y.}~\bibnamefont {Kafri}},
  \bibinfo {author} {\bibfnamefont {A.}~\bibnamefont {Polkovnikov}},\ and\
  \bibinfo {author} {\bibfnamefont {M.}~\bibnamefont {Rigol}},\ }\bibfield
  {title} {\bibinfo {title} {From quantum chaos and eigenstate thermalization
  to statistical mechanics and thermodynamics},\ }\href
  {https://doi.org/10.1080/00018732.2016.1198134} {\bibfield  {journal}
  {\bibinfo  {journal} {Advances in Physics}\ }\textbf {\bibinfo {volume}
  {65}},\ \bibinfo {pages} {239} (\bibinfo {year} {2016})}\BibitemShut
  {NoStop}%
\bibitem [{\citenamefont {Brantut}\ \emph {et~al.}(2012)\citenamefont
  {Brantut}, \citenamefont {Meineke}, \citenamefont {Stadler}, \citenamefont
  {Krinner},\ and\ \citenamefont {Esslinger}}]{Brantut12}%
  \BibitemOpen
  \bibfield  {author} {\bibinfo {author} {\bibfnamefont {J.-P.}\ \bibnamefont
  {Brantut}}, \bibinfo {author} {\bibfnamefont {J.}~\bibnamefont {Meineke}},
  \bibinfo {author} {\bibfnamefont {D.}~\bibnamefont {Stadler}}, \bibinfo
  {author} {\bibfnamefont {S.}~\bibnamefont {Krinner}},\ and\ \bibinfo {author}
  {\bibfnamefont {T.}~\bibnamefont {Esslinger}},\ }\bibfield  {title} {\bibinfo
  {title} {Conduction of ultracold fermions through a mesoscopic channel},\
  }\href {https://doi.org/10.1126/science.1223175} {\bibfield  {journal}
  {\bibinfo  {journal} {Science}\ }\textbf {\bibinfo {volume} {337}},\ \bibinfo
  {pages} {1069} (\bibinfo {year} {2012})},\ \Eprint
  {https://arxiv.org/abs/https://www.science.org/doi/pdf/10.1126/science.1223175}
  {https://www.science.org/doi/pdf/10.1126/science.1223175} \BibitemShut
  {NoStop}%
\bibitem [{\citenamefont {Brantut}\ \emph {et~al.}(2013)\citenamefont
  {Brantut}, \citenamefont {Grenier}, \citenamefont {Meineke}, \citenamefont
  {Stadler}, \citenamefont {Krinner}, \citenamefont {Kollath}, \citenamefont
  {Esslinger},\ and\ \citenamefont {Georges}}]{Brantut13}%
  \BibitemOpen
  \bibfield  {author} {\bibinfo {author} {\bibfnamefont {J.-P.}\ \bibnamefont
  {Brantut}}, \bibinfo {author} {\bibfnamefont {C.}~\bibnamefont {Grenier}},
  \bibinfo {author} {\bibfnamefont {J.}~\bibnamefont {Meineke}}, \bibinfo
  {author} {\bibfnamefont {D.}~\bibnamefont {Stadler}}, \bibinfo {author}
  {\bibfnamefont {S.}~\bibnamefont {Krinner}}, \bibinfo {author} {\bibfnamefont
  {C.}~\bibnamefont {Kollath}}, \bibinfo {author} {\bibfnamefont
  {T.}~\bibnamefont {Esslinger}},\ and\ \bibinfo {author} {\bibfnamefont
  {A.}~\bibnamefont {Georges}},\ }\bibfield  {title} {\bibinfo {title} {A
  thermoelectric heat engine with ultracold atoms},\ }\href
  {https://doi.org/10.1126/science.1242308} {\bibfield  {journal} {\bibinfo
  {journal} {Science}\ }\textbf {\bibinfo {volume} {342}},\ \bibinfo {pages}
  {713} (\bibinfo {year} {2013})},\ \Eprint
  {https://arxiv.org/abs/https://www.science.org/doi/pdf/10.1126/science.1242308}
  {https://www.science.org/doi/pdf/10.1126/science.1242308} \BibitemShut
  {NoStop}%
\bibitem [{\citenamefont {Husmann}\ \emph {et~al.}(2015)\citenamefont
  {Husmann}, \citenamefont {Uchino}, \citenamefont {Krinner}, \citenamefont
  {Lebrat}, \citenamefont {Giamarchi}, \citenamefont {Esslinger},\ and\
  \citenamefont {Brantut}}]{Husmann15}%
  \BibitemOpen
  \bibfield  {author} {\bibinfo {author} {\bibfnamefont {D.}~\bibnamefont
  {Husmann}}, \bibinfo {author} {\bibfnamefont {S.}~\bibnamefont {Uchino}},
  \bibinfo {author} {\bibfnamefont {S.}~\bibnamefont {Krinner}}, \bibinfo
  {author} {\bibfnamefont {M.}~\bibnamefont {Lebrat}}, \bibinfo {author}
  {\bibfnamefont {T.}~\bibnamefont {Giamarchi}}, \bibinfo {author}
  {\bibfnamefont {T.}~\bibnamefont {Esslinger}},\ and\ \bibinfo {author}
  {\bibfnamefont {J.-P.}\ \bibnamefont {Brantut}},\ }\bibfield  {title}
  {\bibinfo {title} {Connecting strongly correlated superfluids by a quantum
  point contact},\ }\href {https://doi.org/10.1126/science.aac9584} {\bibfield
  {journal} {\bibinfo  {journal} {Science}\ }\textbf {\bibinfo {volume}
  {350}},\ \bibinfo {pages} {1498} (\bibinfo {year} {2015})},\ \Eprint
  {https://arxiv.org/abs/https://www.science.org/doi/pdf/10.1126/science.aac9584}
  {https://www.science.org/doi/pdf/10.1126/science.aac9584} \BibitemShut
  {NoStop}%
\bibitem [{\citenamefont {Krinner}\ \emph {et~al.}(2015)\citenamefont
  {Krinner}, \citenamefont {Stadler}, \citenamefont {Husmann}, \citenamefont
  {Brantut},\ and\ \citenamefont {Esslinger}}]{Krinner15}%
  \BibitemOpen
  \bibfield  {author} {\bibinfo {author} {\bibfnamefont {S.}~\bibnamefont
  {Krinner}}, \bibinfo {author} {\bibfnamefont {D.}~\bibnamefont {Stadler}},
  \bibinfo {author} {\bibfnamefont {D.}~\bibnamefont {Husmann}}, \bibinfo
  {author} {\bibfnamefont {J.-P.}\ \bibnamefont {Brantut}},\ and\ \bibinfo
  {author} {\bibfnamefont {T.}~\bibnamefont {Esslinger}},\ }\bibfield  {title}
  {\bibinfo {title} {Observation of quantized conductance in neutral matter},\
  }\href {https://doi.org/10.1038/nature14049} {\bibfield  {journal} {\bibinfo
  {journal} {Nature}\ }\textbf {\bibinfo {volume} {517}},\ \bibinfo {pages}
  {64} (\bibinfo {year} {2015})}\BibitemShut {NoStop}%
\bibitem [{\citenamefont {H\"ausler}\ \emph {et~al.}(2017)\citenamefont
  {H\"ausler}, \citenamefont {Nakajima}, \citenamefont {Lebrat}, \citenamefont
  {Husmann}, \citenamefont {Krinner}, \citenamefont {Esslinger},\ and\
  \citenamefont {Brantut}}]{Hausler17}%
  \BibitemOpen
  \bibfield  {author} {\bibinfo {author} {\bibfnamefont {S.}~\bibnamefont
  {H\"ausler}}, \bibinfo {author} {\bibfnamefont {S.}~\bibnamefont {Nakajima}},
  \bibinfo {author} {\bibfnamefont {M.}~\bibnamefont {Lebrat}}, \bibinfo
  {author} {\bibfnamefont {D.}~\bibnamefont {Husmann}}, \bibinfo {author}
  {\bibfnamefont {S.}~\bibnamefont {Krinner}}, \bibinfo {author} {\bibfnamefont
  {T.}~\bibnamefont {Esslinger}},\ and\ \bibinfo {author} {\bibfnamefont
  {J.-P.}\ \bibnamefont {Brantut}},\ }\bibfield  {title} {\bibinfo {title}
  {Scanning gate microscope for cold atomic gases},\ }\href
  {https://doi.org/10.1103/PhysRevLett.119.030403} {\bibfield  {journal}
  {\bibinfo  {journal} {Phys. Rev. Lett.}\ }\textbf {\bibinfo {volume} {119}},\
  \bibinfo {pages} {030403} (\bibinfo {year} {2017})}\BibitemShut {NoStop}%
\bibitem [{\citenamefont {Lebrat}\ \emph {et~al.}(2018)\citenamefont {Lebrat},
  \citenamefont {Gri\ifmmode~\check{s}\else \v{s}\fi{}ins}, \citenamefont
  {Husmann}, \citenamefont {H\"ausler}, \citenamefont {Corman}, \citenamefont
  {Giamarchi}, \citenamefont {Brantut},\ and\ \citenamefont
  {Esslinger}}]{Lebrat18}%
  \BibitemOpen
  \bibfield  {author} {\bibinfo {author} {\bibfnamefont {M.}~\bibnamefont
  {Lebrat}}, \bibinfo {author} {\bibfnamefont {P.}~\bibnamefont
  {Gri\ifmmode~\check{s}\else \v{s}\fi{}ins}}, \bibinfo {author} {\bibfnamefont
  {D.}~\bibnamefont {Husmann}}, \bibinfo {author} {\bibfnamefont
  {S.}~\bibnamefont {H\"ausler}}, \bibinfo {author} {\bibfnamefont
  {L.}~\bibnamefont {Corman}}, \bibinfo {author} {\bibfnamefont
  {T.}~\bibnamefont {Giamarchi}}, \bibinfo {author} {\bibfnamefont {J.-P.}\
  \bibnamefont {Brantut}},\ and\ \bibinfo {author} {\bibfnamefont
  {T.}~\bibnamefont {Esslinger}},\ }\bibfield  {title} {\bibinfo {title} {Band
  and correlated insulators of cold fermions in a mesoscopic lattice},\ }\href
  {https://doi.org/10.1103/PhysRevX.8.011053} {\bibfield  {journal} {\bibinfo
  {journal} {Phys. Rev. X}\ }\textbf {\bibinfo {volume} {8}},\ \bibinfo {pages}
  {011053} (\bibinfo {year} {2018})}\BibitemShut {NoStop}%
\bibitem [{\citenamefont {Schweigler}\ \emph {et~al.}(2017)\citenamefont
  {Schweigler}, \citenamefont {Kasper}, \citenamefont {Erne}, \citenamefont
  {Mazets}, \citenamefont {Rauer}, \citenamefont {Cataldini}, \citenamefont
  {Langen}, \citenamefont {Gasenzer}, \citenamefont {Berges},\ and\
  \citenamefont {Schmiedmayer}}]{Schweigler17}%
  \BibitemOpen
  \bibfield  {author} {\bibinfo {author} {\bibfnamefont {T.}~\bibnamefont
  {Schweigler}}, \bibinfo {author} {\bibfnamefont {V.}~\bibnamefont {Kasper}},
  \bibinfo {author} {\bibfnamefont {S.}~\bibnamefont {Erne}}, \bibinfo {author}
  {\bibfnamefont {I.}~\bibnamefont {Mazets}}, \bibinfo {author} {\bibfnamefont
  {B.}~\bibnamefont {Rauer}}, \bibinfo {author} {\bibfnamefont
  {F.}~\bibnamefont {Cataldini}}, \bibinfo {author} {\bibfnamefont
  {T.}~\bibnamefont {Langen}}, \bibinfo {author} {\bibfnamefont
  {T.}~\bibnamefont {Gasenzer}}, \bibinfo {author} {\bibfnamefont
  {J.}~\bibnamefont {Berges}},\ and\ \bibinfo {author} {\bibfnamefont
  {J.}~\bibnamefont {Schmiedmayer}},\ }\bibfield  {title} {\bibinfo {title}
  {Experimental characterization of a quantum many-body system via higher-order
  correlations},\ }\href {https://doi.org/10.1038/nature22310} {\bibfield
  {journal} {\bibinfo  {journal} {Nature}\ }\textbf {\bibinfo {volume} {545}},\
  \bibinfo {pages} {323} (\bibinfo {year} {2017})}\BibitemShut {NoStop}%
\bibitem [{\citenamefont {Pigneur}\ \emph {et~al.}(2018)\citenamefont
  {Pigneur}, \citenamefont {Berrada}, \citenamefont {Bonneau}, \citenamefont
  {Schumm}, \citenamefont {Demler},\ and\ \citenamefont
  {Schmiedmayer}}]{Pigneur18}%
  \BibitemOpen
  \bibfield  {author} {\bibinfo {author} {\bibfnamefont {M.}~\bibnamefont
  {Pigneur}}, \bibinfo {author} {\bibfnamefont {T.}~\bibnamefont {Berrada}},
  \bibinfo {author} {\bibfnamefont {M.}~\bibnamefont {Bonneau}}, \bibinfo
  {author} {\bibfnamefont {T.}~\bibnamefont {Schumm}}, \bibinfo {author}
  {\bibfnamefont {E.}~\bibnamefont {Demler}},\ and\ \bibinfo {author}
  {\bibfnamefont {J.}~\bibnamefont {Schmiedmayer}},\ }\bibfield  {title}
  {\bibinfo {title} {Relaxation to a phase-locked equilibrium state in a
  one-dimensional bosonic josephson junction},\ }\href
  {https://doi.org/10.1103/PhysRevLett.120.173601} {\bibfield  {journal}
  {\bibinfo  {journal} {Phys. Rev. Lett.}\ }\textbf {\bibinfo {volume} {120}},\
  \bibinfo {pages} {173601} (\bibinfo {year} {2018})}\BibitemShut {NoStop}%
\bibitem [{\citenamefont {Sommer}\ \emph {et~al.}(2011)\citenamefont {Sommer},
  \citenamefont {Ku}, \citenamefont {Roati},\ and\ \citenamefont
  {Zwierlein}}]{Sommer11f}%
  \BibitemOpen
  \bibfield  {author} {\bibinfo {author} {\bibfnamefont {A.}~\bibnamefont
  {Sommer}}, \bibinfo {author} {\bibfnamefont {M.}~\bibnamefont {Ku}}, \bibinfo
  {author} {\bibfnamefont {G.}~\bibnamefont {Roati}},\ and\ \bibinfo {author}
  {\bibfnamefont {M.~W.}\ \bibnamefont {Zwierlein}},\ }\bibfield  {title}
  {\bibinfo {title} {Universal spin transport in a strongly interacting {Fermi}
  gas},\ }\href {https://doi.org/10.1038/nature09989} {\bibfield  {journal}
  {\bibinfo  {journal} {Nature}\ }\textbf {\bibinfo {volume} {472}},\ \bibinfo
  {pages} {201} (\bibinfo {year} {2011})}\BibitemShut {NoStop}%
\bibitem [{\citenamefont {Luciuk}\ \emph {et~al.}(2017)\citenamefont {Luciuk},
  \citenamefont {Smale}, \citenamefont {B\"ottcher}, \citenamefont {Sharum},
  \citenamefont {Olsen}, \citenamefont {Trotzky}, \citenamefont {Enss},\ and\
  \citenamefont {Thywissen}}]{Luciuk17}%
  \BibitemOpen
  \bibfield  {author} {\bibinfo {author} {\bibfnamefont {C.}~\bibnamefont
  {Luciuk}}, \bibinfo {author} {\bibfnamefont {S.}~\bibnamefont {Smale}},
  \bibinfo {author} {\bibfnamefont {F.}~\bibnamefont {B\"ottcher}}, \bibinfo
  {author} {\bibfnamefont {H.}~\bibnamefont {Sharum}}, \bibinfo {author}
  {\bibfnamefont {B.~A.}\ \bibnamefont {Olsen}}, \bibinfo {author}
  {\bibfnamefont {S.}~\bibnamefont {Trotzky}}, \bibinfo {author} {\bibfnamefont
  {T.}~\bibnamefont {Enss}},\ and\ \bibinfo {author} {\bibfnamefont {J.~H.}\
  \bibnamefont {Thywissen}},\ }\bibfield  {title} {\bibinfo {title}
  {Observation of quantum-limited spin transport in strongly interacting
  two-dimensional fermi gases},\ }\href
  {https://doi.org/10.1103/PhysRevLett.118.130405} {\bibfield  {journal}
  {\bibinfo  {journal} {Phys. Rev. Lett.}\ }\textbf {\bibinfo {volume} {118}},\
  \bibinfo {pages} {130405} (\bibinfo {year} {2017})}\BibitemShut {NoStop}%
\bibitem [{\citenamefont {Nichols}\ \emph {et~al.}(2019)\citenamefont
  {Nichols}, \citenamefont {Cheuk}, \citenamefont {Okan}, \citenamefont
  {Hartke}, \citenamefont {Mendez}, \citenamefont {Senthil}, \citenamefont
  {Khatami}, \citenamefont {Zhang},\ and\ \citenamefont
  {Zwierlein}}]{Nichols19}%
  \BibitemOpen
  \bibfield  {author} {\bibinfo {author} {\bibfnamefont {M.~A.}\ \bibnamefont
  {Nichols}}, \bibinfo {author} {\bibfnamefont {L.~W.}\ \bibnamefont {Cheuk}},
  \bibinfo {author} {\bibfnamefont {M.}~\bibnamefont {Okan}}, \bibinfo {author}
  {\bibfnamefont {T.~R.}\ \bibnamefont {Hartke}}, \bibinfo {author}
  {\bibfnamefont {E.}~\bibnamefont {Mendez}}, \bibinfo {author} {\bibfnamefont
  {T.}~\bibnamefont {Senthil}}, \bibinfo {author} {\bibfnamefont
  {E.}~\bibnamefont {Khatami}}, \bibinfo {author} {\bibfnamefont
  {H.}~\bibnamefont {Zhang}},\ and\ \bibinfo {author} {\bibfnamefont {M.~W.}\
  \bibnamefont {Zwierlein}},\ }\bibfield  {title} {\bibinfo {title} {Spin
  transport in a mott insulator of ultracold fermions},\ }\href
  {https://doi.org/10.1126/science.aat4387} {\bibfield  {journal} {\bibinfo
  {journal} {Science}\ }\textbf {\bibinfo {volume} {363}},\ \bibinfo {pages}
  {383} (\bibinfo {year} {2019})},\ \Eprint
  {https://arxiv.org/abs/https://www.science.org/doi/pdf/10.1126/science.aat4387}
  {https://www.science.org/doi/pdf/10.1126/science.aat4387} \BibitemShut
  {NoStop}%
\bibitem [{\citenamefont {Brown}\ \emph {et~al.}(2019)\citenamefont {Brown},
  \citenamefont {Mitra}, \citenamefont {Guardado-Sanchez}, \citenamefont
  {Nourafkan}, \citenamefont {Reymbaut}, \citenamefont {Hébert}, \citenamefont
  {Bergeron}, \citenamefont {Tremblay}, \citenamefont {Kokalj}, \citenamefont
  {Huse}, \citenamefont {Schauß},\ and\ \citenamefont {Bakr}}]{Brown19}%
  \BibitemOpen
  \bibfield  {author} {\bibinfo {author} {\bibfnamefont {P.~T.}\ \bibnamefont
  {Brown}}, \bibinfo {author} {\bibfnamefont {D.}~\bibnamefont {Mitra}},
  \bibinfo {author} {\bibfnamefont {E.}~\bibnamefont {Guardado-Sanchez}},
  \bibinfo {author} {\bibfnamefont {R.}~\bibnamefont {Nourafkan}}, \bibinfo
  {author} {\bibfnamefont {A.}~\bibnamefont {Reymbaut}}, \bibinfo {author}
  {\bibfnamefont {C.-D.}\ \bibnamefont {Hébert}}, \bibinfo {author}
  {\bibfnamefont {S.}~\bibnamefont {Bergeron}}, \bibinfo {author}
  {\bibfnamefont {A.-M.~S.}\ \bibnamefont {Tremblay}}, \bibinfo {author}
  {\bibfnamefont {J.}~\bibnamefont {Kokalj}}, \bibinfo {author} {\bibfnamefont
  {D.~A.}\ \bibnamefont {Huse}}, \bibinfo {author} {\bibfnamefont
  {P.}~\bibnamefont {Schauß}},\ and\ \bibinfo {author} {\bibfnamefont {W.~S.}\
  \bibnamefont {Bakr}},\ }\bibfield  {title} {\bibinfo {title} {Bad metallic
  transport in a cold atom fermi-hubbard system},\ }\href
  {https://doi.org/10.1126/science.aat4134} {\bibfield  {journal} {\bibinfo
  {journal} {Science}\ }\textbf {\bibinfo {volume} {363}},\ \bibinfo {pages}
  {379} (\bibinfo {year} {2019})},\ \Eprint
  {https://arxiv.org/abs/https://www.science.org/doi/pdf/10.1126/science.aat4134}
  {https://www.science.org/doi/pdf/10.1126/science.aat4134} \BibitemShut
  {NoStop}%
\bibitem [{\citenamefont {Kinoshita}\ \emph {et~al.}(2006)\citenamefont
  {Kinoshita}, \citenamefont {Wenger},\ and\ \citenamefont
  {Weiss}}]{Kinoshita06}%
  \BibitemOpen
  \bibfield  {author} {\bibinfo {author} {\bibfnamefont {T.}~\bibnamefont
  {Kinoshita}}, \bibinfo {author} {\bibfnamefont {T.}~\bibnamefont {Wenger}},\
  and\ \bibinfo {author} {\bibfnamefont {D.~S.}\ \bibnamefont {Weiss}},\
  }\bibfield  {title} {\bibinfo {title} {A quantum {Newton}'s cradle},\ }\href
  {https://doi.org/10.1038/nature04693} {\bibfield  {journal} {\bibinfo
  {journal} {Nature}\ }\textbf {\bibinfo {volume} {440}},\ \bibinfo {pages}
  {900} (\bibinfo {year} {2006})}\BibitemShut {NoStop}%
\bibitem [{\citenamefont {Mazets}\ \emph {et~al.}(2008)\citenamefont {Mazets},
  \citenamefont {Schumm},\ and\ \citenamefont {Schmiedmayer}}]{Mazets08}%
  \BibitemOpen
  \bibfield  {author} {\bibinfo {author} {\bibfnamefont {I.~E.}\ \bibnamefont
  {Mazets}}, \bibinfo {author} {\bibfnamefont {T.}~\bibnamefont {Schumm}},\
  and\ \bibinfo {author} {\bibfnamefont {J.}~\bibnamefont {Schmiedmayer}},\
  }\bibfield  {title} {\bibinfo {title} {Breakdown of integrability in a
  quasi-1d ultracold bosonic gas},\ }\href
  {https://doi.org/10.1103/PhysRevLett.100.210403} {\bibfield  {journal}
  {\bibinfo  {journal} {Phys. Rev. Lett.}\ }\textbf {\bibinfo {volume} {100}},\
  \bibinfo {pages} {210403} (\bibinfo {year} {2008})}\BibitemShut {NoStop}%
\bibitem [{\citenamefont {Tan}\ \emph {et~al.}(2010)\citenamefont {Tan},
  \citenamefont {Pustilnik},\ and\ \citenamefont {Glazman}}]{Tan10}%
  \BibitemOpen
  \bibfield  {author} {\bibinfo {author} {\bibfnamefont {S.}~\bibnamefont
  {Tan}}, \bibinfo {author} {\bibfnamefont {M.}~\bibnamefont {Pustilnik}},\
  and\ \bibinfo {author} {\bibfnamefont {L.~I.}\ \bibnamefont {Glazman}},\
  }\bibfield  {title} {\bibinfo {title} {Relaxation of a high-energy
  quasiparticle in a one-dimensional bose gas},\ }\href
  {https://doi.org/10.1103/PhysRevLett.105.090404} {\bibfield  {journal}
  {\bibinfo  {journal} {Phys. Rev. Lett.}\ }\textbf {\bibinfo {volume} {105}},\
  \bibinfo {pages} {090404} (\bibinfo {year} {2010})}\BibitemShut {NoStop}%
\bibitem [{\citenamefont {Marquardt}(1963)}]{Marquardt63}%
  \BibitemOpen
  \bibfield  {author} {\bibinfo {author} {\bibfnamefont {D.~W.}\ \bibnamefont
  {Marquardt}},\ }\bibfield  {title} {\bibinfo {title} {An algorithm for
  least-squares estimation of nonlinear parameters},\ }\href
  {https://doi.org/10.1137/0111030} {\bibfield  {journal} {\bibinfo  {journal}
  {Journal of the Society for Industrial and Applied Mathematics}\ }\textbf
  {\bibinfo {volume} {11}},\ \bibinfo {pages} {431} (\bibinfo {year} {1963})},\
  \Eprint {https://arxiv.org/abs/https://doi.org/10.1137/0111030}
  {https://doi.org/10.1137/0111030} \BibitemShut {NoStop}%
\bibitem [{\citenamefont {Lanczos}(1950)}]{Lanczos50}%
  \BibitemOpen
  \bibfield  {author} {\bibinfo {author} {\bibfnamefont {C.}~\bibnamefont
  {Lanczos}},\ }\bibfield  {title} {\bibinfo {title} {An iteration method for
  the solution of the eigenvalue problem of linear differential and integral
  operators},\ }\href@noop {} {\bibfield  {journal} {\bibinfo  {journal}
  {Journal of research of the National Bureau of Standards}\ }\textbf {\bibinfo
  {volume} {45}},\ \bibinfo {pages} {255} (\bibinfo {year} {1950})}\BibitemShut
  {NoStop}%
\bibitem [{\citenamefont {Clenshaw}(1955)}]{Clenshaw55}%
  \BibitemOpen
  \bibfield  {author} {\bibinfo {author} {\bibfnamefont {C.~W.}\ \bibnamefont
  {Clenshaw}},\ }\bibfield  {title} {\bibinfo {title} {A note on the summation
  of chebyshev series},\ }\href@noop {} {\bibfield  {journal} {\bibinfo
  {journal} {Mathematics of computation}\ }\textbf {\bibinfo {volume} {9}},\
  \bibinfo {pages} {118} (\bibinfo {year} {1955})}\BibitemShut {NoStop}%
\bibitem [{\citenamefont {Fulton}\ and\ \citenamefont
  {Harris}(2004)}]{Fulton04}%
  \BibitemOpen
  \bibfield  {author} {\bibinfo {author} {\bibfnamefont {W.}~\bibnamefont
  {Fulton}}\ and\ \bibinfo {author} {\bibfnamefont {J.}~\bibnamefont
  {Harris}},\ }\href {https://doi.org/10.1007/978-1-4612-0979-9} {\emph
  {\bibinfo {title} {Representation {Theory}}}},\ \bibinfo {series} {Graduate
  {Texts} in {Mathematics}}, Vol.\ \bibinfo {volume} {129}\ (\bibinfo
  {publisher} {Springer},\ \bibinfo {address} {New York, NY},\ \bibinfo {year}
  {2004})\BibitemShut {NoStop}%
\bibitem [{\citenamefont {Lando}\ and\ \citenamefont
  {Zvonkin}(2004)}]{Lando04}%
  \BibitemOpen
  \bibfield  {author} {\bibinfo {author} {\bibfnamefont {S.~K.}\ \bibnamefont
  {Lando}}\ and\ \bibinfo {author} {\bibfnamefont {A.~K.}\ \bibnamefont
  {Zvonkin}},\ }\href {https://doi.org/10.1007/978-3-540-38361-1} {\emph
  {\bibinfo {title} {Graphs on {Surfaces} and {Their} {Applications}}}},\
  edited by\ \bibinfo {editor} {\bibfnamefont {R.~V.}\ \bibnamefont
  {Gamkrelidze}}\ and\ \bibinfo {editor} {\bibfnamefont {V.~A.}\ \bibnamefont
  {Vassiliev}},\ \bibinfo {series} {Encyclopaedia of {Mathematical}
  {Sciences}}, Vol.\ \bibinfo {volume} {141}\ (\bibinfo  {publisher}
  {Springer},\ \bibinfo {address} {Berlin, Heidelberg},\ \bibinfo {year}
  {2004})\BibitemShut {NoStop}%
\bibitem [{\citenamefont {Devroye}(1986)}]{Devroye86}%
  \BibitemOpen
  \bibfield  {author} {\bibinfo {author} {\bibfnamefont {L.}~\bibnamefont
  {Devroye}},\ }\href {https://doi.org/10.1007/978-1-4613-8643-8} {\emph
  {\bibinfo {title} {Non-{Uniform} {Random} {Variate} {Generation}}}}\
  (\bibinfo  {publisher} {Springer},\ \bibinfo {address} {New York, NY},\
  \bibinfo {year} {1986})\BibitemShut {NoStop}%
\bibitem [{\citenamefont {Araki}(1969)}]{Araki69}%
  \BibitemOpen
  \bibfield  {author} {\bibinfo {author} {\bibfnamefont {H.}~\bibnamefont
  {Araki}},\ }\bibfield  {title} {\bibinfo {title} {Gibbs states of a one
  dimensional quantum lattice},\ }\href {https://doi.org/10.1007/BF01645134}
  {\bibfield  {journal} {\bibinfo  {journal} {Communications in Mathematical
  Physics}\ }\textbf {\bibinfo {volume} {14}},\ \bibinfo {pages} {120}
  (\bibinfo {year} {1969})}\BibitemShut {NoStop}%
\bibitem [{\citenamefont {Kliesch}\ \emph {et~al.}(2014)\citenamefont
  {Kliesch}, \citenamefont {Gogolin}, \citenamefont {Kastoryano}, \citenamefont
  {Riera},\ and\ \citenamefont {Eisert}}]{Kliesch14}%
  \BibitemOpen
  \bibfield  {author} {\bibinfo {author} {\bibfnamefont {M.}~\bibnamefont
  {Kliesch}}, \bibinfo {author} {\bibfnamefont {C.}~\bibnamefont {Gogolin}},
  \bibinfo {author} {\bibfnamefont {M.~J.}\ \bibnamefont {Kastoryano}},
  \bibinfo {author} {\bibfnamefont {A.}~\bibnamefont {Riera}},\ and\ \bibinfo
  {author} {\bibfnamefont {J.}~\bibnamefont {Eisert}},\ }\bibfield  {title}
  {\bibinfo {title} {Locality of temperature},\ }\href
  {https://doi.org/10.1103/PhysRevX.4.031019} {\bibfield  {journal} {\bibinfo
  {journal} {Phys. Rev. X}\ }\textbf {\bibinfo {volume} {4}},\ \bibinfo {pages}
  {031019} (\bibinfo {year} {2014})}\BibitemShut {NoStop}%
\bibitem [{\citenamefont {Deza}\ and\ \citenamefont {Deza}(2016)}]{Deza16}%
  \BibitemOpen
  \bibfield  {author} {\bibinfo {author} {\bibfnamefont {M.~M.}\ \bibnamefont
  {Deza}}\ and\ \bibinfo {author} {\bibfnamefont {E.}~\bibnamefont {Deza}},\
  }\href {https://doi.org/10.1007/978-3-662-52844-0} {\emph {\bibinfo {title}
  {Encyclopedia of {Distances}}}}\ (\bibinfo  {publisher} {Springer},\ \bibinfo
  {address} {Berlin, Heidelberg},\ \bibinfo {year} {2016})\BibitemShut
  {NoStop}%
\bibitem [{\citenamefont {Boas}(1954)}]{Boas54}%
  \BibitemOpen
  \bibfield  {author} {\bibinfo {author} {\bibfnamefont {R.}~\bibnamefont
  {Boas}},\ }\href {https://books.google.pl/books?id=TexQAAAAMAAJ} {\emph
  {\bibinfo {title} {Entire Functions}}},\ \bibinfo {series} {Entire
  Functions}\ No.\ \bibinfo {number} {t. 5}\ (\bibinfo  {publisher} {Academic
  Press},\ \bibinfo {year} {1954})\BibitemShut {NoStop}%
\bibitem [{\citenamefont {Taylor}(1965)}]{Taylor65}%
  \BibitemOpen
  \bibfield  {author} {\bibinfo {author} {\bibfnamefont {A.}~\bibnamefont
  {Taylor}},\ }\href {https://books.google.pl/books?id=Gfd84tA7PXcC} {\emph
  {\bibinfo {title} {General Theory of Functions and Integration}}},\ Blaisdell
  book in pure and applied mathematics\ (\bibinfo  {publisher} {Blaisdell
  Publishing Company},\ \bibinfo {year} {1965})\BibitemShut {NoStop}%
\bibitem [{\citenamefont {Beale}(2011)}]{Beale11}%
  \BibitemOpen
  \bibfield  {author} {\bibinfo {author} {\bibfnamefont {P.}~\bibnamefont
  {Beale}},\ }\href {https://books.google.pl/books?id=KdbJJAXQ-RsC} {\emph
  {\bibinfo {title} {Statistical Mechanics}}}\ (\bibinfo  {publisher} {Academic
  Press},\ \bibinfo {year} {2011})\BibitemShut {NoStop}%
\bibitem [{\citenamefont {Dembo}\ and\ \citenamefont
  {Zeitouni}(2010)}]{Dembo10}%
  \BibitemOpen
  \bibfield  {author} {\bibinfo {author} {\bibfnamefont {A.}~\bibnamefont
  {Dembo}}\ and\ \bibinfo {author} {\bibfnamefont {O.}~\bibnamefont
  {Zeitouni}},\ }\href {https://doi.org/10.1007/978-3-642-03311-7} {\emph
  {\bibinfo {title} {Large {Deviations} {Techniques} and {Applications}}}},\
  \bibinfo {series} {Stochastic {Modelling} and {Applied} {Probability}},
  Vol.~\bibinfo {volume} {38}\ (\bibinfo  {publisher} {Springer},\ \bibinfo
  {address} {Berlin, Heidelberg},\ \bibinfo {year} {2010})\BibitemShut
  {NoStop}%
\bibitem [{\citenamefont {Audenaert}(2007)}]{Audenaert07}%
  \BibitemOpen
  \bibfield  {author} {\bibinfo {author} {\bibfnamefont {K.~M.~R.}\
  \bibnamefont {Audenaert}},\ }\bibfield  {title} {\bibinfo {title} {A sharp
  continuity estimate for the von neumann entropy},\ }\href
  {https://doi.org/10.1088/1751-8113/40/28/S18} {\bibfield  {journal} {\bibinfo
   {journal} {Journal of Physics A: Mathematical and Theoretical}\ }\textbf
  {\bibinfo {volume} {40}},\ \bibinfo {pages} {8127} (\bibinfo {year}
  {2007})}\BibitemShut {NoStop}%
\end{thebibliography}
%

\end{document}